\documentclass[preprint,aps,amsmath,nofootinbib,tightenlines]{revtex4}
\usepackage{graphicx}
\usepackage{bm}
\usepackage{epsfig}

\usepackage[dvips]{hyperref}

\newif\ifpdf
\ifx\pdfoutput\undefined
\pdffalse 
\else
\pdfoutput=1 
\pdftrue
\fi

\def\Aslash{A\!\!\!\!\slash}
\def\Bslash{B\!\!\!\!\slash}
\def\Bcslash{{\cal B}\!\!\!\slash}
\def\Dslash{D\!\!\!\!\slash}
\def\cDslash{{\cal D}\!\!\!\!\slash}

\def\nslash{n\!\!\!\slash}
\def\bnslash{\bar n\!\!\!\slash}

\def\Aslash{A\!\!\!\slash}

\def\vslash{v\!\!\!\slash}
\def\vepslash{\varepsilon\!\!\!\slash}
\def\Delslash{\Delta\!\!\!\!\slash}
\def\OMIT#1{}

\newcommand{\nn}{\nonumber} 

\newcommand{\bn}{{\bar n}}
\newcommand{\bea}{\begin{eqnarray}}
\newcommand{\eea}{\end{eqnarray}}

\newcommand{\bnP}{\bar {\cal P}}
\newcommand{\ppP}{{\cal P}_\perp}

\newcommand{\cP}{{\cal P}}
\newcommand{\cPslash}{ {\cal P}\!\!\!\!\slash}

\newcommand{\mcdot}{\!\cdot\!}

\newcommand{\cD}{{\cal D}}

\newcommand{\SCETa}{\mbox{${\rm SCET}_{\rm I}$ }}

\newcommand{\DvppP}{{{iv\mcdot D}}_c^\perp}

\newcommand{\DSppP}{{{D\!\!\!\!\hspace{0.04cm}\slash}}_c^\perp}
\newcommand{\DSppPl}{\,i\!\overleftarrow {D\!\!\!\!\hspace{0.04cm}\slash}_c^\perp}
\newcommand{\DSppPus}{{{D\!\!\!\!\hspace{0.04cm}\slash}}_{us}^\perp}
\newcommand{\DSppPlus}{\,i\!\overleftarrow {D\!\!\!\!\hspace{0.04cm}\slash}_{us}^\perp}

\newcommand{\cDgppPl}{\,\overleftarrow {\cal D}{}_{\!c\perp}^{\alpha}}

\newcommand{\cDpusLeft}[1]{\,i\!\overleftarrow {\cal D}_{us}^{\perp\, #1}}
\newcommand{\cDpusRight}[1]{\,i\!\overrightarrow {\cal D}_{us}^{T\, #1}}

\newcommand{\lD}{\overleftarrow D{}_c}

\begin{document}
\setlength\baselineskip{15pt}

\ifpdf
\DeclareGraphicsExtensions{.pdf, .jpg}
\else
\DeclareGraphicsExtensions{.eps, .jpg}
\fi


\preprint{ \vbox{ \hbox{hep-ph/0409045} \hbox{MIT-CTP-3521} }}

\title{\phantom{x}
\vspace{0.5cm}
Factorization for Power Corrections to $B\to X_s\gamma$ and 
$B\to X_u \ell\bar\nu$\\ 
\vspace{0.6cm}
}

\author{Keith S. M. Lee}
\affiliation{Center for Theoretical Physics, Massachusetts Institute of\\ 
Technology,
Cambridge, MA 02139\footnote{Electronic address: ksml@mit.edu, iains@mit.edu}
\vspace{0.2cm}}
\author{Iain W. Stewart\vspace{0.4cm}}
\affiliation{Center for Theoretical Physics, Massachusetts Institute of\\
Technology,
Cambridge, MA 02139\footnote{Electronic address: ksml@mit.edu, iains@mit.edu}
\vspace{0.2cm}}

\vspace{0.2cm}

\begin{abstract}
\vspace{0.3cm}

We derive factorization theorems for $\Lambda_{\rm QCD}/m_b$ power corrections
to inclusive $B$-decays in the endpoint region, where $m_X^2\sim
m_b\,\Lambda_{\rm QCD}$. In $B\to X_u\ell\bar\nu$ our results are for the full
triply differential rate. A complete enumeration of $\Lambda_{\rm QCD}/m_b$
corrections is given.  We point out the presence of new $\Lambda_{\rm
  QCD}/m_b$-suppressed shape functions, which arise at tree level with a
$4\pi$-enhanced coefficient, and show that these previously neglected terms
induce an additional significant uncertainty for current inclusive methods of
measuring $|V_{ub}|$ that depend on the endpoint region of phase space.

\end{abstract}

\maketitle

\section{Introduction} \label{sect_intro}

The study of $B$-decays enables precision measurements of Standard Model
parameters as well as searches for new physics. For inclusive measurements, the
large mass of the $b$-quark makes a rigorous theoretical treatment in QCD
possible. A well-known example is the semileptonic decay $B\to
X_c\ell\bar\nu$, which allows measurements of $|V_{cb}|$, $m_b$ and $m_c$ through
moments of the decay spectra.  Analyses of this type are performed at
{\sc BaBar}~\cite{Aubert:2004aw,Aubert:2004te}, Belle~\cite{Abe:2002du}, and
CLEO~\cite{Csorna:2004kp}, where simultaneously fitting a few fundamental
hadronic parameters keeps the theoretical uncertainties under
control~\cite{Gremm:1996yn}.  From measurements of $B\to X_u\ell\bar\nu$ we can
also determine $|V_{ub}|$~\cite{Aubert:2003zw,Gibbons:2004dg,Kakuno:2003fk}.
Finally, in channels involving flavor-changing neutral currents, such as 
$B\to X_s\gamma$ and $B\to X_s \ell^+\ell^-$, we have the luxury of searching 
for new physics with accurate Standard Model calculations of the expected
rates~\cite{Buras:2002er,Hurth:2003vb}.

In inclusive $B$-decays there is often a trade-off between theory and
experiment, because cuts are necessary experimentally, but these less inclusive
spectra make the theory more complicated. For sufficiently inclusive
measurements we can use quark-hadron duality, together with a local operator
product expansion (OPE) in $\Lambda_{\rm
  QCD}/m_b$~\cite{Shifman:1985wx,Chay:1990da,Bigi:1993fe,Blok:1994va,
  Manohar:1994qn,Falk:1993dh}.  In these decays, non-perturbative matrix
elements are defined with the help of the Heavy Quark Effective Theory
(HQET)~\cite{Georgi:1990um,Grinstein:1990mj}, and perturbative corrections are
straightforward to incorporate, although often still difficult to compute. For
$B\to X_c\ell\bar\nu$ decays, this formalism has been worked out to order
$1/m_b^3$ in the power expansion~\cite{Bigi:1995ga,Gremm:1997df} and the
perturbative corrections are in most cases known to
$\alpha_s^2\beta_0$~\cite{Luke:1994du,Luke:1994yc,Gremm:1997gg,Falk:1997jq} and
in a few cases at order $\alpha_s^2$~\cite{Czarnecki:1998kt,Czarnecki:1997hc}.
A simultaneous fit with the hadronic parameters appearing up to order
$1/m_b^2$~\cite{Bauer:2002sh,Benson:2003kp} has provided us with a $\sim 2\%$
determination of $|V_{cb}|$, making it one of the best determined elements of
the CKM matrix.

In $B\to X_u\ell\bar\nu$ decays the situation is more complicated because of the
large $b\to c$ background. Since a model-independent analysis must avoid
extrapolating the experimental data, the computation of decay rates in a
restricted region of phase space becomes important (see 
e.g.~\cite{Falk:1998jq,Bauer:2001rc}).  Phase-space cuts are also important for
$B\to X_s\gamma$ to eliminate softer photons, and for $B\to X_s\ell^+\ell^-$ to
deal with the intermediate $J/\Psi$ and $\Psi'$ resonances.  In many cases,
these restrictions on the phase space can put us in a situation where the local
operator product expansion is no longer applicable, since we are limited to a
region where the allowed masses of the hadronic states satisfy $m_X^2\ll m_b^2$.
To distinguish different cases we use the terminology
\begin{eqnarray}
 && \Delta m_X^2 \sim m_b^2, \qquad\qquad \mbox{totally inclusive}
     \qquad \mbox{(local OPE, HQET)} \nn \,, \\
 && \Delta m_X^2 \sim m_b\Lambda_{\rm QCD}, \qquad \! \mbox{endpoint region}
     \qquad \mbox{(Factorization, SCET)}  \,,\\
 && \Delta m_X^2 \sim \Lambda_{\rm QCD}^2, \qquad\quad \mbox{resonance region}
     \qquad \mbox{(exclusive methods)}  \,,\nn
\end{eqnarray}
where $\Delta m_X^2$ denotes the region of $m_X^2$ extending out from $m_{X_{\rm
    min}}^2$, and the applicable theoretical methods are shown in brackets.

When $m_X^2\sim m_b\Lambda_{\rm QCD}$, the outgoing hadronic states become
jet-like and we enter the so-called endpoint or shape function
region~\cite{Bigi:1994ex,Neubert:1994um,Mannel:1994pm}, which will be the main
focus of this paper.  For $b\to u$ decays this region is important, because of 
cuts on $E_\ell$ or $m_X^2$, which are used to eliminate $b\to c$ events. In 
this region both the perturbative expansion and the power expansion become more
complicated. In particular, there is the usual perturbative expansion at the
scale $\mu^2\simeq m_b^2$, as well as a second perturbative expansion at the
smaller scale $\mu^2\simeq m_X^2$. The rates also exhibit double Sudakov
logarithms. In addition, instead of depending on non-perturbative parameters 
($\lambda_1,\lambda_2, \ldots$) that are matrix elements of local operators, 
the decay rates depend on non-perturbative shape functions.  We shall
refer to the expansion parameter for this region as $\lambda^2\sim
m_X^2/m_B^2\sim \Lambda_{\rm QCD}/m_b$ to distinguish it from the $1/m_b$
expansion for the local OPE. In the endpoint region, the standard OPE no longer
completely justifies the separation of short- and long-distance contributions.
Instead, we must consider a more involved derivation of a QCD factorization
theorem, as is the case in processes such as Drell-Yan and DIS as $x\to
1$~\cite{Mueller:1981sg,Sterman:1995fz}.

For $B\to X_s\gamma$ and $B\to X_u\ell\bar\nu$ at leading order (LO) in
$\lambda$, the factorization theorem for the endpoint decay rates was determined
in Ref.~\cite{Korchemsky:1994jb}. It separates QCD contributions that are hard
($H$), collinear (${\cal J}^{(0)}$) and soft ($f^{(0)}$), so that, 
schematically, a differential decay rate takes the form
\begin{eqnarray}
  d\Gamma = H \times {\cal J}^{(0)} \otimes f^{(0)} \,,
\end{eqnarray}
where $\times$ is normal multiplication and $\otimes$ is a one-parameter
convolution.  Here the hard contributions are perturbative at the scale
$\mu^2\sim m_b^2$, the collinear contributions in ${\cal J}^{(0)}$ are
associated with the inclusive $X$ jet and are treated perturbatively at the
scale $\mu^2\sim m_b\Lambda_{\rm QCD}$, and the soft contributions are factored
into a forward $B$-meson matrix element giving the non-perturbative shape
function $f^{(0)}$~\cite{Neubert:1994um}.  In Ref.~\cite{Bauer:2001yt} this
factorization theorem was rederived using the Soft-Collinear Effective Theory
(SCET)~\cite{Bauer:2000ew,Bauer:2000yr,Bauer:2001ct,Bauer:2001yt}. The
attraction of the effective-theory method is that it provides a formalism for
extending the derivation of factorization theorems beyond LO in the power
expansion.  The main goal of this paper is to derive a factorization theorem for
$B\to X_s\gamma$ and $B\to X_u\ell\bar\nu$ at subleading order, i.e.\ ${\cal
  O}(\lambda^2)$, using SCET. This factorization theorem allows us to separate
perturbative and non-perturbative corrections to all orders in $\alpha_s$.

One method for studying the endpoint region is to start with the local OPE and
sum up the infinite series of the operators that are most singular as we
approach the $m_X^2\sim m_b\Lambda_{\rm QCD}$ region.  This technique was used
in
Refs.~\cite{Neubert:1994um,Neubert:1994ch,Bigi:1994ex,Mannel:1994pm,Falk:1993vb,DeFazio:1999sv},
and provides a method of defining the non-perturbative functions. At LO the
result is the shape function\footnote{We arbitrarily use the term ``shape
  function'' for $f(\ell^+)$ rather than ``distribution function''.  Sometimes
  in the literature the term ``shape function'' is reserved for the distribution
  that enters $d\Gamma(B\to X_u\ell\bar\nu)/dE_\ell$, which is an integral over
  $f$.}
\begin{eqnarray} \label{localexpn}
 \frac12 \sum_{k=0}^\infty  \frac{(-1)^k}{k!}
   \: \delta^{(k)}(\ell^+)\: \langle B_v | 
    \bar h_v  (i n\mcdot D)^k h_v  | B_v \rangle 
  &=& \frac12 \langle B_v | 
    \bar h_v\, \delta(\ell^+ - i n\mcdot D)\, h_v  | B_v \rangle  \nn \\
  &=& f^{(0)}(\ell^+) \,.
\end{eqnarray}
The result is simply the matrix element of a non-local HQET operator, where the
states $|B_v\rangle$ and heavy b-quark fields $h_v$ are defined in HQET, and
$n^\mu$ is a light-like vector along the axis of the jet.  This approach allows
direct contact with the extensive calculations made with the local OPE, which
give terms in the power series.  There are several reasons for considering an
approach where $f(\ell^+)$ is obtained without a summation.  In particular, it is
difficult to go beyond tree level with the summation approach.  Also, owing to the
presence of a kinked Wilson line~\cite{Korchemsky:1993xv}, the renormalization
of the local operators in the sum and final delta-function operator are not
identical~\cite{Bauer:2003pi,Bosch:2004th} (see also~\cite{Korchemsky:1994jb}),
essentially because the moment integrals introduce additional UV
divergences.\footnote{ From the point of view of effective field theory this
  makes sense, since the summation in Eq.~(\ref{localexpn}) attempts to connect
  one effective theory (HQET) to a region with a different expansion parameter
  that is described by a different effective theory (SCET).  Generically the
  renormalization in two EFTs is not interconnected.} For this reason the
expansion in Eq.~(\ref{localexpn}) should be considered to be formal, and care
must be taken in drawing conclusions from operators in the expanded version,
such as the fact that they have trivial dependence on $n^\mu$.
 \OMIT{\footnote{In the
  context of $\bar B^0\to D^0\pi^0$ in Ref.~\cite{Mantry:2003uz} the definition
  with a sum over moments and trivial $n$ dependence would imply that the
  complex soft function appearing there is necessarily real~\cite{private1}. The
  resolution here is again that the moments are insufficient to define the soft
  function fully.}}  Care must also be taken in calculating the hard factor $H$, as
pointed out recently~\cite{Bauer:2003pi,Bosch:2004th}, since the one-loop matrix
element for $f^{(0)}(\ell^+)$ has finite pieces in pure dimensional
regularization. This implies that the quark-level QCD computation does not
directly give the hard contribution, unlike factorization theorems involving
massless quarks such as in DIS. The matching calculation in SCET handles this in
a simple manner because matrix elements of the effective-theory graphs are
necessarily subtracted from the full-theory graphs in order to compute $H$.

For any precision calculation, perturbative corrections play an important role,
and both the resummation of large logarithms and fixed-order calculations
need to be considered.  The position-space version of Eq.~(\ref{localexpn}) has
a kinked Wilson line along $v$-$n$-$v$, which leads to double Sudakov
logarithms~\cite{Korchemsky:1993xv,Korchemsky:1994jb}. These occur between both
the $m_b^2\to m_b\Lambda_{\rm QCD}$ and the $m_b\Lambda_{\rm QCD}\to \Lambda_{\rm
  QCD}^2$ scales and can be summed using renormalization-group techniques. In
moment space the leading and next-to-leading anomalous dimensions can be found
in Ref.~\cite{Korchemsky:1994jb}. For phenomenological purposes, formulae for the
differential rates with resummed logarithms are of practical importance and were
obtained using inverse Mellin transformations in
Refs.~\cite{Leibovich:1999xf,Leibovich:2000ig,Leibovich:2000ey,Akhoury:1995fp}.
These resummations have also been considered in SCET, both in moment
space~\cite{Bauer:2000ew,Bauer:2003pi} and for the differential
rates~\cite{Bosch:2004th}. Finite-order perturbative corrections are currently
known to order $\alpha_s$ for the $H$ and ${\cal J}$
functions~\cite{Bauer:2000yr,Bauer:2003pi,Bosch:2004th}.

Since in the endpoint region $\Lambda_{\rm QCD}/Q\sim 1/5$ to $1/10$, it is
important to consider the effect of power corrections for precision
phenomenology.  By matching from QCD at tree level, contributions of subleading
NLO shape functions have been derived for $B\to X_s\gamma$~\cite{Bauer:2001mh}
and for $B\to X_u\ell\bar\nu$~\cite{Leibovich:2002ys,Bauer:2002yu}, followed by
further analysis in Refs.~\cite{Kraetz:2002rv,Burrell:2003cf,Neubert:2002yx}.  A
single NNLO contribution has also been considered, corresponding to the
``annihilation'' contribution, which is phase space enhanced by
$16\pi^2$~\cite{Voloshin:2001iq,Leibovich:2002ys}. These power corrections
provide the dominant source of theoretical uncertainty in current measurements
of $|V_{ub}|$ and are the focus of our paper, so we discuss them in more detail.

To build some intuition, it is useful to contrast the power expansion in the
endpoint region with the expansion for the local OPE. For the local OPE, all
contributions can be assigned a power of $(\Lambda_{\rm QCD}/m_b)^{k-3}$. The
power of $\Lambda_{\rm QCD}$ and thus $k$ is simply determined by the dimension
of the operator, and the $-3$ accounts for the dimension of the HQET states,
$\langle B_v | \cdots | B_v\rangle$.  For example, the set of local
operators up to dimension 6 is
\begin{eqnarray}
 &&  
  O^3 = \overline h_v \:  h_v\,,\qquad 
  O^{5a} = \overline h_v (i D_T)^2 h_v \,,\qquad
  O^{5b} = g\, \overline h_v \sigma_{\alpha\beta} G^{\alpha\beta} h_v \,,\\[5pt]
 && 
   O^{6a} = \overline h_v (i D^T_\alpha)(i v\mcdot D)(i D_T^\alpha) h_v\,,
  \qquad\qquad\ 
   O^{6b} = i\epsilon^{\alpha\beta\gamma\delta} v_\delta\:
   \overline h_v (i D_\alpha)(i v\mcdot D)(i D_\beta)\gamma_\gamma
   \gamma_5\,  h_v \,,\nn\\[5pt]
 &&
   O^{6c} = (\overline h_v \gamma^\alpha q_L) \, 
                (\overline q_L \gamma_\alpha  h_v) \,, 
   \qquad\qquad\qquad\quad \!\! 
   O^{6d} = (\overline h_v \: q_L ) \, 
                (\overline q_L \:\,   h_v) \,, 
   \nn\\[5pt]
  &&
   O^{6e} = (\overline h_v T^a \gamma^\alpha q_L) \, 
                (\overline q_L T^a \gamma_\alpha  h_v) \,, 
   \qquad\qquad\ 
   O^{6f} = (\overline h_v T^a q_L ) \, 
                (\overline q_L \:T^a \,   h_v) \,, 
    \nn
\end{eqnarray}
where dimensions are shown as superscripts, a superscript/subscript $T$ means
transverse to the HQET velocity parameter $v^\mu$, and an $L$ means
left-handed.\footnote{We write $O^3$ in terms of HQET fields, although strictly
  speaking at lowest order this is not necessary.}  Dimension-$4$ operators are
absent so there are no $1/m_b$ corrections, except the trivial ones that may be
induced by switching to hadronic variables.  For dimension-5 and 6 operators
there are two naming conventions in common use. For $\langle \bar B_v |
\{O^{5a}, O^{5b}, O^{6a}, O^{6b} \} | \bar B_v \rangle$, the parameters are
$\{\lambda_1, \lambda_2, \rho_1, \rho_2 \}$ or $\{\mu_\pi,\mu_G, \rho_D^3,
\rho^3_{LS} \}$.
\begin{figure}[!t]
 \centerline{
  \mbox{\epsfysize=2.5truecm \hbox{\epsfbox{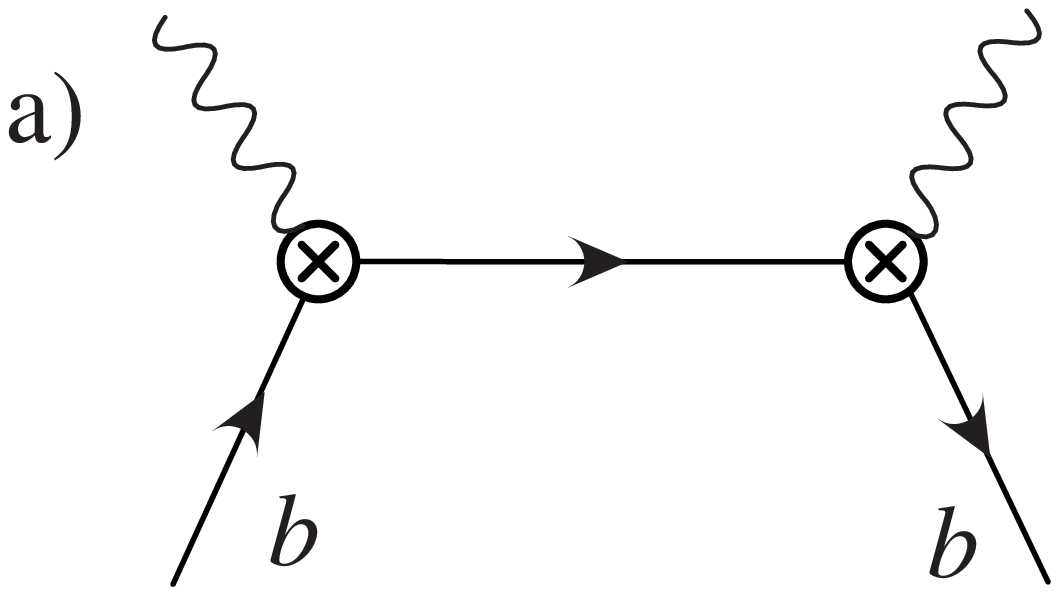}} }\hspace{0.6cm}
  \mbox{\epsfysize=2.5truecm \hbox{\epsfbox{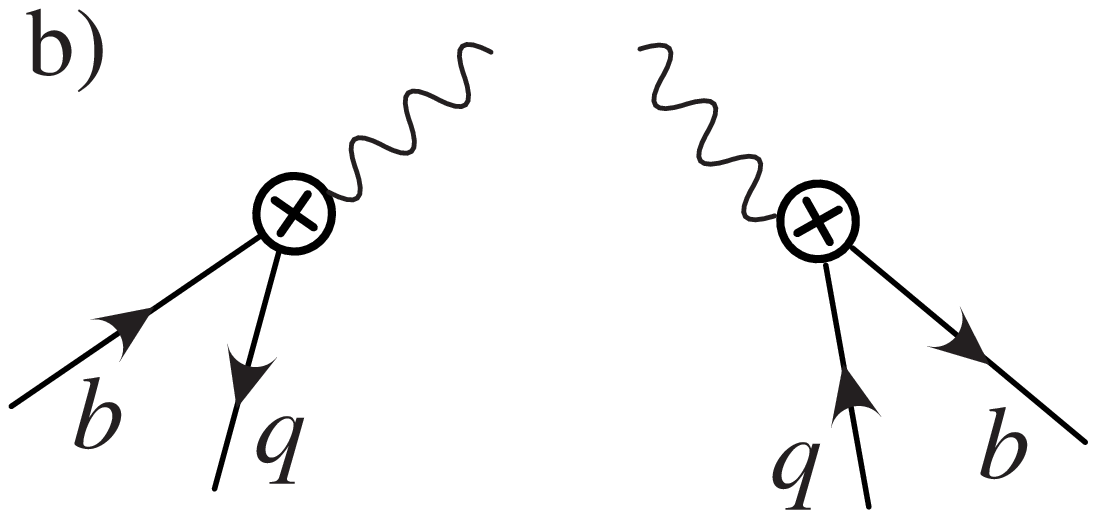}} }\hspace{0.4cm}
  \mbox{\epsfysize=2.5truecm \hbox{\epsfbox{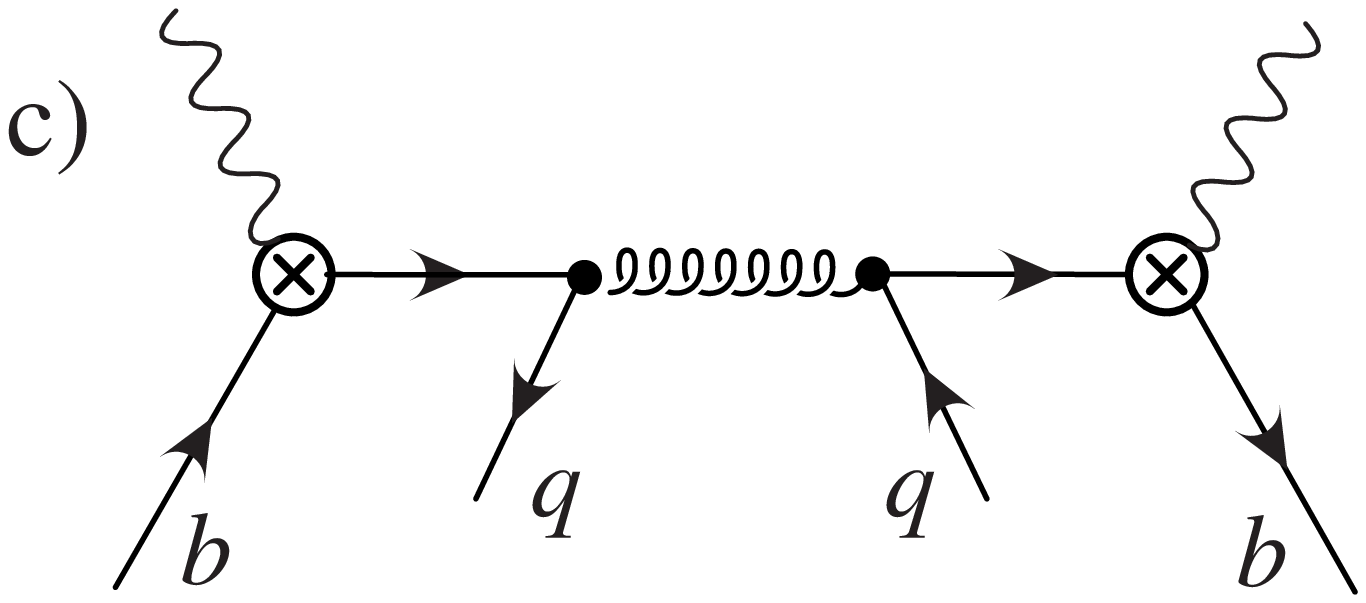}} }
  }
 \vskip -.1cm 
 {\caption[1]{Comparison of the ratio of annihilation contributions to the
     lowest-order result. In the total decay rate, b) is $\sim
     16\pi^2(\Lambda^3/m_b^3) \Delta B\simeq 0.02$, while c) is $\sim
     4\pi\alpha_s(m_b)(\Lambda^3/m_b^3)\simeq 0.003$ when compared to a). In the
     endpoint region, b) is $\sim 16\pi^2 (\Lambda^2/m_b^2)\Delta B\simeq 0.16$,
     a large correction, while c) becomes $\sim
     4\pi\alpha_s(\mu_J)(\Lambda/m_b)\epsilon' \simeq 0.6 \epsilon'$, a
     potentially large correction. }
\label{fig:ann} }
 \vskip -0.1cm
\end{figure}
These operators are generated by connected graphs from the time-ordered product
of two currents, as in Fig.~\ref{fig:ann}a. On the other hand, the four-quark
operators $O^{6c,6d}$ give parameters $f_B^2 B_{1,2}$ and are disconnected (or
rather connected by leptons or photons only), as shown in Fig.~\ref{fig:ann}b,
and thus exhibit a phase-space enhancement relative to Fig.~\ref{fig:ann}a. The
simplest way to see this is to note that for the total rate to $B\to
X_u\ell\bar\nu$, we would cut a two-loop graph for Fig.~\ref{fig:ann}a, while
Fig.~\ref{fig:ann}b would be at one-loop level (the $\ell$--$\bar\nu$ loop).  For
later convenience, we also consider the perturbative correction to the
four-quark operators shown in Fig.~\ref{fig:ann}c, which is suppressed by
$\alpha_s/(4\pi)$ relative to Fig.~\ref{fig:ann}b, and gives the operators
$O^{6e,6f}$. In the total decay rate, if we normalize so that
Fig.~\ref{fig:ann}a $\sim 1$ then
\begin{eqnarray} \label{ope_estimate}
  \mbox{Fig.~1b} \sim 16\pi^2\: \frac{\Lambda^3}{m_b^3}\, \Delta B\: \sim 0.02\,,
  \qquad
  \mbox{Fig.~1c} \sim 4\pi\alpha_s(m_b)\: \frac{\Lambda^3}{m_b^3} \epsilon\: \sim 0.003 \epsilon\,.
\end{eqnarray}
Here $\Delta B=B_2-B_1\sim 0.1$ accounts for the fact that the matrix elements
of the operators generated by Fig.~\ref{fig:ann}b vanish in the factorization
approximation. The factor of $\epsilon$ accounts for any dynamical suppression
of Fig.~\ref{fig:ann}c. The definitions of $B_{1,2}$ are
\begin{eqnarray} 
 \big\langle B_v \big| \big[\bar h_v \gamma_\sigma q_L\big]
   \big[ \bar q_L \gamma_\tau  h_v \big] \big| B_v \big\rangle
  = \frac{f_B^2 m_B}{12} \Big[ (B_1-B_2) g_{\sigma\tau}\: 
    + (4 B_2-B_1) v_\sigma v_\tau\:   \Big] \,.
\end{eqnarray}
Without the $\Delta B$ suppression factor, Fig.~\ref{fig:ann}b would dominate
over other $1/m_b^2$ operators rather than just competing with them.  The ${\cal
  O}(\alpha_s)$ corrections to annihilation are still a small contribution in
the local OPE, for any $\epsilon\le 1$.  In particular possible
enhancements of these contributions have been shown to cancel for the total
$b\to u$ decay rate~\cite{Bigi:1991ir}.

In the endpoint region there are extra enhancement factors and the dimensions of
the operators no longer determine the size of their contributions. The fact that
annihilation effects are larger in the endpoint was first pointed out in
Ref.~\cite{Bigi:1991ir}. The power counting in SCET organizes these
contributions in a systematic fashion and allows us to be more quantitative
about how large these contributions are. Since some background material is
required, we postpone this power counting until Sec.~\ref{sect_J}. The
derivation given here is more heuristic, but leads to the same results.  For
Fig.~1a the intermediate quark propagator becomes collinear, giving an
$m_b/\Lambda$ enhancement.  This explains why a larger portion of the decay rate
is concentrated in the endpoint region.  For Fig.~1b there is no
quark-propagator enhancement but also no reduction from the phase space.  A
numerical estimate for this contribution was made in
Ref.~\cite{Voloshin:2001iq}.  Finally, for Fig.~1c in the endpoint region there
can be {\em three} collinear propagators, giving a large $m_b^3/\Lambda^3$
enhancement to this diagram. In Sec.~\ref{sect_NLO} we show that this graph
contains the maximum possible enhancement.  In summary, 
if we consider the rate  integrated only over the endpoint region then 
Fig.~1a $\sim 1$ and
\begin{eqnarray}
   \mbox{Fig.~1b} \sim 16\pi^2\: \frac{\Lambda^2}{m_b^2} \Delta B\: \sim 0.2\,,
  \qquad
  \mbox{Fig.~1c} \sim 4\pi\alpha_s(1.4\,{\rm GeV})\: \frac{\Lambda}{m_b}\: 
  \sim 0.6\: \epsilon'
  \,.
\end{eqnarray}
The non-perturbative function that gives $\epsilon'$ differs from the local
operators that give $\epsilon$ in Eq.~(\ref{ope_estimate}).  Since $\epsilon'
\sim 0.3$ is possible\footnote{The only rigorous scaling argument for a
  dynamical suppression that we are aware of is the large-$N_c$ limit, where
  $\Delta B\sim 1/N_c$. For Fig.\ref{fig:ann}c the $q\bar q$ contraction gives a
  matrix element that is leading order in $N_c$. If these contributions are
  small, then a rough estimate is $\epsilon'\sim 1/N_c\sim 0.3$.  }, we conclude
that the contribution from Fig.~1c may actually give one of the largest
uncertainties in extracting $V_{ub}$ with methods such as $E_\ell$ or $m_X^2$
cuts that depend on the endpoint region. It has not been considered in recent
error estimates in the literature. The main phenomenological outcome of our
analysis is a proper consideration of this term for endpoint spectra.

Theoretically, the main result of our analysis is a complete theoretical
description for the NLO term, $\Gamma^{(2)}$, in the power expansion of decay
spectra in the endpoint region,
\begin{eqnarray}\label{series}
  \frac{d\Gamma}{dZ_i} \bigg|_{\rm endpoint} = 
     \frac{d\Gamma^{(0)}}{dZ_i} + 
      \frac{d\Gamma^{(2)}}{dZ_i} + \ldots  \,.
\end{eqnarray}
Here $Z_i$ denotes a generic choice of the possible spectrum variables,
$\{P^+,P^-,E_\gamma, q^2, s_H, m_b, \ldots \}$.  At NLO we use SCET to determine
the contributions to the spectra. These contributions are tabulated in the body
of the paper, but the generic structure of a term in $({1}/{\Gamma_0})
{d\Gamma^{(2)}}/{dZ_i}$ is
\begin{eqnarray}\label{Tlo}
  \int [dz_{n'}] H^{(j_1)}(z_{n'},m_b,Z_i) 
  \int [dk^+_n]\ {\cal J}^{(j_2)}(z_{n'},k^+_n,P^\pm)\: 
    f^{(j_3)}(k^+_n) \,,
\end{eqnarray}
where the number of convolution parameters varies from $n=1$ to $n=3$ and $n'=1$
or $2$, and for $n=2$ $[dk^+_n]=dk^+_1 dk^+_2$ etc. The dependence on the
$z_{n'}$ parameters appears only in jet functions that vanish at tree level. In
Eq.~(\ref{Tlo}) the $(j_1)$, $(j_2)$, $(j_3)$ powers indicate whether the power
suppression occurs in the hard, jet or soft regions respectively. The power
corrections start at ${\cal O}(\lambda^2)$, which is $\sim 1/m_b$, and so
$j_1+j_2+j_3=2$. Here $j_{1,3}\ge 0$ while $j_2$ can be negative.  Phase-space
and kinematic corrections give an $H^{(2)}$ with the same jet and shape
functions as at leading order. Other more dynamic power corrections involve new
hard $H^{(0)}$ functions, and obtain their power suppression from the product of
jet and soft factors. We show that the operators at NLO allow $-4 \le j_2 \le 2$
and $0 \le j_3 \le 6$.  The largest jet function ($j_2=-4$) occurs for exactly
the endpoint contribution generated by the four-quark operators ($j_3=6$) from
Fig.~1c.

0ur analysis can be compared with the closely related physical problem of deep
inelastic scattering with $Q^2\gg \Lambda^2$, in the limit where Bjorken
$x\sim 1-\Lambda/Q$. With no parametric scaling for $x$, the power corrections
in DIS at twist $4$ were computed in Refs.~\cite{Jaffe:1981td,
  Jaffe:1982pm,Ellis:1982cd,Ellis:1982wd,Shuryak:1981pi,Shuryak:1981kj}. As
$x\to 1$ the relative importance of the power-suppressed operators changes and
the importance of contributions from four-quark operators has been discussed in
Ref.~\cite{Gardi:2004ia}.

The outline of the remainder of this paper is as follows.  In Sec.~II we give
the basic ingredients needed for our computations, including the weak
Hamiltonian (Sec.~II~A) and expressions for the hadronic tensors and decay rates
(Sec.~II~B). In Sec.~II~C we give a detailed discussion of the endpoint
kinematics and light-cone variables, and in Sec.~II~D we briefly summarize a few
results obtained using the optical theorem for the forward scattering amplitude,
and the procedure for switching between partonic and hadronic variables that is
relevant in the endpoint region.  In Sec.~III we turn to the discussion of the
SCET formalism. Many of the ingredients necessary for our computation are
readily available in the literature.  Of particular note are expressions for the
heavy-to-light currents at ${\cal O}(\lambda^2)$~\cite{Beneke:2002ni}, which we
have verified. In Sec.~IV we review the derivation of the factorization theorem
at LO, but do so in a way that makes the extension beyond LO more accessible.
We consider power corrections of ${\cal O}(\lambda)$ in Sec.~V, and show that
they vanish. In Sec.~VI we discuss the true NLO factorization theorem, which is
${\cal O}(\lambda^2)$. In particular, in Sec.~VI~A we switch to hadronic
variables and re-expand the LO result, in Sec.~VI~B we enumerate all the
time-ordered products that occur at this order, and in Sec.~VI~C we show that
the tree-level matching is simplified by using the SCET formalism. In Sec.~VI~D
we give definitions for the non-perturbative shape functions that appear, and
then in Sec.~VI~E we derive the factorization theorems for the most important
contributions in some detail. Finally, in Sec.~VI~F we summarize the hard
coefficient functions for the subleading time-ordered products. Next, in
Sec.~VII we present a useful summary of the NLO decay-rate results, including
the phase-space corrections. We also compare with results in the literature
where they are available.  Our conclusions and discussion are given in
Sec.~VIII.  Further details are relegated to appendices, including the expansion
of the heavy quark field and calculation of the power-suppressed heavy-to-light
currents at tree level in Appendix~A, and a review of constraints on the
currents from reparameterization invariance in Appendix~B.  For the reader
interested in getting an overview of our results while skipping the details, we
suggest reading Secs.~II, III and IV, the introduction to Sec.~VI, and
Secs.~VI~A, VI~B and VII.  A reader interested only in final results may skip
directly to the summary in Sec.~VII.


\section{Basic Ingredients}

In this section we give the ingredients necessary for studying the decays $B\to
X_s\gamma$ and $B\to X_u \ell \bar\nu$ in the endpoint region to NLO. A proper
treatment requires a separation of the scales $m_W^2\gg m_b^2\gg m_b\Lambda_{\rm
  QCD} \gg \Lambda_{\rm QCD}^2$ in the form of a factorization theorem. This is
accomplished by the following steps:
\begin{itemize}
\item[1)] Match on to the weak Hamiltonian, $H_W$, at $\mu^2=m_W^2$ and
  run down to $\mu^2=m_b^2$, just as in the standard local OPE.
\item[2)] Match $H_W$ at $\mu^2\simeq m_b^2$ on to SCET, with collinear and 
  usoft degrees of freedom and an expansion in 
  $\lambda=\sqrt{\Lambda_{\rm QCD}/m_b}$.
  Run from $\mu^2=m_b^2$ to $\mu^2=\mu_0^2\simeq m_b\Lambda_{\rm QCD}$.
\item[3)] At $\mu^2 = \mu_0^2$ integrate out the collinear modes, which, given a
  complete factorization in step 2), is trivial. Then run from $\mu^2=\mu_0^2$ to
  $\mu^2\sim 1\,{\rm GeV^2}\gtrsim \Lambda_{\rm QCD}^2$.
\end{itemize} 
In Sec.~\ref{sect_Hw} we discuss the weak Hamiltonian. The kinematics and
differential decay rates for the endpoint region are given in
Sec.~\ref{sect_kin}.  Then in Sec.~\ref{sect_J} we give the necessary
effective-theory Lagrangians and currents to ${\cal O}(\lambda^2)$.

\subsection{Weak Effective Hamiltonians} \label{sect_Hw}

For $B\to X_u\ell\bar\nu$ the effective Hamiltonian is simply
\begin{eqnarray}
  {\cal H}_{\rm eff}^u &=&  -\frac{4G_F}{\sqrt2}\, V_{ub}
    (\bar{u}\gamma_\mu P_L b)
    (\bar{\ell}\gamma^\mu P_L \nu_\ell) \,,
\end{eqnarray}
and the current $\bar u \gamma_\mu P_L b$ is the basis for our analysis of the
QCD part of the problem.  The Hamiltonian for the weak radiative decay $B\to
X_s\gamma$ is more involved:
\begin{eqnarray}
  {\cal H}_{\rm eff}^s &=& 
   -\frac{4G_F}{\sqrt2}\, V^{\phantom{*}}_{tb} V^*_{ts}\, 
  \Big[ C_{7\gamma}\, {\cal O}_{7\gamma}  
   + C_{8g}\, {\cal O}_{8g} 
   + \sum_{i=1}^6 C_i\, {\cal O}_i \Big]
  \,,
\end{eqnarray}
where $P_{R,L} = \frac12 (1 \pm \gamma_5)$ and
\begin{eqnarray}
 {\cal O}_{7\gamma} 
  &=& \frac{e}{16\pi^2}\, m_b\: \bar s\,\sigma_{\mu\nu}F^{\mu\nu}\,P_R\,b\, 
  \,,\qquad\quad
 {\cal O}_{8g}
  = \frac{g}{16\pi^2}\, m_b\: \bar s\,\sigma_{\mu\nu}G^{\mu\nu}\,P_R\,b\, \,,
   \\[5pt]
 {\cal O}_1 &=& (\bar{c}\gamma_\mu P_L b) (\bar{d}\gamma^\mu P_L c) \,,
   \qquad\qquad\quad
 {\cal O}_2 = (\bar c \gamma_\mu P_L b) (\bar s \gamma^\mu P_L c) \,,
   \nn\\
 {\cal O}_{\{{3,5}\}}  &=&  (\bar{d} \gamma_\mu P_L  b) 
   \sum_q (\bar q\, \gamma^\mu P_{L,R}\: q)\,,
   \qquad\quad\!\!
 {\cal O}_{\{{4, 6}\}} =  (\bar{d}_{\beta} \gamma^\mu P_L b_{\alpha})
  \sum_q (\bar{q}_{\alpha} \gamma^\mu P_{L,R}\: q_{\beta}) \,,
 \nn
\end{eqnarray}
with $F_{\mu\nu}$ and $G_{\mu\nu}$ the electromagnetic and QCD field strengths.
In the expression for ${\cal H}_{\rm eff}^s$, we have used unitarity,
$V_{cb}^{\phantom{*}}V_{cs}^* = -V_{tb}^{\phantom{*}} V_{ts}^* -
V_{ub}^{\phantom{*}} V_{us}^*$, and have dropped the small ($\sim 2\%$)
corrections from $V_{ub}^{\phantom{*}} V_{us}^*$ and $m_s/m_b$.

For the total $B\to X_s\gamma$ rate the perturbative corrections are known at
NLO~\cite{Buras:2002er,Hurth:2003vb}.  Effective scheme-independent coefficients
$C_{7,8}^{\rm eff}$ are defined in a way that includes contributions from the 
penguin operators ($C_{3-6}$). A totally inclusive analysis is considerably
simplified by the fact that at leading order in $1/m_b$ the matrix
elements can be evaluated directly in full QCD rather than first having to
match on to HQET. For an endpoint analysis, the matching at $m_W$ and running 
to $\mu\sim m_b$ is the same. However, at the scale $\mu\simeq m_b$ the operators 
in $H_W$ need to be matched on to operators in SCET before the OPE is performed.
In performing the matching, the only subtle complication is the treatment of 
the charm mass. For simplicity, the approach we take here is formally to let 
$m_c\sim m_b$, so that charm-mass effects are all hard and are integrated out 
in matching on to SCET. This agrees with the treatment of the $O_{i\ne 7}$ 
advocated in Ref.~\cite{Neubert:2001sk} for the endpoint region. Since 
numerically $m_c^2 \sim m_b\Lambda$, perhaps a better alternative would be to 
keep charm-mass effects in the operators of SCET until below the jet scale 
$m_b\Lambda$. This second approach is more involved, and in particular it is 
clear from Ref.~\cite{Buras:2001mq} that it would necessitate introducing two 
types of collinear charm quark, as well as soft and ultrasoft charm quarks. 
For this reason, we stick to the former approach and leave the latter for 
future investigation.

At lowest order in the $\lambda$ power expansion, there is only the SCET analog of
the $\bar s \sigma_{\mu\nu}P_R b$ current called $J^{(0)}$ (cf. Eq.~(\ref{J0})),
and ${\cal O}_{1-8}$ can make contributions to its Wilson coefficient. At NLL
order in $\alpha_s$ the effect of the other operators can be taken into
account by using~\cite{Greub:1996tg,Chetyrkin:1996vx}
\begin{eqnarray}
  C_\gamma(\mu,z) = C_7^{\rm eff(0)}(\mu) + \frac{\alpha_s}{4\pi} 
    C_7^{\rm eff(1)}(\mu)
    + \sum_{k} C_k^{\rm eff (0)}(\mu) \Big[ r_k(\rho) + \gamma_{k7}^{\rm eff
    (0)} \ln\Big(\frac{m_b}{\mu}\Big) \Big]\,,
\end{eqnarray}
in place of $C_7$ when matching on to SCET.  Here the dependence on
$\rho=m_c^2/m_b^2$ enters from the four-quark operators with charm quarks. In the
endpoint region the ${\cal O}(\alpha_s)$ effects from the process 
$b\to s\gamma g$ all appear in the jet and shape functions. 
For later convenience, we define
\begin{eqnarray} \label{Deltas}
  \Delta_\gamma(\mu,\rho) = \frac{C_\gamma(\mu,\rho)}{C_7^{\rm eff(0)}(\mu)} - 1\,.
\end{eqnarray}

The photon in $B\to X_s\gamma$ is collinear in the opposite direction to the jet
$X_s$, so propagators connecting the two are hard. Thus, beyond LO in the power
expansion, the photon will typically be emitted by effective-theory currents
$J^{(i)}$ (which could be four-quark operators). We shall discuss the matching
on to these subleading currents for ${\cal O}_7$ only. Some of the contributions
from the other ${\cal O}_i$ will just change the Wilson coefficients of the
subleading currents and thus not modify the structure of the power-suppressed
factorization theorems (indeed some of them are already known since they are
fixed by reparameterization invariance). These other operators may also induce
time-ordered products that would involve operators with quarks collinear to the
photon direction, but these are not considered here.

\subsection{Hadronic Tensors and Decay Rates} \label{sect_decay}

In this subsection, we summarize general results for the hadronic tensors and
decay rates, without restricting ourselves to the endpoint region.  For both
decays $B\to X_s\gamma$ and $B\to X_u \ell \bar\nu$,  momentum conservation 
for the hadrons gives
\begin{eqnarray} \label{kin1}
  p_B^\mu = m_B v^\mu = p_X^\mu + q^\mu \,,
\end{eqnarray}
where $p_X^\mu$ is the sum of the four-momenta of all the hadrons in $X$,
$q^\mu$ is the momentum of the $\gamma$ or the pair of leptons $(\ell\bar\nu)$,
and the velocity $v^\mu$ satisfies $v^2=1$. For $B\to X_s\gamma$, $q^\mu =
n\mcdot q\, \bn^\mu/2=E_\gamma \bn^\mu$, where $v\cdot q=E_\gamma $ is the
photon energy in the rest frame of the $B$ meson. In this case $q^2=0$ and
\begin{eqnarray} \label{kin2}
  m_{X_s}^2 = m_B (2E_{X_s}-m_B) = m_B(m_B-2 E_\gamma) \,,
\end{eqnarray} 
so the differential rate involves only one variable, $m_X$ or $E_\gamma$.  For
$B\to X_u \ell \bar\nu$, Eq.~(\ref{kin1}) implies
\begin{eqnarray} \label{kin3}
  E_X = v\mcdot p_X = \frac{m_B^2 + m_X^2 - q^2}{2m_B} \,, \qquad
  v\cdot q = m_B - E_X  = E_{\ell}+ E_{\nu}\,,
\end{eqnarray}
where $p_X^2=m_X^2$. The differential decay rate involves three variables,
several common choices of which are $\{E_\ell,E_\nu,q^2\}$, 
$\{E_\ell, v\mcdot q, q^2\}$, or $\{E_\ell, m_X^2, q^2\}$.

To derive the inclusive decay rates for $\bar B\to X_s\gamma$ and
$\bar B\to X_u \ell \bar\nu$, the matrix elements are separated into a
leptonic/photonic part $L_{\mu\nu}$ and a hadronic part $W_{\mu\nu}$. Here
\begin{eqnarray} \label{defnW}
  W_{\mu\nu} &=& \frac{1}{2m_B} \sum_X (2\pi)^3 \delta^4(p_B-q-p_X) 
  \langle \bar B | J_\mu^\dagger | X \rangle \langle X | J_\nu | \bar B \rangle
  \nn\\
  &=& - g_{\mu\nu} W_1 + v_\mu v_\nu W_2 + i\epsilon_{\mu\nu\alpha\beta}
  v^\alpha q^\beta W_3 + q_\mu q_\nu W_4 + (v_\mu q_\nu + v_\nu q_\mu) W_5 
  \,,
\end{eqnarray}
in which we use the hadronic current $J$ and relativistic normalization for the 
$|\bar B\rangle$ states.  For convenience we define projection tensors 
$P_i^{\mu\nu}$ so that
\begin{eqnarray}
   W_i = P_i^{\mu\nu}\: W_{\mu\nu} \,.
\end{eqnarray}
They are
\begin{eqnarray} \label{ProjW}
  P_1^{\mu\nu} &=& -\frac{1}{2} g^{\mu\nu}
  + \frac{q^2\, v^{\mu}v^\nu + q^\mu q^\nu  - v\mcdot q (v^\mu q^\nu+v^\nu
    q^\mu)} {2 [q^2-(v\mcdot q)^2]}\,,\\[5pt]
  P_2^{\mu\nu} &=& \frac{3 q^2\, P_1^{\mu\nu} + q^2 g^{\mu\nu} -q^\mu q^\nu}
    { [q^2-(v\mcdot q)^2]} \,,\qquad
  P_3^{\mu\nu} = \frac{-i \epsilon^{\mu\nu\alpha\beta} q_\alpha v_\beta}
    { 2[q^2-(v\mcdot q)^2]} \,,\nn \\[5pt]
  P_4^{\mu\nu} &=& \frac{g^{\mu\nu} - v^\mu v^\nu + 3 P_1^{\mu\nu}}
    { [q^2-(v\mcdot q)^2]} \,, \qquad
  P_5^{\mu\nu} = \frac{g^{\mu\nu} +  4 P_1^{\mu\nu}-P_2^{\mu\nu} - q^2
    P_4^{\mu\nu}} {2 v\mcdot q} \,.\nn
\end{eqnarray}

Contracting the lepton/photon tensor $L^{\mu\nu}$ with $W^{\mu\nu}$ and 
neglecting the mass of the leptons gives the differential decay rates
\begin{eqnarray} \label{dGamma}
  \frac{d\Gamma^s}{dE_\gamma} 
    &=& \Gamma_0^s\: \frac{8 E_\gamma}{m_B^3} (4 W_1^s -W_2^s -2 E_\gamma W_5^s) 
    \,, \\[5pt]
 \frac{d^3\Gamma^u}{ dE_\ell dq^2 dE_\nu}  
    &=& \Gamma_0^u\: \frac{96}{m_B^5} \Big[ q^2 W_1^u + 
    (2 E_e E_\nu -q^2/2) W_2^u  + q^2 (E_e-E_\nu) W_3^u \Big]
    \theta(4 E_e E_\nu-q^2) \,, \nn\\[-5pt] \nn
%
\end{eqnarray}
where $W_i=W_i(q^2,v\mcdot q)$ and the normalization factors are
\begin{eqnarray}
  \Gamma_0^s = \frac{G_F^2\, m_B^3}{32\pi^4}\,
  |V_{tb} V_{ts}^*|^2\, \alpha_{\rm em}\, [\overline m_b(m_b)]^2
  |C_7^{\rm eff(0)}(m_b)|^2 \,,\qquad
 \Gamma_0^u = \frac{G_F^2\, m_B^5}{192\pi^3}\,
  |V_{ub}|^2\,.
\end{eqnarray}
Here $\overline m_b(\mu)$ is the $\overline {\rm MS}$ mass. For convenience, we
have pulled out a Wilson coefficient $C_7^{\rm eff(0)}$ so that contributions
from other coefficients appear in ratios $C_i/C_7^{\rm eff(0)}$ in the SCET
Wilson coefficients (for example the quantity $\Delta_\gamma$ in
Eq.~(\ref{Deltas})).  Note that we have chosen to stick with hadronic variables
here (using $m_B$ rather than $m_b$).  When we eventually compute the $W_i$, we
shall have to deal with switching between partonic and hadronic variables.
However, we shall see that the situation is quite different from that in the 
local OPE (as we discuss further in Sec.~\ref{sect_partonic} below). 
In particular, it is the hadronic phase space that turns out to be required.

In Eq.~(\ref{dGamma}), $0\le E_\ell , E_\nu \le (m_B^2-m_\pi^2)/(2 m_B)$. A set
of useful dimensionless hadronic variables is
\begin{eqnarray}
  x_H^\gamma = \frac{2 E_\gamma}{m_B} \,, \qquad
  x_H = \frac{2 E_{\ell}}{m_B} \,, \qquad
  \quad y_H = \frac{q^2}{m_B^2}\,, \qquad 
  s_H = \frac{m_X^2}{m_B^2} \,.
\end{eqnarray}
In terms of these variables,
\begin{eqnarray}
 \frac{2 E_\nu}{m_B} = 1-s_H+y_H-x_H \,,\qquad
 \frac{2 E_X}{m_B} = 1 + s_H - y_H \,,
\end{eqnarray}
and $W_i=W_i(y_H,s_H)$.  
For $B\to X_s\gamma$,
\begin{eqnarray} \label{dGamma2t}
\frac{d\Gamma^s}{dx_H^\gamma} 
    &=& \Gamma_0^s\: \frac{2 x^H_\gamma}{m_B} 
     \Big\{ 4 W_1^s -W_2^s -m_B\, x_H^\gamma W_5^s \Big\} 
    \,,
\end{eqnarray}
with $0\le x_H^\gamma \le 1 - m_{K^*}^2/m_B^2$. For
$B\to X_u\ell\bar\nu$,
\begin{eqnarray}  \label{dGamma2u}
 \frac{d^3\Gamma^u}{ dx_H\, dy_H\, ds_H}  
    &=& \Gamma_0^u\: {24m_B} \Big\{ y_H W_1^u + 
    \frac12 \big[(1\!-\!x_H)(x_H\!-\!y_H) \!-\! x_H s_H\big] W_2^u \\
 && 
   + \frac{1}{2}\, m_B y_H \big(2x_H\!+\!s_H\!-\!y_H\!-\! 1\big) W_3^u \Big\}
     \theta\big[ (1\!-\!x_H)(x_H\!-\!y_H) \!-\! x_H s_H \big]\,,
 \nn
\end{eqnarray} 
and, depending on the order of integration, there are several
useful combinations of the limits, which are shown in Table~\ref{table_limits}.
\begin{table}[t!]
\[
\begin{array}{llll}
 \hline
i)\hspace{1cm} &  
0 \le x_H \le 1- {r_\pi^2}   & 
  r_\pi^2 \le s_H \le 1-x_H  & 
  0 \le y_H \le x_H -\frac{s_H\, x_H}{1-x_H}  
 \\[7pt]
ii) &  
 0 \le x_H \le 1- r_\pi^2    &
 0 \le y_H \le x_H -\frac{r_\pi^2 x_H}{1-x_H}  & 
 r_\pi^2 \le s_H \le \frac{1-x_H}{x_H}(x_H-y_H)  
 \\[7pt]
iii) &  
 r_\pi^2 \le s_H \le 1  \hspace{2cm} &  
  0 \le x_H \le 1-s_H  \hspace{2cm} &  
  0 \le y_H \le x_H -\frac{s_H\, x_H}{1-x_H} 
  \\[7pt]
iv) &  
 r_\pi^2 \le s_H \le 1  &
  0 \le y_H \le \big(1-\sqrt{s_H}\big)^2  & 
  x_H^{\rm min} \le x_H \le x_H^{\rm max}    
 \\[7pt]
v) &  
   0 \le y_H \le \big(1\!-\! r_\pi\big)^2  & 
 r_\pi^2 \le s_H \le \big(1\!-\!\sqrt{y_H}\big)^2  &
  x_H^{\rm min} \le x_H \le x_H^{\rm max}   
 \\[7pt]
vi) &  
 0 \le y_H \le \big(1\!-\! r_\pi \big)^2  \,\, &
x_H^{\rm min \, *} \le x_H \le x_H^{\rm max \, *}  \,\, &
r_\pi^2 \le s_H \le 1 + y_H - \frac{y_H}{x_H} - x_H \\[5pt]
 \hline
\end{array}
\]
\vskip -.1cm
\caption{
 Limits for different orders of integration in $B\to X_u\ell\bar\nu$ with
 variables $\{x_H, s_H, y_H\}$. Here
 $r_\pi=m_\pi/m_B$, while 
 $\{x_H^{\rm max},x_H^{\rm min}\} = \big[(1+y_H-s_H)\pm
\sqrt{(1+y_H-s_H)^2-4y_H}\,\big]/2$ and $\{x_H^{\rm max \,*},x_H^{\rm min \,*}\} =
\{x_H^{\rm max},x_H^{\rm min}\}|_{s_H=r_\pi^2}$.
 Results for the
phase-space limits of partonic variables are obtained by dropping the
$H$-subscripts and setting $r_{\pi}=0$. \label{table_limits}} 
\end{table}
If we integrate over all values of $x_H$ (cases iv) \& v)), the rate becomes
\begin{eqnarray}  \label{dGamma2up}
 \frac{d^2\Gamma^u}{  dy_H\, ds_H}  
    &=& \Gamma_0^u\: {2m_B} \sqrt{(1\!-\!y_H\!+\! s_H)^2\!-\! 4\, s_H}\:
   \Big\{ 12 y_H W_1^u  \\
 && 
   +\big[(1\!-\!y_H\!+\! s_H)^2 \!-\! 4 s_H\big] W_2^u \Big\}
 \,.
 \nn
\end{eqnarray}

\subsection{Light-cone Hadronic Variables and Endpoint Kinematics} 
\label{sect_kin}

We are interested in the jet-like region corresponding to $\Lambda_{\rm QCD} \ll
m_X \ll E_X$ for both $B\to X_s\gamma$ and $B\to X_u \ell\bar\nu$.  In this
region, the hadrons in the $X$ occur in a jet in the $B$ rest frame with 
$E_X\sim m_B$ and $m_X^2\lesssim m_B\Lambda_{\rm QCD}$.\footnote{In
  Refs.~\cite{Bauer:2003pi,Bosch:2004th} an intermediate situation,
  $m_X^2\lesssim \Delta^2$ with $\sqrt{m_B\Lambda_{\rm QCD}}\ll \Delta\ll m_B$,
  was also considered, but we do not consider it here.}  The momentum of the
states $X$ is therefore restricted, but they still form a complete set for
Eq.~(\ref{defnW}).  The width of the jet is determined by noting that the
typical perpendicular momentum between any two final-state hadrons is $\Delta
p_\perp \lesssim \sqrt{m_B\Lambda}\sim 1.6\,{\rm GeV}$, where we use
$\Lambda\sim 0.5\,{\rm GeV}$ to denote a typical hadronic scale for $B$-mesons
(examples being $\Lambda_{\rm QCD}$ and $\overline\Lambda$). We also assume that
there are enough states with $m_X^2\lesssim m_B\Lambda$ that the endpoint region
is still inclusive and Eq.~(\ref{WT}) is valid. As we shall see below, the
factorization in this region gives non-perturbative shape functions rather than
just local operators.

For jet-like kinematics it is natural to introduce a light-like vector $n$ along
the jet direction and an auxiliary vector $\bn$ such that $n^2=\bn^2=0$ and
$n\cdot \bn =2$, and refer to components $(p^+,p^-,p_\perp)=(n\mcdot p,
\bn\mcdot p, p_\perp^\mu)$.  Some useful decompositions are
\begin{eqnarray}
 && g^{\mu\nu}_\perp = g^{\mu\nu} - \frac12 (n^\mu \bn^\nu+ n^\nu \bn^\mu), 
  \qquad 
   g^{\mu\nu}_T = g^{\mu\nu}-v^{\mu} v^\nu ,
  \qquad
  \epsilon_\perp^{\mu\nu} = \frac12 \epsilon^{\mu\nu\alpha\beta}\bn_\alpha
  n_\beta \,,
   \qquad\nn\\
 && p^\mu_\perp = p^\mu - \frac{\bn\mcdot p}{2}\, n^\mu 
  - \frac{n\mcdot p}{2}\, \bn^\mu
  \,,\qquad \ \ 
  p^\mu_T = p^\mu - v^\mu v\mcdot p \,,
\end{eqnarray}
where we take $\epsilon^{0123}=1$. Note that the subscript $T$ means transverse
to $v^\mu$, so $p_T^\mu \ne p_\perp^\mu$.  For the final factorization theorem
for the differential decay rates we shall use a frame where
$q_\perp^\mu=v_\perp^\mu=0$, and $v^\mu=(n^\mu+\bn^\mu)/2$.\footnote{If one desires,
  he or she can take $v=(1,0,0,0)$, $n=(1,0,0,-1)$, and $\bn=(1,0,0,1)$.  A more
  general frame is required only for working out the constraints from
  reparameterization invariance.} Thus $q^\mu = \bn\mcdot q\, n^\mu/2+n\mcdot
q\, \bn^\mu/2$ and
\begin{eqnarray} \label{qs}
  n\mcdot q = m_B - n\mcdot p_X \,,\qquad 
  \bn\mcdot q = m_B -\bn\mcdot p_X \,.
\end{eqnarray}
For $B\to X_s\gamma$ the photon momentum is taken along the $\bn$ light-like
direction, i.e.\ $q_\mu=E_\gamma \bn_\mu$, and
\begin{eqnarray}
  \bn\cdot p_X = m_B\,, \qquad 
   n\cdot p_X = m_B -2 E_\gamma = m_B (1-x_H^\gamma) \,.
\end{eqnarray}
For $B\to X_u \ell\bar\nu$ the phase space is more complicated and for
convenience we define the dimensionless variables
\begin{eqnarray} \label{ybaru}
  \overline y_H = \frac{\bn\mcdot p_X}{m_B} \,,\qquad \qquad
  u_H = \frac{n\mcdot p_X}{m_B} \,.
\end{eqnarray}
Now $m_X^2 = n\mcdot p_X\, \bn\mcdot p_X$ and $n\mcdot p_X+\bn\mcdot p_X= 
(m_B^2-q^2+m_X^2)/m_B$, so 
\begin{eqnarray}
   s_H = u_H \overline y_H  \,,\qquad
    y_H = (1-u_H)(1-\overline y_H) \,,
\end{eqnarray}
and, making the choice $\overline y_H \ge u_H$, we have
\begin{eqnarray}
 \{ \overline y_H , u_H\} &=& \frac{1}{2} \Big[ 1-y_H+s_H 
    \pm \sqrt{(1-y_H+s_H)^2-4 s_H}\,\Big] \,.
\end{eqnarray}
\begin{table}[t!]
\[
\begin{array}{llll}
 \hline \\[-8pt]
 i) \hspace{0.4cm} & 
 0 \le x_H \le 1 - r_\pi^2 \hspace{0.7cm} &
 r_\pi^2 \le u_H \le 1- x_H &
 {\rm max}\big\{ 1-x_H \,, \frac{r_\pi^2}{u_H} \big\}
    \le \overline y_H \le 1
 \\[7pt]
 ii) \hspace{0.8cm} & 
  0 \le x_H \le 1 - r_\pi^2 \hspace{0.8cm} &
  {\rm max}\big\{ 1-x_H, \frac{r_\pi^2}{1-x_H} \big\} \le \overline y_H \le 1 
    \hspace{0.8cm} &
  \frac{r_\pi^2}{\overline y_H} \le u_H \le 
      1-x_H
 \\[7pt]
 iii) \hspace{0.4cm} & 
 r_\pi^2 \le u_H \le 1 &
  0 \le x_H \le 1- u_H &
 {\rm max}\big\{ 1-x_H , \frac{r_\pi^2}{u_H}\big\}
    \le \overline y_H \le 1
 \\[7pt]
 iv) \hspace{0.4cm} &
 r_\pi^2 \le u_H \le 1  &
 \max\big\{\frac{r_\pi^2}{u_H}, u_H
  \big\} \le \overline y_H \le 1  &
  1-\overline y_H \le x_H \le 1-u_H
 \\[7pt]
 v) \hspace{0.4cm} &
  r_\pi \le \overline y_H  
     \le 1 \hspace{0.7cm} & 
  \frac{r_\pi^2}{\overline y_H}\: \le u_H \le\: \overline y_H
   \hspace{1cm} &
  1-\overline y_H \le x_H \le 1-u_H
 \\[7pt]
 vi)  \hspace{0.4cm} & 
   r_\pi \le \overline y_H \le 1 &
   1-\overline y_H \le  x_H \le 1-\frac{r_\pi^2}{\overline y_H} &
  \frac{r_\pi^2}{\overline y_H} \le u_H \le 1-x_H
 \\[7pt]
 \hline
\end{array}
\]
\caption{Full phase-space limits for $B\to X_u\ell\bar\nu$ with variables $x_H$, 
$\overline y_H$, and $u_H$. The parameter $r_\pi=m_\pi/m_B$. Results for the
phase-space limits of partonic variables are obtained by dropping the
$H$-subscripts and setting $r_{\pi}=0$.\label{table_limits2}}
\end{table}

So far we have not made any restriction to the endpoint. The variables
$\overline y_H$ and $u_H$ provide an equally good description of the full $B\to
X_u\ell\bar\nu$ phase space as $y_H$ and $s_H$, namely
\begin{eqnarray} \label{dGamma3u}
 \frac{1}{\Gamma_0^u} \frac{d^3\Gamma^u}{ dx_H\, d\overline y_H\, du_H}  
    &=& \: {24m_B} (\overline y_H\!-\! u_H)
   \Big\{(1\!-\! u_H)(1\!-\! \overline y_H) W_1^u + 
    \frac12 (1\!-\!x_H\!-\!u_H)(x_H\!+\!\overline y_H\!-\!1) W_2^u \nn\\
 && \hspace{-2cm}
  + \frac{m_B}{2}\,(1\!-\! u_H)(1\!-\! \overline y_H)
     \big(2x_H\!+\! u_H \!+\!\overline y_H\!-\! 2\big) W_3^u \Big\}
    \,, 
\end{eqnarray}
where $W_i=W_i(u_H,\overline y_H)$ and we have suppressed the theta function
from Eq.~(\ref{dGamma2u}). Integrating over $x_H$ from Table~\ref{table_limits2}
(cases iv) \& v)) gives
\begin{eqnarray} \label{dGamma4u}
  \frac{1}{\Gamma_0^u} \frac{d^2\Gamma^u}{  d\overline y_H\, du_H}  
    &=& {24m_B} (\overline y_H\!-\! u_H)^2
   \Big\{(1\!-\! u_H)(1\!-\! \overline y_H)  W_1^u + 
    \frac{1}{12} (\overline y_H\!-\! u_H)^2 W_2^u 
  \Big\}
     \,.
\end{eqnarray}
The full limits for $\overline y_H$ and $u_H$ are given in
Table~\ref{table_limits2}.

\begin{figure}[!t]
 \centerline{
  \mbox{\epsfysize=5.4truecm \hbox{\epsfbox{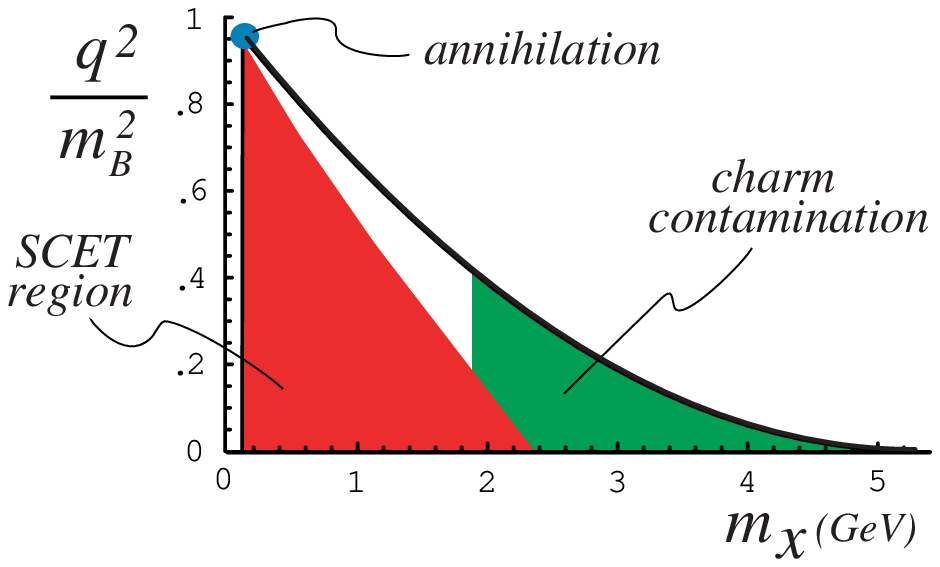}} }\ \ 
  \mbox{\epsfysize=5.4truecm \hbox{\epsfbox{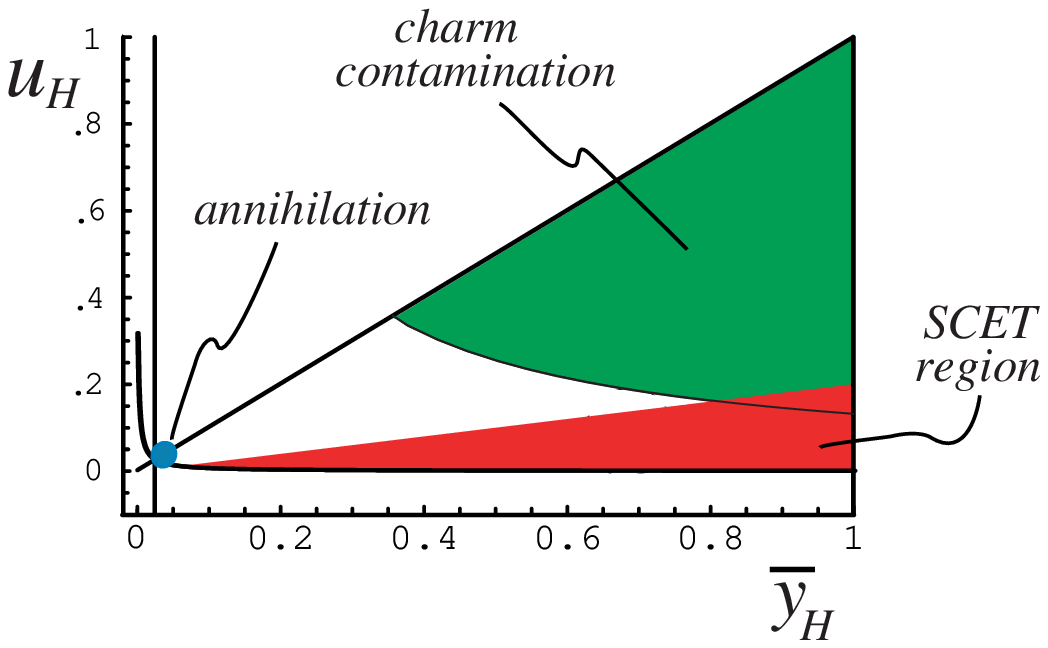}} }
  }
 \vskip -.5cm 
 {\caption[1]{Allowed phase space for $B\to X_u \ell\bar \nu$, where $m_\pi <
     m_X < m_B$. The second figure shows the same regions using the
     variables defined in Eq.~(\ref{ybaru}). We indicate the region where charm
     contamination enters, $m_X > m_D$, and the region of phase space where
     annihilation contributions enter.  Also shown is the region where the SCET
     expansion converges, which is taken to be $u_H/\overline y_H\le 0.2$ and
     corresponds to $m_X^2/(4E_X^2) \lesssim 0.14$.  }
\label{fig:phase} }
 \vskip -0.1cm
\end{figure}

In Ref.~\cite{Bosch:2004bt}, it was pointed out that a natural set of variables
in the endpoint region consists of the hadronic variable $n\cdot p_X$ and
partonic variable $\bn\cdot p$, where
\begin{eqnarray}
  \bn\mcdot p &=& \bn\mcdot p_X + m_b -m_B = \bn\mcdot p_X - \overline\Lambda
  + \ldots\,, \nn\\
  p_\omega^\mu &=& \frac{n^\mu}{2} \bn\mcdot p + \frac{\bn^\mu}{2} n\mcdot p_X
  \,.
\end{eqnarray}
They are natural because the LO factorization theorem dictates that the
kinematic variables appear in the jet functions and soft functions only as
\begin{eqnarray}
  {\cal J}(\bn\mcdot p\, k^+) \,, \qquad  f(n\mcdot p_X-k^+) \,,
\end{eqnarray}
where $k^+$ is the convolution parameter (cf. Sec.~\ref{LOfact}).  We shall see
that this remains true of the shape functions and jet functions at subleading
order in the power expansion. For a dimensionless version of $\bn\mcdot p$ we
use $\overline y=\bn\mcdot p/m_b$. A comparison of the phase space with the
variables $\{y_H=q^2/m_B^2, \sqrt{s_H}=m_X/m_B\}$ and $\{u_H,\overline y_H\}$ is
shown in Fig.~\ref{fig:phase} and corresponds to the limits shown in columns iv)
and v) of Tables~\ref{table_limits} and \ref{table_limits2}.  The figure on the
right is the analog of Fig.~1 in Ref.~\cite{Bosch:2004bt} with dimensionless
variables.

For a strict SCET expansion we want 
\begin{eqnarray} \label{endpt}
  \frac{n\mcdot p_X}{ \bn\mcdot p_X} \le \lambda_H^2 \ll 1 \,,
\end{eqnarray}
where the expansion is in the parameter $\lambda_H$. Eq.~(\ref{endpt}) is
equivalent to $u_H/ \overline y_H\le \lambda_H^2$.  For $B\to X_s\gamma$,
Eqs.~(\ref{kin2}) and (\ref{endpt}) imply that the endpoint region is
\begin{eqnarray}
 E_\gamma \gtrsim (m_B/2 - \Lambda)\sim 2.1\,{\rm GeV} \,.
\end{eqnarray}
For $B\to X_u \ell\bar\nu$, satisfying the criterion in Eq.~(\ref{endpt}) with
values  $\lambda_H^2\simeq 0.2$ is equivalent to
\begin{eqnarray} \label{endpt2}
  \frac{m_X}{2E_X}\lesssim \frac{\lambda_H}{1+\lambda_H^2} = 0.37 \,,
\end{eqnarray}
or $y_H \lesssim 1 - {2.68} \sqrt{s_H} + s_H $.  Not exceeding a given expansion
parameter in Eq.~(\ref{endpt}) corresponds to specifying a triangular region of
phase space (shown for $u_H/\overline y_H\le 0.2$ in Fig.~\ref{fig:phase}).  We
refer to this as the SCET region of phase space.\footnote{Note that the SCET
  expansion here is actually in powers of $\lambda^2$ since odd terms in \SCETa
  tend to be absent~\cite{Beneke:2003pa}.  In Sec.~\ref{sect_NLOa} we show
  that the ${\cal O}(\lambda)$ contributions for inclusive decays indeed vanish.
  Thus $\lambda^2=0.2$ is not a large expansion parameter. A smaller value
  $\lambda_H^2=0.1$ could be chosen if desired. } As can be seen from
Fig.~\ref{fig:phase}, the simpler restriction $y_H \le 1 - 2.44 \sqrt{s_H}$ 
gives a very good approximation to the SCET region since the boundary is roughly
linear in the $q^2$ and $m_X$ variables.

In calculating decay rates at subleading order, it is important to define
carefully how the phase-space integrals are treated once we compute doubly
differential or singly differential decay rates.  The philosophy we adopt for
$B\to X_u\ell\bar\nu$ is that we use SCET to compute the $W_i$, and hence the
triply differential decay rate, for the SCET region in Eq.~(\ref{endpt}).
In general, one may wish to integrate this rate over a larger region of phase
space, and thus need to construct the full $W_i$'s. This could be done using
\begin{eqnarray} \label{combine}
  W_i^{\rm full} &=&  W_i^{\rm SCET} \theta( \lambda_H^2 \overline y_H - u_H)
  + W_i^{\rm OPE} \theta(u_H - \lambda_H^2 \overline y_H) \,,
\end{eqnarray}
where the SCET expansion in $\lambda$ is used for the first term and the
standard local OPE in $\Lambda/m_b$ is used for the second term. Thus,
$\lambda_H$ does not play the role of a strict expansion parameter, but rather
provides us with a means of interpolating any differential spectrum between the
full OPE and full SCET results by varying $\lambda_H$ between $0$ and $1$. We
only consider the first term in Eq.~(\ref{combine}) here. Depending on the final
spectrum that one looks at and the other cuts imposed, the error in including a
larger region of phase space than the SCET region may be power suppressed.  The
parameter $\lambda_H$ provides us with a way of testing this by considering the
difference between taking $\lambda_H=0.2$ and $\lambda_H=1$.  We present our
final results in a manner that makes it easy to take the $\lambda_H\to 1$ limit
for situations where a large enough region has been smeared over  that this
is the case.

One can also refer to a shape-function region, corresponding to the region $0\le
u_H \lesssim 0.1$ where the non-perturbative function $f$ is important. The
expansion for $B\to X_s\gamma$ to second order in $\lambda$, where
$1-x_H^\gamma\sim \lambda^2$, is
\begin{eqnarray} \label{dGamma3t}
\frac{d\Gamma^s}{dx_H^\gamma} 
    &=& \Gamma_0^s\: \frac{2}{m_B}  
     \bigg[ \Big\{ 4 W_1^s -W_2^s -m_B\, W_5^s \Big\}
   - (1\!-\!x_H^\gamma) \Big\{   
   4 W_1^{s} \!-\! W_2^{s} \!-\!2 m_B\, W_5^{s} \Big\} \bigg] 
    \,.
\end{eqnarray}
For $B\to X_u\ell\bar\nu$, if we integrate over all $x_H$ and expand in $u_H\sim
\lambda^2$, then from Eq.~(\ref{dGamma4u}) the first two orders in the
expansion are
\begin{eqnarray} \label{dGamma5u}
  \frac{1}{\Gamma_0^u}  \frac{d^2\Gamma^u}{  d\overline y_H\, du_H}  
    &=& {24m_B} \bigg[
   \Big\{ \overline y_H^{\,2}(1\!-\! \overline y_H) W_1^{u} +
    \frac{\overline y_H^{\,4}}{12}\, W_2^{u} \Big\}
  + u_H \Big\{ 
    \overline y_H (\overline y_H^{\,2}\!+\!\overline y_H\!-\!2) W_1^{u}
   -  \frac{\overline y_H^{\,3}}{3}    W_2^{u} 
  \Big\}  \bigg]
     \,. \nn\\
\end{eqnarray}
By the endpoint region in $x_H$ we mean $x_H^c \le x_H\le 1-r_\pi^2$,
where $1-x_H^c\sim 0.1$ corresponds to making a cut on the lepton's energy
spectrum.  The limit on $u_H$ forces it to be small, $u_H\le 1-x_H^c$, so 
shape-function effects are important here. This cut still allows a large range for
$\overline y_H$. Expanding the triply differential rate in Eq.~(\ref{dGamma3u})
and keeping the first two orders in the expansion for $1-x_H\sim u_H\sim
\lambda^2$ gives
\begin{eqnarray} \label{dGamma6u}
 \frac{1}{\Gamma_0^u} 
  \frac{d^3\Gamma^u}{ dx_H^{cut}\, d\overline y_H\, du_H}  
    &=& {24m_B} \bigg[
   \Big\{ \overline y_H (1\!-\! \overline y_H) 
  \big( W_1^{u} + \frac{m_B\overline y_H}{2} W_3^{u}\big) \Big\} \\
 && \hspace{-0cm} 
   - u_H\Big\{   (1\!-\!\overline y_H^{\,2})W_1^{u} + 
   \frac{m_B}{2} \overline y_H (2\!-\!\overline y_H\!-\!\overline y_H^{\,2})
    W_3^{u} \Big\} \nn\\
 && \hspace{-0cm} 
    + (1\!-\! x_H \!-\! u_H)\Big\{ \frac{\overline y_H^{\,2}}{2}\, W_2^{u}
   -{m_B} \overline y_H (1\!-\! \overline y_H)  W_3^{u} \Big\}
   \bigg]
    \,. \nn
\end{eqnarray}
  Finally, to consider the
$d\Gamma/(dq^2dm_X^2)$ spectrum in the endpoint region we let $\zeta=1-y_H+s_H$
and expand in $s_H/\zeta^2$, which to linear order gives
\begin{eqnarray} \label{dGamma5sy}
  \frac{1}{\Gamma_0^u}  \frac{d^2\Gamma^u}{  dy_H\, ds_H}  
    &=& {2m_B} \bigg[ 
   \Big\{ 12 \zeta (1\!-\! \zeta) W_1^{u} + \zeta^3 W_2^{u} \Big\}
  + \frac{6 s_H}{\zeta} \Big\{ 2 (\zeta^2\!+\!2\zeta\!-\!2) W_1^u - \zeta^2 W_2^u
  \Big\}  \bigg]
     \,. \nn\\
\end{eqnarray}
In each of 
Eqs.~(\ref{dGamma3t})--(\ref{dGamma5sy})
there will also be an expansion of the $W_i$ themselves, which we discuss later
on.

The results in this section can easily be extended to any desired order in
$\Lambda/m_B$ by expanding to higher order in the phase space.

\subsection{OPE and Partonic Variables}   \label{sect_partonic}

We have not yet made use of quark-hadron duality or formulated
the method for computing the $W_i$. The usual procedure to compute the $W_i$ is
to use an operator product expansion and calculate the forward scattering
amplitude
\begin{eqnarray} \label{T}
  T_{\mu\nu} &=& \frac{1}{2 m_B}  \langle \bar B| \hat T_{\mu\nu}
  |\bar B\rangle \\
   &=& - g_{\mu\nu} T_1 + v_\mu v_\nu T_2 + i\epsilon_{\mu\nu\alpha\beta}
  v^\alpha q^\beta T_3 + q_\mu q_\nu T_4 + (v_\mu q_\nu + v_\nu q_\mu) T_5 
  \,,\nn
\end{eqnarray}
where
\begin{eqnarray} \label{T2}
 \hat T_{\mu\nu} &=& -i \int\!\! d^4x\, e^{-iq\cdot x}\, 
   T J_\mu^\dagger(x) J_\nu(0) \,,
\end{eqnarray} 
and for the $T_i^{(s,u)}(q^2,v\mcdot q)$ we use the corresponding hadronic
currents, which are
\begin{eqnarray}
 J_\mu^s &=& \bar s\,i\sigma_{\mu\nu} q^\nu P_R\, b 
 \,,
 \qquad\qquad
 J_\mu^u = \bar u\, \gamma^\mu P_L b \,.
%
%
\end{eqnarray} 
Here $J_\mu^s$ comes from the operator ${\cal O}_7$.  
The operator product expansion relates the $W_i^{(u,s)}$ to the forward
scattering amplitudes through
\begin{eqnarray} \label{WT}
  W_i = -\frac{1}{\pi} \:{\rm Im}\, T_i \,.
\end{eqnarray}
When we compute the $W_i$ with an OPE, the partonic variables depending on $m_b$
and the hadronic variables involving $m_B$ will need to be related order by
order in the $1/m_b$ expansion. In particular, the heavy meson mass to second
order is
\begin{eqnarray}
  m_B = m_b + \overline \Lambda - \frac{\lambda_1}{2m_b} -
  \frac{3c_F(\mu)\lambda_2(\mu)}{2 m_b} + \ldots \,,
\end{eqnarray}
where we shall also use $\lambda_2=c_F(\mu) \lambda_2(\mu)$ as a definition of
the non-perturbative matrix element that has the perturbative coefficient
$c_F(\mu)$ absorbed.  

In applying the local OPE, part of the expansion involved in switching to
hadronic variables occurs because the phase-space limits are partonic. In fact,
if we calculate the triply differential rate for $B\to X_u\ell\bar\nu$ with the
local OPE, and then consider integrating it over the hadronic phase space, then
the integrand has support over the partonic phase space only, so the limits are
reduced to this more restricted case. In the local OPE the signal for this is
the occurrence of factors like
\begin{eqnarray} \label{deltaOPE}
   \delta^{(n)}[(m_b v-q)^2 ] \,,
\end{eqnarray} 
which must be smeared sufficiently so that quark-hadron duality can be used.
For the $\delta[(m_b v-q)^2 ]$ that occurs at LO, integrating once to get the
doubly differential rate gives a theta function that imposes partonic limits.

On the other hand, with the SCET expansion in the endpoint region the support of
the triply differential rate is larger. We never encounter singular
distributions like the one in Eq.~(\ref{deltaOPE}), but instead obtain a
non-trivial forward $B$-hadronic matrix element that gives $f^{(0)}(\ell^+)$.
This function knows about the difference between the hadronic and partonic phase
space already at leading order in the power counting, and more generally the LO
factorization result with ${\cal O}(\alpha_s)$ corrections (cf.
Eq.~(\ref{Wfact00}) below) does not cause a restriction of the hadronic phase
space.  Therefore, we shall use the full hadronic phase-space limits in our
computation.\footnote{Note that this might imply that direct calculations of
  less differential subleading spectra must be treated with care. For example,
  if one directly computes a singly differential rate by tying up lepton lines
  then the partonic phase-space restrictions might appear to creep back in if
  one is not sufficiently careful about the structure of the factorization
  theorem.}

\OMIT{
It is useful to define a few partonic variables, namely 
\begin{eqnarray}
  x^\gamma = \frac{2E_\gamma}{m_b} \,,\qquad\quad
  x = \frac{2E_\ell}{m_b} \,,\qquad\quad 
  \overline y \equiv \frac{\bn\mcdot p}{m_b}  \,.
\end{eqnarray}
The variable $\overline y$ will appear in the calculation of hard coefficients
in Secs.~\ref{LOfact} and~\ref{sect_Summary}.
}

\section{SCET ingredients} \label{sect_J}



To derive the expansion of the $T_i$ in the endpoint region, we determine the
SCET currents and Lagrangians in a power expansion in $\lambda$, and use this
expansion to separate the collinear jet-like effects from the non-perturbative
ultrasoft shape functions.  This will allow us to determine results for the
$T_i$ order by order in the expansion. We write
\begin{eqnarray}
   T_i &=& T_i^{(0)} + T_i^{(1)} +  T_i^{(2)} + \ldots \,.
\end{eqnarray}
The ingredients required for this include expansions of the 
Lagrangians and currents to second order in $\lambda$,
\begin{eqnarray}
 {\cal L} = {\cal L}^{(0)} +  {\cal L}^{(1)} +  {\cal L}^{(2)} \,,\qquad
 J = J^{(0)} + J^{(1)} + J^{(2)} \,,
\end{eqnarray}
and are described in this section.

The LO Lagrangian for usoft light quarks and usoft gluons, 
${\cal L}_{us}^{(0)}$, is identical to full QCD. For usoft heavy quarks 
we have the HQET Lagrangian, which has an expansion in 
$\Lambda/m_b\sim \lambda^2$, namely
\begin{eqnarray}
  {\cal L}_h^{(0)} &=& \bar h_v iv\mcdot D_{us} h_v \,, \\
  {\cal L}_h^{(2)} &=& \frac{1}{2m_b}\, O_h \,,\qquad
   O_h = \bar h_v (i D_T)^2 h_v 
   + \frac{1}{2}\, c_F(\mu)\, \bar h_v \sigma_{\mu\nu}\, gG^{\mu\nu}_{us}\, h_v 
   \,, \nn
\end{eqnarray}
where $c_F(\mu)$ is the Wilson coefficient of the chromomagnetic operator.
At leading order, the SCET$_{\rm I}$ Lagrangian for collinear
quarks is~\cite{Bauer:2000yr,Bauer:2001ct}
\begin{eqnarray}\label{L0}
{\cal L}_{\xi\xi}^{(0)} 
&=& \bar \xi_n \left[ i n\mcdot D
+ i \Dslash^\perp_c  W \frac{1}{\bnP} W^\dagger 
  i \Dslash^\perp_c\right] \frac{\bnslash}{2}\, \xi_n  \,,
\end{eqnarray}
where the collinear covariant derivative is $iD_c^\mu = {\cal P}^\mu +
gA_n^\mu$, with label operator $\cP^\mu$, the full derivative is $in\mcdot
D=in\mcdot\partial + g n\mcdot A_{us} + gn\mcdot A_n$, and the Wilson line $W$
is built out of $\bn\mcdot A_n$ fields: 
\begin{eqnarray} \label{W}
   W =  \Big[ \sum_{\rm perms} \exp\Big( -\frac{g}{\bnP}\: 
  \bn\mcdot  A_{n,q}(x) \ \Big) \Big] \,.
\end{eqnarray}
The collinear covariant derivative satisfies 
$f(i\bn\mcdot D_c) = W f(\bnP) W^\dagger$. 
In the leading-order collinear Lagrangians for both quarks and gluons, ${\cal
  L}_{\xi\xi}^{(0)}+{\cal L}_{cg}^{(0)}$, we can factorize the usoft gluons with
the field redefinitions~\cite{Bauer:2001yt}
\begin{eqnarray}  \label{fr}
  && \xi_n \to Y\xi_n \,,\qquad A_n^\mu \to Y A_n^\mu Y^\dagger \,,\\
  && Y = Y(x) = P\exp\Big(ig \int_{-\infty}^0\!\!\!\! ds\:
     n\mcdot A_{us}(x\!+\! ns)\Big) \,, \nn 
\end{eqnarray}
where the new collinear fields no longer transform under usoft gauge
transformations. For convenience, we also define
\begin{eqnarray} \label{calB}
  {\cal H}_v &=& Y^\dagger h_v\,,\qquad
   \psi_{us} = Y^\dagger q_{us} \,, \qquad\qquad
    {\cal D}_{us} = Y^\dagger D_{us} Y 
  \nn\\
    \chi_n &=& W^\dagger \xi_n \,, \qquad
  {\cal D}_c = W^\dagger D_c W \,,\qquad
  ig {\cal B}_c^\mu
   = \Big[\frac{1}{\bnP} W^\dagger  [i\bn\mcdot D_c , iD_c^\mu] W\Big] \,.\qquad
\end{eqnarray} 
The leading-order collinear Lagrangians become
\begin{eqnarray} \label{L00}
{\cal L}_{\xi\xi}^{(0)} &=& \bar \chi_n \left[ i n\mcdot {\cal D}_c
  + i \cDslash^\perp_c   \frac{1}{\bnP} 
  i \cDslash^\perp_c\right] \frac{\bnslash}{2} \: \chi_n  \,, \qquad
{\cal L}_{cg}^{(0)}
   = -\frac{1}{2} {\rm tr}\big[ G_c^{\mu\nu} G^c_{\mu\nu} \big] + \mbox{g.f.}
  \,,
\end{eqnarray}
where $ig G_c^{\mu\nu}= [iD_c^\mu, i D_c^\nu]$ and g.f. stands for 
gauge-fixing terms.
Thus the field redefinition removes the coupling of $n\mcdot A_{us}$ gluons to
collinear fields at LO. At subleading order it organizes the usoft gluons into
gauge invariant blocks, like $(Y^\dagger \ldots Y)$, which are inserted only a 
finite number of times. The subleading Lagrangians that we require are all
available in the literature~\cite{Beneke:2002ph,Pirjol:2002km}. After one makes
the above field redefinition, the subleading quark Lagrangians we require
are~\cite{Bauer:2003mg}
\begin{eqnarray} \label{Lxxnew}
{\cal L}_{\xi\xi}^{(1)} 
  &=&  \bar \chi_n\,  i\cDslash^\perp_{us}\: 
  \frac{1}{\bnP}\,  i \cDslash^\perp_c\:  \frac{\bnslash}{2} \,\chi_n 
  + {\rm h.c.}\,, \\
{\cal L}_{\xi q}^{(1)} 
 &=&  \bar\chi_n \:   ig\Bcslash_c^\perp
  \psi_{us} \mbox{ + h.c.} 
  \,,\nn\\
  {\cal L}_{\xi\xi}^{(2a)} &=&   \bar \chi_n \:  
 i\cDslash_{us}^\perp\,  i\! \cDslash_{us}^\perp \: \frac{1}{\bnP}
  \frac{\bnslash}{2}\, \chi_n 
  \,,\nn\\
 {\cal L}_{\xi\xi}^{(2b)} &=&   
    \bar \chi_n i\cDslash^\perp_c  i\bn\mcdot {\cal D}_{us} 
  \frac{1}{\bnP^2}\, \frac{\bnslash}{2}\:  i\cDslash^\perp_c \: \chi_n
   \,. \nn
\end{eqnarray} 
We also need the subleading terms in the mixed usoft-collinear gluon action,
\begin{eqnarray}
 {\cal L}_{cg}^{(1)} &=& 
  - {2}\: {\rm tr} \big[ {\cal G}_c^{\mu\nu} {\cal H}_{\mu}^{\:\,\tau}
     \big]  g^\perp_{\nu\tau}  + \mbox{g.f.}\,,\\
 {\cal L}_{cg}^{(2a)} &=& 
     {\rm tr} \big[ {\cal H}^{\mu\nu} {\cal H}^{\sigma\tau}
     \big]  g^\perp_{\mu\tau}\: g^\perp_{\nu\sigma}
    -  {\rm tr} \big[ {\cal H}^{\mu\nu} {\cal H}_{\mu}^{\:\,\tau}
     \big]  g^\perp_{\nu\tau}
    -  {\rm tr} \big[ {\cal G}_c^{\mu\nu} {\cal G}_{u}^{\sigma\tau}
     \big]  g^\perp_{\mu\sigma}\: g^\perp_{\nu\tau} + \mbox{g.f.}\,,\nn\\
  {\cal L}_{cg}^{(2b)} &=& 
    - {\rm tr} \big[ {\cal G}_c^{\mu\nu} {\cal H}_{\mu}^{\:\,\tau}
     \big] n_\nu \bn_\tau
   + \mbox{g.f.}\,,\nn 
\end{eqnarray}
where 
\begin{eqnarray}
&&  ig\, {\cal G}_c^{\mu\nu}  = W^\dagger [i D_c^\mu, iD_c^\nu ] W \,,\qquad
  ig\, {\cal G}_u^{\mu\nu}    = Y^\dagger [i D_{us}^\mu, iD_{us}^\nu ] Y \,,
  \nn\\
&&   ig\, {\cal H}^{\mu\nu} = [ W^\dagger iD_c^\mu W, Y^\dagger i D_{us}^\nu Y]
  \,,
\end{eqnarray}
and g.f. denotes gauge-fixing terms that are required by reparameterization
invariance.

Next consider the heavy-to-light currents. For simplicity, we work in a
frame where $v_\perp=0$. Expanding the $J^u$ current to ${\cal O}(\lambda^2)$
gives
\begin{eqnarray}\label{Jeff0}
 J^{u} &=& e^{i( \bnP \frac{n}{2}+\ppP - m_b v)\cdot x }\, \bigg\{ 
 \sum_{j} \! \int\!\! d\omega  C_j^{(v)}(\omega) J^{u(0)}_{j}(\omega) 
 + \sum_{j} \! \int\!\! d\omega  B_j^{(v)}(\omega) J^{u(1)}_{j}(\omega) \nn\\
 && \hspace{3cm}
  + \sum_{j} \! \int\!\! d\omega  A_j^{(v)}(\omega) J^{u(2)}_{j}(\omega)\bigg\}
  \,, 
\end{eqnarray}
where the superscript $(0)$, $(1)$, $(2)$ indicates the order in $\lambda$. We
have an analogous result for $J^s$ with $(v)\to (t)$ and $u\to s$. After 
one makes the field redefinition in Eq.~(\ref{fr}), the leading-order SCET 
heavy-to-light current is~\cite{Bauer:2000yr}
\begin{eqnarray} \label{J0}
 J^{(0)}_{j}(\omega) &=&  (\bar\xi_n W)_{\omega} \Gamma_j^{(0)} 
  (Y^\dagger h_v) 
  = \bar\chi_{n,\omega} \Gamma_j^{(0)} {\cal H}_v \,,
\end{eqnarray}
where $j=1$-$3$ for $u$ and $j=1$-$4$ for $s$. Despite the fact that the
operator in $J^{(0)}$ is ${\cal O}(\lambda^4)$, we use the superscript $(0)$, to
indicate that it is the lowest-order current.  At the first subleading order, we
have the ${\cal O}(\lambda^5)$
currents~\cite{Chay:2002vy,Beneke:2002ph,Bauer:2002aj,Pirjol:2002km}
\begin{eqnarray} \label{J1}
  J^{(1a)}_j(\omega) &=&  \frac{1}{\omega} \big( \bar\chi_n\, i\!\! \cDgppPl  
    \big)_{\omega} 
    \Upsilon_j^{(1a)} \: {\cal H}_v \,, \\
  J^{(1b)}_j(\omega_1,\omega_2) 
    &=&  \frac{1}{m_b\: n\mcdot v} \: \bar\chi_{n,\omega_1} 
    \, (ig {\cal B}_{c\perp}^\alpha)_{\omega_2}\,
    \Upsilon_j^{(1b)}
    \: {\cal H}_v \,. \nn 
\end{eqnarray}
The subscript $\omega$ notation in Eqs.~(\ref{J0}) and (\ref{J1}) indicates that
the field carries momentum $\omega$, for example
\begin{eqnarray}
  \bar\chi_{n,\omega} = \bar\chi_n\: \delta(\omega-\bnP^\dagger) \,,\qquad
  (ig {\cal B}_{c\perp}^\alpha)_{\omega_2} = (ig {\cal B}_{c\perp}^\alpha)
    \: \delta(\omega_2-\bnP^\dagger) \,.
\end{eqnarray}

Finally, at second order we find that the basis of ${\cal O}(\lambda^6)$ 
currents consists of
\begin{eqnarray} \label{J2}
 J^{(2a)}_j(\omega) &=&  \frac{1}{2m_b}\: \bar\chi_{n,\omega} \,
   \Upsilon^{(2a)}_{j\,\alpha}  
    {\cDpusRight{\alpha} } \, {\cal H}_v  
   \,, \nn\\
 J^{(2b)}_j(\omega) &=& \frac{1}{\omega}\: \bar\chi_{n,\omega} \,
   \Upsilon^{(2b)}_{j\,\alpha}  \,
    \cDpusLeft{\alpha}\, {\cal H}_v \,,\\  
 J^{(2c)}_j(\omega) &=& - \Big(\bar\chi_n  \frac{\bn\mcdot v}{\bnP}\, 
  \frac{ ig n\mcdot {\cal B}_c}{n\mcdot v} \Big)_{\omega} 
   \Upsilon^{(2c)}_{j}\: {\cal H}_v  
    \,, \nn\\
 J^{(2d)}_j(\omega_1,\omega_2) &=&  \frac{-1}{m_b} 
   \bar\chi_{n,\omega_1}\, 
   \Big(\frac{ig n\mcdot {\cal B}_c}{n\mcdot v} \Big)_{\omega_2}
   \Upsilon^{(2d)}_{j}\:  {\cal H}_v  \,,
   \nn\\
 J^{(2e)}_j(\omega_1,\omega_2) &=& \frac{-1}{m_b\, n\mcdot v\,\omega_1} 
   \big( \bar\chi_n i\! \cDgppPl \big)_{\omega_1}
   ( ig {\cal B}_{c\perp}^\beta)_{\omega_2} \Upsilon_{j\alpha\beta}^{(2e)}\:
   {\cal H}_v \,, \nn\\
 J^{(2f)}_j(\omega_1,\omega_2) &=&  \frac{-1}{m_b\,} 
   \bar\chi_{n,\omega_1}
    \Big(\frac{1}{ n\mcdot v\,\bnP}  i {\cal D}_{c\perp}^{\alpha} 
   \: ig {\cal B}_{c\perp}^{\beta} \Big)_{\omega_2} 
   \Upsilon^{(2f)}_{j\,\alpha\beta}\:
   {\cal H}_v \,.
   \nn
\end{eqnarray}
These currents were first derived in Ref.~\cite{Beneke:2002ni} (see also
\cite{Beneke:2002ph}) and we agree with their results. Two differences are that
in Eq.~(\ref{J2}) we have determined the functional dependence and the number of
parameters $\omega_i$, and that terms from the separation of label and residual
momenta (or equivalently the multipole expansion of the fields) are treated
differently here from Ref.~\cite{Beneke:2002ni}. In Appendix~\ref{app_J} we give
a more detailed comparison, along with the details of our calculation.

For the LO Dirac structures we use the basis from
Refs.~\cite{Chay:2002vy,Pirjol:2002km}, which is the most convenient for
considering reparameterization invariance, i.e.\
\begin{eqnarray} \label{Jeff3}
 \Gamma_{1-3}^{u(0)} &=&  P_R \Big\{ \gamma^\mu\,, v^\mu \,, 
   \frac{n^\mu}{n\mcdot v} \Big\}\,,\qquad
 \Gamma_{1-4}^{s(0)} =  P_R\  q_\tau\: \Big\{ i\sigma^{\mu\tau} \,, 
   \gamma^{[\mu,} v^{\tau]}\,, \frac{\gamma^{[\mu,} n^{\tau]} }{n\mcdot v} \,,  
   \frac{n^{[\mu,}v^{\tau]}}{n\mcdot v}  \Big\} \,.
\end{eqnarray}
Here we slightly reorganize the basis in Ref.~\cite{Pirjol:2002km} to 
reflect the constraints from RPI and to use subscripts $j>11$ for the 
currents that vanish when $v_\perp=0$. This gives
\begin{eqnarray} \label{Upsilon1}
  \Upsilon_{1-3}^{u(1a)} \!&=&\! 
    P_R \Big\{ \gamma^\alpha_\perp \frac{\bnslash}{2} \gamma^\mu 
     ,  \gamma^\alpha_\perp \frac{\bnslash}{2}  v^\mu
     ,  \gamma^\alpha_\perp \frac{\bnslash}{2}  n^\mu 
    \!-\!2 g^{\alpha\mu}_\perp \Big\} \,, \ \
  \Upsilon_{4-7}^{u(1b)} \!=\!\!
    P_R \Big\{  \gamma^\mu \gamma^\alpha_\perp
    ,   v^\mu \gamma^\alpha_\perp\: , n^\mu \gamma^\alpha_\perp
    ,  g_\perp^{\mu\alpha} \Big\} \,, \nn\\
  \Upsilon_{1-4}^{s(1a)} \!&=&\!\! 
    P_R \Big\{i\gamma^\alpha_\perp \frac{\bnslash}{2} \sigma^{\mu\nu}\,,\, 
    \gamma^\alpha_\perp \frac{\bnslash}{2}
     \gamma^{\mbox{\tiny $[$}\mu,} v^{\nu\mbox{\tiny$]$}}\,,\, 
    \gamma^\alpha_\perp \frac{\bnslash}{2}
     \gamma^{\mbox{\tiny$[$}\mu,} n^{\nu\mbox{\tiny$]$}}
    +2 g_\perp^{\alpha \mbox{\tiny$[$}\mu\,,} \gamma^{\nu\mbox{\tiny$]$}}\,,\,
    \gamma^\alpha_\perp \frac{\bnslash}{2}
     n^{\mbox{\tiny$[$}\mu,} v^{\nu\mbox{\tiny$]$}}
    -2 g_\perp^{\alpha\mbox{\tiny$[$}\mu\,,} v^{\nu\mbox{\tiny$]$}} 
     \Big\}\,, \nn \\
  \Upsilon_{5-11}^{s(1b)} \!&=&\!\!
    P_R \Big\{ i \sigma^{\mu\nu}\gamma^\alpha_\perp   \,,\,
    \gamma^{\mbox{\tiny $[$}\mu,} v^{\nu\mbox{\tiny$]$}}\gamma^\alpha_\perp\,,\, 
    \gamma^{\mbox{\tiny$[$}\mu,} n^{\nu\mbox{\tiny$]$}}\gamma^\alpha_\perp\,,\,
    n^{\mbox{\tiny$[$}\mu,} v^{\nu\mbox{\tiny$]$}} \gamma^\alpha_\perp\,,\,
    g_\perp^{\alpha \mbox{\tiny$[$}\mu\,,} \gamma^{\nu\mbox{\tiny$]$}} \,,\,
    g_\perp^{\alpha\mbox{\tiny$[$}\mu\,,} v^{\nu\mbox{\tiny$]$}}  \,, 
    g_\perp^{\alpha\mbox{\tiny$[$}\mu\,,} {n^{\nu\mbox{\tiny$]$}}}
    \bigg\}\,. 
\end{eqnarray}
Note that we do not list evanescent Dirac structures that can become necessary
when computing perturbative corrections in dimensional regularization (see for
example Refs.~\cite{Beneke:2004rc,Becher:2004kk}).  For our purposes, the Dirac
structures for the $J^{(2)}$ currents that appear at lowest order in
$\alpha_s(m_b)$ will suffice; these are
\begin{eqnarray} \label{Upsilon2}
    \Upsilon_{1}^{u(2a)} \!\! &=& \!\! 
    P_R\: \gamma^\mu  \gamma^\alpha_T
     \,, \ \
 \Upsilon_{1}^{u(2b)} \!=\!
    P_R\: \gamma^\alpha_\perp \frac{\bnslash}{2} \gamma^\mu 
     \,, \ \ 
  \Upsilon_{1}^{u(2c)} \!=\! 
    P_R\:  \gamma^\mu 
     \,, \ \quad 
   \Upsilon_{1}^{u(2d)} \!=\! 
    P_R\: \gamma^\mu    \frac{\nslash\bnslash}{4}
     \,, \ \nn\\
  \Upsilon_{1}^{u(2e)} \!\! &=& \!\! 
    P_R\:  \gamma^\perp_\alpha \frac{\bnslash}{2} \gamma^\mu \frac{\nslash}{2}
     \gamma^\perp_\beta
     \,, \ \qquad
  \Upsilon_{1}^{u(2f)} \! = \! 
    P_R\:  \gamma^\mu \gamma^\perp_\alpha \gamma^\perp_\beta
     \,, \nn\\
 \Upsilon_{1}^{s(2a)} \!\!&=&\!\! 
    P_R\: i\sigma^{\mu\tau} \gamma^\alpha_T q_\tau
     \,, \ \ 
\Upsilon_{1}^{s(2b)} \!=\! 
    P_R\: \gamma^\alpha_\perp \frac{\bnslash}{2}\, i\sigma^{\mu\tau} q_\tau
     \,, \ 
\Upsilon_{1}^{s(2c)} \!=\! 
    P_R\: i\sigma^{\mu\tau} q_\tau
     \,, \ \ 
 \Upsilon_{1}^{s(2d)} \!=\! 
    P_R\: i\sigma^{\mu\tau} q_\tau \frac{\nslash\bnslash}{4}
     \,, \ \nn\\
 \Upsilon_{1}^{s(2e)} \!& =& \! 
    P_R\:  \gamma^\perp_\alpha \frac{\bnslash}{2} i\sigma^{\mu\tau}q_\tau
     \frac{\nslash}{2} \gamma^\perp_\beta \,,\ \quad 
  \Upsilon_{1}^{s(2f)} \!=\! 
    P_R\:  i\sigma^{\mu\tau}q_\tau \gamma^\perp_\alpha \gamma^\perp_\beta
  \,.
\end{eqnarray}
The complete tree-level currents for arbitrary $v_\perp$ are given in
Appendix~\ref{app_J}, and in Appendix~\ref{app_RPI} we show how the $v_\perp$
dependence is necessary to satisfy the full constraints from RPI at this order.
When the ${\cal O}(\lambda^2)$ currents that show up at order $\alpha_s(m_b)$ 
are determined, the number of possible $\Upsilon_j^{(2)}$ structures will
increase, as occurred in Eq.~(\ref{Upsilon1}).  However, this will not affect 
the form of the factorization theorems we derive, since they are expressed in 
terms of traces of the $\Upsilon_j^{(2)}$ factors; this makes it trivial to
incorporate new Dirac structures that arise beyond tree level.

At lowest order in $\alpha_s(m_b)$, the non-zero coefficients in 
Eq.~(\ref{Jeff0}) are
\begin{eqnarray}
  C_1^{(v)}=C_1^{(t)}= B_1^{(v)} = B_{1,7}^{(t)} = -B_6^{(v)} = 1 
 \,.
\end{eqnarray}
The one-loop results and RGE-improved coefficients $C_i$
can be found in Ref.~\cite{Bauer:2000yr}. Our coefficients are linear
combinations of these:
\begin{eqnarray} \label{matchC}
 C_{1}^{(v)}(\hat\omega,1) 
 &=& 1 - \frac{\alpha_s(m_b)C_F}{4\pi} \bigg\{ 2\!\ln^2(\hat\omega) 
  + 2 {\rm Li}_2(1\!-\!\hat\omega) 
  + \ln(\hat\omega) \Big( \frac{3\hat\omega-2}{1-\hat\omega}\Big)
  + \frac{\pi^2}{12} + 6\bigg\} , \nn \\*
C_{2}^{(v)} (\hat\omega,1) 
 &=& \frac{\alpha_s(m_b)C_F}{4\pi}\:  \bigg\{ \frac{2}{(1-\hat\omega)} 
  + \frac{2\hat\omega\ln(\hat\omega)} {(1-\hat\omega)^2} \bigg\} \,, \nn \\
C_{3}^{(v)}(\hat\omega,1) 
 &=& \frac{\alpha_s(m_b)C_F}{4\pi} \bigg\{ 
  \frac{(1-2\hat\omega)\hat\omega \ln(\hat\omega) }{(1-\hat\omega)^2}
  - \frac{\hat\omega}{1-\hat\omega} \bigg\} 
 \,, \nn \\*
C_{1}^{(t)}(\hat\omega,1) 
 &=& 1 + \Delta_\gamma(m_b,\rho) 
  - \frac{\alpha_s(m_b)C_F}{4\pi} \bigg\{ 2\!\ln^2(\hat\omega) 
  + 2 {\rm Li}_2(1\!-\!\hat\omega) 
  +  \ln(\hat\omega) \Big( \frac{4\hat\omega-2}{1-\hat\omega} \Big) 
  + \frac{\pi^2}{12} + 6 \bigg\} , \nn \\
C_{2}^{(t)}(\hat\omega,1)  
 &=& 0  \,, \quad 
C_{3}^{(t)}(\hat\omega,1) 
 = \frac{\alpha_s(m_b)C_F}{4\pi} \bigg\{ 
    \frac{-2\hat\omega \ln(\hat\omega)}{1-\hat\omega} \bigg\} 
 \,, \quad 
C_{4}^{(t)}(\hat\omega,1)  
 = 0 \,,
\end{eqnarray}
where $\hat\omega=\omega/m_b$ and $\Delta_\gamma(m_b,\rho)$ was given in
Eq.~(\ref{Deltas}).  Thus, non-zero values for $C_{2,3}^{(v)}$ and $C_3^{(t)}$
are generated at one-loop order, while $C_{2,4}^{(t)}$ are still zero at this
order.  For the $J^{(1b)}$ currents, the one-loop matching coefficients were
derived in Ref.~\cite{Beneke:2004rc}, and the anomalous dimensions and
RGE-improved coefficients were computed in Ref.~\cite{Hill:2004if}.

By reparameterization invariance $B_{1-3}^{(v)}(\omega)=C_{1-3}^{(v)}(\omega)$
and $B_{1-4}^{(t)} (\omega) = C_{1-4}^{(t)}(\omega)$, so their one-loop matching
coefficients are in Eq.~(\ref{matchC}). For the ${\cal O}(\lambda^2)$ current
coefficients the RPI constraints are
\begin{eqnarray}
  A_1^{(v)} = A_2^{(v)} = -\frac12 A_3^{(v)} = C_1^{(v)}(\omega) \,,\qquad
  A_4^{(v)} = A_5^{(v)} = A_6^{(v)} = B_7^{(v)}(\omega_1,\omega_2) \,,\nn\\
  A_1^{(t)} = A_2^{(t)} = -\frac12 A_3^{(t)} = C_1^{(t)}(\omega) \,,\qquad
  A_4^{(t)} = A_5^{(t)} = A_6^{(t)} = B_7^{(t)}(\omega_1,\omega_2) \,.
\end{eqnarray}

Note that we have not included effects associated with a non-zero strange-quark
mass in our basis of subleading SCET operators, and it would be interesting 
from a formal view-point to consider the possibility of such power-suppressed 
terms in the future (cf.~\cite{Leibovich:2003jd}).


\section{Leading-Order Factorization}   \label{LOfact}


In this section we review the leading-order factorization
theorem~\cite{Korchemsky:1994jb} as derived using SCET~\cite{Bauer:2001yt}. The
result is discussed in detail, which will allow us to refer back to this section
when some of the steps are repeated at NLO. Throughout this section we shall
suppress the $u$ and $s$ superscripts, since all the manipulations are identical
in both cases, $B\to X_u\ell\bar\nu$ and $B\to X_s\gamma$.

To derive the leading order $T_i^{(0)}$ we need only consider $\hat
T^{(0)}_{\mu\nu}$, given by taking two LO currents $J^{(0)}$, as shown in
Table~\ref{table_LOT}.
\begin{table} 
\[
\begin{array}{ccccccc}
\hbox{T-product} & & \hbox{Example Diagram} & \hspace{-0cm}
  \begin{array}{c}\mbox{Hard, Jet, and} \\ \mbox{Shape Functions}\end{array} 
  \hspace{-0cm} &  
  \hbox{Usoft Operator}   
  \\ \hline\\
 T^{(0)} 
  & & 
     \epsfxsize=6cm \lower25pt\hbox{\epsfbox{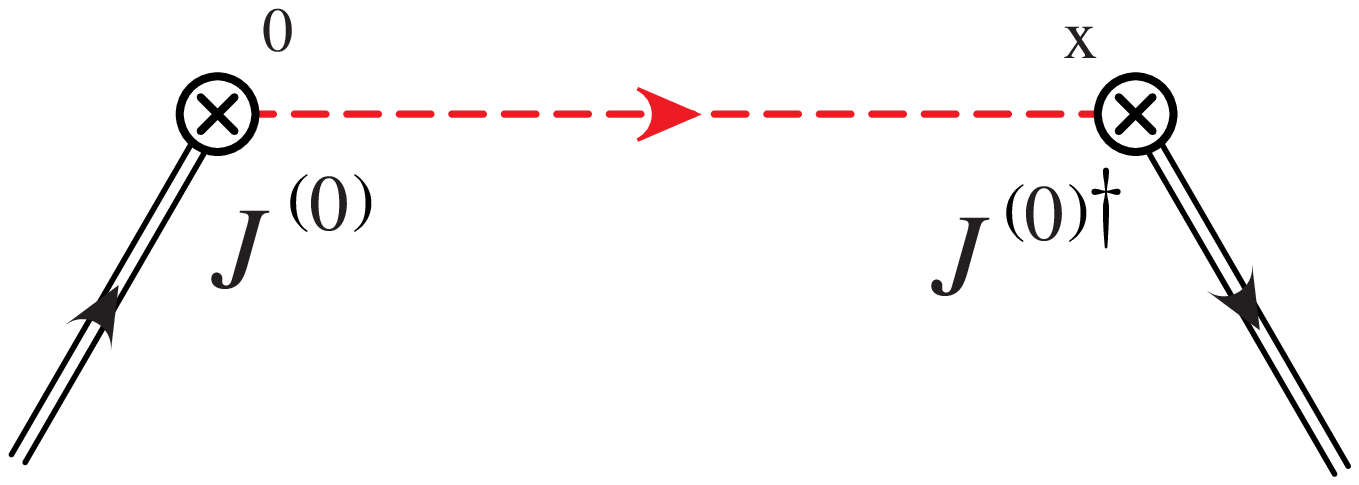}} &  
  h^{[0]}\, {\cal J}^{(0)}\, f^{(0)} 
  &  
  \bar h_v(x) h_v(0)
\end{array}
\]
\vskip -.1cm
\caption{
Lowest-order insertion of SCET currents. The double lines are heavy
  quarks and the dashed line is a collinear light quark.
 \label{table_LOT}} 
\end{table}
The phase factor in Eq.~(\ref{Jeff0}) gives  
\begin{eqnarray} \label{phase}
  e^{-i (q+ \frac{n}{2} \bnP - m_b v)\,\cdot\, x} 
    \equiv e^{-i r\,\cdot\, x} \, 
\end{eqnarray}
and $\exp(-i \ppP\cdot x)$. The current has $q_\perp=0$, so this $\ppP$ term
will contribute only to fixing the perpendicular momentum of the jet function to
be zero. The large label momentum in the phase in Eq.~(\ref{Jeff0}) also gets
fixed by momentum conservation: 
\begin{eqnarray}
 \bnP = m_b-\bn\cdot q = m_b-m_B + \bn\cdot p_X \equiv \bn\cdot p\,,
\end{eqnarray}
where $\bn\mcdot p_X$ is the large momentum in the jet $X$. Then the remaining
momentum $r^\mu\sim \Lambda$ since $\bn\cdot r = \bn\mcdot q +\bnP - m_b = 0$
and
\begin{eqnarray} \label{rdef}
   n\cdot r &=& n\mcdot q - m_b = m_B - m_b - n\mcdot p_X \,.
\end{eqnarray}
At lowest order
\begin{eqnarray} \label{rdef0}
 \bn\mcdot p = \bn\cdot p_X \,,\qquad\qquad
  n\cdot r =   \overline \Lambda - n\mcdot p_X  \,,
\end{eqnarray}
where both $\overline\Lambda, n\cdot p_X \sim \Lambda$ (and higher-order terms
in $m_B-m_b$ will be needed only when we go beyond LO). For the time being we
stick to the partonic variables $\bn\mcdot p$ and $n\mcdot r$; later, we shall 
perform the expansion involved in switching to hadronic variables. 
Using the states defined with HQET, we get
\begin{eqnarray} \label{Tlo1}
   W_{\mu\nu}^{(0)} &=&  \frac{(-1)}{\pi} {\rm Im}\: 
   \frac12 \langle \bar B_v | \hat T_{\mu\nu}^{(0)}| \bar B_v \rangle \,,\\
  \hat T_{\mu\nu}^{(0)} &=& -i\!
    \int\!\! d^4x\,   e^{-i r\cdot x}\  
   T\: J^{(0)\dagger}_{j'\,\mu}(x)\: J^{(0)}_{j\,\nu}(0) \,. \nn
\end{eqnarray}
Separating out the hard Wilson coefficients, we have
\begin{eqnarray} \label{LOhard}
 \hat T_{\mu\nu}^{(0)} \!\! &=&\!\! \sum_{j,j'} \! \int\!\! d\omega d\omega' \: 
   C_{j'}(\omega')  C_j(\omega) \delta(\omega'\!-\!\bn\mcdot p)(-i)\!
    \int\!\! d^4x\,   e^{-i r\cdot x}\  
   T\: J^{(0)\dagger}_{j'\,\mu}(\omega',x)\: J^{(0)}_{j\,\nu}(\omega,0) .\ \  
\end{eqnarray}
The effective-theory currents in the remaining time-ordered product depend only
on collinear and usoft fields describing momenta $p^{2}\ll m_b^2$, i.e.\
\begin{eqnarray} \label{Tlo2}
  T\, J^{(0)\dagger}_{j'\,\mu}(\omega',x)\: J^{(0)}_{j\,\nu}(\omega,0)
  &=& \big[ \bar {\cal H}_v  \overline \Gamma_{j'\mu}^{(0)} 
    \chi_{n,\omega'} \big](x)\ 
   \big[\bar \chi_{n,\omega} \Gamma_{j\nu}^{(0)} {\cal H}_v \big](0)\,,
\end{eqnarray}
where  
$ \bar\Gamma \equiv \gamma^0\Gamma^\dagger\gamma^0$.
It is useful to group the collinear and usoft fields into common brackets by
using a Fierz rearrangement.  For spin and color we can use
\begin{eqnarray} \label{Fierz}
 1 \otimes 1 &=& \frac{1}{2}\ \sum_{k=1}^{6} F_k^\bn \otimes F_k^n \\
 &=& \frac{1}{2} \Big[ 
  \big(\frac{\bnslash}{2N_c}\big)\!\otimes\! \big(\frac{\nslash}{2} \big)
 + \big(\frac{-\bnslash\gamma_5}{2N_c}\big)\!\otimes\!
   \big(\frac{\nslash\gamma_5}{2}\big)
 + \big(\frac{-\bnslash\gamma_\perp^\alpha}{2N_c}\big) \!\otimes\! 
     \big(\frac{\nslash\gamma^\perp_\alpha}{2}\big) \nn\\
 &&\ \ \ 
 +  \big({\bnslash T^a}\big) \!\otimes\! \big(\frac{\nslash T^a}{2}\big) 
 + \big({-\bnslash\gamma_5 T^a}\big) \!\otimes\! 
   \big(\frac{\nslash\gamma_5 T^a}{2}\big) 
 + \big( {-\bnslash\gamma_\perp^\alpha T^a}\big) \!\otimes\! 
  \big(\frac{\nslash\gamma^\perp_\alpha T^a}{2} \big)
  \Big] \,.\nn
\end{eqnarray}
Equation~(\ref{Fierz}) is valid as long as the identity matrices on the LHS are
inserted such that on the RHS the $F_k^\bn$ appear as $\bar\xi_n A F_k^\bn B
\xi_n$, where $A$ and $B$ do not contain $\bnslash$ factors. In Eq.~(\ref{Tlo2})
we insert identity matrices to the right of the $\overline \Gamma^{(0)}_{j'\mu}$
and to the left of the $\Gamma^{(0)}_{j\nu}$, which gives
\begin{eqnarray} \label{melt1}
 \langle \bar B_v | T\, J^{(0)\dagger}_{j'\,\mu}(\omega',x)\: 
  J^{(0)}_{j\,\nu}(\omega,0) | \bar B_v \rangle 
  &=& \frac{(-1)}{2}\
   \langle \bar B_v | T \big[ \bar {\cal H}_v(x) \overline
    \Gamma_{j'\mu}^{(0)}\, {F_k^n}\, \Gamma_{j\nu}^{(0)} \,
    {\cal H}_v(0) \big] | \bar B_v \rangle \nn\\
  && \times \langle 0 | T \big[ \bar\chi_{n,\omega}(0)\, 
     {F_k^\bn}\,
    \chi_{n,\omega'}(x) \big] | 0 \rangle \,.
\end{eqnarray}
Here the states $|\bar B_v\rangle$ have HQET normalization, $\langle B_v(k') |
B_v(k) \rangle = 2 v^0 (2\pi)^3 \delta^3({\bf k'\!-\!k})$, and are defined as
energy eigenstates of the LO usoft Hamiltonian generated from ${\cal
  L}_{us}^{(0)}$.

Only the vacuum matrix element with $k=1$ is non-zero:\footnote{Note that in
  Eq.~(\ref{Jmelt}) $\delta^2(x_\perp)\sim \lambda^{2}$, since from the
momentum-space view-point it is obtained by combining integrals over both the
label momentum $p_\perp\sim \lambda$ and residual momenta $k_\perp\sim
\lambda^2$, rather than just the latter.}
\begin{eqnarray} \label{Jmelt}
  &&\hspace{-0.9cm}
 \big\langle 0 \big| T \big[ \bar\chi_{n,\omega}(0) \frac{\bnslash}{2N_c}
    \chi_{n,\omega'}(x) \big] \big| 0 \big\rangle 
 =
  \big\langle 0 \big| T \big[ (\bar \xi_n W)_{\omega}(0) \frac{\bnslash}{2N_c}
    (W^\dagger \xi_n)_{\omega'}(x) \big] \big| 0 \big\rangle \nn\\
  &=& 
   (-2i)\, \delta(\omega\!-\!\omega') \delta^2(x_\perp) \delta(x^+) 
    \int\!\! \frac{dk^+}{2\pi}\: e^{-i k^+ x^-/2}\ {\cal J}^{(0)}_\omega(k^+) \,.
\end{eqnarray}
This definition of ${\cal J}_\omega^{(0)}$ agrees with
Ref.~\cite{Bauer:2001yt}.\footnote{To check this, one must note that here we
  contract color indices on the LHS and have an extra minus sign, since the
  $\chi_n$ fields are in the opposite order to Ref.~\cite{Bauer:2001yt}.} Owing
to the forward matrix element, the momentum-conserving delta function gives the
$\delta(\omega'-\omega)$, which can be used to eliminate the $\omega'$ integral
in Eq.~(\ref{Tlo1}), and the delta function in Eq.~(\ref{LOhard}) then sets
\begin{eqnarray}
  \omega=\bn\mcdot p \,.
\end{eqnarray} 
The appearance of the $\delta(x^+)$ ensures that in the remaining usoft matrix
element in Eq.~(\ref{melt1}) time ordering is the same as path ordering along
$x^-$. Up to order $\alpha_s$, the one-loop diagrams plus counterterms give
\begin{eqnarray} \label{Jexpn}
  {\cal J}_{\omega}^{(0)}(k^+) = \frac{\omega}{\omega k^+ \!+\! i\epsilon} 
  \bigg\{ 
  1\!+\! \frac{\alpha_s(\mu)C_F}{4\pi} \bigg[ 
  2 \ln^2\Big(\frac{-\omega k^+ \!-\!i\epsilon}{\mu^2}\Big) 
  \!-\! 3 \ln\Big(\frac{-\omega k^+ \!-\!i\epsilon}{\mu^2}\Big) 
  \!+\! 7 \!-\!\frac{\pi^2}{3} \bigg]
  \bigg\}\,,
\end{eqnarray}
which agrees with Refs.~\cite{Bauer:2003pi,Bosch:2004th}.

\begin{figure}[t!]
 \centerline{
  \mbox{\epsfysize=3.truecm \hbox{\epsfbox{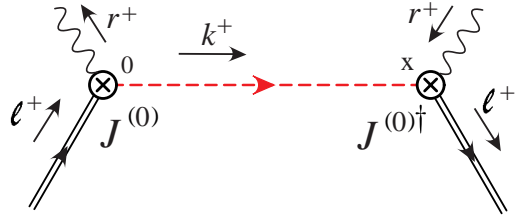}} }
  }
 \vskip -.2cm 
{\caption[1]{Momentum routing for leading-order insertion of currents.}
\label{fig:LOa} }
 \vskip -0.1cm
\end{figure}

Next, we simplify the spin structure of the $B$ matrix element in
Eq.~(\ref{melt1}) using the formula 
\begin{equation}
  P_v \Gamma P_v 
  =   P_v  {\rm Tr}\Big[ \frac{1}{2} P_v \Gamma\Big]
                 +  s_\mu {\rm Tr}\Big[-\frac{1}{2} s^\mu\Gamma\Big] 
  \equiv \sum_{m=1}^2 P^h_m\, {\rm Tr} [P^H_m \Gamma] 
   \,,
   \label{eq:projform}
\end{equation}
which is valid between heavy quark fields.
Here, $P_1^h = P_v$, $P_1^H=P_v/2$, $P_2^h=s_\mu$, $P_2^H = -s^\mu/2$ for $P_v =
(1 + \vslash)/2$ and $s_\mu \equiv P_v \gamma_\mu^T \gamma_5 P_v$. For our LO
matrix element the $s^\mu$ term vanishes and, taking into account the delta
functions in Eq.~(\ref{Jmelt}), we have only a function of
\begin{eqnarray} \label{xtdefn}
 \tilde x^\mu=\bn\mcdot x\:  n^\mu/2 \,,
\end{eqnarray}
namely
\begin{eqnarray} \label{Tr0}
  &&\hspace{-2cm}
  \langle \bar B_v | T \big[ \bar {\cal H}_v(\tilde x)\, \overline
    \Gamma_{j'\mu}^{(0)}\, \frac{\nslash}{2}\, \Gamma_{j\nu}^{(0)} \,
    {\cal H}_v(0) \big] | \bar B_v \rangle 
  =
   {\rm Tr} \Big[ \frac{P_v}{2}\, \overline \Gamma_{j'\mu}^{(0)} \,
  \frac{\nslash}{2}\, \Gamma_{j\nu}^{(0)} \Big] \langle \bar B_v | \bar
  h_v(\tilde x) 
   Y(\tilde x,0) h_v(0) | \bar B_v \rangle \,,
\end{eqnarray}
where $Y(\tilde x,\tilde y) \equiv Y(\tilde x)Y^\dagger(\tilde y) = {\rm P}\,
\exp\left( ig\int_{y^-}^{x^-} \mbox{d}s\: n\mcdot A_{us}(s n/2) \right)$.
Combining the phases from Eqs.~(\ref{Tlo1}) and (\ref{Jmelt}) and noting that
$d^4x=(dx^+ dx^- d^2x_\perp)/2$, we see that the matrix element in
Eq.~(\ref{melt1}) is now
\begin{eqnarray} \label{Tsimpler}
   {i}\: {\rm Tr} \Big[ \frac{P_v}{2}\, \overline \Gamma_{j'\mu}^{(0)} \,
  \frac{\nslash}{2}\, \Gamma_{j\nu}^{(0)} \Big]
   \int\!\! {dk^+}\!\! \int\!\! \frac{dx^-}{4\pi} e^{-i x^- (r^+ + k^+)/2} 
     {\cal J}_{\omega}^{(0)}(k^+) 
     \langle \bar B_v | \bar h_v(\tilde x) Y(\tilde x,0) h_v(0) | \bar B_v 
   \rangle \,.
\end{eqnarray}
After we pull out the large phases, the residual momenta are as shown in
Fig.~\ref{fig:LOa}, where $\ell^+=r^+ + k^+$.
Equation~(\ref{Tsimpler}) involves just the leading-order shape function
\begin{eqnarray} \label{f0}
  f^{(0)}(\ell^+) &=&
  \frac{1}{2} \int\!\! \frac{dx^-}{4\pi} e^{-i x^- \ell^+/2} 
   \ \langle \bar B_v | \bar h_v(\tilde x) Y(\tilde x,0) h_v(0) 
  | \bar B_v \rangle \nn\\
  &=&\frac{1}{2}\: \langle \bar B_v | \bar h_v
  \delta(\ell^+\!-\! in\mcdot D)  h_v | \bar B_v \rangle \,.
\end{eqnarray}
Note that we have taken $m_b\gg \Lambda$, and that Eq.~(\ref{f0}) depends only
on the residual momentum and masses that are of order $\Lambda$. Since we are
free to integrate by parts in the forward matrix element, it is evident that
$f^{(0)}(\ell^+)$ is real. Momentum conservation implies that the residual
momentum of the $b$-quark should not be larger than the residual mass
$\bar\Lambda$ of the $B$-meson, so $f(\ell^+)$ is non-zero only for $\ell^+ \le
m_B-m_b = \overline \Lambda$. Inserting $r^+$ from Eq.~(\ref{rdef}), this
implies that $k^+ \le p_X^+$.  In $\overline {\rm MS}$ the renormalized shape
function depends only on ratios of the dimensional regularization scale $\mu$
and momentum scales $\sim\Lambda$.

To derive the leading-order jet function we need the imaginary part of $(-i)$
times the result in Eq.~(\ref{Tsimpler}), which is the imaginary part of ${\cal
  J}^{(0)}_\omega(k^+)\: f^{(0)}(\ell^+)$.  Although the $B$ matrix element
integrated over $x^-$ is real, it does provide a theta function that 
restricts $k^+< p_X^+$ and therefore cannot be ignored in taking the imaginary 
part. The function ${\cal J}_\omega^{(0)}(k^+)$ has an imaginary part for 
$k^+>0$ only, so we have support for the imaginary part over a finite interval 
only.  The crucial point is that these theta functions are generated by 
dynamics within the effective theory and occur even though we have taken the 
$m_b\to\infty$ limit.
For the leading-order jet function for $B\to X_{u,s}$ we therefore have
[$\omega=\bn\mcdot p$]
\begin{eqnarray} \label{jetdef}
  {\cal J}^{(0)}(\omega\,k^+) = -\frac{1}{\pi}\: {\rm Im} 
   \Big[\, 
   {\cal J}^{(0)}_{\omega}(k^+)\  \theta(p_X^+ - k^+)\theta(k^+)\, \Big] \,.
\end{eqnarray}
By RPI-III invariance~\cite{Manohar:2002fd}, this ${\cal J}^{(0)}$ is a
non-trivial function of only the invariant product $(\omega k^+)$, 
times an overall factor of $\omega$ and the theta functions. 
For notational convenience, we suppress these extra dependences when writing 
the arguments of ${\cal J}^{(0)}$ as $\omega k^+$. 

Since $r^+ = n\mcdot q -m_b \leq \ell^+ \leq \overline\Lambda$, or equivalently
$0 \leq k^+ \leq p_X^+$, it is convenient to use the variables
$p_\omega^2=\bn\mcdot p\, n\mcdot p_X$ and $z$, where
\begin{eqnarray}
  k^+ = z\ p_X^+ \,,
\end{eqnarray} 
so that $\omega k^+ = z\, \bn\mcdot p\, n\mcdot p_X = z\, p_\omega^2$, and
$J_\omega^{(0)}(k^+)$ is non-zero only for $0\le z \le 1$. At order $\alpha_s$,
taking the imaginary part of Eq.~(\ref{Jexpn}) gives
\begin{eqnarray} \label{J01loop}
  {\cal J}^{(0)}(z,p_\omega^2,\mu) &=& \frac{\omega}{p_\omega^2} 
   \bigg\{ \delta(z) \bigg[
   1 + \frac{\alpha_s(\mu)C_F}{4\pi} \Big( 2 \ln^2\frac{p_\omega^2}{\mu^2} - 
   3 \ln \frac{p_\omega^2}{\mu^2} + 7-\pi^2 \Big) \bigg] \\
 && +\frac{\alpha_s(\mu)C_F}{4\pi} \bigg[ \Big(\frac{4\ln z}{z}\Big)_+ + 
  \Big(4 \ln\frac{p_\omega^2}{\mu^2}-3 \Big)\: \frac{1}{(z)_+} \bigg] \bigg\}
   \theta(1\!-\!z)\,\theta(z) \,,\nn
\end{eqnarray}
with the standard definitions for the plus functions. Our definition for the
jet function agrees with Refs.~\cite{Bauer:2003pi,Bosch:2004th} once we
compensate for the extra $\theta(p_X^+-k^+)$ that we included in
Eq.~(\ref{jetdef}). The jet function depends only on the dimensionless parameter
$z$ and ratios of $\mu^2$ and the momentum at the jet scale, $p_\omega^2\sim
Q\Lambda$.

At this point we have separated momentum scales in \SCETa, and, combining
Eqs.~(\ref{Tlo1}), (\ref{melt1}), (\ref{Jmelt}), (\ref{jetdef}), (\ref{Tr0}),
and (\ref{f0}), we arrive at the LO factorization theorem in terms of partonic
variables,
\begin{eqnarray} \label{Wfact000}
W_i^{(0)} &=&  
  h_i(\bn\mcdot p,m_b,\mu)\: 
  \int_{r^+}^{\bar \Lambda} \! d\ell^+\: 
   {\cal J}^{(0)}(\bn\mcdot p(\ell^+ \!-\! r^+),\mu )\: 
   f^{(0)}(\ell^+,\mu)\\ 
&=&
  h_i(\bn\mcdot p,m_b,\mu) \:
  \int_{0}^{\overline\Lambda-r^+} \! \! dk^+\: 
   {\cal J}^{(0)}(\bn\mcdot p\, k^+,\mu )\: 
   f^{(0)}(k^+  \!+\! r^+,\mu)
 \,.\nn
\end{eqnarray}
Using Eq.~(\ref{rdef0}), the final LO result in terms of hadronic variables can
be written as
\begin{eqnarray} \label{Wfact00}
W_i^{(0)} &=& 
  h_i(p_X^-,m_b,\mu)\: 
  \int_{\overline\Lambda- p_X^+}^{\bar \Lambda} \! d\ell^+\: 
   {\cal J}^{(0)}(p_X^-(\ell^+ \!-\! \overline\Lambda\!+\! p_X^+),\mu )\: 
   f^{(0)}(\ell^+,\mu)
 \nn\\
 &=& 
  h_i(p_X^-,m_b,\mu) \:
  \int_{0}^{p_X^+} \! \! dk^+\: 
   {\cal J}^{(0)}(p_X^-\, k^+,\mu )\: 
   f^{(0)}(k^+  \!+\!\overline\Lambda\!-\! p_X^+,\mu)
 \\ 
 &=& 
  h_i(p_X^-,m_b,\mu)\ 
  p_X^+\!
  \int_{0}^{1} \!\! dz\ 
   {\cal J}^{(0)}(z\, m_X^2,\mu )\:
   f^{(0)}\big(\overline \Lambda - p_X^+(1\!-\! z),\mu\big) \,. \nn
\end{eqnarray}
For practical applications we would also make use of a short-distance mass
definition for $m_b$ (the cancellation of infrared renormalon ambiguities for
the shape function was demonstrated recently in Ref.~\cite{Gardi:2004ia}).
Here, the dependence on whether it is $X_s$ or $X_u$ occurs only in $h_i^{(0)}$
and the values of the kinematic variables $\bn\cdot p$ and $n\cdot p_X$.  The
functional forms of ${\cal J}^{(0)}$ and $f^{(0)}$ are independent of which
process we consider.  The hard coefficients are given by
\begin{eqnarray} \label{h0}
  h_i(\omega,m_b,\mu) &=& \sum_{j,j'} \!
   C_{j'}(\omega,m_b,\mu)  C_j(\omega,m_b,\mu) \ 
  {\rm Tr} \Big[ \frac{P_v}{2}\, \overline \Gamma_{j'\mu}^{(0)} \,
  \frac{\nslash}{2}\, \Gamma_{j\nu}^{(0)} \Big] P_i^{\mu\nu}\,,
\end{eqnarray}
with the projectors $P_i^{\mu\nu}$ defined in Eq.~(\ref{ProjW}).

Taking the traces in Eq.~(\ref{h0}) for $B\to X_u\ell\bar\nu$, we find that
\begin{eqnarray} \label{hiuLOf}
  h_1^{u} &=& \frac{1}{4} \big[ C_1^{(v)} \big]^2  \,,\qquad\\
  h_2^{u} &=&
   \frac{n\mcdot q\, \big[ (C_1^{(v)})^2 + C_1^{(v)} C_2^{(v)}+
     C_2^{(v)} C_3^{(v)} \big] }{(n\mcdot q\! -\! \bn\mcdot q)}
   + \frac{(C_2^{(v)})^2}{4} 
   + \frac{(n\mcdot q)^2\, \big[ (C_3^{(v)})^2 +  2 C_1^{(v)} C_3^{(v)} \big]}
    {(n\mcdot q\! -\! \bn\mcdot q)^2}    \,,\nn\\
  h_3^{u} &=& \frac{(C_1^{(v)})^2}{2(n\mcdot q\! -\! \bn\mcdot q)} \,,\qquad
  h_4^{u} = \frac{C_3^{(v)}\, (2C_1^{(v)}+C_3^{(v)})}
    {(n\mcdot q\! -\! \bn\mcdot q)^2}\:  \,,\nn\\
  h_5^{u} &=& -\frac{\big[ (C_1^{(v)})^2 + C_1^{(v)} C_2^{(v)} +
     C_2^{(v)} C_3^{(v)}\big]  }{2(n\mcdot q\! -\! \bn\mcdot q)} 
   - \frac{\big[ (C_3^{(v)})^2 + 2 C_1^{(v)} C_3^{(v)} 
   \big] n\mcdot q } {(n\mcdot q\! -\! \bn\mcdot q)^2} 
    \,.\nn
\end{eqnarray}
For $B\to X_s\gamma$, where $\bn\mcdot q=0$, the traces give
\begin{eqnarray} \label{hisLOf}
  h_1^{s} &=& \frac{(n\mcdot q)^2}{4}
    \big( C_1^{(t)}-\frac12 C_2^{(t)} - C_3^{(t)} \big)^2 \,,\quad
  h_2^{s} = 0  \,,\quad
  h_3^{s} = \frac{n\mcdot q}{2} \big( C_1^{(t)}-\frac12 C_2^{(t)} 
   - C_3^{(t)} \big)^2\,,\nn\\
  h_4^{s} &=& -\frac{1}{4} \big( 3C_1^{(t)}-2 C_2^{(t)} -2 C_3^{(t)}
    - C_4^{(t)}\big) 
   \big( C_1^{(t)}-2 C_3^{(t)} + C_4^{(t)} \big)\,,\nn\\
  h_5^{s} &=& \frac{n\mcdot q}{2} \big( C_1^{(t)}-\frac12 C_2^{(t)} 
    - C_3^{(t)} \big)^2 \,.
\end{eqnarray}
Here the  $h_i^{0u}$'s depend on $\ln(\mu/m_b)$ and in addition $\overline
y=\bn\mcdot p/m_b$ through the ${\cal O}(\alpha_s)$ corrections to the $C_i$'s
(see Eqs.~(\ref{matchC}) for results at $\mu=m_b$).

The factors of $n\mcdot q=m_B x_\gamma^H$ (for $B\to X_s\gamma$) and $\bn\mcdot
q=m_B(1-\overline y_H)$, $n\mcdot q=m_B(1-u_H)$ (for $B\to X_u\ell\bar\nu$)
are purely kinematic 
and so can be directly replaced by these hadronic variables. 
Expanding to leading order about $x_\gamma^H=1$ and $u_H=0$ gives 
the results that should be used at LO in the SCET expansion. For notational
convenience, we write
\begin{eqnarray}
  h_i^{u} = h_i^{0u} + h_i^{0'u} + \ldots \,,\qquad
  h_i^{s} = h_i^{0s} + h_i^{0's} + \ldots \,.
\end{eqnarray}
For $B \rightarrow X_u \ell \bar\nu$ we have
\begin{eqnarray} \label{hiuLO}
  h_1^{0u} &=& \frac{1}{4} \big[ C_1^{(v)} \big]^2  \,,\qquad\\
  h_2^{0u} &=& \frac{1}{\overline y_H} \big[ (C_1^{(v)})^2
   \!+\! C_1^{(v)}\, C_2^{(v)} \!+\! C_2^{(v)}\, C_3^{(v)} \big]
   + \frac{1}{4} (C_2^{(v)})^2
   +  \frac{1}{\overline y_H^{\,2}} \big[ 
   (C_3^{(v)})^2 \!+\! 2 C_1^{(v)}\, C_3^{(v)}\big] \,,\nn\\
  h_3^{0u} &=& \frac{1}{2m_B \overline y_H} (C_1^{(v)})^2\,,\qquad
  h_4^{0u} = \frac{1}{m_B^2\,\overline y_H^{\,2}}\: 
    C_3^{(v)}\, (2C_1^{(v)}+C_3^{(v)})
    \,,\nn\\
  h_5^{0u} &=& -\frac{1}{2m_B \overline y_H} \big[ (C_1^{(v)})^2
   \!+\! C_1^{(v)}\, C_2^{(v)} \!+\! C_2^{(v)}\, C_3^{(v)} \big]
   - \frac{1}{m_B\overline y_H^{\,2}}\, C_3^{(v)} \big[ 
    2 C_1^{(v)}+ C_3^{(v)}\big] 
     \,, \nn
\end{eqnarray}
while for $B\to X_s\gamma$ we have
\begin{eqnarray} \label{hisLO}
  h_1^{0s} &=& \frac{m_B^2}{4}
    \big( C_1^{(t)}\!-\!\frac12\, C_2^{(t)} \!-\! C_3^{(t)} \big)^2 \,,\qquad
  h_2^{0s} = 0  \,,\qquad
  h_3^{0s} = \frac{m_B}{2} 
   \big( C_1^{(t)}\!-\!\frac12 C_2^{(t)} \!-\! C_3^{(t)} \big)^2\,,\nn\\
  h_4^{0s} &=& -\frac{1}{4} \big( 3C_1^{(t)}\!-\!2 C_2^{(t)} \!-\! 2 C_3^{(t)}
    \!-\! C_4^{(t)}\big) 
   \big( C_1^{(t)}\!-\! 2 C_3^{(t)} \!+\! C_4^{(t)} \big)\,,\nn\\
  h_5^{0s} &=& \frac{m_B}{2} \big( C_1^{(t)}\!-\!\frac12 C_2^{(t)} 
   \!-\! C_3^{(t)}
   \big)^2 \,. 
\end{eqnarray}
The results in Eqs.~(\ref{hiuLO}) and (\ref{hisLO}) agree with
Refs.~\cite{Bauer:2003pi,Bosch:2004th}. At NLO in the power expansion, it will
be necessary to keep the next term $h_i^{0'f}$ from the expansion of the
kinematic prefactors in Eqs.~(\ref{hiuLOf}) and (\ref{hisLOf}).  For $B\to
X_u\ell\bar\nu$ we find
\begin{eqnarray} \label{hiuLOp}
  h_1^{0'u} &=& 0  \,,\qquad\\
  h_2^{0'u} &=& u_H\bigg\{
   \frac{(1\!-\!\overline y_H )}{\overline y_H^2} \big[ (C_1^{(v)})^2
    \!+\! C_1^{(v)}\, C_2^{(v)} \!+\! C_2^{(v)}\, C_3^{(v)} \big]
   +  \frac{2(1\!-\!\overline y_H)}{\overline y_H^{\,3}} \big[ 
   (C_3^{(v)})^2 \!+\! 2 C_1^{(v)}\, C_3^{(v)}\big]\bigg\} \,,\nn\\
  h_3^{0'u} &=& \frac{u_H}{2m_B \overline y_H^2} (C_1^{(v)})^2\,,\qquad
  h_4^{0'u} = \frac{2u_H}{m_B^2\,\overline y_H^{\,3}}\: 
    C_3^{(v)}\, (2C_1^{(v)}+C_3^{(v)})
    \,,\nn\\
  h_5^{0'u} &=& - u_H \bigg\{ \frac{1}{2m_B \overline y_H^2} \big[ (C_1^{(v)})^2
   \!+\! C_1^{(v)}\, C_2^{(v)} \!+\! C_2^{(v)}\, C_3^{(v)} \big]
   + \frac{(2\!-\!\overline y_H)}{m_B\overline y_H^{\,3}}\, C_3^{(v)} \big[ 
    2 C_1^{(v)}+ C_3^{(v)}\big]  \bigg\}
     \,, \nn
\end{eqnarray}
while for $B\to X_s\gamma$ we have
\begin{eqnarray} \label{hisLOp}
  h_1^{0's} &=& \frac{m_B^2(x_\gamma^H\!-\! 1)}{2}
    \big( C_1^{(t)}\!-\!\frac12\, C_2^{(t)} \!-\! C_3^{(t)} \big)^2 \,,\qquad
  h_2^{0's} = 0  \,,\\
  h_3^{0's} &=& \frac{m_B(x_\gamma^H\!-\! 1)}{2} 
   \big( C_1^{(t)}\!-\!\frac12 C_2^{(t)} \!-\! C_3^{(t)} \big)^2\,,\qquad
  h_4^{0's} = 0\,,\nn\\
  h_5^{0's} &=& \frac{m_B(x_\gamma^H\!-\! 1)}{2} 
   \big( C_1^{(t)}\!-\!\frac12 C_2^{(t)} 
   \!-\! C_3^{(t)}
   \big)^2 \,. \nn
\end{eqnarray}

In Ref.~\cite{Korchemsky:1994jb}, the LO triply differential rate for $B\to
X_u\ell\bar\nu$ was found to satisfy $d^3\Gamma^u/ dx dy dy_0 \propto
(x-y)(y_0-x)$.  In terms of the $h_i^{0u}$, this corresponds to the relations
$h_2^{0u} = 4 h_1^{0u}/ {\bar y_H}$ and $m_B h_3^{0u} = 2 h_1^{0u}/ {\bar y_H}$
(where at LO we can set $y = (1-u_H)(1-\bar y_H)$, $y_0 = 2 - u_H - \bar y_H$).
Eqs.~(\ref{hiuLO}) agree with this result at tree level in the hard functions,
but give non-zero corrections to the first ($h_2^{0u}$) relation of order
$\alpha_s(m_b)$ from the Wilson coefficients $C_2^{(v)}$ and $C_3^{(v)}$.


\section{Vanishing Time-Ordered Products at ${\cal O}(\lambda)$} \label{sect_NLOa}

%
%

To work out the factorization beyond LO is now simply a matter of going to
higher order in $\lambda$ in SCET. At order $\lambda$ the time-ordered products
are
\begin{eqnarray}
 \hat T^{1a} &=& -i \int\!\! d^4x \: e^{-i r\cdot x}\ 
      {\rm T} \big[ 
       J^{(0)\dagger}(x)\: J^{(1a)}(0)+J^{(1a)\dagger}(x)\: J^{(0)}(0)  \big] 
       \,,\\
 \hat T^{1b} &=& -i \int\!\! d^4x \: e^{-i r\cdot x}\ 
    {\rm T} \big[ 
     J^{(0)\dagger}(x)\: J^{(1b)}(0) +  J^{(1b)\dagger}(x)\: J^{(0)}(0)\big]
      \,,\nn\\
 \hat T^{1L} &=& -i \int\!\! d^4x\, d^4y \: e^{-i r\cdot x}\ 
      {\rm T} \big[ J^{(0)\dagger}(x)\: i{\cal L}_{\xi\xi}^{(1)}(y) J^{(0)}(0) 
       \big] \,.\nn
\end{eqnarray}
We shall see that these time-ordered products give vanishing matrix elements for
$B\to X_s\gamma$ and $B\to X_u\ell \bar\nu$. The result follows almost directly
from chirality and the fact that the currents in Eqs.~(\ref{J0}), (\ref{J1}) and
(\ref{J2}) all involve left-handed collinear quark fields. However, we must be
careful to check that terms proportional to the chiral condensate do not
contribute.  We follow an approach similar to the previous section: by using the
Fierz transformation in Eq.~(\ref{Fierz}), we can collect the usoft and collinear
fields into separate parts ${\cal J}\otimes {\cal S}$, and the matrix
element then factorizes to give $\langle 0 | {\cal J} | 0 \rangle \otimes
\langle B_v | {\cal S} | B_v \rangle$. These parts are still connected by
indices and spacetime integrals, which are represented here by the $\otimes$.

For $T^{1a}$ and $T^{1b}$, we take the expressions for the currents from
Eqs.~(\ref{J0}) and (\ref{J1}). The same usoft fields appear as in the 
leading-order $T^{(0)}$, and so the T-products involve soft matrix elements 
that are similar to the one in Eq.~(\ref{melt1}), namely
\begin{eqnarray} \label{Jnlo1}
&& \big\langle \bar B_v \big| T \big[ \bar {\cal H}_v(x) \, \overline
     \Gamma_{j\mu}^{(0)} \, {F_k^n}\, \Upsilon_{j'\nu\alpha}^{(1a)}\, 
    {\cal H}_v(0) \big] \big| \bar B_v \big\rangle \,,\nn\\
&& \big\langle \bar B_v \big| T \big[ \bar {\cal H}_v(x) \overline
    \Upsilon_{j'\mu\alpha}^{(1a)}\, {F_k^n}\, \Gamma_{j\nu}^{(0)} \,
    {\cal H}_v(0) \big] \big| \bar B_v \big\rangle \,,
\end{eqnarray}
with similar matrix elements for $T^{1b}$ but with $\Upsilon_{j'\mu}^{(1a)}\to
\Upsilon_{j'\mu}^{(1b)}$.  The first term in Eq.~(\ref{Jnlo1}) is multiplied by
the collinear matrix element
\begin{eqnarray} \label{Jc1a}
 {\cal J}^{(1a)}(0,x) &=& \frac{1}{\omega'}\: \Big\langle 0 \Big| T \Big[ 
     (\bar \chi_n^L\, i\! \cDgppPl  )_{\omega'}(0)\, {F_k^\bn}\,
    \chi^L_{n,\omega}(x) \big] \Big| 0 \Big\rangle  \,,
\end{eqnarray}
while the second term is multiplied by $[{\cal J}^{(1a)}(x,0)]^\dagger$. For
$T^{1b}$ the analogous result is
\begin{eqnarray}
 {\cal J}^{(1b)}(0,x) &=& \frac{-1}{m_b n\mcdot v} \Big\langle 0 \Big| T \big[ 
     \bar \chi^L_{n,\omega'}(0)\, (ig\, {\cal
       B}_{c\perp}^\alpha)_{\omega_2}(0)\,
    {F_k^\bn}\,
    \chi^L_{n,\omega}(x) \big] \Big| 0 \Big\rangle  \,.
\end{eqnarray}
In both ${\cal J}^{(1a)}$ and ${\cal J}^{(1b)}$, the fact that we need an 
overall color singlet eliminates the possibilities $k=4,5,6$. Owing to the 
presence of the $\alpha$ $\perp$-index, rotational invariance also eliminates 
$k=1,2$, leaving only $F_3^{\bn} = -\bnslash\gamma_\perp^\beta/(2N_c)$. 
However, this term has the wrong chiral structure and also vanishes, i.e.\
\begin{eqnarray} \label{vanish1}
 \bar\chi_n^L F_3^\bn \chi_n^L \propto
   \bar\chi_n P_R \bnslash \gamma_\perp^\beta P_L \chi_n = 0 \,,
\end{eqnarray}
so both $T^{1a}$ and $T^{1b}$ give vanishing corrections.

For $T^{1L}$ we need to use the Fierz identity in Eq.~(\ref{Fierz}) twice to
group together all usoft and collinear factors. This gives the usoft matrix
element
\begin{eqnarray}
 \big\langle B_v \big|
  \bar{\cal H}_v(x) \overline\Gamma^{(0)}_{j'} F_{k'}^n \big(i \cDslash_{us}^\perp
  \gamma_\perp^\beta \frac{\bnslash}{2} \big)(y) F_k^n \Gamma^{(0)}_j 
  {\cal H}_v(0) \big| B_v\big \rangle
\end{eqnarray}
multiplied by the collinear matrix element of a four-quark operator, which we
claim satisfies
\begin{eqnarray} \label{vanish2}
  \Big\langle 0 \Big| 
   \Big[\bar\chi_n^L (y) F_{k'}^\bn \chi_n^L(x)\Big]
   \Big[\bar\chi_n^L(0) F_k^\bn ( i {\cal D}_{c\perp}^\beta \chi_n^L)(y) \Big]
   \Big| 0 \Big\rangle =0\,.
\end{eqnarray}
To prove Eq.~(\ref{vanish2}) we begin by noting that color now allows $T^a$
terms as long as they occur in both $F$'s, so either $k,k' \in \{1,2,3\}$ or
$k,k' \in \{4,5,6\}$. For either possibility the argument is the same, so for
convenience we take the former case.  Rotational invariance now demands that
one of $k,k'$ is equal to $3$ and the other is $1$ or $2$. 
If $k'=3$, then the first pair of collinear quark fields vanishes, just as 
in Eq.~(\ref{vanish1}), while if $k=3$, then the second pair of collinear 
quark fields vanishes.

The results in Eqs.~(\ref{vanish1}) and (\ref{vanish2}) rely on the 
underlying assumption that there is no structure in the
vacuum that can flip the chirality. In QCD we know that this is not the case,
since the chiral condensate (and instantons) take $L\leftrightarrow R$. The
above argument is valid because we have used chirality only at the $\mu^2\sim
Q\Lambda$ scale where we are matching perturbatively, and not for $\mu^2\sim
\Lambda^2\ll Q\Lambda $, where it is badly broken. The same argument applies
when perturbatively matching on to the weak Hamiltonian at $\mu\simeq m_W \gg
\Lambda$.



\section{Factorization at Next-to-Leading Order} \label{sect_NLO}

Since the order $\lambda$ contributions vanish, the first power corrections
occur at order $\lambda^2=\Lambda/Q$ and will be referred to as NLO corrections.
The NLO contributions to the decay rates have several sources, including the
following:
\begin{itemize}
\item[i)] expansion of kinematic factors occurring in front of the $W_i$ in
  the decay rates given by Eqs.~(\ref{dGamma3u}) and (\ref{dGamma4u})\,,
\item[ii)] expansion of the kinematic factors appearing in the $h_i$ at LO,
  i.e.\ in Eqs.~(\ref{hiuLOf}) and (\ref{hisLOf})\,,
\item[iii)] expansion associated with the conversion from partonic to hadronic 
  variables, given in Eq.~(\ref{Wi0rNLO})\,,
\item[iv)] expansion from higher-order operators contributing to the
  time-ordered products, given in Eq.~(\ref{Wfact00b}).
\end{itemize}  
Once the region of phase space where the SCET expansion is valid is properly
defined, as in Sec.~\ref{sect_kin}, the corrections in i) and ii) are
straightforward to compute. The corrections from i) do depend on the choice of
how the $x_H$ charged-lepton variable is treated, for example whether we
integrate over all of $x_H$ or instead look at an $x_H$ spectrum with a cut. 
Results for i)
are given in Eqs.~(\ref{dGamma5u}), (\ref{dGamma6u}) and (\ref{dGamma3t}). 
For ii) the required terms are derived by keeping one more term in the Taylor 
series in passing from Eqs.~(\ref{hiuLOf}) and (\ref{hisLOf}) to 
Eqs.~(\ref{hiuLO}) and (\ref{hisLO}).
We give the results for i) and ii) in Sec.~\ref{sectKin}. Note that these
terms are already in terms of hadronic variables and are therefore unaffected by
the conversion in iii).

The NLO terms from category iii) are discussed in Sec.~\ref{Trexpn} below.
The contributions from category iv) require several sections.  In
Sec.~\ref{Tlam2} we give a complete list of the time-ordered products arising at
second order in SCET (category iv)), along with a summary of the jet and shape
functions they generate.  In Sec.~\ref{sectJet2} we carry out the tree-level
matching calculations for these time-ordered products and define the jet
functions at leading order in $\alpha_s$.  In Sec.~\ref{sectshape} we give
operator definitions for the shape functions.  Detailed derivations of the NLO
factorization theorems for these time-ordered products are presented in
Sec.~\ref{sectT2s}. Finally, results for the computation of the traces that
give the NLO hard coefficients are given in Sec.~\ref{sectHard}.

Note that the final results for contributions from i), ii), iii) and iv) are
summarized in Sec.~\ref{sect_Summary}.

\subsection{Switching to hadronic variables at order $\lambda^2$}
\label{Trexpn}

At LO the factorization theorem in terms of partonic variables was given in
Eq.~(\ref{Wfact000}):
\begin{eqnarray}  \label{Wi0remember}
W_i^{(0)} 
&=&
  h_i^{0}(\bn\mcdot p,m_b,\mu) \:
  \int_{0}^{\overline\Lambda-r^+} \! \! dk^+\: 
   {\cal J}^{(0)}(\bn\mcdot p\, k^+,\mu )\: 
   f^{(0)}(k^+  \!+\! r^+,\mu)
 \,.
\end{eqnarray}
In this expression, the variables $\bn\cdot p$ and $n\cdot r$ have a series
expansion once we switch to hadronic variables. For the accuracy needed at NLO
we have
\begin{eqnarray} \label{rdefNLO}
   \bn\cdot p = \bn\cdot p_X - \overline\Lambda\,,
  \qquad\qquad
   n\cdot r = (\overline \Lambda - n\mcdot p_X) 
    - \frac{(\lambda_1+3\lambda_2)}{2m_b} \,.
\end{eqnarray}
Expanding Eq.~(\ref{Wi0remember}) gives the LO term in Eq.~(\ref{Wfact00}) plus
the NLO terms
\begin{eqnarray} \label{Wi0rNLO}
  \big[ W_i^{(0)} \big]^{\rm NLO}
&=&
  - \frac{(\lambda_1+3\lambda_2)}{2m_b}\:  h_i^{0}(p_X^-,m_b) \:
  \int_{0}^{p_X^+} \! \! dk^+\: 
   {\cal J}^{(0)}(p_X^-\, k^+)\: 
  \\[5pt]
 &&\quad
  \times \bigg\{ \frac{df^{(0)}(k^+  \!+\! \overline\Lambda\!-\!p_X^+)}{dk^+}
  - \delta(k^+ \!-\! p_X^+) f^{(0)}(k^+  \!+\!
  \overline\Lambda\!-\!p_X^+)\bigg\}\nn\\
 && - \overline\Lambda \int_{0}^{p_X^+} \! \! dk^+\: 
  \frac{d}{d\bn\mcdot p}\Big\{ 
 h_i^0(\bn\mcdot p,m_b) \: {\cal J}^{(0)}(\bn\mcdot p\, k^+)  \Big\}
  \Big|_{\bn\cdot p=p_X^-}\ 
  f^{(0)}(k^+  \!+\! \overline\Lambda\!-\!p_X^+)
 \, \nn\\[5pt]
 &=& 
   - \frac{(\lambda_1+3\lambda_2)}{2m_B} h_i^{0}(p_X^-,m_b)  
  \big[{{\cal J}^{(0)}\otimes \widetilde{f^{(0)}}}\big](p_X^+,p_X^-)
 -\overline\Lambda\  
  \big[{(h_i^{0f}{\cal J}^{(0)})}'\otimes f^{(0)}\big](p_X^+,p_X^-) \,.\nn
\end{eqnarray}
Note that in taking the derivative with respect to $\bn\mcdot p$ we 
differentiate only the $C_j(\bn\mcdot p)$'s in $h_i^0$ and not the prefactors
depending on $\bn\mcdot q$. For later convenience, we switched to hadronic
variables in the final line of Eq.~(\ref{Wi0rNLO}) and defined 
\begin{eqnarray} \label{J0tildef0}
   \big[{{\cal J}^{(0)}\otimes \widetilde{f^{(0)}}}\big](p_X^+,p_X^-)
  & = & \int_{0}^{p_X^+} \! \! dk^+\:
   {\cal J}^{(0)}(p_X^-\, k^+)\:
   \bigg\{ \frac{df^{(0)}(k^+  \!+\! \overline\Lambda\!-\!p_X^+)}{dk^+}
   \\[5pt]
  && \qquad
  - \delta(k^+ \!-\! p_X^+) f^{(0)}(k^+  \!+\!
  \overline\Lambda\!-\!p_X^+)\bigg\},\nn\\
   \big[{(h^{0f}{\cal J}^{(0)})}'\otimes f^{(0)}\big](p_X^+,p_X^-)
  & = & \int_{0}^{p_X^+} \! \! dk^+\:
  \frac{d}{d\bn\mcdot p}\Big\{
 h^{0f}(\bn\mcdot p,m_b) \: {\cal J}^{(0)}(\bn\mcdot p\, k^+)  
  \Big\}\Big|_{\bn\cdot p=p_X^-}\nn\\
&& \qquad \times  f^{(0)}(k^+  \!+\! \overline\Lambda\!-\!p_X^+). \nn
\end{eqnarray}

\subsection{Time-Ordered Products at order $\lambda^2$}  \label{Tlam2}

\begin{table} 
\[
\begin{array}{ccccccc}
\hbox{T-product} & & \hbox{Example Diagram} & \hspace{-0.cm}
  \begin{array}{c}\mbox{Hard, Jet, and} \\ \mbox{Shape Functions}\end{array} 
  \hspace{-0.cm} &  
  \hbox{Usoft Operator}   
  \\ \hline\\
  \hat T^{(2H)}
   & & 
     \epsfxsize=6.2cm \lower25pt\hbox{\epsfbox{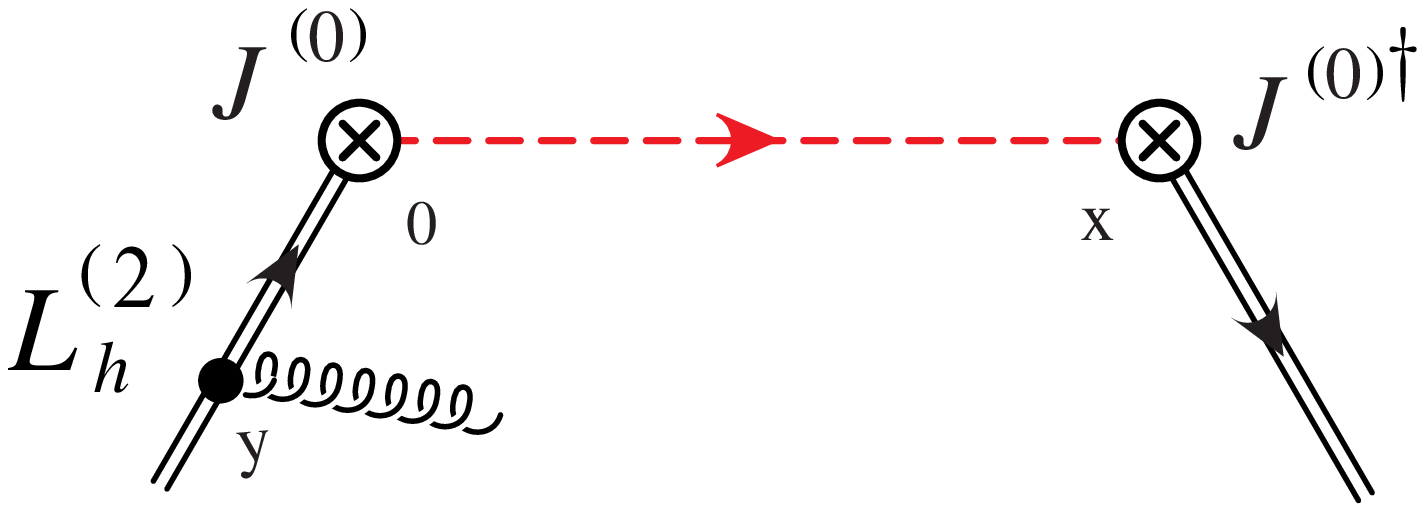}} &  
  h^{0} {\cal J}^{(0)}\, f_0^{(2)} & 
 \bar h_v(x)  h_v(0) i{\cal L}^{(2)}_{\rm h}(y)
  \\ \\
 \hat T^{(2a)} 
  & & 
     \epsfxsize=6cm \lower25pt\hbox{\epsfbox{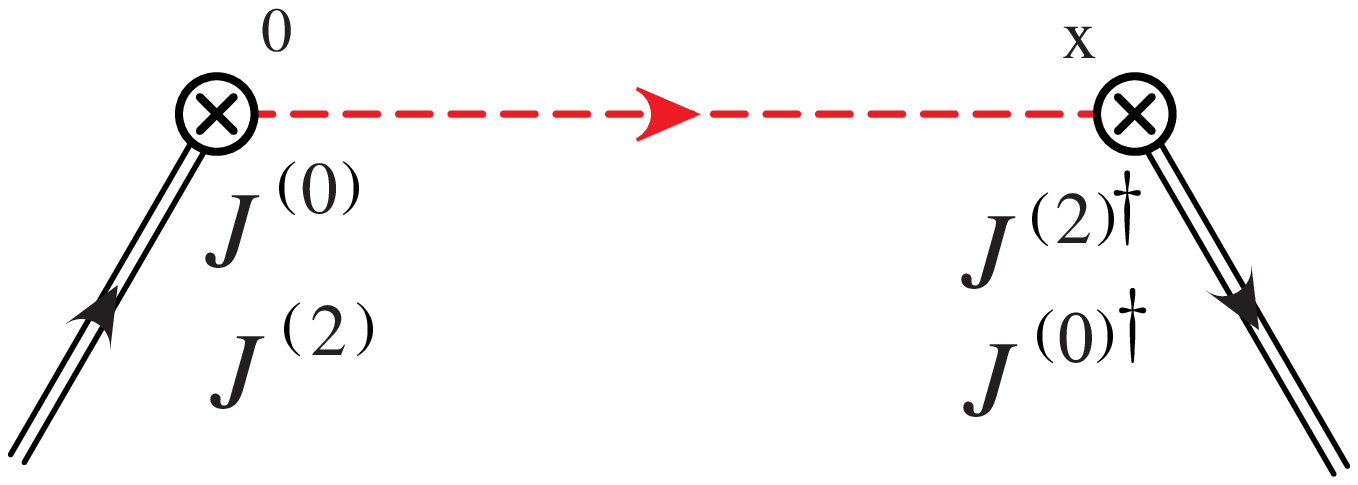}} &  
  h^{1,2} {\cal J}^{(0)}\, f^{(2)}_{1,2} &  
  \begin{array}{c} 
  \bar h_v(x) (D_{T,\perp} h_v)(0)\\ (\bar h_v D_{T,\perp})(x) h_v(0)
  \end{array}
  \\ \\
%
  \hat T^{(2L)}
  & & 
     \epsfxsize=6cm \lower25pt\hbox{\epsfbox{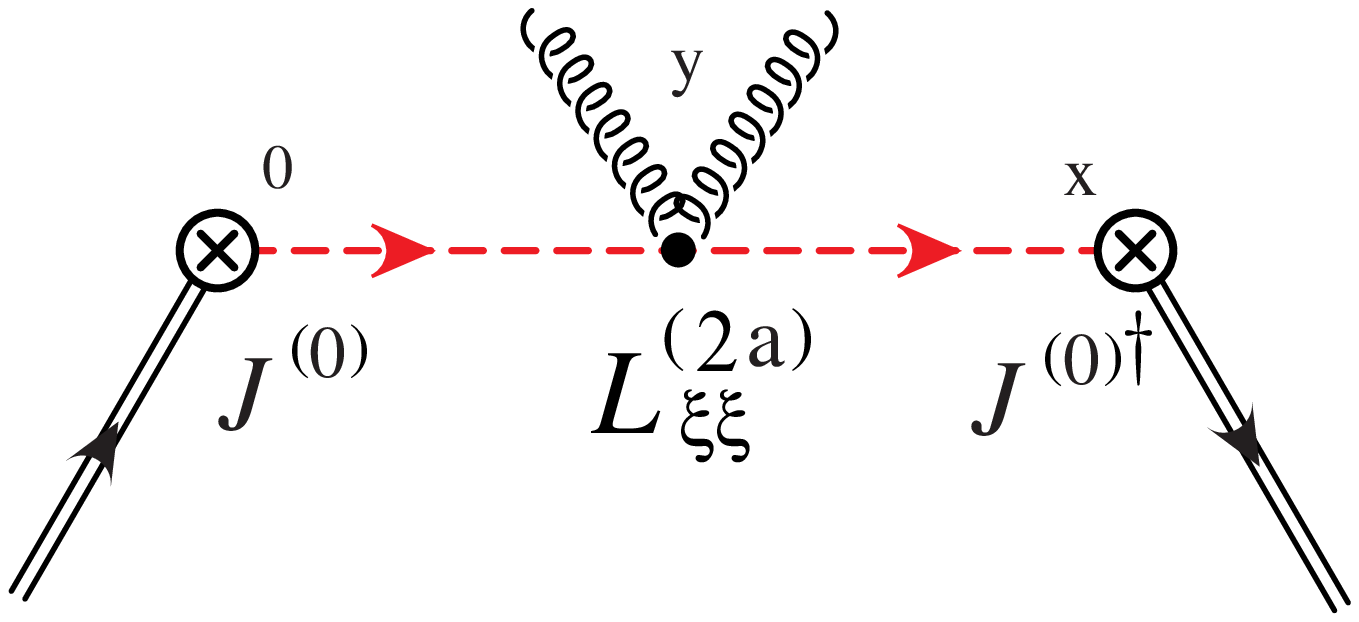}} &  
  \begin{array}{c}
  h^{3,4}\, {\cal J}_{1,2}^{(-2)}\, f^{(4)}_{3,4} \\[8pt]
  h^{3,4}\, {\cal J}_{3,4}^{(-2)}\, g^{(4)}_{3,4}
  \end{array}
  &  
  \bar h_v(x) (D_\perp D_\perp)(y) h_v(0)    
  \\ \\
 \hat T^{(2q)}
& & 
     \epsfxsize=6cm \lower25pt\hbox{\epsfbox{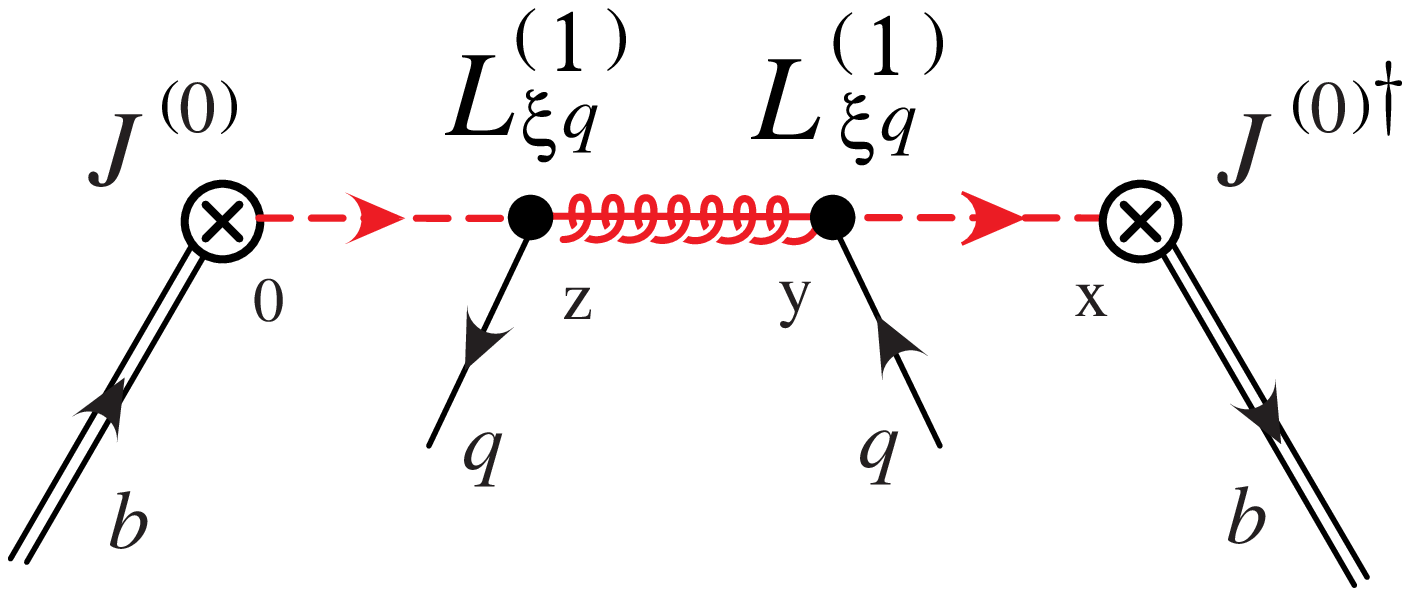}} &  
  \begin{array}{c}
  h^{5,6}\,{\cal J}^{(-4)}_1\, f_{5,6}^{(6)} \\[8pt]
   h^{5-8}\,{\cal J}^{(-4)}_{2-4}\, g_{5-10}^{(6)}
  \end{array}
  & 
  \begin{array}{c} \bar h_v(x) q(y) \bar q(z) h_v(0)
  \end{array}
\end{array}
\]
\vskip -.1cm
\caption{
Time-ordered products that are of order $\lambda^2=\Lambda/m_b$ overall, 
and that are non-zero at tree level.  The power of $\lambda^2$ is obtained by
multiplying the powers from the jet functions ${\cal J}$ by those from the shape
functions $f$ or $g$. We suppress color and Dirac structure in the usoft operators
listed, which can be found in the text.  The time-ordered product
in the last row has not been considered in the literature and is enhanced
relative to the other entries by a prefactor of $4\pi\alpha_s(E_X\Lambda)\sim 5$.
 \label{table_T}} 
\end{table}

\begin{table} 
\[
\begin{array}{ccccccc}
\hbox{T-product} &  \hbox{Example Diagram} & \hspace{-0.cm}
  \begin{array}{c}\mbox{Hard, Jet, and} \\ \mbox{Shape Functions}\end{array} 
  \hspace{-0.cm} & 
  \hbox{Usoft Operator}   
  \\ \hline\\
  \hat T^{(2b)} 
  &  
     \epsfxsize=5.2cm \lower35pt\hbox{\epsfbox{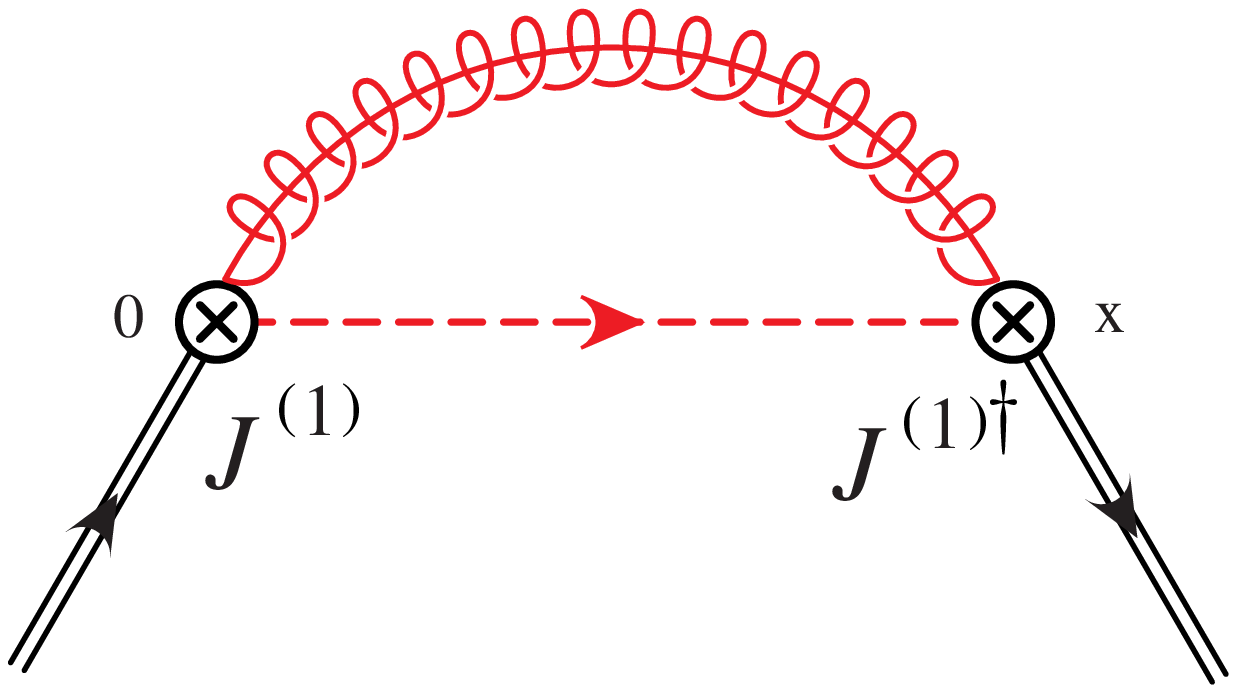}} &  
   h^{[2b]}\, {\cal J}_{1,2}^{(2)} \, f^{(0)}
  &   \bar h_v(x) h_v(0)
  \\ 
  \hat T^{(2c)}
& 
     \epsfxsize=6cm \lower30pt\hbox{\epsfbox{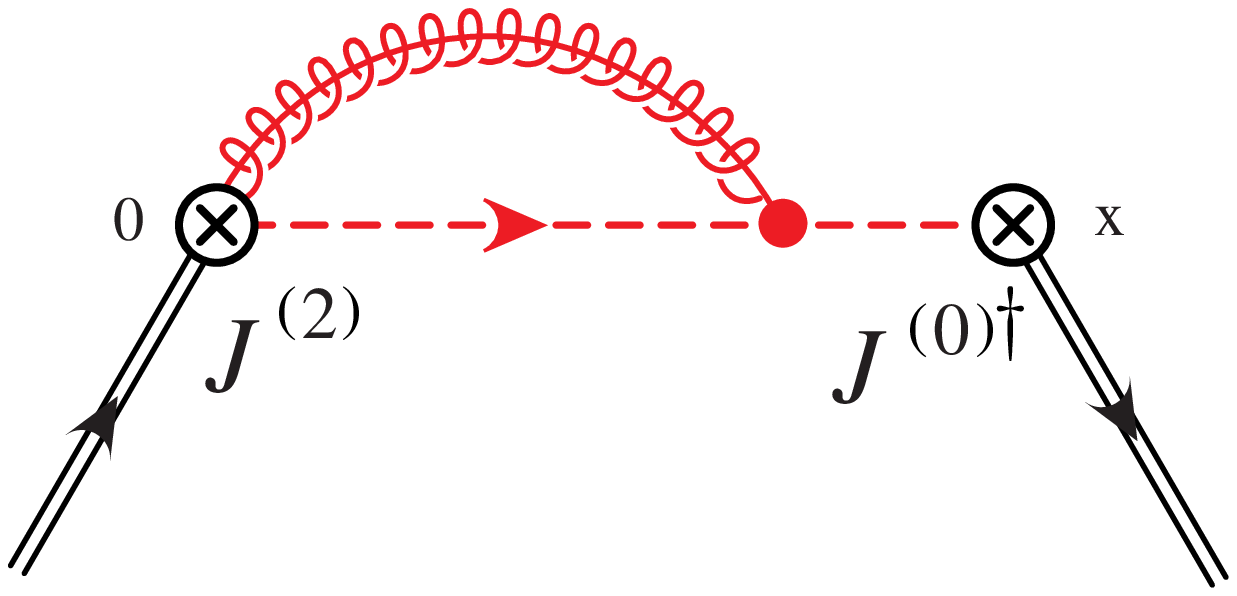}} &  
  h^{[2c]}\, {\cal J}_{3-10}^{(2)}\, f^{(0)}
  &  \bar h_v(x) h_v(0)
  \\ 
  \hat T^{(2La)}
& 
     \epsfxsize=6cm \lower30pt\hbox{\epsfbox{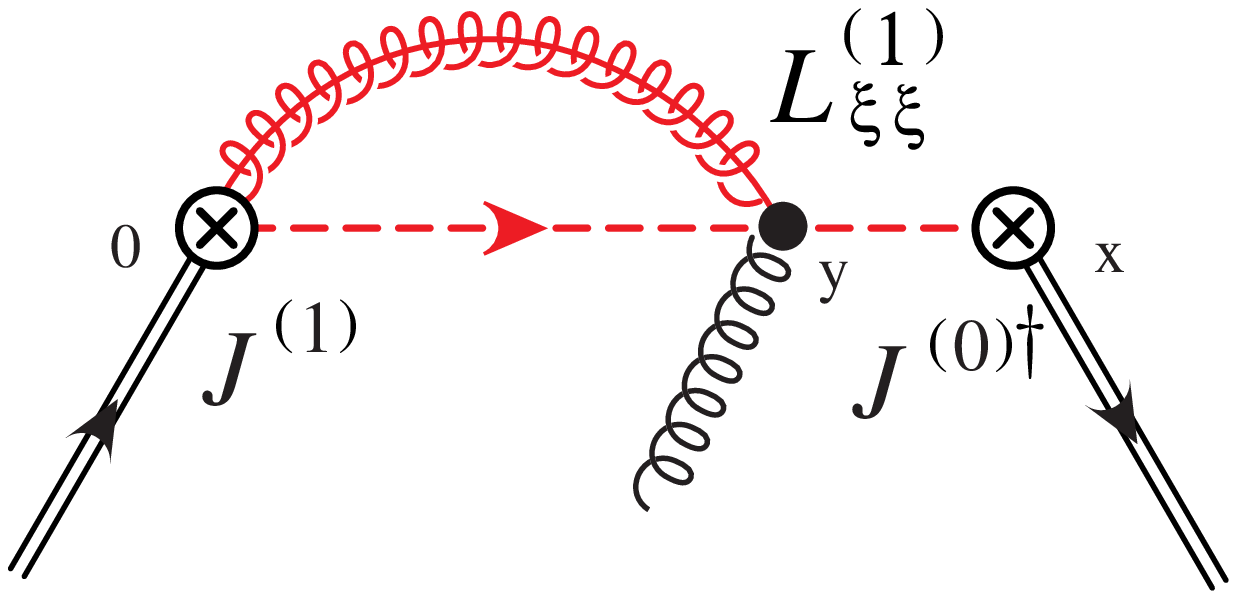}} &  
  h^{[2La]}\, {\cal J}_{j'}^{(0)} \, g_{11,12}^{(2)} & 
  \begin{array}{c} 
   \bar h_v(x) D_\perp(y) h_v(0)
  \end{array}
  \\ \\
  \hat T^{(2Lb)} 
& 
     \epsfxsize=6cm \lower30pt\hbox{\epsfbox{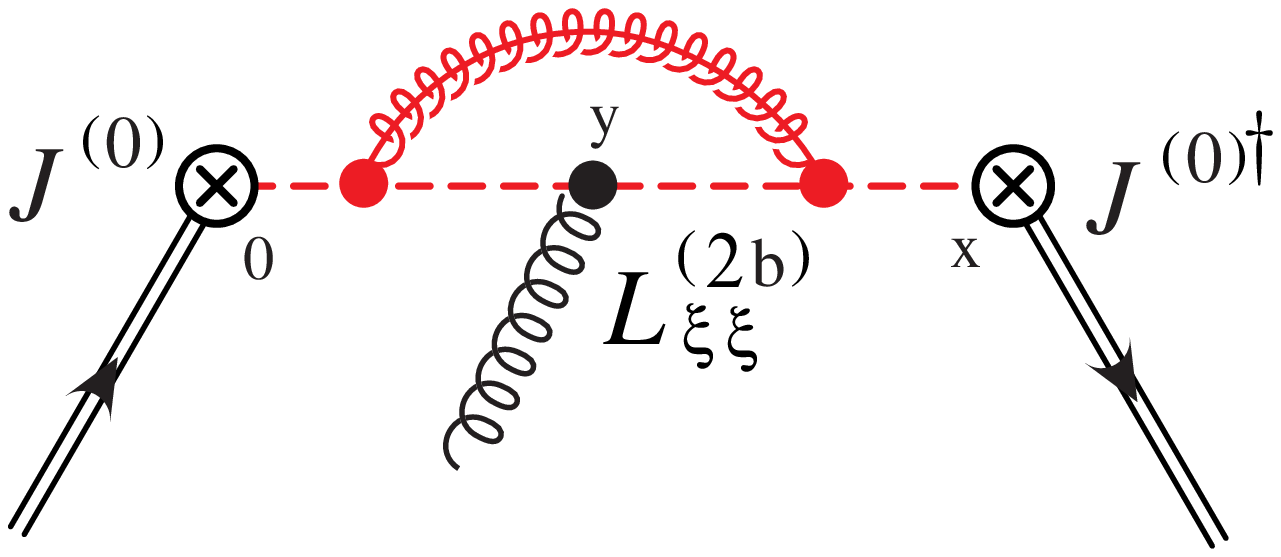}} &  
  h^{[2Lb]}\, {\cal J}_{j'}^{(0)} \, g_{13,14}^{(2)} &
  \begin{array}{c} 
  \bar h_v(x) \bn\mcdot D(y) h_v(0)
  \end{array}
  \\ \\
  \hat T^{(2LL)}
& 
     \epsfxsize=6.5cm \lower25pt\hbox{\epsfbox{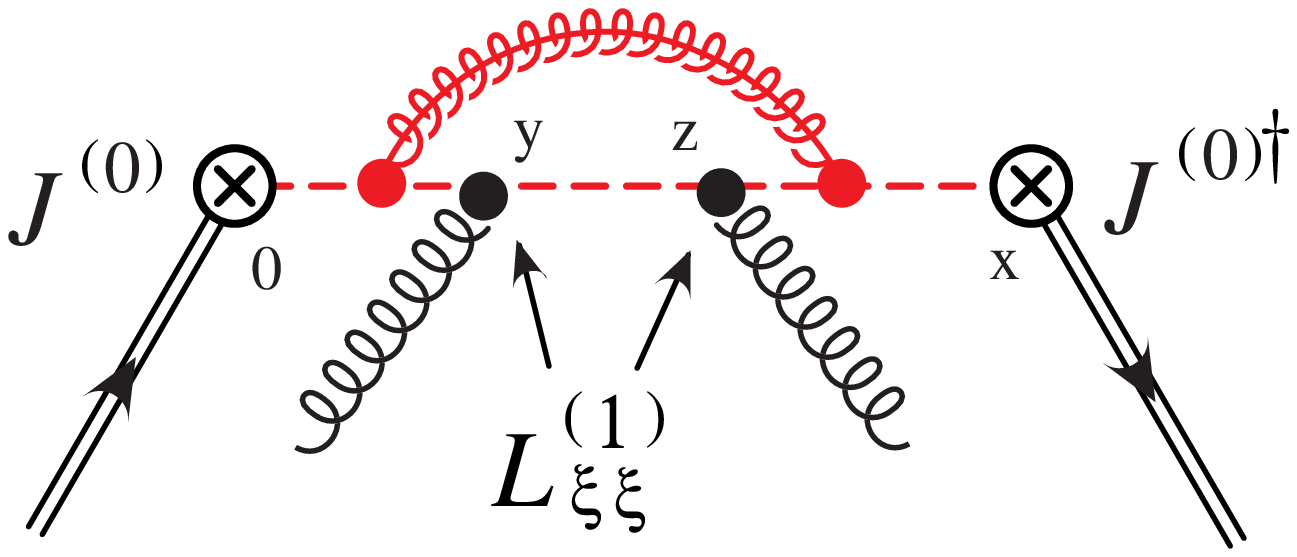}} &  
  h^{[2LL]}\,{\cal J}_{j'}^{(-2)}\, g_{15-26}^{(4)} & 
  \begin{array}{c} \bar h_v(x) D_\perp(y) D_\perp(z) h_v(0)
  \end{array}
\end{array}
\]
\caption{Time-ordered products that are of order $\Lambda/m_b$, but have
  jet functions that start at one-loop order. The last three rows introduce 
  new shape functions that were not present at tree level. Vertices that are 
  not labeled are from ${\cal L}^{(0)}_{\xi\xi}$. \label{table_T2}} 
\end{table}
\begin{table} 
\[
\begin{array}{ccccccc}
\hbox{T-product} & & \hbox{Example Diagram} & & \hspace{-0.4cm}
  \begin{array}{c}\mbox{Hard, Jet, and}\\ \mbox{Shape Functions}\end{array} 
  \hspace{-0.4cm} & 
  \hbox{Usoft Operator}   
  \\ \hline\\
  \hat T^{(2Ga)} & & 
     \epsfxsize=6.5cm \lower25pt\hbox{\epsfbox{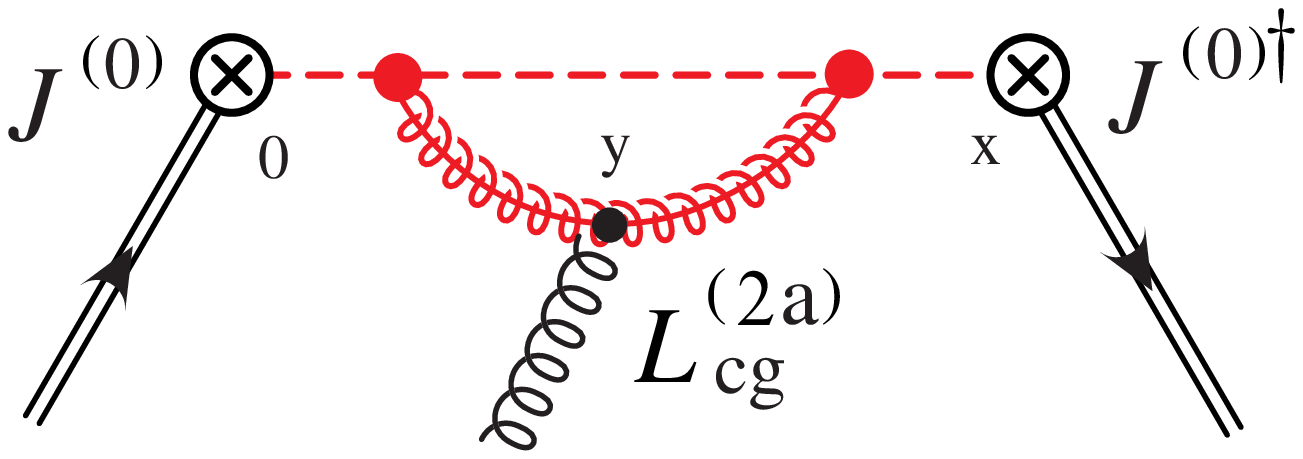}} &  &
  \begin{array}{c}
    h^{[2Ga]}\, {\cal J}_{j'}^{(-2)}\, f_{3,4}^{(4)} 
  \end{array}
  & 
  \begin{array}{l} 
    \bar h_v(x) (D_\perp D_\perp)(y)  h_v(0)
  \end{array}
  \\ \\
 \hat T^{(2Lb)} & & 
     \epsfxsize=6.5cm \lower25pt\hbox{\epsfbox{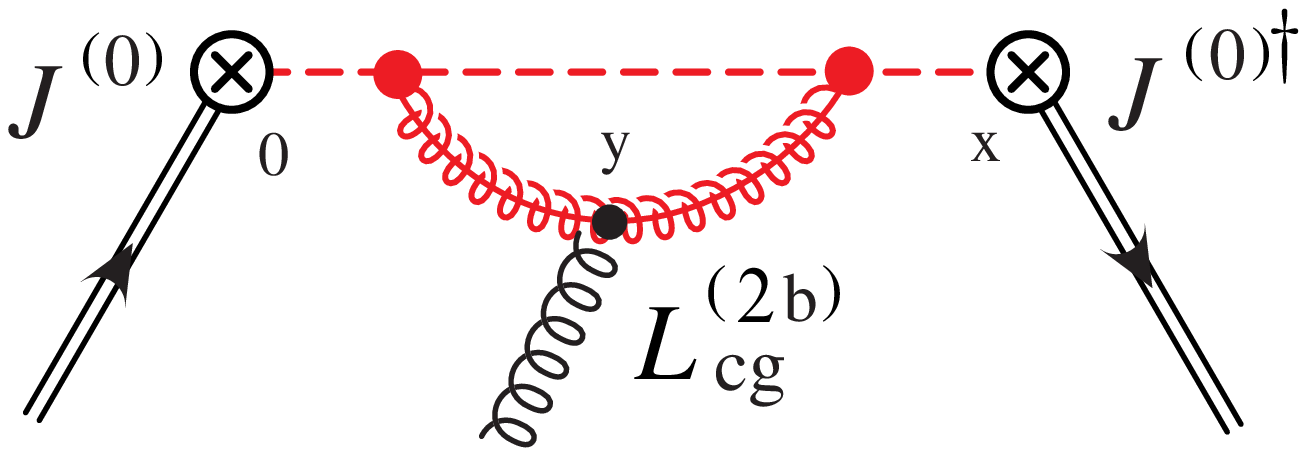}} &  &
  \begin{array}{c}
    h^{[2Lb]}\, {\cal J}_{j'}^{(0)} \, g_{13,14}^{(2)}
  \end{array}
  & 
  \begin{array}{l} 
    \bar h_v(x) \bn\mcdot D(y)  h_v(0)
  \end{array}
  \\ \\
  \hat T^{(2La)} & & 
     \epsfxsize=6.5cm \lower25pt\hbox{\epsfbox{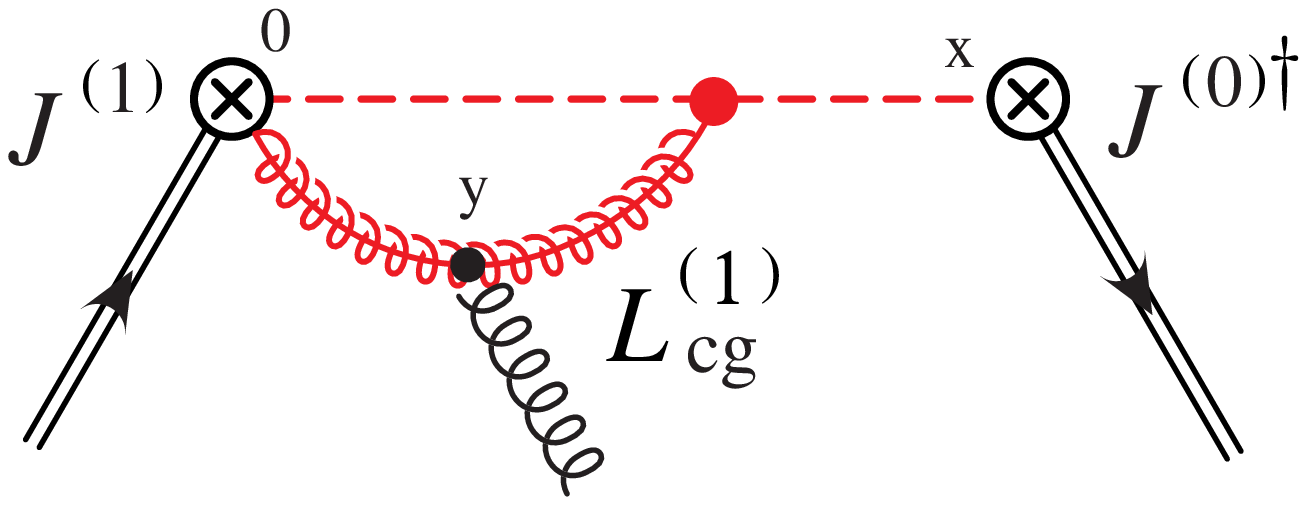}} &  &
  h^{[2La]}\, {\cal J}_{j'}^{(0)}\, g_{11,12}^{(2)} & 
  \begin{array}{c} \bar h_v(x) D_\perp(y)  h_v(0)
  \end{array}
\end{array}
\]
\caption{Examples of non-abelian terms in time-ordered products that are of 
  order $\Lambda/m_b$ and have
  jet functions that start at one-loop order. 
 \label{table_T3}} 
\end{table}

To enumerate all the possible time-ordered product contributions at this order
we consider all possible combinations of SCET currents and Lagrangians from
Sec.~\ref{sect_J},
\begin{eqnarray}
  J^{(n_1)\dagger} {\cal L}^{(n_2)} \cdots {\cal L}^{(n_{j-1})} J^{(n_j)} \,,
\end{eqnarray}
where $n_1+\ldots + n_j=2$ for NLO, i.e. ${\cal O}(\lambda^2)$.  It is useful to
divide these time-ordered products into two categories, those that have a jet
function that starts at tree level and those whose jet function starts at
one-loop order only. To determine into which category a time-ordered product 
falls, we first note that the jet functions are vacuum-to-vacuum matrix 
elements, so all collinear fields are contracted.  Since there are no external 
$\perp$-momenta, at tree level all collinear lines have no $\perp$-momentum 
and factors of ${\cal P}_\perp$ all vanish. (This is also true beyond tree 
level for factors of ${\cal P}_\perp$ that act on all collinear fields in a $J$ or $J^\dagger$.)
Thus, for example, the product $J^{(1b)\dagger} J^{(1b)}$ has a jet function
that starts at one-loop order since we must contract both the collinear quark 
and gluon lines. A second example consists of time-ordered products that 
involve a ${\cal L}_{\xi\xi}^{(1)}$ insertion, 
in which neither the ${\cal P}_\perp$ nor the $A_{n}^\perp$ in the $D_\perp^c$
can contribute at tree level.

The time-ordered products that appear already at tree level will be most
important phenomenologically, and include
\begin{eqnarray} \label{T2tree}
  \hat T^{(2H)} &=& -i \int\!\! d^4x\, d^4y \: e^{-i r\cdot x}\ 
      {\rm T} \big[ J^{(0)\dagger}(x)\:  i{\cal L}_h^{(2)}(y)\: J^{(0)}(0) 
        \big] \,,\\
  \hat T^{(2a)} &=& -i  \int\!\! d^4x \: e^{-i r\cdot x}\ 
      \sum_{\kappa=a,b} {\rm T} \big[ J^{(2\kappa)\dagger}(x)\: J^{(0)}(0) 
       +  J^{(0)\dagger}(x)\: J^{(2\kappa)}(0)
 \big] \,,
 \nn \\
 \hat T^{(2L)} &=& -i \int\!\! d^4x\, d^4y \: e^{-i r\cdot x}\ 
      {\rm T} \big[ J^{(0)\dagger}(x)\: i{\cal L}_{\xi\xi}^{(2a)}(y) J^{(0)}(0) 
        \big] \,,\nn \\
 \hat T^{(2q)} &=& - \frac{i}{2} \int\!\! d^4x\, d^4y\, d^4z \: e^{-ir\cdot x} \
      {\rm T} \big[ J^{(0)\dagger}(x)\: i{\cal L}_{\xi q}^{(1)}(y)
     \: i{\cal L}_{\xi q}^{(1)}(z) J^{(0)}(0) 
        \big] \,.\nn
\end{eqnarray}
These time-ordered products give jet functions that are either the same as at
lowest order, namely $J^{(0)}$, or enhanced by two or four powers of
$\lambda$, namely $J^{(-2,-4)}$. The corresponding shape functions are power 
suppressed relative to $f^{(0)}$ by two, four and six powers respectively. 

The most pertinent information is summarized in Table~\ref{table_T}, along with
examples of Feynman diagrams and the forms of the non-perturbative usoft
operators.  The overall power of $\lambda^2$ is obtained by simply adding the
superscripts in the operator insertions.  Dividing this overall power into that
of the jet and usoft terms is slightly more involved since it depends on the
individual operators. At tree level it is simply that every additional collinear
propagator enhances the corresponding jet function by $\lambda^{-2}$. Generating
additional propagators requires inserting subleading operators with additional
usoft fields that produce power-suppressed shape functions. The power counting
in \SCETa restricts us to consider at most two ${\cal O}(\lambda)$ operator
insertions, so this guarantees that the greatest enhancement from the jet 
function is ${\cal O}(\lambda^{-4})$ and the set of possible terms is finite.

From Table~\ref{table_T}, we see that for $T^{(2a)}$ and $T^{(2H)}$ the 
collinear fields give a jet function that is identical to ${\cal J}^{(0)}$ 
defined in Eq.~(\ref{jetdef}), but have $\Lambda/Q$-suppressed shape functions 
$f^{(2)}_{1,2}$ and $f_0^{(2)}$.  
For $T^{(2L)}$ we find an enhancement of $Q/\Lambda$ in the jet function 
${\cal J}^{(-2)}$, but shape functions that are further suppressed, namely
$f_{3,4}^{(4)}$, which are down by $\Lambda^2/Q^2$.\footnote{Note that in the
  subleading Lagrangian ${\cal L}_{\xi\xi}^{(2)}$ the two $\perp$ derivatives
  appear at the same spacetime point, providing a simple explanation for why
  this occurred to all orders in the twist expansion in
  Ref.~\cite{Bauer:2001mh}.} Finally for $T^{(2q)}$ we have the biggest
enhancement, $Q^2/\Lambda^2$, through the jet function ${\cal J}^{(-4)}$, and 
a shape function $f^{(6)}_{5,6}$ involving a four-quark operator, which is down
by $\Lambda^3/Q^3$.

In Table~\ref{table_T} we also notice that it's only the $J^{(0)}$ currents that
appear, plus the subleading currents $J^{(2a)}$ and $J^{(2b)}$ whose Wilson
coefficients are related to the coefficients in $J^{(0)}$ by reparameterization
invariance (see Appendix~\ref{app_RPI}). This implies that the logarithms
encoded in the running of the subleading hard functions $h^{1-8}$ are the same
as the logarithms that can be resummed in $h^0$. (Here the term ``logarithms''
includes Sudakov double logarithms as well as the usual single logarithms.)
This result depends on the fact that the Lagrangians that are inserted do not
run.  This implies that the logarithms resummed between the scales $m_b^2$ and
$m_b\Lambda$ are universal for these terms (besides the simple HQET running from
$c_F(\mu)$ in ${\cal L}_h^{(2)}$, which is easily taken into account).
Additional logarithms occur below the $m_b\Lambda$ scale and there is no reason
to expect that they are universal. A computation of the anomalous dimensions of
the corresponding soft operators would allow these additional logarithms to be
resummed. Finally, for the time-ordered products that appear at one-loop order
in the jet functions, even the logarithms between $m_b^2$ and $m_b\Lambda$ are
not the same as at LO. They are, however, entirely determined by the running of
the $J^{(1)}$ currents that were calculated in Ref.~\cite{Hill:2004if}.

At one-loop order in the jet function, the remaining time-ordered products start
contributing. They are
\begin{eqnarray} \label{T2loop}
 \hat T^{(2b)} &=& -i \int\!\! d^4x\: e^{-i r\cdot x}\!\!\!
     \sum_{\kappa=a,b,\, \ell=a,b} \!\!
      {\rm T} \big[ J^{(1\kappa)\dagger}(x)\: J^{(1\ell)}(0) 
        \big] \,,
    \\
  \hat T^{(2c)} &=& -i \int\!\! d^4x \: e^{-i r\cdot x}
     \sum_{\ell=c,d,e,f} 
     {\rm T} \big[ J^{(2\ell)\dagger}(x)\: J^{(0)}(0) 
       +  J^{(0)\dagger}(x)\: J^{(2\ell)}(0) \big] \,,\nn
     \\
   \hat T^{(2La)} &=& -i \int\!\! d^4x\,d^4y \: e^{-i r\cdot x}\!
      \sum_{\kappa=a,b}\! {\rm T} \big[ 
       J^{(1\kappa)\dagger}(x)\: i{\cal L}_{\xi\xi}^{(1)}(y)  J^{(0)}(0) 
       + J^{(0)\dagger}(x)\: i{\cal L}_{\xi\xi}^{(1)}(y)  J^{(1\kappa)}(0)
       \big]
        \,\nn\\
    && -i \int\!\! d^4x\,d^4y \: e^{-i r\cdot x}\!
      \sum_{\kappa=a,b}\!
      {\rm T} \big[ J^{(1\kappa)\dagger}(x)\: i{\cal L}_{cg}^{(1)}(y) J^{(0)}(0) 
       \!+\! J^{(0)\dagger}(x)\: i{\cal L}_{cg}^{(1)}(y)  J^{(1\kappa)}(0) \big]
     \,, \nn\\
   \hat T^{(2Lb)} &=& -i \int\!\! d^4x\, d^4y \: e^{-i r\cdot x}\ 
      {\rm T} \big[ J^{(0)\dagger}(x)\: i{\cal L}_{\xi\xi}^{(2b)}(y) J^{(0)}(0) 
        \big]  \,\nn\\
   && -i \int\!\! d^4x\, d^4y \: e^{-i r\cdot x}\ 
      {\rm T} \big[ J^{(0)\dagger}(x)\: i{\cal L}_{cg}^{(2b)}(y) J^{(0)}(0) 
        \big]
      \,,\nn 
\end{eqnarray}
\begin{eqnarray}
   \hat T^{(2LL)} &=& - \frac{i}{2} \int\!\! d^4x\, d^4y \, d^4z \: e^{-i r\cdot x}\ 
      {\rm T} \big[ J^{(0)\dagger}(x)\: i{\cal L}_{\xi\xi}^{(1)}(y)
     \: i{\cal L}_{\xi\xi}^{(1)}(z) J^{(0)}(0) 
        \big] \nn \\
 && - {i} \int\!\! d^4x\, d^4y \, d^4z \: e^{-i r\cdot x}\ 
      {\rm T} \big[ J^{(0)\dagger}(x)\: i{\cal L}_{\xi\xi}^{(1)}(y)
     \: i{\cal L}_{cg}^{(1)}(z) J^{(0)}(0) 
        \big] \nn \\
 && - \frac{i}{2} \int\!\! d^4x\, d^4y \, d^4z \: e^{-i r\cdot x}\ 
      {\rm T} \big[ J^{(0)\dagger}(x)\: i{\cal L}_{cg}^{(1)}(y)
     \: i{\cal L}_{cg}^{(1)}(z) J^{(0)}(0) 
        \big] \,,\nn \\
   \hat T^{(2Ga)} &=& -i \int\!\! d^4x\, d^4y \: e^{-i r\cdot x}\ 
      {\rm T} \big[ J^{(0)\dagger}(x)\: i{\cal L}_{cg}^{(2a)}(y) J^{(0)}(0) 
        \big]
       \,.\nn
\end{eqnarray}
Information about how these time-ordered products contribute to the
factorization theorems at ${\cal O}(\lambda^2)$ is summarized in
Tables~\ref{table_T2} and~\ref{table_T3}.  Note that $T^{(2b)}$ and $T^{(2c)}$
match on to subleading jet functions and the leading-order shape function, and 
so these ${\cal O}(\alpha_s)$ corrections can be calculated without 
introducing new non-perturbative information.  The remaining T-products give 
new subleading shape functions that were not present at tree level.  
Together the terms in Eqs.~(\ref{T2tree}) and (\ref{T2loop}) provide the 
complete set of time-ordered products that contribute at this order in 
$\lambda$ and at any order in $\alpha_s$.

Adding up the contributions from the various subleading time-ordered products, we
find that
\begin{eqnarray} \label{Wfact00b}
W_i^{(2)f}  &=& 
  \frac{h_i^{0f}(\bn\mcdot p)}{2m_b} \: 
  \int_{0}^{p_X^+} \!\! dk^+\ 
  {\cal J}^{(0)}(\bn\mcdot p\, k^+,\mu )\:
   f^{(2)}_0\big(k^+ + r^+,\mu\big) \,
  \\
 &+&
  \sum_{r=1}^2
  \frac{h_i^{rf}(\bn\mcdot p)}{m_b} \: 
  \int_{0}^{p_X^+} \!\! dk^+\ 
  {\cal J}^{(0)}(\bn\mcdot p\, k^+,\mu )\:
   f^{(2)}_r\big( k^+ + r^+,\mu\big) \,
  \nn\\
 &+& 
  \sum_{r=3}^4
  \frac{ h_i^{rf}(\bn\mcdot p)}{m_b} 
  \int\!\! dk_1^+\: dk_2^+ \
   {\cal  J}^{(-2)}_{1\pm 2}(\bn\mcdot p\, k_j^+,\mu )\:
   f^{(4)}_r\big(k_j^+ + r^+,\mu\big) \,
 \nn\\
 &+&
  \sum_{r=5}^6
    \frac{h_i^{rf}(\bn\mcdot p)}{\bn\mcdot p} 
  \int\!\! dk_1^+ dk_2^+ dk_3^+  \ 
 {\cal J}^{(-4)}_1(\bn\mcdot p\, k_{j'}^+,\mu )\:
   f^{(6)}_r\big(k_{j'}^+ + r^+,\mu\big) \, \nn\\[5pt]
 &+&
  \frac{h_i^{00f}(\bn\mcdot p)}{m_b} \: 
  \int_{0}^{p_X^+} \!\! dk^+\ 
  {\cal J}^{(0)}(\bn\mcdot p\, k^+,\mu )\:
   g^{(2)}_0\big( k^+ + r^+,\mu\big) \,
  \nn\\
 &+& 
  \sum_{r=3}^4
  \frac{ h_i^{rf}(\bn\mcdot p)}{m_b} 
  \int\!\! dk_1^+\: dk_2^+ \
   {\cal  J}^{(-2)}_{3\pm 4}(\bn\mcdot p\, k_j^+,\mu )\:
   g^{(4)}_r\big(k_j^+ + r^+,\mu\big) \, \nn\\
 &+&
  \sum_{r=5}^6
    \frac{h_i^{rf}(\bn\mcdot p)}{\bn\mcdot p} 
  \int\!\! dk_1^+ dk_2^+ dk_3^+  \ 
 {\cal J}^{(-4)}_2(\bn\mcdot p\, k_{j'}^+,\mu )\:
   g^{(6)}_r\big(k_{j'}^+ + r^+,\mu\big) \, \nn\\
 &+&
  \sum_{r=7}^8
    \frac{h_i^{rf}(\bn\mcdot p)}{\bn\mcdot p} 
  \int\!\! dk_1^+ dk_2^+ dk_3^+  \ 
 \big[ {\cal J}^{(-4)}_3(\bn\mcdot p\, k_{j'}^+,\mu )\:
   g^{(6)}_r\big(k_{j'}^+ + r^+,\mu\big) \, \nn\\
  &&\qquad\qquad\qquad\qquad\qquad\qquad
   + {\cal J}^{(-4)}_4(\bn\mcdot p\, k_{j'}^+,\mu )\:
   g^{(6)}_{r+2}\big(k_{j'}^+ + r^+,\mu\big) \big] \nn \\
 &+& \sum_{m=1,2} \int\!\! dz_1 dz_2\ 
  \frac{h_i^{[2b]m+8}(z_1,z_2,\bn\mcdot p)}{m_b} \:
  \int_{0}^{p_X^+} \! \!\! dk^+\: 
   {\cal J}^{(2)}_m(z_{1},z_2,p_X^-\, k^+ )\: 
   f^{(0)}(k^+  \!+\!\overline\Lambda\!-\! p_X^+) 
 \nn 
\end{eqnarray}
\begin{eqnarray}
 &+&  \sum_{m=3,4} 
  \frac{h_i^{[2c]m+8}(\bn\mcdot p)}{m_b} \:
  \int_{0}^{p_X^+} \! \!\! dk^+\: 
   {\cal J}^{(2)}_m(p_X^-\, k^+ )\: 
   f^{(0)}(k^+  \!+\!\overline\Lambda\!-\! p_X^+) 
 \nn \\  
 &+& \sum_{m=5}^{10} \int\!\! dz_1  \:
  \frac{h_i^{[2c]m+8}(z_1,\bn\mcdot p)}{m_b} \:
  \int_{0}^{p_X^+} \! \!\! dk^+\: 
   {\cal J}^{(2)}_m(z_{1},p_X^-\, k^+ )\: 
   f^{(0)}(k^+  \!+\!\overline\Lambda\!-\! p_X^+)
  \nn\\
 &+& W_i^{[2La]f}[\, g_{11,12}^{(2)}\, ] \nn\\
 &+&  W_i^{[2Lb]f}[\, g_{13,14}^{(2)}\, ] \nn\\
 &+&  W_i^{[2LL]f}[\, g_{15-26}^{(4)}\, ]\nn\\
 &+&  W_i^{[2Ga]f}[\, f_{3,4}^{(4)}\, ]   \nn \,,
\end{eqnarray}
where $j = 1,2$ and $j' = 1,2,3$ and ${\cal J}_{1\pm 2}^{(-2)}={\cal
  J}_1^{(-2)}\pm {\cal J}_2^{(-2)}$ for $r=3,4$ respectively. Recall that 
$f=s$ for $B\to X_s\gamma$ and $f=u$ for $B\to X_u\ell\bar\nu$. 
If we work at tree level in the jet functions, then only the first four lines 
of Eq.~(\ref{Wfact00b}) are required (with shape functions $f_{0-6}$). 
In our notation, the $f_i$ denote shape functions that appear from tree-level 
matching, while the $g_i$ denote shape functions that show up only when
one-loop corrections are considered for the jet functions.

\subsection{Tree-Level Matching Calculations}   \label{sectJet2}

In this section, we show how computing the four tree-level diagrams in
Table~\ref{table_T} immediately allows us to obtain the tree-level jet 
functions, properly convoluted with the non-perturbative shape functions that 
show up at NLO order. Note that no summation of operators is necessary. 
An operator-based derivation of the factorization theorems at all orders in 
the jet functions is given later, in Sec.~\ref{sectT2s}.

First consider the computation of Fig.~\ref{fig:LOa} using the LO jet function
from the time-ordered product in Eq.~(\ref{Tlo1}), with currents as in
Eq.~(\ref{Tlo2}). Working entirely in momentum space and contracting the
collinear quark fields in the diagram gives
\begin{eqnarray} \label{calc1}
 && (-i) \overline {\cal H}_v(\ell^+) \overline \Gamma_{\mu}^{(0)} 
    i\frac{\nslash}{2} \frac{\bn\mcdot p}{(\bn\mcdot p k^+ + i\epsilon)}
   \Gamma_{\nu}^{(0)} {\cal H}_v(\ell^+) \,,
\end{eqnarray}
where $\bn\mcdot p$ is the large momentum flowing through the collinear quark
propagator. Taking $(-1/\pi)$ times the imaginary part gives 
\begin{eqnarray} \label{Jtree0}
{\cal J}^{(0)}(k^+) &=& 
   \delta(k^+) + {\cal O}(\alpha_s) \,,
\end{eqnarray}
for the tree-level jet function, which is equal to $\delta(\ell^+-r^+)$ by
momentum conservation 
since
\begin{eqnarray}
  \ell^+ = k^+ + r^+ \,.
\end{eqnarray}
Thus, from Eq.~(\ref{calc1}) we can
directly write the convolution of jet and soft factors as 
\begin{eqnarray}
 && {\rm Tr} \big[P_v \overline \Gamma_{\mu}^{(0)} 
    \frac{\nslash}{2}  \Gamma_{\nu}^{(0)}\big] 
     \int d\ell^+ \Big\{ {\cal J}^{(0)}(\ell^+-r^+) \Big\}\ \Big\{
    \bar { h}_v Y
    \delta(\ell^+ \!-\! in\mcdot\partial) Y^\dagger {h}_v  \Big\} \,,
\end{eqnarray}
where we have used ${\cal H}_v = Y^\dagger h_v$. Performing the integral at tree
level gives the well-known fact that the decay spectrum is determined by the
shape function evaluated at $r^+$. Note that in SCET we needed to compute only
one diagram rather than resumming an infinite sum of operators.

\begin{figure}[t!]
 \centerline{
  \mbox{\epsfysize=3.truecm \hbox{\epsfbox{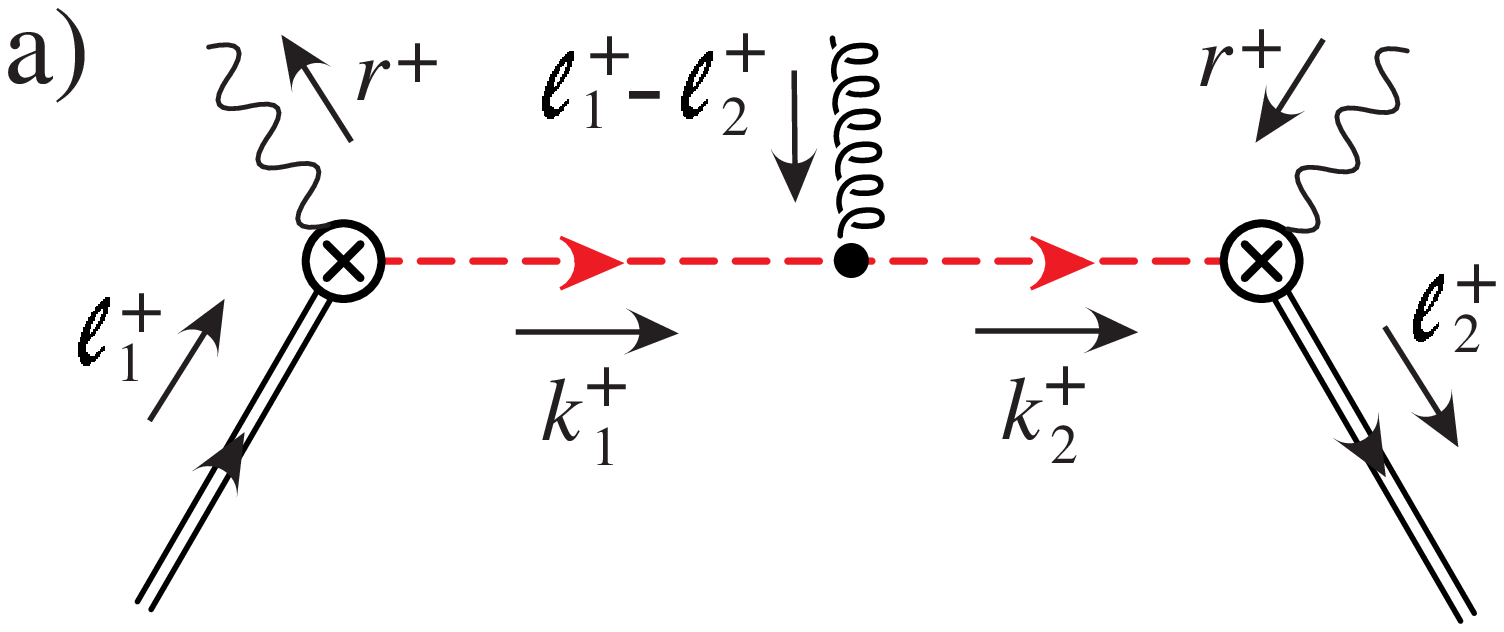}} }\qquad\qquad
  \mbox{\epsfysize=3.truecm \hbox{\epsfbox{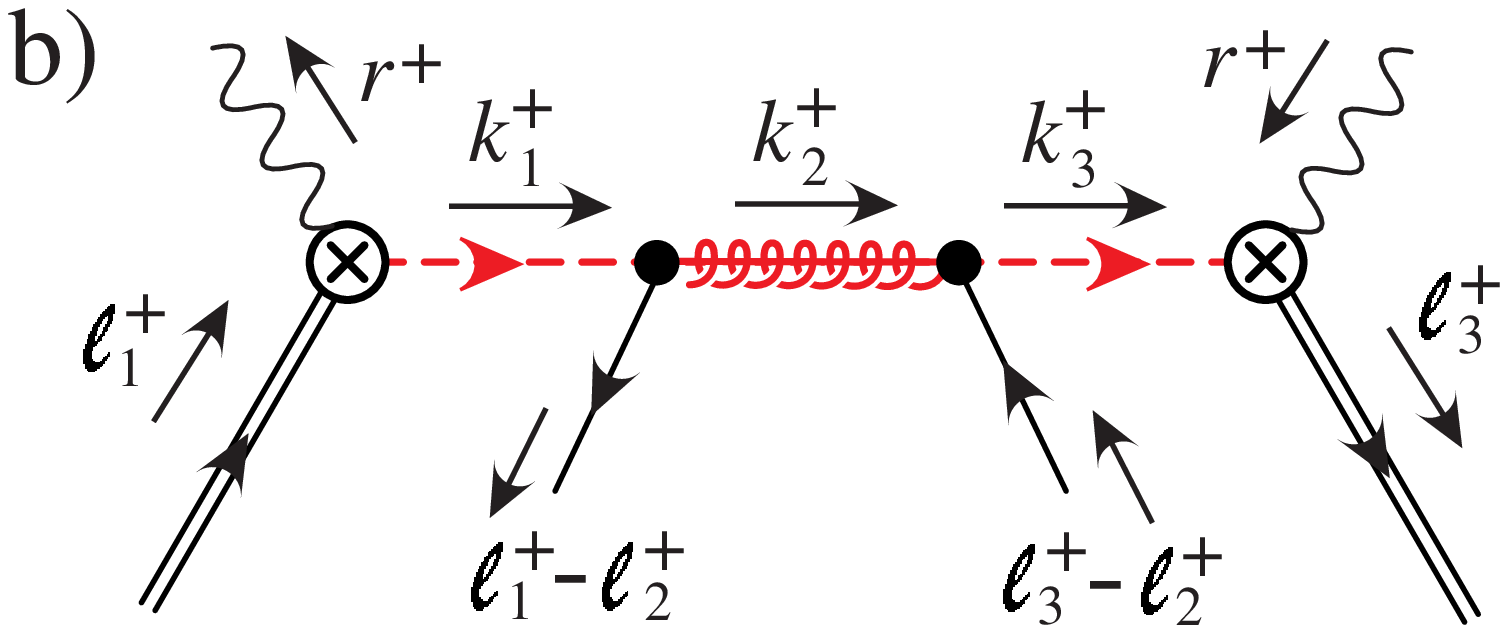}} }
  }
 \vskip -.2cm 
 {\caption[1]{Momentum routing for a) $T^{(2L)}$ and b) $T^{(2q)}$. In a), the
     operator with one gluon comes from $[i\cDslash_{us}^\perp
     i\cDslash_{us}^\perp]$.}
\label{fig:NLOa} }
 \vskip -0.1cm
\end{figure}
At NLO the tree-level diagrams are shown in Table~\ref{table_T}. The graphs for
$T^{(2H)}$ and $T^{(2a)}$ involve the same calculation as in Eq.~(\ref{calc1})
but just leave different soft operators. 

For $T^{(2L)}$ the momentum routing is shown in Fig.~\ref{fig:NLOa}a.
Contracting the collinear quark propagators in the graph gives
\begin{eqnarray}
&& (-i)(i)^3 \overline {\cal H}_v(\ell_2^+) \overline \Gamma_{\mu}^{(0)} 
    \frac{\nslash}{2} \frac{\bn\mcdot p}{(\bn\mcdot p\, k_2^+ \!+\! i\epsilon)}
    [i\cDslash_{us}^\perp\,  i \cDslash_{us}^\perp](\ell_2^+\!-\!\ell_1^+)
   \: \frac{1}{\bn\mcdot p}
  \frac{\bnslash}{2}\,
    \frac{\nslash}{2} \frac{\bn\mcdot p}{(\bn\mcdot p\, k_1^+ \!+\! i\epsilon)}
   \Gamma_{\nu}^{(0)} {\cal H}_v(\ell_1^+)  \nn\\
 &&= - \frac{1}{\bn\mcdot p} {\rm Tr}\Big[P_v \overline \Gamma_{\mu}^{(0)} 
    \frac{\nslash}{2} \gamma_\perp^\alpha\gamma_\perp^\beta \Gamma_{\nu}^{(0)}
    \Big]
    \frac{1}{(k_1^+\!+\!i\epsilon)(k_2^+\!+\!i\epsilon)}
   \overline {\cal H}_v(\ell_2^+) 
    [i\cD^{\perp\alpha}_{us}\,  i \cD^{\perp\beta}_{us}](\ell_2^+\!-\!\ell_1^+)
   {\cal H}_v(\ell_1^+) \,,\nn\\
\end{eqnarray}
where momentum conservation sets $k_1^+=\ell_1^+ - r^+$ and $k_2^+=\ell_2^+-r^+$. 
Taking $(-1/\pi)$ times the imaginary part gives the tree-level jet function
\begin{eqnarray} \label{Jtreem2}
 {\cal J}^{(-2)}_1(k_j^+)  & = & 
  \frac{\delta(k_1^+)-\delta(k_2^+)}
       {k_2^+ - k_1^+} + {\cal O}(\alpha_s) \,, 
\end{eqnarray}
and the convolutions
\begin{eqnarray}
 && - \frac{1}{\bn\mcdot p} {\rm Tr}\Big[P_v \overline \Gamma_{\mu}^{(0)} 
    \frac{\nslash}{2} \gamma_\perp^\alpha\gamma_\perp^\beta \Gamma_{\nu}^{(0)}
    \Big] \int\!\! d\ell_1^+ d\ell_2^+ 
    \Big\{ {\cal J}^{(-2)}(\ell_j^+-r^+) \Big\} \nn\\
 && \qquad \times \Big\{ \bar h_v Y
    \delta(\ell_2^+-in\mcdot \partial) Y^\dagger i D_{us}^{\perp\alpha} 
     i D_{us}^{\perp\beta}  Y \delta(\ell_1^+-in\mcdot\partial) Y^\dagger h_v
   \Big\} \,.
\end{eqnarray}

Finally, we consider the NLO time-ordered product $T^{(2q)}$ with the momentum
routing in Fig.~\ref{fig:NLOa}b. This graph gives
\begin{eqnarray}
 && 
  (-i)^2(i)^4 g^2 \Big[\overline {\cal H}_v(\ell_3^+) \overline \Gamma_{\mu}^{(0)} 
    \frac{\nslash}{2} \frac{\bn\mcdot p}{(\bn\mcdot p\, k_3^+ \!+\! i\epsilon)}
    \gamma_\perp^\tau T^A \psi_{us}(\ell_3^+\!-\!\ell_2^+)\Big] 
   \frac{1}{\bn\mcdot p\,k_2^+ \!+\! i\epsilon}\\
 && \times \Big[
   \bar \psi_{us}(\ell_1^+\!-\!\ell_2^+)\gamma^\perp_\tau T^A
    \frac{\nslash}{2} \frac{\bn\mcdot p}{(\bn\mcdot p\, k_1^+ \!+\! i\epsilon)}
   \Gamma_{\nu}^{(0)} {\cal H}_v(\ell_1^+) \Big]  \nn\\
 &&
  = - \frac{1}{\bn\mcdot p} 
  \frac{g^2}{(k_1^+\!+\!i\epsilon)(k_2^+\!+\!i\epsilon)(k_3^+\!+\!i\epsilon)}
 \Big[\overline {\cal H}_v \overline \Gamma_{\mu}^{(0)} 
    \frac{\nslash}{2} 
    \gamma_\perp^\tau T^A \psi_{us}\Big] 
   \Big[\bar \psi_{us}T^A \gamma^\perp_\tau  \frac{\nslash}{2}
   \Gamma_{\nu}^{(0)} {\cal H}_v \Big] \nn \,,
\end{eqnarray} 
where $k_1^+=\ell_1^+ - r^+$, $k_2^+=\ell_2^+-r^+$, and $k_3^+=\ell_3^+-r^+$.
Taking $(-1/\pi)$ times the imaginary part gives the tree-level jet function
\begin{eqnarray} \label{Jtreem4}
 {\cal J}^{(-4)}_1(k_j^+)  & = & 
  {g^2}
  \left[\frac{\delta(k_1^+)}{(k_2^+)(k_3^+)} 
      + \frac{\delta(k_2^+)}{(k_1^+)(k_3^+)} 
      + \frac{\delta(k_3^+)}{(k_1^+)(k_2^+)} 
      - \pi^2 \delta(k_1^+)\delta(k_2^+)\delta(k_3^+)
      \right] \,. 
\end{eqnarray}
Writing out the convolutions for these four-quark operators, we find
\begin{eqnarray}
 && - \frac{1}{\bn\mcdot p}\int\!\! d\ell_1^+ d\ell_2^+ d\ell_3^+ 
    \Big\{ {\cal J}^{(-4)}(\ell_j^+-r^+) \Big\}
   \Big[\bar { h}_v Y \delta(\ell_3^+\!-\!in\mcdot\partial)
   \overline \Gamma_{\mu}^{(0)} 
    \frac{\nslash}{2} 
    \gamma_\perp^\tau T^A Y^\dagger q_{us}\Big] \nn\\
&&\qquad \times  \delta(\ell_2^+\!-\!in\mcdot\partial) 
   \Big[\bar q_{us} Y T^A \gamma^\perp_\tau  \frac{\nslash}{2}
   \Gamma_{\nu}^{(0)} \delta(\ell_1^+\!-\!in\mcdot\partial)  Y^\dagger h_v
   \Big] \,.
\end{eqnarray}

\subsection{NLO Soft Operators and Shape Functions}  \label{sectshape}

In this section, we define the shape functions that can appear in the NLO
factorization theorem. At tree level, many of these functions have already been
defined in Ref.~\cite{Bauer:2001mh} (we use a sequential enumeration of these
functions, and translate to the notation in Ref.~\cite{Bauer:2001mh} in
Table~\ref{table:notatn}).  We also define the additional shape functions with
four-quark operators that are not given in Ref.~\cite{Bauer:2001mh}, and later
in section~\ref{sectT2s} the shape functions that appear only beyond tree level
in the subleading jet functions.

We begin by enumerating the usoft operators. For the time-ordered product of the
LO shape-function operator and the subleading HQET Lagrangian we define
\begin{eqnarray} 
 O_0^{(2)}(\ell^+) &=&
  \int\!\! \frac{dx^-}{\!8\pi}\: e^{-\frac{i}{2} x^- \ell^+ } \!\! 
  \int\!\! d^4y \: 
   T\: \big[\bar h_v(\tilde x) Y(\tilde x,0) h_v(0)\ i O_h(y) \big]
  \,.
\end{eqnarray}
With $ig G_{us\perp}^{\mu\nu}=[iD_{us}^{\perp\mu}, iD_{us}^{\perp\nu}]$, the
remaining ultrasoft operators that play an important role include
\begin{eqnarray} \label{softops1}
   O_{1}^{\beta}(\ell^+)\: 
 &=& \int\!\! \frac{dx^-}{\!8\pi}\,  e^{-\frac{i}{2}\ell^+ x^-}
  \Big[ 
  \overline {\cal H}_v(\tilde x)   
   (  i\! {\overrightarrow {\cal D}_{\!us}^{\beta}} {\cal H}_v)(0) 
   + \mbox{(h.c., $\tilde x\leftrightarrow 0$)} \Big] \nn
   \\ 
 && 
  = \frac12 
   \bar h_v 
   \big\{ i D^{\beta}_{\rm us}, \delta(\ell^+ \!\! -\! in\mcdot D_{\rm us})\big\} 
      h_v 
  \,, \\
 P^\beta_{1\lambda}(\ell^+)
 &=& \int\!\! \frac{dx^-}{\!8\pi}\, e^{-\frac{i}{2}\ell^+ x^-} 
  \Big[ 
   \overline {\cal H}_v(\tilde x)   \gamma^T_{\lambda}\gamma_5
   (  i\! {\overrightarrow {\cal D}_{\!us}^{\beta}} {\cal H}_v)(0) 
  + \mbox{(h.c., $\tilde x\leftrightarrow 0$)} \Big]\nn\\
 &=& \frac12 \:
    \bar h_v \big\{  i D^{\beta}_{\rm us}, 
   \delta(\ell^+ \!\! -\! in\mcdot D_{\rm us})\big\}
    \gamma_T^\lambda \gamma_5  h_v 
  \,, \nn\\
  O^\beta_{2}(\ell^+)
 &=& \int\!\! \frac{dx^-}{\!8\pi}\, e^{-\frac{i}{2}\ell^+ x^-} 
  \Big[\!-\!i 
   {\rm T}\,   \overline {\cal H}_v(\tilde x)   
   (  i\! {\overrightarrow {\cal D}_{\!us}^{\beta}} {\cal H}_v)(0) 
  + \mbox{(h.c., $\tilde x\leftrightarrow 0$)} \Big]\nn\\
 &=& \frac{i}2 \:
    \bar h_v  \big[  i D^{\beta}_{\rm us}, 
   \delta(\ell^+ \!\! -\! in\mcdot D_{\rm us})\big] 
    h_v 
  \,, \nn\\
  P^\beta_{2\lambda}(\ell^+)
 &=& \int\!\! \frac{dx^-}{\!8\pi}\, e^{-\frac{i}{2}\ell^+ x^-} 
  \Big[\!-\!i 
   {\rm T}\,   \overline {\cal H}_v(\tilde x)   \gamma^T_{\lambda}\gamma_5
   (  i\! {\overrightarrow {\cal D}_{\!us}^{\beta}} {\cal H}_v)(0) 
  + \mbox{(h.c., $\tilde x\leftrightarrow 0$)} \Big]\nn\\
 &=& \frac{i}2 \:
    \bar h_v  \big[  i D^{\beta}_{\rm us}, 
   \delta(\ell^+ \!\! -\! in\mcdot D_{\rm us})\big] 
   \gamma_T^\lambda \gamma_5  h_v 
  \,, \nn
\end{eqnarray}
\begin{eqnarray}
  O_3^{\alpha\beta}(\ell^+_1,\ell^+_2) \! &=& 
   \int\!\! \frac{dx^- \, dy^-\,}{32\pi^2} 
    e^{-\frac{i}{2} \ell_2 x^-}  e^{-\frac{i}{2}(\ell^+_1 - \ell^+_2) y^-}
   \big[\overline {\cal H}_v(\tilde x) \{i D_{us}^{\perp\alpha}, i
   D_{us}^{\perp\beta}\}(\tilde y) {\cal H}_v(0) \big]
   \nonumber \\
  &=& \frac{1}{2}\:
  \bar h_v \delta(\ell_2^+ \!-\! in \mcdot D_{us} )\left\{i\cD^{\perp\alpha}_{us},
   i\cD^{\perp\beta}_{us}\right\}\delta(\ell_1^+ \!-\! i n \mcdot D_{us}) h_v\,,
    \nonumber \\
%
%
%
P_{4\lambda}^{\alpha\beta}(\ell^+_1,\ell^+_2)\! &=&
  i\int\!\! \frac{dx^- \, dy^-\,}{32\pi^2} 
    e^{-\frac{i}{2} \ell_2 x^-}  e^{-\frac{i}{2}(\ell^+_1 - \ell^+_2) y^-}
   \big[\overline {\cal H}_v(\tilde x) [i \cD_{us}^{\perp\alpha}, i
   \cD_{us}^{\perp\beta}](\tilde y) \gamma^T_\lambda\gamma_5\,
   {\cal H}_v(0) \big] \nn\\
 &=& -\frac{1}2 \:
 \bar h_v \delta(\ell^+_2\!-\! in \mcdot D_{us}  )g G_{us \perp}^{\alpha\beta}
  \delta(\ell^+_1\!-\! i n \mcdot D_{us}  )\gamma_\lambda^T \gamma_5  h_v 
  \,, \nn 
\end{eqnarray}
and the four-quark operators
\begin{eqnarray}
O_5^{\alpha\beta}(\ell^+_{1,2,3})  &=& \!\!\!
   \int\!\! \frac{dx^-  dy^- dz^-}{128\pi^3} 
    e^{-\frac{i}{2} \{ \ell^+_3 x^- +(\ell^+_2 - \ell^+_3) 
    y^- +(\ell^+_1 - \ell^+_2) z^- \} }  
   \big[\overline {\cal H}_v(\tilde x) \gamma^\beta P_L T^A \psi_{us}^\bn(\tilde
   y)
 \\
 &&\quad \times  
   \bar\psi_{us}^\bn(\tilde z) \gamma^\alpha P_L T^A {\cal H}_v(0) \big] 
 \nn\\
 &=& \frac{1}{2} \:
 \big\{\bar h_v \delta( \ell^+_3\!-\! in \mcdot D_{us} )\gamma^\beta
   P_L T^A q^\bn\big\}\: 
 \delta(\ell_2^+-in\mcdot \partial)\: \big\{\bar q^\bn \gamma^\alpha P_L 
  \delta(\ell^+_1\!-\! i n \mcdot D_{us}  ) T^A h_v \big\}\,, \nonumber \\
O_6^{\alpha\beta}(\ell^+_{1,2,3})  \!\! &=&  \!\!
   \int\!\! \frac{dx^- dy^- dz^-}{128\pi^3} 
    e^{-\frac{i}{2} \{ \ell^+_3 x^- +(\ell^+_2 - \ell^+_3) 
    y^- +(\ell^+_1 - \ell^+_2) z^- \} }  
   \big[\overline {\cal H}_v(\tilde x) \gamma^\beta P_L 
   \psi_{us}^\bn(\tilde y) 
   \nn\\
  && \quad 
  \times\bar\psi_{us}^\bn(\tilde z) \gamma^\alpha 
  P_L  {\cal H}_v(0) \big]
  \nn\\
 &=& \frac{1}{2} \:
 \big[\bar h_v \delta( \ell^+_3\!-\! in \mcdot D_{us} ) \gamma^\beta
 P_L 
 q^\bn\big] \: \delta(\ell_2^+\!-\!in\mcdot \partial)\: \big[
 \bar q^\bn \gamma^\alpha P_L \delta(\ell^+_1\!-\! i n \mcdot D_{us}  ) 
  h_v\big]\,.
  \nonumber 
\end{eqnarray}
Here $\psi_{us}^\bn=(\bnslash\nslash)/4\, \psi_{us}$ and the flavor of the 
quarks in these operators is $s$ for $B\to X_s\gamma$
and $u$ for $B\to X_u\ell\bar\nu$.  Note that a minimal basis
of Dirac structures for the bilinears is $\bar h_v \{1,\gamma_T^\mu\gamma_5\}
h_v$.  In the second line for each operator we have used our freedom to
integrate by parts since only the forward part of these usoft operators is
required. The operators in Eq.~(\ref{softops1}) were obtained by matching with
tree-level quark propagators in Ref.~\cite{Bauer:2001mh} and we agree with 
these.\footnote{It seems that Ref.~\cite{Bauer:2001mh} has a typographical error
in the phase factor of $O_3^{\mu\nu}$ and $P_{4\alpha}^{\mu\nu}$ and that it should
read $e^{i(\omega_1-\omega_2)t_1} e^{-i\omega_1 t_2}$, in which case the
variables in that paper are related to ours by $\omega_i = -\ell_i$.} 
The four-quark
operator $O_5^{\alpha\beta}$ appears when gluon propagators are included at tree
level, as we discuss further below. The operator $\int d\ell_1^+ d\ell_3^+
O_6^{\alpha\beta}$ occurs from the disconnected annihilation contribution, as
shown in Ref.~\cite{Leibovich:2002ys}. However, since this contribution is of
order~$(\Lambda/m_b)^2$ in the endpoint region, we do not include it in our
analysis here.

When ${\cal O} (\alpha_s)$ corrections are included in the subleading 
factorization theorems, then in general we require additional usoft 
operators. For example,
at ${\cal O}(\alpha_s)$ there are two additional soft operators generated by
$T^{(2L)}$, and six additional operators from $T^{(2q)}$. Definitions for these
operators can be found in Sec.~\ref{sectT2s}.

The subleading shape functions of the operators in Eq.~(\ref{softops1}) are
defined by the matrix elements
\begin{eqnarray} \label{defnfs1}
  \langle \bar B_v | O_0(\ell^+) | \bar B_v\rangle 
    &=& f_0^{(2)}(\ell^+) \,,\\[3pt]
  \langle \bar B_v | O_{1}^{\beta}(\ell^+) | \bar B_v\rangle 
    &=& \Big(v^\beta \!-\! \frac{n^\beta}{n\mcdot v}\Big) \: 
     f_1^{(2)}(\ell^+) \,,\nn\\[3pt]
 \langle \bar B_v | P^{\beta\lambda}_{1}(\ell^+) | \bar B_v\rangle 
    &=& \: \epsilon_\perp^{\beta\lambda} \:
     g_0^{(2)}(\ell^+) \,,\nn\\[3pt]
  \langle \bar B_v | P_{2}^{\beta\lambda}(\ell^+) | \bar B_v\rangle 
    &=& \: \epsilon_\perp^{\beta\lambda} \:
     f_2^{(2)}(\ell^+) \,,\nn\\[3pt] 
  \langle \bar B_v | O_{3}^{\alpha\beta}(\ell^+) | \bar B_v\rangle 
    &=& \: g_\perp^{\alpha\beta} \:
     f_3^{(4)}(\ell_1^+,\ell_2^+) \,,\nn\\[3pt] 
  \langle \bar B_v | P_{4\lambda}^{\alpha\beta}(\ell^+) | \bar B_v\rangle 
    &=& \: - \epsilon_\perp^{\alpha\beta} 
     \Big(v_\lambda \!-\! \frac{n_\lambda}{n\mcdot v}\Big) \:
     f_4^{(4)}(\ell_1^+,\ell_2^+) \,,\nn
\end{eqnarray}
where we take $v_\perp=0$ and $\epsilon_\perp^{\beta\lambda} =
\epsilon^{\beta\lambda\sigma\tau} v_\sigma n_\tau/(n\mcdot v)$. For the 
four-quark operators we need
\begin{eqnarray}  \label{defnfs2}
 n_\alpha n_\beta
 \langle \bar B_v | O_{5}^{\alpha\beta}(\ell_{1,2,3}^+) | \bar B_v\rangle 
    &=&  f_5^{(6)}(\ell_1^+,\ell_2^+,\ell_3^+) \:
      \,,\\[3pt] 
  (g^\perp_{\alpha\beta}- i \epsilon^\perp_{\alpha\beta})
 \langle \bar B_v | O_{5}^{\alpha\beta}(\ell_{1,2,3}^+) | \bar B_v\rangle 
    &=&  f_6^{(6)}(\ell_1^+,\ell_2^+,\ell_3^+) \:
      \,.\nn
\end{eqnarray}
Recall that we use the notation $f_i$ to indicate shape functions that appear 
from tree-level matching and $g_i$ to indicate shape functions 
that show up only when higher-order perturbative corrections for the jet 
functions are considered. 
Definitions for all the $g_i$ functions are given with the
derivation of the factorization theorems in Sec.~\ref{sectT2s}. For the
$f_i$, we give the translation to the notation in Ref.~\cite{Bauer:2001mh} in
Table~\ref{table:notatn}.
 
In Ref.~\cite{Bauer:2001mh}, it was shown that the equations of
motion for $h_v$ imply that
\begin{eqnarray}
  f_1^{(2)}(\omega) = 2\omega f^{(0)}(\omega) \,,
\end{eqnarray}
which reduces the number of NLO unknowns, and that the $B$ matrix element of the
operator $O_{2\beta}$ vanishes, so we shall not need it for our results.  The
same is true of two additional operators, $O_4^{\alpha\beta}$ and
$P_{3\lambda}^{\alpha\beta}$, which are the analogs of $P_4^{\alpha\beta}$ but
with anticommutators. Ref.~\cite{Bauer:2001mh} also removes $P_{1\lambda}^\beta$
from the basis by noting that all the moments of the corresponding shape
function vanish. However, we keep this operator since it is unclear whether the
moments are sufficient to define the function completely beyond tree level in
the jet function.  (We do show, however, that this operator is not matched on to
at tree level in the hard function.)

\begin{table}[t!]
\begin{center}
\begin{tabular}{cc}
 Our Notation \hspace{0.2cm} &  Notation in Ref.~\cite{Bauer:2001mh} \\
\hline \\[-8pt]
$f^{(0)}(\ell^+)$ & $f(-\ell^+)$ \\
$f_0^{(2)}(\ell^+)$ & $t(-\ell^+)$ \\
$f_1^{(2)}(\ell^+)$ & $g_1(-\ell^+)$ \\
$f_2^{(2)}(\ell^+)$ & $h_1(-\ell^+)$ \\
$f_3^{(4)}(\ell^+_1,\ell_2^+)$ 
             & $g_2(-\ell^+_1,-\ell_2^+)$ \\
\;\;$f_4^{(4)}(\ell^+_1,\ell_2^+) \;\;$ 
             & $h_2(-\ell^+_1,-\ell_2^+)$ \\
$f_5^{(6)}(\ell^+_1,\ell_2^+,\ell_3^+)$ 
             & --- \\
\;\;$f_6^{(6)}(\ell^+_1,\ell_2^+,\ell_3^+) \;\;$ 
             & --- \\[7pt]
\hline
\end{tabular}
\caption{Relation of our $f_i$ functions to the notation in
Ref.~\cite{Bauer:2001mh}.}
   \label{table:notatn}
\end{center}
\end{table}



\subsection{Factorization Calculations at ${\cal O}(\lambda^2)$} \label{sectT2s}


We now derive the factorization theorems for the subleading 
${\cal O}(\lambda^2)$ terms.
We show that it is exactly the soft functions defined in the previous
section that appear in the decay rates, and give operator definitions
for the jet functions that can be used beyond tree level. 

\subsubsection{Calculation of $\hat T^{(2H)}$}  

For $T^{(2H)}$ the factorization is identical to that at LO in
Sec.~\ref{LOfact}, except that the final usoft matrix element is the
time-ordered product
\begin{eqnarray} 
  &&
  \int\!\! \frac{dx^-}{\!8\pi} e^{-\frac{i}{2} x^- \ell^+ } \int d^4y
   \: \Big\langle \bar B_v \Big| T\: \Big[\bar h_v(\tilde x) Y(\tilde x,0) h_v(0)\ 
   \frac{i\, O_h(y)}{2m_b} \Big]
   \Big| \bar B_v \Big\rangle 
   = \frac{f_0^{(2)}(\ell^+)}{2m_b} \,.
\end{eqnarray}
The factorization theorem for $\hat T^{(2H)}$ is therefore
\begin{eqnarray} \label{WfactQ2}
W_i^{(2H)} &=& 
   h_i^{0}(\bn\mcdot p,m_b,\mu)
  \int_{n\cdot q -m_b}^{\bar \Lambda} \!\!\!\!\!\! d\ell^+\: 
   {\cal J}^{(0)}(\bn\mcdot p(\ell^+ \!+\! m_b \!-\! n\mcdot q),\mu )\: 
   \frac{f_0^{(2)}(\ell^+,\mu)}{2m_b} \,.
\end{eqnarray}
As noted in Ref.~\cite{Bauer:2001mh}, the function $f_0^{(2)}$ can simply be
absorbed into the LO $f^{(0)}$, since at this order they always appear
together in the combination
\begin{eqnarray}
  f^{(0)}(\ell^+) + \frac{1}{2m_b}\, f_0^{(2)}(\ell^+) \,. 
\end{eqnarray}

\subsubsection{Calculation of $\hat T^{(2a)}$}

For the remaining time-ordered products, the factorization is more complicated.
However, some aspects of the LO analysis in Sec.~\ref{LOfact} remain the
same at subleading order. In particular, we still have the $\exp(-ir\cdot x)$
phase factor in Eq.~(\ref{phase}), a $\delta(\omega'-\bn\mcdot p)$ appears in
the separation of hard Wilson coefficients, as it does in Eq.~(\ref{LOhard}), 
and the vacuum matrix element of collinear fields still gives a
$\delta(\omega-\omega')$. For convenience, we integrate over $\omega$ and
$\omega'$ and remove these delta functions right from the start when
considering $T^{(2a)}$, $T^{(2L)}$, and $T^{(2q)}$.

For $\hat T^{(2a)}$ the analog of Eq.~(\ref{LOhard}) is
\begin{eqnarray} \label{Tnlo2a}
  \hat T_{\mu\nu}^{(2a)} &=& -i \sum_{j,j'} \sum_{\kappa=a,b}
   C_{j'}(\bn\mcdot p)  A_j^{(2\kappa)}(\bn\mcdot p) \!
    \int\!\! d^4x\,   e^{-i r\cdot x}\  \\
  &&\times   {\rm T}\: \big[ J^{(0)\dagger}_{j'\,\mu}(\bn\mcdot p,x)\:
   J^{(2\kappa)}_{j\,\nu}(\bn\mcdot p,0) 
   + J^{(2\kappa)\dagger}_{j\,\mu}(\bn\mcdot p,x)\:
   J^{(0)}_{j'\,\nu}(\bn\mcdot p,0) \big]  \,,
  \nn 
\end{eqnarray}
where $A_j^{(2\kappa)}(\omega,\mu)$ are the Wilson coefficients for the $J^{(2k)}$
currents, as in Eq.~(\ref{Jeff0}). The products of currents in square brackets
are
\begin{eqnarray}
J^{(0)\dagger}_{j'} J^{(2a)}_{j} + J^{(2a)\dagger}_{j} J^{(0)}_{j'}
 & = &  \frac{1}{2m_b}  \big[ \bar {\cal H}_v \overline \Gamma^{(0)}_{j'} 
   \chi_{n\omega'}\big](x)
   \big[\bar \chi_{n,\omega} \Upsilon^{(2a)}_{j\alpha} 
    {\,i \overrightarrow {\cal D}_{us}^{T\alpha}}\, {\cal H}_v \big](0) \nn \\
 & & \hspace{0cm}+   \frac{1}{2m_b}
    \big[\bar {\cal H}_v (-i) {\overleftarrow {\cal D}_{us}^{T\alpha}} 
     \overline \Upsilon^{(2a)}_{j\alpha}
     \chi_{n,\omega} \big](x)\;
    \big[\bar \chi_{n,\omega'} \Gamma^{(0)}_{j'} {\cal H}_v \big](0) \,,
    \nn\\[5pt]
J^{(0)\dagger}_{j'} J^{(2b)}_{j} + J^{(2b)\dagger}_{j} J^{(0)}_{j'}
 & = & \frac{1}{\omega} \big[ \bar {\cal H}_v   \overline\Gamma^{(0)}_{j'}
   \chi_{n,\omega'} \big](x)
  \big[ \bar\chi_{n,\omega} \Upsilon^{(2b)}_{j\alpha}
   i \overleftarrow {\cal D}_{us}^{\perp\alpha} {\cal H}_v \big](0)  \nn \\ 
 & & \hspace{0cm} +
   \frac{1}{\omega} \big[ (\bar {\cal H}_v  
    (-i) \overrightarrow {\cal D}_{us}^{\perp\alpha} ) 
   \overline \Upsilon_{j\alpha}^{(2b)} \chi_{n,\omega} \big](x)\;
   \big[ \bar\chi_{n,\omega'} \Gamma^{(0)}_{j'} {\cal H}_v \big](0) \,. 
\end{eqnarray}
Next we insert identity matrices to Fierz transform using Eq.~(\ref{Fierz}) 
and take the $\langle B_v |\cdots | B_v\rangle/2$ matrix element. These two
terms then become
\begin{eqnarray}   \label{eq:J02a}
 && \big\langle J^{(0)\dagger}_{j'} J^{(2a)}_{j} 
  \!+\! J^{(2a)\dagger}_{j} J^{(0)}_{j'} \big\rangle
 =
  \frac{(-1)}{8m_b}\,  \Big\{
   \big\langle \bar B_v \big|  \bar {\cal H}_v(x) 
    \overline \Gamma_{j'}^{(0)} {F_k^n} \Upsilon^{(2a)}_{j\alpha} 
    (  i \overrightarrow {\cal D}_{us}^{T\alpha} {\cal H}_v)(0) 
     \big| \bar B_v \big\rangle 
   \ \\
  &&\ \ -
   \big\langle \bar B_v \big| 
   (\bar {\cal H}_v i\! {\overleftarrow {\cal D}_{us}^{T\alpha}} )(x)
   \overline \Upsilon^{(2a)}_{j\alpha} {F_k^n} \Gamma_{j'}^{(0)}  
   {\cal H}_v(0)  \big| \bar B_v \big\rangle 
 \Big\} \big\langle 0 \big| 
   \, \bar \chi_{n,\omega}(0) 
    {F_k^\bn}  \chi_{n,\omega'}(x) 
    \big| 0 \big\rangle \,,
    \nn \\[5pt]
 && \big\langle J^{(0)\dagger}_{j'} J^{(2b)}_{j} 
  \!+\! J^{(2b)\dagger}_{j} J^{(0)}_{j'}\big\rangle 
  =  \frac{1}{4 \bn\mcdot p}\,
 \Big\{   \big\langle \bar B_v \big| 
    \, \bar {\cal H}_v(x) 
    \overline \Gamma_{j'}^{(0)} {F_k^n} \Upsilon^{(2b)}_{j\alpha} 
     (i \overrightarrow {\cal D}_{us}^{\perp\alpha} {\cal H}_v)(0) 
     \big| \bar B_v \big\rangle 
   \ \nn\\
  &&\ \  -
   \big\langle \bar B_v \big| 
   (\bar {\cal H}_v i\! {\overleftarrow {\cal D}_{us}^{\perp\alpha}} )(x)
   \overline \Upsilon^{(2b)}_{j\alpha} {F_k^n} \Gamma_{j'}^{(0)}  
    {\cal H}_v(0)\,  \big| \bar B_v \big\rangle 
  \Big\} \big\langle 0 \big| 
    \bar \chi_{n,\omega}(0) 
    {F_k^\bn} \chi_{n,\omega'}(x) 
    \big| 0 \big\rangle \,, \nn
\end{eqnarray}
where it is understood that we have $\omega=\omega'=\bn\mcdot p$, and we
have used $q_\perp = 0$ in order to integrate the $iD^\perp$ derivatives by
parts in the $J^{(2b)}$ terms. From Eq.~(\ref{eq:J02a}) we see that the
collinear vacuum matrix elements are the same as in the LO case. Hence they are
non-zero only for $k=1$ and are given by the LO jet function ${\cal
  J}_\omega^{(0)}(k^+)$ from Eq.~(\ref{Jmelt}). 

Using the trace formula in Eq.~(\ref{eq:projform}), we can simplify the spin
structures in the usoft matrix elements in Eq.~(\ref{eq:J02a}), where we can now
set $F_1^n=\nslash/2$. Here both the $P_v$ and $s^\mu$ terms in the $F^H_{k'}$
matrices will give non-zero contributions, so $k'=1,2$ and we have
\begin{eqnarray} \label{trace2a}
 &&\hspace{-0.3cm}
 -{\rm Tr}\Big\{ P^H_{k'} \overline \Gamma_{j'}^{(0)}\, \frac{\nslash}{2}
    \Big[ \frac{ g_{\alpha\beta}^T}{8m_b} \Upsilon^{(2a)}_{j\alpha} A^{(2a)}_j
     \!-\! \frac{g_{\alpha\beta}^\perp}{4\bn\mcdot p}  \Upsilon^{(2b)}_{j\alpha} 
      A^{(2b)}_j  \Big]\Big\} 
    \big\langle \bar B_v \big| {\rm T} \, \bar {\cal H}_v(x)  {P_{k'}^h} 
    (  i \overrightarrow {\cal D}_{us}^{\beta} {\cal H}_v)(0) 
     \big| \bar B_v \big\rangle   \nn\\
 &&\hspace{-0.3cm}
 + {\rm Tr}\Big\{ P^H_{k'} \Big[ 
    \frac{ g_{\alpha\beta}^T}{8m_b} \overline \Upsilon^{(2a)}_{j\alpha} A^{(2a)}_j
   \!-\!\frac{g_{\alpha\beta}^\perp}{4\bn\mcdot p} \overline
   \Upsilon^{(2b)}_{j\alpha}  A^{(2b)}_j 
   \Big]  \frac{\nslash}{2}\,   \Gamma_{j'}^{(0)}\Big\}
  \big\langle \bar B_v \big| {\rm T}\,  
   (\bar {\cal H}_v i\! {\overleftarrow {\cal D}_{us}^{\beta}} )(x) P_{k'}^h
   {\cal H}_v(0)  \big| \bar B_v \big\rangle 
  \nn\\
&& \hspace{-0.3cm}
  = -\frac14 {\rm Tr}\big\{ X_\beta^{k'=1} \big\}
   \langle \bar B_v |(  {O}_1^\beta + i  { O}_2^\beta )| \bar B_v \rangle
    +\frac14 {\rm Tr}\big\{ \overline X_\beta^{\,k'=1} \big\}  
    \langle \bar B_v |( - { O}_1^\beta + i  { O}_2^\beta )| \bar B_v \rangle\nn\\
&& \hspace{-0.25cm}
   -\frac14 {\rm Tr}\big\{ X_{\beta\lambda}^{k'=2} \big\}
   \langle \bar B_v |(  {P}_1^{\beta\lambda} 
    + i  {P }_2^{\beta\lambda} )| \bar B_v \rangle
    +\frac14 {\rm Tr}\big\{ \overline X_{\beta\lambda}^{\,k'=2} \big\}  
    \langle \bar B_v |( - { P}_1^{\beta\lambda}
     + i  { P}_2^{\beta\lambda} )| \bar B_v \rangle\,, 
\end{eqnarray}
where 
\begin{eqnarray} \label{Xdefn}
  X^{k'} &=& P^H_{k'} \overline \Gamma_{j'\mu}^{(0)}\, \frac{\nslash}{2}
    \Big[ \frac{ g_{\alpha\beta}^T}{4m_b} \Upsilon^{(2a)}_{j\alpha\nu} A^{(2a)}_j
     \!-\! \frac{g_{\alpha\beta}^\perp}{2\bn\mcdot p}  
     \Upsilon^{(2b)}_{j\alpha\nu} A^{(2b)}_j  \Big]\,,\nn \\
  \overline X^{\,k'} &=& P^H_{k'} 
    \Big[ \frac{g_{\alpha\beta}^T}{4m_b} \overline \Upsilon^{(2a)}_{j\alpha\mu} 
     A^{(2a)}_j \!-\! \frac{g_{\alpha\beta}^\perp}{2\bn\mcdot p}  
     \overline \Upsilon^{(2b)}_{j\alpha\mu} A^{(2b)}_j  \Big] 
   \frac{\nslash}{2} \Gamma_{j'\nu}^{(0)}\, .
\end{eqnarray}
Note that the second trace is the complex conjugate of the first with
$(\mu\leftrightarrow \nu)$, since $\gamma_0 (F_{k'}^H)^\dagger \gamma_0 =
F_{k'}^H$ and ${\rm Tr}[ \overline X ] = {\rm Tr}[ X^\dagger] = {\rm Tr}[ X^* ]
= ({\rm Tr} X )^*$.  These usoft matrix elements give the shape functions $f_1$,
$g_0$, and $f_2$ in Eq.~(\ref{defnfs1}) at any matching order in $\alpha_s$. The
index $\beta$ is transverse to $v$, as follows from Eq.~(\ref{trace2a}) with
$v_\perp=0$. Combining Eq.~(\ref{trace2a}) with the delta functions and
prefactors from the jet function and the $(-i)$ from $\hat T^{(2a)}$ gives
\begin{eqnarray} \label{T2aresult}
  && \int\! dk^+ {\cal J}_{\omega}^{(0)}(k^+) \Big[
   {\rm Tr}\big\{ X_\beta^{k'=1} \!+\! \overline X_\beta^{\,k'=1} \big\}
    \Big(-\! \frac{n^\beta}{n\mcdot v}\Big) \: 
     f_1^{(2)}(k^+\!+\!r^+)
    \\
&& + {\rm Tr}\big\{ X_{\beta\lambda}^{k'=2} \!
    +\! \overline X_{\beta\lambda}^{\,k'=2}\big\}
     \epsilon_\perp^{\beta\lambda} \: g_0^{(2)}(k^+\!+\!r^+)
   + {\rm Tr}\big\{ X_{\beta\lambda}^{k'=2} \!
    -\! \overline X_{\beta\lambda}^{\,k'=2}\big\}
   {i}\epsilon_\perp^{\beta\lambda} \: f_2^{(2)}(k^+\!+\!r^+)
   \Big] \,. \nn
\end{eqnarray}
We include these traces in the definition of the hard coefficients at
subleading order, as was done in Eq.~(\ref{h0}), defining
\begin{eqnarray}\label{h2ab}
  h_i^{[2a]1} &=& \frac{1}{8} \sum_{j,j'} C_{j'}\,A^{(2a)}_j\:
     {\rm Tr}\Big\{ P_v \overline \Gamma_{j'\mu}^{(0)}\, \frac{\nslash}{2}
    \Big(-\!\frac{n^\alpha}{n\mcdot v}\Big)
      \Upsilon^{(2a)}_{j\alpha\nu} + \mbox{(h.c. $\mu\leftrightarrow\nu$)}
     \Big\}\: P_i^{\mu\nu} \,, \\
  h_i^{[2a]2} &=& \frac{1}{2} \sum_{j,j'} C_{j'}\:
     {\rm Tr}\Big\{ P_v  \gamma_5 \gamma_\lambda^T P_v
    \overline \Gamma_{j'\mu}^{(0)}\, \frac{\nslash}{2}
    \Big[ \frac{i\epsilon^{\alpha\lambda}_\perp}{4} 
     \Upsilon^{(2a)}_{j\alpha\nu} A^{(2a)}_j
     - \frac{i\epsilon^{\alpha\lambda}_\perp\, m_b}{2\bn\mcdot p}  
     \Upsilon^{(2b)}_{j\alpha\nu} 
      A^{(2b)}_j  \Big]  \nn\\
 && + \mbox{(h.c. $\mu\leftrightarrow\nu$)}\Big\}\: 
   P_i^{\mu\nu} \,,\nn\\
  h_i^{[2a]00} &=& \frac{1}{2} \sum_{j,j'} C_{j'}\:
     {\rm Tr}\Big\{ P_v  \gamma_5 \gamma_\lambda^T P_v
    \overline \Gamma_{j'\mu}^{(0)}\, \frac{\nslash}{2}
    \Big[ \frac{\epsilon^{\alpha\lambda}_\perp}{4} 
     \Upsilon^{(2a)}_{j\alpha\nu} A^{(2a)}_j
     - \frac{\epsilon^{\alpha\lambda}_\perp\, m_b}{2\bn\mcdot p}  
     \Upsilon^{(2b)}_{j\alpha\nu} 
      A^{(2b)}_j  \Big]  \nn\\
 && + \mbox{(h.c. $\mu\leftrightarrow\nu$)}\Big\}\: 
   P_i^{\mu\nu} \,.\nn
\end{eqnarray}
For the Dirac structures $\Upsilon^{(2a)}$ and $\Upsilon^{(2b)}$ that are
present at tree level $h_i^{[2a]00}$ is zero, so $g_0^{(2)}$ comes in with
$\alpha_s(m_b)$ suppression. We were unable to prove our suspicion that
$g_0^{(2)}$ would not appear beyond this level.

Using Eq.~(\ref{h2ab}) with Eq.~(\ref{T2aresult}) and taking $(-1/\pi)$ times 
the imaginary part gives the NLO factorization theorem for $W_i^{(2a)}$: 
\begin{eqnarray}
  W_i^{(2a)}
  &=& \frac{h_i^{[2a]1}(\bn\mcdot p,m_b,\mu)}{m_b}\:
  \int_{0}^{p_X^+} \! dk^+ \
   {\cal J}^{(0)}(\bn\mcdot p\: k^+\!, \mu)\:
   f_1^{(2)}(k^+ \!+\! r^+,\mu)  
  \\
 &+& \frac{h_i^{[2a]2}(\bn\mcdot p,m_b,\mu)}{m_b}\:
  \int_{0}^{p_X^+} \! dk^+ \
   {\cal J}^{(0)}(\bn\mcdot p\: k^+\!,\mu)\:
   f_2^{(2)}(k^+ \!+\! r^+,\mu)  \nn\\
 &+& \frac{h_i^{[2a]00}(\bn\mcdot p,m_b,\mu)}{m_b}\:
  \int_{0}^{p_X^+} \! dk^+ \
   {\cal J}^{(0)}(\bn\mcdot p\: k^+\!,\mu)\:
   g_0^{(2)}(k^+ \!+\! r^+,\mu) 
 \nn \,.
\end{eqnarray}
Explicit results for the coefficients $h_i^{[2a]}$ are summarized in 
Sec.~\ref{sectHard}.

\subsubsection{Calculation of $T^{(2L)}$}

For $T^{(2L)}$ the heavy-to-light currents are identical to those in $T^{(0)}$,
and so the same Wilson coefficients appear as in Eq.~(\ref{LOhard}), i.e.\
\begin{eqnarray} \label{Tnlo2L}
  \hat T_{\mu\nu}^{(2L)} &=& -i \sum_{j,j'} 
   C_{j'}(\bn\mcdot p)  C_j(\bn\mcdot p) \!
    \int\!\! d^4x\! \int\!\! d^4y\:   e^{-i r\cdot x}\:
   {\rm T}\: \big[ J^{(0)\dagger}_{j'\,\mu}(\bn\mcdot p,x)\:
    {\cal L}_{\xi\xi}^{(2a)}(y)
   J^{(0)}_{j\,\nu}(\bn\mcdot p,0) \big]  \,.
  \nn \\
\end{eqnarray}
The difference is the extra Lagrangian insertion with mixed collinear and usoft
fields, namely
\begin{eqnarray} \label{T2Lfierz}
&& \hspace{-1cm} J_{j'}^{(0)\dagger}(x) {\cal L}_{\xi\xi}^{(2a)}(y) J_j^{(0)}(0)
   \\
& = & \big[\bar {\cal H}_v \overline\Gamma_{j'}^{(0)}
       \chi_{n,\omega'}^L \big](x)
      \Big[ \bar \chi_{n}
        i\cDslash_{us}^\perp\,  i\! \cDslash_{us}^\perp 
      \frac{\bnslash}{2}  \frac{1}{\bnP} \chi_n \Big](y)\;
      \big[ \bar\chi_{n,\omega}^L  \Gamma_j^{(0)} {\cal H}_v \big](0)
    \nn\\
&=& \frac14 \sum_{k,k'=1}^4
      \big[\bar {\cal H}_v(x) \overline\Gamma_{j'}^{(0)} F_{k'}^{L n}
      \Big(  i\cDslash_{us}^\perp\,  i\! \cDslash_{us}^\perp 
      \frac{\bnslash}{2} \Big)(y) 
      F_k^{n R} \Gamma_j^{(0)} {\cal H}_v(0) \Big]\nn\\
 &&\qquad \times
     \big[ \bar\chi_n(y) F_{k'}^{\bn L} \chi^L_{n,\omega'}(x) \big]
      \big[ \bar\chi^L_{n,\omega}(0) 
     F_k^{R \bn}  \frac{1}{\bnP} \chi_n(y) \big]
    \nn  \,,
\end{eqnarray}
where in the second line we have used analogs of the Fierz formula in
Eq.~(\ref{Fierz}) that are appropriate with left- and right-handed projectors,
\begin{eqnarray} \label{FierzLR}
 P_L \otimes 1 &=& \frac{1}{2}\ \sum_{k=1}^{4} F_k^{\bn L} \otimes F_k^{L n}
 = \frac{1}{2} \Big[ 
  \big(\frac{\bnslash}{N_c} P_L\big)\!\otimes\! \big(P_L \frac{\nslash}{2} \big)
 + \big(\frac{-\bnslash\gamma_\perp^\alpha P_L}{2N_c}\big) \!\otimes\! 
     \big(P_L \frac{\nslash\gamma^\perp_\alpha}{2}\big) \nn\\
 &&\ \ \ 
 +  \big(2 {\bnslash P_L T^a}\big) \!\otimes\! \big(P_L \frac{\nslash}{2} T^a\big) 
 + \big( {-\bnslash\gamma_\perp^\alpha P_L T^a}\big) \!\otimes\! 
  \big( P_L \frac{\nslash\gamma^\perp_\alpha }{2} T^a \big)
  \Big] \,,\nn\\
 P_R \otimes 1 &=& \frac{1}{2}\ \sum_{k'=1}^{4} F_{k'}^{R \bn}\otimes F_{k'}^{n R}
 = \frac{1}{2} \Big[ 
  \big(P_R\frac{\bnslash}{N_c} \big)\!\otimes\! \big( \frac{\nslash}{2} P_R\big)
 + \big(P_R\frac{-\bnslash\gamma_\perp^\beta }{2N_c}\big) \!\otimes\! 
     \big( \frac{\nslash\gamma^\perp_\beta}{2} P_R\big) \nn\\
 &&\ \ \ 
 +  \big(2  P_R {\bnslash T^b}\big) \!\otimes\!
   \big( \frac{\nslash}{2} P_R T^b\big) 
 + \big( {- P_R\bnslash\gamma_\perp^\beta T^b}\big) \!\otimes\! 
  \big( \frac{\nslash\gamma^\perp_\beta }{2} P_R T^b \big)
  \Big] \,.
\end{eqnarray}

The vacuum matrix element of the collinear four-quark operator in
Eq.~(\ref{T2Lfierz}) must be a color singlet, implying that either 
$k,k'\in \{1,2\}$, so that both $F^\bn$'s have no $T^a$'s, 
or $k,k'\in \{3,4\}$ and the color structure gets reduced when we take the 
matrix element, using $T^a\otimes T^b = T^A\otimes T^A\:
\delta^{ab}/(N_c^2\!-\!1)$.  Rotational invariance then implies that the matrix
element gives $\delta_{k,k'}$ so we do the sum over $k'$ and set $k=k'$. This
leaves four terms in 
$[\bar\chi_n F_{k'}^{\bn L} \chi_n^L][\bar\chi_n^L F_k^{R\bn} \chi_n]$:
\begin{eqnarray} \label{collin4q}
  &&\hspace{-0.2cm}
 \big[F_{k}^{\bn L}  \big] \otimes \big[   F_k^{R \bn}   \big] 
 = \delta_{k1}
  \big[ \frac{\bnslash}{N_c} P_L \big]\otimes
      \big[  \frac{\bnslash}{N_c} P_L \big] 
  + \delta_{k2}
    \frac{(g^\perp_{\alpha\beta}\!+\! i\epsilon^\perp_{\alpha\beta})}{2}
  \big[ \frac{\bnslash\gamma_\perp^\lambda}{2N_c}  P_L \big]\otimes
      \big[  \frac{\bnslash\gamma^\perp_\lambda}{2N_c}  P_R \big] \\
&& +
  \delta_{k3}  \frac{4\delta^{ab}}{(N_c^2\!-\!1)}
  \big[ {\bnslash} P_L T^A \big]\otimes \big[  {\bnslash} P_L T^A \big]
  +\delta_{k4}
  \frac{(g^\perp_{\alpha\beta}\!+\! i\epsilon^\perp_{\alpha\beta})}{2} 
    \frac{4\delta^{ab}}{N_c^2\!-\! 1}
   \big[ \frac{\bnslash\gamma_\perp^\lambda}{2}  P_L T^A \big]\otimes
      \big[  \frac{\bnslash\gamma^\perp_\lambda}{2}  P_R T^A \big]
  .\nn
\end{eqnarray}
When we use Eq.~(\ref{eq:projform}) to reduce the Dirac structure between the
heavy quark fields in Eq.~(\ref{T2Lfierz}), and take into account the
$\delta^{ab}$'s from Eq.~(\ref{collin4q}), the usoft operators become
\begin{eqnarray} \label{L2trace}
  && \frac14 \sum_{k''=1}^2\: {\rm Tr}\Big[ P^H_{k''}\,
      \overline\Gamma_{j'}^{(0)} F_{k}^{L n}
      \gamma_\perp^\sigma \gamma_\perp^\tau  \frac{\bnslash}{2} 
       F_k^{n R} \Gamma_j^{(0)}  \Big]
      \Big[\bar {\cal H}_v(x)  F_{k''}^h
      \big(  i\cD_{us}^{\perp\sigma}  i \cD_{us}^{\perp\tau} \big)(y) 
      {\cal H}_v(0) \Big]\,, \\
 && \frac14 \sum_{k''=1}^2\: {\rm Tr}\Big[ P^H_{k''}\,
      \overline\Gamma_{j'}^{(0)} F_{k}^{L n}
      \gamma_\perp^\sigma \gamma_\perp^\tau  \frac{\bnslash}{2} 
       F_{k}^{n R} \Gamma_j^{(0)}  \Big]
   \Big[\bar {\cal H}_v(x)  F_{k''}^h
   T^A\big(  i\cD_{us}^{\perp\sigma}  i \cD_{us}^{\perp\tau} \big)(y) T^A
      {\cal H}_v(0) \Big]\nn ,
\end{eqnarray}
for $k\in \{1,2\}$ and $k\in \{3,4\}$, respectively. The spin-trace that occurs
in both terms can be simplified with $ \gamma_\perp^\sigma
\gamma_\perp^\tau\bnslash\, = (g_\perp^{\sigma\tau}\! -\! i
\epsilon_\perp^{\sigma\tau}\gamma_5)\bnslash $ and $\nslash \gamma_\perp^\alpha
\gamma_\perp^\beta\, = \nslash (g_\perp^{\alpha\beta}\! -\! i
\epsilon_\perp^{\alpha\beta}\gamma_5) $:
\begin{eqnarray} \label{simpTr}
  &&{\rm Tr}\Big[ P^H_{k''}\,
      \overline\Gamma_{j'}^{(0)} F_{k}^{L n}
      \gamma_\perp^\sigma \gamma_\perp^\tau  \frac{\bnslash}{2} 
       F_k^{n R} \Gamma_j^{(0)}  \Big] 
  = \Big[ (\delta_{k1}+\delta_{k3})
   (g_\perp^{\sigma\tau}\!-\! i\epsilon_\perp^{\sigma\tau}) 
  \nn\\
  &&\qquad  + (\delta_{k2}+\delta_{k4})
   (g_\perp^{\sigma\tau}\!+\! i\epsilon_\perp^{\sigma\tau}) 
   (g_\perp^{\alpha\beta}\!-\! i\epsilon_\perp^{\alpha\beta}) 
  \Big] {\rm Tr}\Big[ P^H_{k''}\,
      \overline\Gamma_{j'}^{(0)}   \frac{\nslash}{2} P_R \Gamma_j^{(0)}  \Big]
  \,.
\end{eqnarray}
The $(g_\perp^{\alpha\beta}\!-\! i\epsilon_\perp^{\alpha\beta})$ terms contract
with the $(g^\perp_{\alpha\beta}\!+\!  i\epsilon^\perp_{\alpha\beta})/2$'s in
Eq.~(\ref{collin4q}) to give a $2$.

From Eq.~(\ref{collin4q}) there are four jet functions:
\begin{eqnarray} \label{Jetm2}
 && \Big\langle 0 \Big| \Big[ \bar \chi_n (y) \frac{\bnslash P_L}{N_c}
  \chi^L_{n,\omega'}(x) \Big] \Big[ \bar\chi^L_{n,\omega}(0) 
   \frac{\bnslash P_L}{N_c} \frac{1}{\bnP} \chi_n(y) \Big] 
   \Big| 0 \Big\rangle \\
 &&\qquad =  \frac{-4 \delta(\omega\!-\!\omega')}{\omega}\,
 \delta^2(x_\perp) \delta(x^+) \delta^2(y_\perp) \delta(y^+) 
 \int\!\! \frac{dk^+_1\, dk^+_2}{(2\pi)^2}  \:
  e^{-\frac{i}{2} [k_2^+x^- \!+ (k_1^+\!- k_2^+)y^-]}
  {\cal J}^{(-2)}_{1\omega}(k^+_j)
  \,, \nn\\[5pt]
&& 2\Big\langle 0 \Big| \Big[ \bar \chi_n (y) 
  \frac{\bnslash \gamma_\perp^\lambda}{2N_c}
  \chi^L_{n,\omega'}(x) \Big] \Big[ \bar\chi^L_{n,\omega}(0) 
   \frac{\bnslash \gamma^\perp_\lambda}{2N_c} \frac{1}{\bnP} \chi_n(y) \Big] 
   \Big| 0 \Big\rangle \nn\\
 &&\qquad =  \frac{-4 \delta(\omega\!-\!\omega')}{\omega}\,
 \delta^2(x_\perp) \delta(x^+) \delta^2(y_\perp) \delta(y^+) 
 \int\!\! \frac{dk^+_1\, dk^+_2}{(2\pi)^2}  \:
  e^{-\frac{i}{2} [k_2^+x^- \!+ (k_1^+\!- k_2^+)y^-]}
  {\cal J}^{(-2)}_{2\omega}(k^+_j)
  \,, \nn\\[5pt]
 && \frac{4}{(N_c^2\!-\!1)}\,
 \Big\langle 0 \Big| \Big[ \bar \chi_n(y) {\bnslash P_L T^A}
  \chi^{\,L}_{n,\omega'}(x) \Big] \Big[ \bar\chi^{\,L}_{n,\omega}(0) 
   {\bnslash P_L T^A} \frac{1}{\bnP}  \chi_n(y) \Big] 
   \Big| 0 \Big\rangle \nn\\[5pt]
 && \qquad =  \frac{-4 \delta(\omega\!-\!\omega')}{\omega}\,
 \delta^2(x_\perp) \delta(x^+) \delta^2(y_\perp) \delta(y^+) 
 \int\!\! \frac{dk^+_1\, dk^+_2}{(2\pi)^2}  \:
  e^{-\frac{i}{2} [k_2^+x^- \!+ (k_1^+\!- k_2^+)y^-]}
  {\cal J}^{(-2)}_{3\omega}(k^+_j)
  \,, \nn\\[5pt]
 && \frac{2}{(N_c^2\!-\!1)}\,
 \Big\langle 0 \Big| \Big[ \bar \chi_n(y) {\bnslash\gamma_\perp^\lambda P_L T^A}
  \chi^{\,L}_{n,\omega'}(x) \Big] \Big[ \bar\chi^{\,L}_{n,\omega}(0) 
   {\bnslash \gamma^\perp_\lambda P_R T^A} \frac{1}{\bnP}  \chi_n(y) \Big] 
   \Big| 0 \Big\rangle \nn\\[5pt]
 && \qquad =  \frac{-4 \delta(\omega\!-\!\omega')}{\omega}\,
 \delta^2(x_\perp) \delta(x^+) \delta^2(y_\perp) \delta(y^+) 
 \int\!\! \frac{dk^+_1\, dk^+_2}{(2\pi)^2}  \:
  e^{-\frac{i}{2} [k_2^+x^- \!+ (k_1^+\!- k_2^+)y^-]}
  {\cal J}^{(-2)}_{4\omega}(k^+_j)
  \,.\nn
\end{eqnarray}
At tree level, only ${\cal J}_{1\omega}^{(-2)}$ is non-zero. It has the value
\begin{eqnarray}
  {\cal J}_{1\omega}^{(-2)}(n\mcdot k_1,n\mcdot k_2) =
  \frac{1}{[n\mcdot k_1 +i\epsilon]}
  \frac{1}{[n\mcdot k_2 +i\epsilon]} \,.
\end{eqnarray}
At one-loop order, ${\cal J}_{3\omega}^{(-2)}$ will become non-zero, while 
${\cal J}_{2\omega}^{(-2)}$ and ${\cal J}_{4\omega}^{(-2)}$ are non-zero only
at two-loop order. The expansion here is in $\alpha_s$ at the jet scale, 
$\mu^2\sim m_X^2$. We denote the imaginary parts by 
\begin{eqnarray}
  {\cal J}^{(-2)}_j = \Big(\frac{-1}{\pi}\Big)\: {\rm Im}\ 
    {\cal J}^{(-2)}_{j\omega} \,.
\end{eqnarray} 
Eqs.~(\ref{collin4q}) and (\ref{simpTr}) also show that if we work at any order
in $\alpha_s$ there are four shape functions generated by $T^{(2L)}$, i.e.\
\begin{eqnarray}
 &&  (g^\perp_{\sigma\tau}\!-\! i\epsilon^\perp_{\sigma\tau})
  \int\!\! \frac{dx^- dy^-}{32\pi^2}
   e^{-\frac{i}{2} [x^- \ell_2^+ +y^- \ell_1^+]} \: 
 \langle \bar B_v |  \bar {\cal H}_v(\tilde x) F_{k''}^{h}
 \big(  i\cD_{us}^{\perp\sigma}  i \cD_{us}^{\perp\tau}\big)(\tilde y) 
    {\cal H}_v(0) | \bar B_v \rangle\nn  \\
&& \qquad
 = \delta_{k''1}\:  f_3(\ell_1^+,\ell_2^+)
  +  \delta_{k''2}\: \Big( v_\eta\!-\! \frac{n_\eta}{n\mcdot v}\Big)
   f_4(\ell_1^+,\ell_2^+) \,, \nn\\ 
 &&  (g^\perp_{\sigma\tau}\!-\! i\epsilon^\perp_{\sigma\tau})
  \int\!\! \frac{dx^- dy^-}{32\pi^2}
   e^{-\frac{i}{2} [x^- \ell_2^+ +y^- \ell_1^+]} \: 
 \langle \bar B_v |  \bar {\cal H}_v(\tilde x) F_{k''}^{h} T^A
 \big(  i\cD_{us}^{\perp\sigma}  i \cD_{us}^{\perp\tau}\big)(\tilde y)  T^A
    {\cal H}_v(0) | \bar B_v \rangle\nn  \\
 && \qquad
 = \delta_{k''1}\: g_3(\ell_1^+,\ell_2^+)
  + \delta_{k''2}\: \Big( v_\eta\!-\! \frac{n_\eta}{n\mcdot v}\Big)
   g_4(\ell_1^+,\ell_2^+) \,, \nn\\ 
 &&  (g^\perp_{\sigma\tau}\!+\! i\epsilon^\perp_{\sigma\tau})
  \int\!\! \frac{dx^- dy^-}{32\pi^2}
   e^{-\frac{i}{2} [x^- \ell_2^+ +y^- \ell_1^+]} \: 
 \langle \bar B_v |  \bar {\cal H}_v(\tilde x) F_{k''}^{h}
 \big(  i\cD_{us}^{\perp\sigma}  i \cD_{us}^{\perp\tau}\big)(\tilde y) 
    {\cal H}_v(0) | \bar B_v \rangle\nn  \\
&& \qquad
 = \delta_{k''1}\: f_3(\ell_1^+,\ell_2^+)
  - \delta_{k''2}\: \Big( v_\eta\!-\! \frac{n_\eta}{n\mcdot v}\Big)
  f_4(\ell_1^+,\ell_2^+)  \,, \nn\\ 
 &&  (g^\perp_{\sigma\tau}\!+\! i\epsilon^\perp_{\sigma\tau})
  \int\!\! \frac{dx^- dy^-}{32\pi^2}
   e^{-\frac{i}{2} [x^- \ell_2^+ +y^- \ell_1^+]} \: 
 \langle \bar B_v |  \bar {\cal H}_v(\tilde x) F_{k''}^{h} T^A
 \big(  i\cD_{us}^{\perp\sigma}  i \cD_{us}^{\perp\tau}\big)(\tilde y)  T^A
    {\cal H}_v(0) | \bar B_v \rangle\nn  \\
&& \qquad
 = \delta_{k''1}\: g_3(\ell_1^+,\ell_2^+)
  - \delta_{k''2}\: \Big( v_\eta\!-\! \frac{n_\eta}{n\mcdot v}\Big)
    g_4(\ell_1^+,\ell_2^+)\,.
\end{eqnarray}
Here the index $\eta$ is from $F_2^h=P_v \gamma_\eta^T \gamma_5 P_v$.  At tree
level in the jet function, only the first of these four combinations occurs. 
The functions $f_{3,4}$ were defined in Eq.~(\ref{defnfs1}). The definitions of
$g_{3,4}$ are
\begin{eqnarray}
 \langle \bar B_v | O_{7}^{\alpha\beta}(\ell^+) | \bar B_v\rangle 
    &=& \: g_\perp^{\alpha\beta} \:
     g_3^{(4)}(\ell_1^+,\ell_2^+) \,,\nn\\[3pt] 
  \langle \bar B_v | P_{7\lambda}^{\alpha\beta}(\ell^+) | \bar B_v\rangle 
    &=& \: - \epsilon_\perp^{\alpha\beta} 
     \Big(v_\lambda \!-\! \frac{n_\lambda}{n\mcdot v}\Big) \:
     g_4^{(4)}(\ell_1^+,\ell_2^+) \,,
\end{eqnarray}
where
\begin{eqnarray}
 O_7^{\alpha\beta}(\ell^+_1,\ell^+_2) &=& 
   \int\!\! \frac{dx^- \, dy^-\,}{32\pi^2} 
    e^{-\frac{i}{2} \ell_2 x^-}  e^{-\frac{i}{2}(\ell^+_1 - \ell^+_2) y^-}
   \big[\overline {\cal H}_v(\tilde x) T^A \{i \cD_{us}^{\perp\alpha}, i
   \cD_{us}^{\perp\beta}\}(\tilde y) T^A {\cal H}_v(0) \big] \,,
   \nonumber \\
%
P_{7\lambda}^{\alpha\beta}(\ell^+_1,\ell^+_2) &=&
  i\int\!\! \frac{dx^- \, dy^-\,}{32\pi^2} 
    e^{-\frac{i}{2} \ell_2 x^-}  e^{-\frac{i}{2}(\ell^+_1 - \ell^+_2) y^-}
   \big[\overline {\cal H}_v(\tilde x) T^A [i \cD_{us}^{\perp\alpha}, i
   \cD_{us}^{\perp\beta}](\tilde y) T^A \gamma^T_\lambda\gamma_5\,
   {\cal H}_v(0) \big] \,.\nn\\
%
\end{eqnarray}
The hard coefficients can now be defined to be
\begin{eqnarray} \label{h2L}
   h^{[2L]3}_i &=&  - \frac{m_b}{\bn\mcdot p} \sum_{j,j'} C_{j'}\, C_j\:
     {\rm Tr}\Big\{ \frac{P_v}{2}
     \overline \Gamma_{j'}^{(0)}\, \frac{\nslash}{2} P_R
     \Gamma^{(0)}_{j}  \Big\}\: 
   P_i^{\mu\nu} = - \frac{m_b}{\bn\mcdot p}\: h_i^0 \,, \\
   h^{[2L]4}_i &=&  \frac{m_b}{\bn\mcdot p} \sum_{j,j'} C_{j'}\, C_j\:
     \frac{n_\eta}{n\mcdot v} {\rm Tr}\Big\{ \frac{P_v}{2}
      \gamma_T^\eta\gamma_5 P_v
      \overline \Gamma_{j'}^{(0)}\, \frac{\nslash}{2} P_R
      \Gamma^{(0)}_{j}  \Big\}\: 
   P_i^{\mu\nu} \,. \nn
\end{eqnarray}

Combining equations, we get the final result for the $T^{(2L)}$ contributions:
\begin{eqnarray}
  W_i^{(2L)}
  &=& \frac{h_i^{[2L]3}(\bn\mcdot p,m_b)}{m_b}
  \int \! dk^+_1 dk^+_2 \Big\{
   \big[{\cal J}^{(-2)}_1(\bn\mcdot p\: k^+_j)
   +{\cal J}^{(-2)}_2(\bn\mcdot p\: k^+_j)\big]\:
   f_3(k^+_j \!+\! r^+) \nn\\ 
 && \qquad\qquad + \big[{\cal J}^{(-2)}_3(\bn\mcdot p\: k^+_j)
   +{\cal J}^{(-2)}_4(\bn\mcdot p\: k^+_j)\big]\:
   g_3(k^+_j \!+\! r^+)\nn \Big\}
  \\
 &+& \frac{h_i^{[2L]4}(\bn\mcdot p,m_b)}{m_b}
  \int \! dk^+_1 dk^+_2 \Big\{
   \big[{\cal J}^{(-2)}_1(\bn\mcdot p\: k^+_j)
   -{\cal J}^{(-2)}_2(\bn\mcdot p\: k^+_j)\big]\:
   f_4(k^+_j \!+\! r^+) \nn\\ 
 && \qquad\qquad + \big[{\cal J}^{(-2)}_3(\bn\mcdot p\: k^+_j)
   -{\cal J}^{(-2)}_4(\bn\mcdot p\: k^+_j)\big]\:
   g_4(k^+_j \!+\! r^+) \Big\}
  \,.
\end{eqnarray}
Explicit results for the hard coefficients $h_i^{[2L]}$ are
summarized in 
Sec.~\ref{sectHard}.

\subsubsection{Calculation of $T^{(2q)}$}

For $T^{(2q)}$ the heavy-to-light currents are also identical to those in
Eq.~(\ref{Tlo1}), but now we have two insertions of the Lagrangian ${\cal
  L}_{\xi q}^{(1)}$, i.e.\
\begin{eqnarray} \label{Tnlo2q}
  \hat T_{\mu\nu}^{(2q)} &=& -\frac{i}{2} \sum_{j,j'} 
   C_{j'}(\bn\mcdot p)  C_j(\bn\mcdot p) \!
    \int\!\! d^4x\! \int\!\! d^4y\:   e^{-i r\cdot x}\:
   {\rm T}\: \big[ J^{(0)\dagger}_{j'\,\mu}(\bn\mcdot p,x)\:
    {\cal L}_{\xi q}^{(1)}(y)\:{\cal L}_{\xi q}^{(1)}(z)\:
   J^{(0)}_{j\,\nu}(\bn\mcdot p,0) \big]  \,.
  \nn \\
\end{eqnarray}
Owing to the hermitian-conjugate terms in each ${\cal L}_{\xi q}^{(1)}$, there are
two ways to generate the four-quark operator shown in Table~\ref{table_T}, where
the positions in the figure are labeled either $0-z-y-x$ or $0-y-z-x$. This
average over $z\leftrightarrow y$ cancels the $1/2$ in the definition of
$T^{(2q)}$. Fierz transforming the product of operators with 
Eq.~(\ref{FierzLR}) in order to group the collinear and usoft fields, we have
\begin{eqnarray}
&& \hspace{-1cm} J_{j'}^{(0)\dagger}(x) {\cal L}_{\xi q}^{(1)}(y) 
   {\cal L}_{\xi q}^{(1)}(z) J_j^{(0)}(0)
   \\
& = & \big[\bar{\cal H}_v \overline\Gamma_{j'}^{(0)}
    \chi^L_{n,\omega'} \big](x)
  \Big[ \bar \chi_n ig {\Bcslash}_{c}^\perp 
    \psi_{us}\Big](y)
   \Big[ \bar\psi_{us}   \big(\!-\!ig{\Bcslash}^{\perp}_c\big) 
    \chi_n  \Big](z)
    \big[ \bar\chi^L_{n,\omega} \Gamma_j^{(0)} {\cal H}_v \big](0)
  \nn\\
& = &   \frac14 \sum_{k,k'=1}^4
  \Big[  \Big(\bar \chi_n ig {\Bcslash}_{\perp}^{c}
   \Big)(y) F_{k'}^{\bn L}  \chi^L_{n,\omega'}(x)\Big]
 \bar\chi^L_{n,\omega}(0)  F_{k}^{R\bn} 
  \Big(\!-\! ig{\Bcslash}_{\perp}^{c}  
  \chi_n\Big)(z)
  \nn\\
 &&\times
  \big[\bar {\cal H}_v(x) \overline\Gamma_{j'}^{(0)} F_{k'}^{L n} 
     \psi_{us}(y) \big]
  \big[  \bar \psi_{us}(z) F_{k}^{n R} \Gamma_j^{(0)} {\cal H}_v(0) \big] 
  \mbox{  + $(y\leftrightarrow z)$}
    \nn \,.
\end{eqnarray}
The vacuum matrix element of the collinear operator must be a color singlet, so
either $k,k'\in \{1,2\}$ or $k,k'\in \{3,4\}$, and then rotational invariance
implies that $k=k'$. This leaves four non-trivial collinear operators, which 
are given by
\begin{eqnarray} \label{collin5q}
  \Big[  \Big(\bar \chi_n ig {\Bcslash}_{\perp}^{c}
   \Big)(y) F_{k}^{\bn L}  \chi^L_{n,\omega'}(x)\Big]
 \bar\chi^L_{n,\omega}(0)  F_{k}^{R\bn} 
  \Big(\!-\! ig{\Bcslash}_{\perp}^{c}  
  \chi_n\Big)(z) \,,
\end{eqnarray}
with the decomposition of $F_k^{\bn L}\otimes F_k^{R\bn}$ given by exactly
Eq.~(\ref{collin4q}). For the usoft operators we always have $\nslash
\psi_{us}=\nslash \psi_{us}^\bn$, so this projects on to two components of the
light quark fields given by $\psi_{us}^\bn$ with $\bnslash\psi_{us}^\bn=0$. The
Dirac structure can therefore be simplified by noting that
\begin{eqnarray}
  \bar \psi_{us}^\bn P_{L,R} \Gamma {\cal H}_v
     &=& {\rm Tr}\big[\bnslash P_{L,R} \Gamma P_v \big] 
             \bar \psi_{us}^\bn P_{L,R} \frac{\nslash}{2} {\cal H}_v
     + {\rm Tr}\big[\gamma^\perp_\sigma  \frac{\nslash\bnslash}{4}\Gamma P_v\big] 
             \bar \psi_{us}^\bn P_{L,R} \gamma_\perp^\sigma {\cal H}_v \,, \nn\\
 \bar {\cal H}_v  \Gamma P_{R,L} \psi_{us}^\bn 
     &=& {\rm Tr}\big[\Gamma P_{R,L} \bnslash  P_v \big] 
             \bar  {\cal H}_v  \frac{\nslash}{2} P_{R,L} \psi_{us}^\bn
     + {\rm Tr}\big[\Gamma \frac{\bnslash\nslash}{4} \gamma^\perp_\tau P_v\big] 
             \bar {\cal H}_v \gamma_\perp^\tau P_{R,L}\psi_{us}^\bn  \,, 
\end{eqnarray}
and using rotational invariance of the usoft matrix element to determine that
$\gamma_\perp^\sigma$ and $\gamma^\tau_\perp$ terms are restricted to
appearing together.
Thus $\big[\bar {\cal H}_v \overline\Gamma_{j'}^{(0)} F_{k}^{L n} \psi_{us}
\big] \big[ \bar \psi_{us} F_{k}^{n R} \Gamma_j^{(0)} {\cal H}_v \big] $
together with the $(g^\perp_{\alpha\beta}\!+\!i\epsilon^\perp_{\alpha\beta})/2$
factor from Eq.~(\ref{collin4q}) gives
\begin{eqnarray}
&&\delta_{k1}\bigg\{ {\rm Tr}\Big[ \overline\Gamma_{j'}^{(0)} \frac{\nslash}{2} 
   P_{R} \bnslash  P_v \Big] 
   {\rm Tr}\Big[\bnslash P_{L}\frac{\nslash}{2} \Gamma_j^{(0)} P_v \Big]  
   \big[\bar {\cal H}_v  \frac{\nslash}{2} P_R 
     \psi_{us}^\bn \big]
  \big[  \bar \psi_{us}  \frac{\nslash}{2} P_R {\cal H}_v \big]\nn\\
 &&\qquad
   + {\rm Tr}\Big[ \overline\Gamma_{j'}^{(0)} P_L \frac{\nslash}{2} 
  \gamma_\perp^\tau   P_v \Big] 
   {\rm Tr}\Big[\gamma_\perp^\sigma  \frac{\nslash}{2}P_{R} \Gamma_j^{(0)} P_v 
   \Big]  
   \big[\bar {\cal H}_v  \gamma^\perp_\tau P_R 
     \psi_{us}^\bn \big]
  \big[  \bar \psi_{us}^\bn  \gamma^\perp_\sigma P_R {\cal H}_v \big] \bigg\}
  \nn\\
&& - \delta_{k2}
  \frac{(g^\perp_{\alpha\beta}\!+\!i\epsilon^\perp_{\alpha\beta})}{2}
 \bigg\{ 
  {\rm Tr}\Big[ \overline\Gamma_{j'}^{(0)} \frac{\nslash\bnslash}{2} 
  \gamma_\perp^\alpha P_{R}  P_v \Big] 
   {\rm Tr}\Big[  \frac{\bnslash\nslash}{2}\gamma_\perp^\beta P_{R} 
   \Gamma_j^{(0)} P_v \Big]  
   \big[\bar {\cal H}_v  \frac{\nslash}{2} P_L  \psi_{us}^\bn \big]
  \big[  \bar \psi_{us}^\bn  \frac{\nslash}{2}P_L {\cal H}_v \big]\nn\\
 &&\qquad
   + {\rm Tr}\Big[ \overline\Gamma_{j'}^{(0)} P_L \frac{\nslash}{2} 
  \gamma_\perp^\alpha\gamma_\perp^\tau   P_v \Big] 
   {\rm Tr}\Big[\frac{\nslash}{2} \gamma_\perp^\sigma\gamma_\perp^\beta  P_{R} 
   \Gamma_j^{(0)} P_v \Big]  
   \big[\bar {\cal H}_v  \gamma^\perp_\tau P_L 
     \psi_{us}^\bn \big]
  \big[  \bar \psi_{us}^\bn  \gamma^\perp_\sigma P_L {\cal H}_v \big] \bigg\}
  \nn\\
&& + \delta_{k3}\Big\{ \cdots \Big\} + \delta_{k4}\Big\{ \cdots \Big\} 
 \,,
\end{eqnarray}
where the structures for the $\delta_{k3}$ ($\delta_{k4}$) term are identical to
those of $\delta_{k1}$ ($\delta_{k2}$) except with extra $T^A\otimes T^A$
factors in the operators. Simplifying the Dirac structures and using rotational
invariance of the $\langle \bar B_v |\cdots | \bar B_v \rangle$ matrix element
gives
\begin{eqnarray} \label{T2qdirac}
&& \delta_{k1}\bigg\{ {\rm Tr}\Big[ \overline\Gamma_{j'}^{(0)}
   \frac{\nslash\bnslash}{2} 
   P_{L}   P_v \Big] 
   {\rm Tr}\Big[\frac{\bnslash\nslash}{2} P_{R} \Gamma_j^{(0)} P_v \Big]  
   \big[\bar {\cal H}_v  \frac{\nslash}{2} P_R 
     \psi_{us}^\bn \big]
  \big[  \bar \psi_{us}^\bn  \frac{\nslash}{2} P_R {\cal H}_v \big] \nn\\
 &&\qquad
   + {\rm Tr}\Big[ \overline\Gamma_{j'}^{(0)}  \frac{\nslash}{2} 
   \gamma_\perp^\tau P_L  P_v \Big] 
   {\rm Tr}\Big[\gamma_\perp^\tau \frac{\nslash}{2} P_{R} 
   \Gamma_j^{(0)} P_v \Big]  
   \big[\bar {\cal H}_v  \gamma^\perp_\sigma P_R
     \psi_{us}^\bn \big]
  \big[  \bar \psi_{us}^\bn  \gamma_\perp^\sigma P_R {\cal H}_v \big] 
\bigg\}
  \nn\\
&& \!\!\!\!\! - \delta_{k2}
 \bigg\{ 
  {\rm Tr}\Big[ \overline\Gamma_{j'}^{(0)} \frac{\nslash\bnslash}{2} 
  \gamma_\perp^\tau P_{R}  P_v \Big] 
   {\rm Tr}\Big[  \frac{\bnslash\nslash}{2}\gamma^\perp_\tau P_{R} 
   \Gamma_j^{(0)} P_v \Big]  
   \big[\bar {\cal H}_v  \frac{\nslash}{2} P_L  \psi_{us}^\bn \big]
  \big[  \bar \psi_{us}^\bn  \frac{\nslash}{2}P_L {\cal H}_v \big]\nn\\
 &&\qquad
   + {\rm Tr}\Big[ \overline\Gamma_{j'}^{(0)} \nslash P_R 
    P_v \Big] 
   {\rm Tr}\Big[ {\nslash} P_{R} \Gamma_j^{(0)} P_v \Big]  
   \big[\bar {\cal H}_v  \gamma^\perp_\sigma P_L  \psi_{us}^\bn \big]
  \big[  \bar \psi_{us}^\bn  \gamma_\perp^\sigma P_L {\cal H}_v \big] \bigg\}
  \nn\\
&& \!\!\!\!\!  
 + \delta_{k3}\Big\{ \cdots \Big\} + \delta_{k4}\Big\{ \cdots \Big\}\,.
\end{eqnarray}

From Eq.~(\ref{collin5q}) the jet functions are
\begin{eqnarray}
   && \frac{4}{N_c^2\!-\! 1} \Big\langle 0 \Big| 
 \big[\bar \chi_n g {\Bcslash}_{\perp}^{c}\big](y) 
   \frac{\bnslash \gamma^\perp_\alpha P_L T^A}{2} 
   \chi^L_{n,\omega'}(x)
 \bar\chi^L_{n,\omega}(0)  
   \frac{\bnslash \gamma_\perp^\alpha P_L T^A}{2}
   \big[ g{\Bcslash}_{\perp}^{c} \chi_n\big](z)
  \Big| 0 \Big\rangle \nn \\
 &&\quad =  \frac{4iN_\delta }{\omega}
 \int\!\! \frac{dk^+_1\, dk^+_2\, dk^+_3}{(2\pi)^3}  \:
  e^{-\frac{i}{2} [k_3^+x^- \!+ (k_2^+\!- k_3^+)y^-\!+ (k_1^+\!- k_2^+)z^-]}
 \ {\cal J}^{(-4)}_{1\omega}(k^+_j) 
  \,, \nn\\ 
   && \Big\langle 0 \Big| 
 \big[\bar \chi_n g {\Bcslash}_{\perp}^{c}\big](y) 
   \frac{\bnslash \gamma^\perp_\alpha P_L}{2N_c} 
   \chi^L_{n,\omega'}(x)
 \bar\chi^L_{n,\omega}(0)  
   \frac{\bnslash \gamma_\perp^\alpha P_L}{2N_c}
   \big[g{\Bcslash}_{\perp}^{c} \chi_n\big](z)
  \Big| 0 \Big\rangle\nn \\
 &&\quad =  \frac{4iN_\delta }{\omega}
 \int\!\! \frac{dk^+_1\, dk^+_2\, dk^+_3}{(2\pi)^3}  \:
  e^{-\frac{i}{2} [k_3^+x^- \!+ (k_2^+\!- k_3^+)y^-\!+ (k_1^+\!- k_2^+)z^-]}
 \ {\cal J}^{(-4)}_{2\omega}(k^+_j) 
  \,, \nn 
\end{eqnarray}
\begin{eqnarray}
   && \frac{4}{N_c^2\!-\! 1} \Big\langle 0 \Big| 
 \big[\bar \chi_n g {\Bcslash}_{\perp}^{c}\big](y) 
   \bnslash  P_L T^A
   \chi^L_{n,\omega'}(x)
 \bar\chi^L_{n,\omega}(0)  
    \bnslash P_L T^A
   \big[g{\Bcslash}_{\perp}^{c} \chi_n\big](z)
  \Big| 0 \Big\rangle\nn \\
 &&\quad =  \frac{4iN_\delta }{\omega}
 \int\!\! \frac{dk^+_1\, dk^+_2\, dk^+_3}{(2\pi)^3}  \:
  e^{-\frac{i}{2} [k_3^+x^- \!+ (k_2^+\!- k_3^+)y^-\!+ (k_1^+\!- k_2^+)z^-]}
 \ {\cal J}^{(-4)}_{3\omega}(k^+_j) 
  \,, \nn\\[5pt]
  &&  \Big\langle 0 \Big| 
 \big[\bar \chi_n g {\Bcslash}_{\perp}^{c}\big](y) 
   \frac{\bnslash P_L}{N_c} 
   \chi^L_{n,\omega'}(x)
 \bar\chi^L_{n,\omega}(0)  
   \frac{\bnslash P_L}{N_c}
   \big[g{\Bcslash}_{\perp}^{c} \chi_n\big](z)
  \Big| 0 \Big\rangle \nn \\
 &&\quad =  \frac{4iN_\delta }{\omega}
 \int\!\! \frac{dk^+_1\, dk^+_2\, dk^+_3}{(2\pi)^3}  \:
  e^{-\frac{i}{2} [k_3^+x^- \!+ (k_2^+\!- k_3^+)y^-\!+ (k_1^+\!- k_2^+)z^-]}
 \ {\cal J}^{(-4)}_{4\omega}(k^+_j) 
  \,, 
\end{eqnarray}
where the common prefactor is $N_\delta = \delta(\omega\!-\!\omega')
\delta^2(x_\perp) \delta(x^+) \delta^2(y_\perp) \delta(y^+) \delta^2(z_\perp)
\delta(z^+)$ and the $1/\bnP$ factors in the operators do not act outside the
square brackets.  At tree level, only ${\cal J}^{(-4)}_{1\omega}(k^+_j)$ is 
non-zero. It has the value
\begin{eqnarray}
 {\cal J}^{(-4)}_{1\omega}(n\mcdot k_1,n\mcdot k_2,n\mcdot k_3) =
  \frac{1}{[n\mcdot k_1 +i\epsilon]
  [n\mcdot k_2 +i\epsilon][n\mcdot k_3 +i\epsilon]}
  \,.
\end{eqnarray}
As before, we denote the imaginary parts by
\begin{eqnarray}
  {\cal J}^{(-4)}_j = \Big(\frac{-1}{\pi}\Big)\: {\rm Im}\ 
    {\cal J}^{(-4)}_{j\omega} \,.
\end{eqnarray} 
From Eq.~(\ref{T2qdirac}) there are eight shape functions with different Dirac
and color structures, namely 
\begin{eqnarray}
 && 
  \int\!\! \frac{dx^- dy^- dz^-}{128\pi^3}
 e^{-\frac{i}{2} \{ \ell^+_3 x^- +(\ell^+_2 - \ell^+_3)
    y^- +(\ell^+_1 - \ell^+_2) z^- \} }
 \langle \bar B_v |  \bar {\cal H}_v(\tilde x)\, {\nslash} P_R 
   \psi_{us}^\bn(\tilde y) 
   \bar \psi_{us}^\bn(\tilde z)\, {\nslash} P_R {\cal H}_v(0)
   | \bar B_v \rangle 
  \nn  \\
&& \qquad
  = g_9(\ell_1^+,\ell_2^+,\ell_2^+)
  \,, \nn\\ 
 && 
  2\int\!\! \frac{dx^- dy^-dz^-}{128\pi^3}
 e^{-\frac{i}{2} \{ \ell^+_3 x^- +(\ell^+_2 - \ell^+_3)
    y^- +(\ell^+_1 - \ell^+_2) z^- \} }
 \langle \bar B_v |  \bar {\cal H}_v(\tilde x) 
      \gamma_\perp^\eta P_R 
   \psi_{us}^\bn (\tilde y) 
   \bar \psi_{us}^\bn(\tilde z) 
   \gamma^\perp_\eta P_R {\cal H}_v(0)
   | \bar B_v \rangle 
  \nn  \\
&& \qquad
  = g_{10}(\ell_1^+,\ell_2^+,\ell_2^+)
  \,, \nn\\ 
 && 
  \int\!\! \frac{dx^- dy^- dz^-}{128\pi^3}
 e^{-\frac{i}{2} \{ \ell^+_3 x^- +(\ell^+_2 - \ell^+_3)
    y^- +(\ell^+_1 - \ell^+_2) z^- \} }
 \langle \bar B_v |  \bar {\cal H}_v(\tilde x)\, {\nslash} P_L 
   \psi_{us}^\bn(\tilde y) 
   \bar \psi_{us}^\bn(\tilde z)\, {\nslash} P_L {\cal H}_v(0)
   | \bar B_v \rangle 
   \nn  \\
&& \qquad
  = g_5(\ell_1^+,\ell_2^+,\ell_2^+)
  \,, \nn\\ 
 && 
  2\int\!\! \frac{dx^- dy^-dz^-}{128\pi^3}
 e^{-\frac{i}{2} \{ \ell^+_3 x^- +(\ell^+_2 - \ell^+_3)
    y^- +(\ell^+_1 - \ell^+_2) z^- \} }
 \langle \bar B_v |  \bar {\cal H}_v(\tilde x) 
      \gamma_\perp^\eta P_L 
   \psi_{us}^\bn (\tilde y) 
   \bar \psi_{us}^\bn(\tilde z) 
   \gamma^\perp_\eta P_L {\cal H}_v(0)
   | \bar B_v \rangle 
  \nn  \\
&& \qquad
  = g_6(\ell_1^+,\ell_2^+,\ell_2^+)
  \,, \nn\\
&& 
  \int\!\! \frac{dx^- dy^- dz^-}{128\pi^3}
    e^{-\frac{i}{2}  [\ \cdots\:\, ]} 
 \langle \bar B_v |  \bar {\cal H}_v(\tilde x)\, {\nslash} P_R 
   T^A \psi_{us}^\bn(\tilde y) 
   \bar \psi_{us}^\bn(\tilde z)\, {\nslash} P_R T^A {\cal H}_v(0)
   | \bar B_v \rangle 
  \nn  \\
&& \qquad
  = g_7(\ell_1^+,\ell_2^+,\ell_2^+)
  \,, \nn\\ 
 && 
 2 \int\!\! \frac{dx^- dy^-dz^-}{128\pi^3}
    e^{-\frac{i}{2} [\ \cdots\:\, ]} 
 \langle \bar B_v |  \bar {\cal H}_v(\tilde x) 
      \gamma_\perp^\eta P_R T^A 
   \psi_{us}^\bn (\tilde y) 
   \bar \psi_{us}^\bn(\tilde z) 
   \gamma^\perp_\eta P_R T^A {\cal H}_v(0)
   | \bar B_v \rangle 
  \nn  \\
&& \qquad
  = g_{8}(\ell_1^+,\ell_2^+,\ell_2^+)
  \,, \nn
\end{eqnarray}
\begin{eqnarray}
 && 
  \int\!\! \frac{dx^- dy^- dz^-}{128\pi^3}
    e^{-\frac{i}{2}  [\ \cdots\:\, ]} 
 \langle \bar B_v |  \bar {\cal H}_v(\tilde x)\, \nslash P_L T^A 
   \psi_{us}^\bn(\tilde y) 
   \bar \psi_{us}^\bn(\tilde z)\, {\nslash} P_L T^A {\cal H}_v(0)
   | \bar B_v \rangle 
  \nn  \\
&& \qquad
  = f_5(\ell_1^+,\ell_2^+,\ell_2^+)
  \,, \nn\\ 
 && 
 2 \int\!\! \frac{dx^- dy^-dz^-}{128\pi^3}
    e^{-\frac{i}{2}  [\ \cdots\:\, ]} 
 \langle \bar B_v |  \bar {\cal H}_v(\tilde x) 
      \gamma_\perp^\eta P_L T^A
   \psi_{us}^\bn (\tilde y) 
   \bar \psi_{us}^\bn(\tilde z) 
   \gamma^\perp_\eta P_L T^A {\cal H}_v(0)
   | \bar B_v \rangle 
  \nn  \\
&& \qquad
  = f_6(\ell_1^+,\ell_2^+,\ell_2^+)
  \,, 
\end{eqnarray}
where $\psi_{us}^\bn = (\bnslash\nslash)/4\: \psi_{us}$ and the ellipses
denote exactly the same exponent as in the preceding expressions. 
Only the last two of these shape functions show up when we work at tree level 
in the jet functions.

We define the hard coefficients to be 
\begin{eqnarray} \label{h2q}
  h^{[2q]5}_i &=& \frac{1}{4}  \frac{m_b}{\bn\mcdot p} \sum_{j,j'} C_{j'}\, C_j\:
     {\rm Tr}\Big\{ {P_v}
     \overline \Gamma_{j'}^{(0)}\, \frac{\nslash\bnslash}{2} \gamma_\perp^\tau P_R
      \Big\}\:
   {\rm Tr}\Big\{ {P_v} \frac{\bnslash\nslash}{2}\gamma_\tau^\perp P_R 
    \Gamma_j^{(0)}\Big\}
   P_i^{\mu\nu} \,, \nn \\
   h^{[2q]6}_i &=& \frac{1}{2}  \frac{m_b}{\bn\mcdot p} \sum_{j,j'} C_{j'}\, C_j\:
     {\rm Tr}\Big\{ {P_v}
     \overline \Gamma_{j'}^{(0)}\, {\nslash} P_R
      \Big\}\:
   {\rm Tr}\Big\{ {P_v} \nslash P_R \Gamma_j^{(0)}
   \Big\}
   P_i^{\mu\nu} \,, \nn\\
   h^{[2q]7}_i &=&  \frac{1}{4}  \frac{m_b}{\bn\mcdot p} \sum_{j,j'} C_{j'}\, C_j\:
     {\rm Tr}\Big\{ {P_v}
     \overline \Gamma_{j'}^{(0)}\, \frac{\nslash\bnslash}{2} P_L
      \Big\}\:
   {\rm Tr}\Big\{ {P_v} \frac{\bnslash\nslash}{2} P_R \Gamma_j^{(0)}\Big\}
   P_i^{\mu\nu} \,, \nn\\
   h^{[2q]8}_i &=& \frac{1}{2}  \frac{m_b}{\bn\mcdot p} \sum_{j,j'} C_{j'}\, C_j\:
     {\rm Tr}\Big\{ {P_v}
     \overline \Gamma_{j'}^{(0)}\, \frac{\nslash}{2}\gamma_\tau^\perp P_L
      \Big\}\:
   {\rm Tr}\Big\{ {P_v} \gamma_\perp^\tau \frac{\nslash}{2} P_R \Gamma_j^{(0)}
   \Big\}
   P_i^{\mu\nu} \,,
\end{eqnarray}
and the final result from the $T^{(2q)}$ contributions is then
\begin{eqnarray}
  W_i^{(2q)}
  &=& \frac{h_i^{[2q]7}}{m_b}
  \int \! dk^+_1 dk^+_2 dk^+_2 \Big\{
   {\cal J}^{(-4)}_4(\bn\mcdot p\: k^+_j) g_9(k^+_j \!+\! r^+)
   +{\cal J}^{(-4)}_3(\bn\mcdot p\: k^+_j) \:
   g_7(k^+_j \!+\! r^+) \Big\} \nn\\ 
 &+& \frac{h_i^{[2q]8}}{m_b}
  \int \! dk^+_1 dk^+_2 dk^+_2 \Big\{
   {\cal J}^{(-4)}_4(\bn\mcdot p\: k^+_j) g_{10}(k^+_j \!+\! r^+)
   +{\cal J}^{(-4)}_3(\bn\mcdot p\: k^+_j) \:
   g_{8}(k^+_j \!+\! r^+) \Big\} \nn\\ 
 &+& \frac{h_i^{[2q]5}}{m_b}
  \int \! dk^+_1 dk^+_2 dk^+_2 \Big\{
   {\cal J}^{(-4)}_2(\bn\mcdot p\: k^+_j) g_5(k^+_j \!+\! r^+)
   +{\cal J}^{(-4)}_1(\bn\mcdot p\: k^+_j) \:
   f_{5}(k^+_j \!+\! r^+) \Big\} \nn\\ 
 &+& \frac{h_i^{[2q]6}}{m_b}
  \int \! dk^+_1 dk^+_2 dk^+_2 \Big\{
   {\cal J}^{(-4)}_2(\bn\mcdot p\: k^+_j) g_6(k^+_j \!+\! r^+)
   +{\cal J}^{(-4)}_1(\bn\mcdot p\: k^+_j) \:
   f_{6}(k^+_j \!+\! r^+) \Big\} 
  \,.\nn\\
\end{eqnarray}
Explicit results for the hard coefficients $h_i^{[2q]5,6}$ are summarized in
Sec.~\ref{sectHard}.

\subsubsection{Calculation of $\hat T^{(2b)}$ and $\hat T^{(2c)}$}

For later convenience we set
\begin{eqnarray}
  B_j^{(1a)} J^{(1a)}_j +  B_j^{(1b)} J^{(1b)}_j &=& 
    \int\! d\omega_1 d\omega_2\: B_j^{(1)}(\omega_1,\omega_2) 
    \ J^{(1)}_j(\omega_1,\omega_2)\,, 
%
%
\end{eqnarray}
where
\begin{eqnarray}
 J_j^{(1)}(\omega_1,\omega_2)  &=&
 \frac{1}{m_b} \overline \chi_{n,\omega_1} 
  (i g {\cal B}_{c\perp}^\alpha)_{\omega_2} \: \Upsilon_{j\alpha}^{(1)}\:
  {\cal H}_v 
  \,,
%
%
\end{eqnarray}
and $B_j^{(1)f}(\omega_1,\omega_2)=-B_j^{(1a)f}(\omega_1\!+\!\omega_2)\:
m_b/(\omega_1\!+\!\omega_2)$ for $j=1$--$3$ when $f=u$, and for $j=1$--$4$ when
$f=s$, and
$B_j^{(1)f}(\omega_1,\omega_2)=B_j^{(1b)f}(\omega_1,\omega_2)/(n\mcdot v)$ for
the remaining $j$'s.

Both $\hat T^{(2b)}$ and $\hat T^{(2c)}$ have jet functions that vanish at tree
level but are non-zero at one-loop order. The steps are the same as in previous
sections. Here we have
\begin{eqnarray} \label{T2bstart}
 \hat T^{(2b)} \!\!&=&\!\!
 -i \sum_{j,j'}\int\!\! d\omega_1 d\omega_2 d\omega_3 d\omega_4
 \, B_j^{(1)}(\omega_{1,2})
   B_{j'}^{(1)}(\omega_{3,4}) \delta(\omega_3\!+\!\omega_4\!-\! \bn\mcdot p) 
  \nn\\
 &&\qquad\times
  \int\!\! d^4x\: e^{-i r\cdot x}\
      {\rm T} \big[ J_{j'}^{(1)\dagger}(\omega_{3,4},x)\: 
        J_j^{(1)}(\omega_{1,2},0) \big] \,.
\end{eqnarray}
To Fierz transform we can use
\begin{eqnarray} \label{FierzLxR}
 P_L\otimes P_R=  \big( \frac{\nslash}{2} P_R \big) \otimes 
   \big( \frac{\bnslash}{2N_c} P_L \big) + 
   \big( \frac{\nslash}{2} P_R T^a \big) \otimes 
   \big( {\bnslash} P_L T^a \big) \,,
\end{eqnarray}
and we can drop the second term, since color implies that it does not
contribute in the matrix elements. Thus,
\begin{eqnarray} \label{T2b1}
  J_{j'}^{(1)\dagger} J_j^{(1)} &=&
    \frac{-1}{2m_b^2} {\rm Tr}\Big[ \frac{P_v}{2}
 \, 
   \overline \Upsilon_{j'\alpha'}^{(1)}\frac{\nslash}{2} P_R 
     \Upsilon_{j\alpha}^{(1)} \Big]
   \Big[\bar {\cal H}_v(x)  {\cal H}_v(0) \Big]\\
 &&\qquad \times
   \Big[ \big(\bar \chi_{n,\omega_1} ig {\cal B}_{c,\omega_2}^{\perp\alpha}\big)(0)
    \frac{\bnslash}{N_c} P_L \big(\!-\!ig {\cal B}_{c,\omega_4}^{\perp\alpha'}
     \chi_{n,\omega_3}\big)(x)\Big]\nn\,,
\end{eqnarray}
where we have used the fact that the second term in the reduction in
Eq.~(\ref{eq:projform}) gives a vanishing matrix element. Between $\bar B$
mesons, the soft operator in Eq.~(\ref{T2b1}) gives the leading-order shape
function $f^{(0)}$.  The matrix element of the collinear operator gives the jet
functions
\begin{eqnarray} \label{Jmelt2b}
  &&\hspace{-0.9cm}
 \big\langle 0 \big| \Big[ \big(\bar \chi_{n,\omega_1} 
    ig {\cal B}_{c,\omega_2}^{\perp\alpha}\big)(0)
    \frac{\bnslash}{N_c} P_L \big(\!-\!ig {\cal B}_{c,\omega_4}^{\perp\alpha'}
     \chi_{n,\omega_3}\big)(x)\Big]
   \big| 0 \big\rangle \\
  &=& 
   (-2i)(\omega_1\!+\omega_2)\, \delta(\omega_1\!+\!\omega_2\!-\!\omega_3\!-\!\omega_4)
  \delta^2(x_\perp) \delta(x^+) 
    \int\!\! \frac{dk^+}{2\pi}\: e^{-\frac{i}{2} k^+ x^-}\,\nn\\
 &&\qquad\times
  \big[ g_\perp^{\alpha\alpha'}{\cal J}^{(2)}_{1\omega_j}(k^+) + 
     i \epsilon_\perp^{\alpha\alpha'}{\cal J}^{(2)}_{2\omega_j}(k^+) \big],\nn
\end{eqnarray}
and their imaginary parts are denoted by ${\cal J}_{m}^{(2)}(\omega_j k^+) =
(-1/\pi) {\rm Im}\: {\cal J}_{m\omega_j}(k^+)$.  Noting the delta functions in
Eqs.~(\ref{T2bstart}) and (\ref{Jmelt2b}), we define the hard coefficients to
be 
\begin{eqnarray} \label{h2b}
  h_i^{[2b]9}(z_{1},z_{2},\bn\mcdot p) &=& \sum_{j,j'} \!
   B^{(1)}_{j'}(\omega_{3,4})  B_j^{(1)}(\omega_{1,2}) \ 
  g_\perp^{\alpha\alpha'}\, \frac{\bn\mcdot p}{m_b}\:
  {\rm Tr} \Big[ \frac{P_v}{2}\, \overline \Upsilon^{(1)}_{j'\alpha'\mu} \,
  \frac{\nslash}{2}\,  \Upsilon^{(1)}_{j\alpha\nu} \Big] P_i^{\mu\nu}\,,\\
 h_i^{[2b]10}(z_{1},z_{2},\bn\mcdot p) &=& \sum_{j,j'} \!
   B^{(1)}_{j'}(\omega_{3,4})  B_j^{(1)}(\omega_{1,2}) \ 
  i\epsilon_\perp^{\alpha\alpha'}\, \frac{\bn\mcdot p}{m_b}\:
  {\rm Tr} \Big[ \frac{P_v}{2}\, \overline \Upsilon^{(1)}_{j'\alpha'\mu} \,
  \frac{\nslash}{2}\,  \Upsilon^{(1)}_{j\alpha\nu} \Big] P_i^{\mu\nu}\,,\nn
\end{eqnarray}
where $\omega_1=z_1\bn\mcdot p$, $\omega_2=(1\!-\!z_1)\bn\mcdot p$,
$\omega_3=z_2\bn\mcdot p$, $\omega_4=(1\!-\!z_2)\bn\mcdot p$. This gives the
factorization theorem
\begin{eqnarray} \label{W2bi}
W_i^{(2b)} 
 &=& \int\!\! dz_1 dz_2\ 
  \frac{h_i^{[2b]9}(z_1,z_2,\bn\mcdot p)}{m_b} \:
  \int_{0}^{p_X^+} \! \!\! dk^+\: 
   {\cal J}^{(2)}_1(z_{1},z_2,p_X^-\, k^+ )\: 
   f^{(0)}(k^+  \!+\!\overline\Lambda\!-\! p_X^+) 
 \\  
 &+& \int\!\! dz_1 dz_2\ 
  \frac{h_i^{[2b]10}(z_1,z_2,\bn\mcdot p)}{m_b} \:
  \int_{0}^{p_X^+} \! \!\! dk^+\: 
   {\cal J}^{(2)}_2(z_{1},z_2,p_X^-\, k^+ )\: 
   f^{(0)}(k^+  \!+\!\overline\Lambda\!-\! p_X^+) .\ \ \phantom{x}
  \nn
\end{eqnarray}

%
%

For $\hat T^{(2c)}$ we have
\begin{eqnarray} \label{Tnlo2c}
  \hat T^{(2c)} &=& -i \sum_{j,j'} \sum_{\ell=c,d,e,f} \int\! [d\omega_n]
   \: \delta\big( \omega \!-\!\bn\mcdot p\big)\
    C_{j'}(\omega)  A_j^{(2\ell)}(\omega_n) \!
    \int\!\! d^4x\,   e^{-i r\cdot x}\nn  \\
  &&\times   {\rm T}\: \big[ J^{(0)\dagger}_{j'\,\mu}(\bn\mcdot p,x)\:
   J^{(2\ell)}_{j\,\nu}(\omega_n,0) 
   + J^{(2\ell)\dagger}_{j\,\mu}(\omega_n,x)\:
   J^{(0)}_{j'\,\nu}(\bn\mcdot p,0) \big]  \,,  
\end{eqnarray}
where we have $n=1$ for $\ell=c$ and $n=1,2$ for $\ell=d,e,f$.  For convenience
we set $J^{(2\ell)}_{j}(\omega_n)= \overline {\cal C}^{(2\ell)}_{n,\omega_n}
\Upsilon^{(2\ell)}_j {\cal H}_v$, so that ${\cal C}^{(2\ell)}$ contains the
product of collinear fields in each current.  Fierz transforming 
$\hat T^{(2c)}$ with Eq.~(\ref{FierzLxR}) (again we need the first term only) 
and then reducing the Dirac structure using Eq.~(\ref{eq:projform}) gives
\begin{eqnarray}
&& J^{(0)\dagger}_{j'} J^{(2\ell)}_{j} + J^{(2\ell)\dagger}_{j} J^{(0)}_{j'}\\
 &&=  \frac{-1}{4}\, {\rm tr}\Big[ \frac{P_v}{2} \bar\Upsilon_j^{(2\ell)} 
  \frac{\nslash}{2} P_R \Gamma_{j'}^{(0)} +\mbox{h.c.} \Big] 
   \Big[\bar {\cal H}_v(x)  {\cal H}_v(0) \Big]
   \Big[\bar \chi_{n,\omega}(0) \frac{\bnslash P_L}{N_c}
     {\cal C}^{(2\ell)}_{n,\omega_{1,2}}(x) 
    + \mbox{(h.c., $x\leftrightarrow 0$)}\Big]\nn\\
 &&\ +  \frac{-1}{4}\, {\rm tr}\Big[ \frac{P_v}{2} \bar\Upsilon_j^{(2\ell)} 
  \frac{\nslash}{2} P_R \Gamma_{j'}^{(0)} -\mbox{h.c.} \Big] 
   \Big[\bar {\cal H}_v(x)  {\cal H}_v(0) \Big]
   \Big[\bar \chi_{n,\omega}(0) \frac{\bnslash P_L}{N_c}
     {\cal C}^{(2\ell)}_{n,\omega_{1,2}}(x) 
    - \mbox{(h.c., $x\leftrightarrow 0$)}\Big]\nn\,.
\end{eqnarray}
The usoft operator here gives the leading-order shape function $f^{(0)}$.
Next we define the jet functions
\begin{eqnarray} \label{Jmelt2c}
  &&\hspace{-0.9cm}
 (-1)\big\langle 0 \big| \bar \chi_{n,\omega}(0)
    \frac{\bnslash P_L}{N_c}  \Big(\Big[\frac{1}{\bnP} ig n\mcdot{\cal B}_{c}\Big]
     \chi_{n}\Big)_{\omega_1}\!\!(x) \pm \mbox{(h.c., $x\leftrightarrow 0$)}
   \big| 0 \big\rangle \\
  &&\qquad = 
   \frac{-2i}{\omega}\, \delta(\omega\!-\!\omega_1)
  \delta^2(x_\perp) \delta(x^+) 
    \int\!\! \frac{dk^+}{2\pi}\: e^{-\frac{i}{2} k^+ x^-}
   {\cal J}^{(2)}_{3\omega,4\omega}(k^+) ,\nn\\
&&\hspace{-0.9cm}
 \frac{1}{m_b} \big\langle 0 \big| \bar \chi_{n,\omega}(0)
    \frac{\bnslash P_L}{N_c}  \big[( ig n\mcdot{\cal B}_{c})_{-\omega_2}
     \chi_{n,\omega_1}\big](x) \pm \mbox{(h.c., $x\leftrightarrow 0$)}
   \big| 0 \big\rangle \nn\\
  &&\qquad =
   \frac{-2i}{m_b}\, \delta(\omega\!-\!\omega_1\!-\!\omega_2)
  \delta^2(x_\perp) \delta(x^+) 
    \int\!\! \frac{dk^+}{2\pi}\: e^{-\frac{i}{2} k^+ x^-}
   {\cal J}^{(2)}_{5\omega_{1,2},6\omega_{1,2}}(k^+) ,\nn\\
&&\hspace{-0.9cm}
 \frac{-1}{m_b} \big\langle 0 \big| \bar \chi_{n,\omega}(0)
    \frac{\bnslash P_L}{N_c} \frac{1}{\omega_1}
   \big[ig {\cal B}_{c,-\omega_2}^{\perp\alpha'} 
   \big( i\cD_{c\perp}^\alpha  \chi_{n}\big)_{\omega_1}\big](x) 
   \pm \mbox{(h.c., $x\leftrightarrow 0$)}
   \big| 0 \big\rangle \nn\\
  &&\qquad =
   \frac{-2i}{m_b}\, \delta(\omega\!-\!\omega_1\!-\!\omega_2)
  \delta^2(x_\perp) \delta(x^+) 
    \int\!\! \frac{dk^+}{2\pi}\: e^{-\frac{i}{2} k^+ x^-}
   {\cal J}^{(2)}_{7\omega_{1,2},8\omega_{1,2}}(k^+) ,\nn\\
&&\hspace{-0.9cm}
 \frac{1}{m_b} \big\langle 0 \big| \bar \chi_{n,\omega}(0)
    \frac{\bnslash P_L}{N_c} \frac{1}{\omega_2}
   \big[ \big(ig {\cal B}_{c}^{\perp\alpha'}\, i\!\! \cDgppPl)_{\omega_2}
     \chi_{n,\omega_1}\big](x) 
   \pm \mbox{(h.c., $x\leftrightarrow 0$)}
   \big| 0 \big\rangle \nn\\
  &&\qquad =
   \frac{-2i}{m_b}\, \delta(\omega\!-\!\omega_1\!-\!\omega_2)
  \delta^2(x_\perp) \delta(x^+) 
    \int\!\! \frac{dk^+}{2\pi}\: e^{-\frac{i}{2} k^+ x^-}
   {\cal J}^{(2)}_{9\omega_{1,2},10\omega_{1,2}}(k^+) ,\nn
\end{eqnarray}
and their imaginary parts ${\cal J}_{m}^{(2)}(\omega_j k^+) = (-1/\pi) {\rm
  Im}\: {\cal J}_{m\omega_j}(k^+)$, as well as the hard coefficients
\begin{eqnarray} \label{h2c}
  h_i^{[2c]11,12}(\bn\mcdot p) &=& \frac{m_b}{2\bn\mcdot p} \sum_{j,j'} \!
   C_{j'}(\bn\mcdot p)  A_j^{(2c)}(\bn\mcdot p) \ 
  {\rm Tr} \Big[ \frac{P_v}{2}\, \overline \Upsilon^{(2c)}_{j\mu} \,
  \frac{\nslash}{2}\,  \Gamma^{(0)}_{j'\nu} \pm \mbox{h.c.} \Big] P_i^{\mu\nu}
  \,,\\
 h_i^{[2c]13,14}(z_{1},\bn\mcdot p) &=& \frac12 \sum_{j,j'}  \!
   C_{j'}(\bn\mcdot p)  A_j^{(2d)}(\omega_{1,2}) \ 
  {\rm Tr} \Big[ \frac{P_v}{2}\, \overline \Upsilon^{(2d)}_{j\mu} \,
  \frac{\nslash}{2}\,  \Gamma^{(0)}_{j'\nu} \pm \mbox{h.c.} \Big] P_i^{\mu\nu} 
  \,,\nn\\
 h_i^{[2c]15,16}(z_{1},\bn\mcdot p) &=& \frac12 \sum_{j,j'}  \!
   C_{j'}(\bn\mcdot p)  A_j^{(2e)}(\omega_{1,2}) \ 
  {\rm Tr} \Big[ \frac{P_v}{2}\, \overline \Upsilon^{(2e)}_{j\mu} \,
  \frac{\nslash}{2}\,  \Gamma^{(0)}_{j'\nu} \pm \mbox{h.c.} \Big] P_i^{\mu\nu} 
  \,,\nn\\
 h_i^{[2c]17,18}(z_{1},\bn\mcdot p) &=& \frac12 \sum_{j,j'}  \!
   C_{j'}(\bn\mcdot p)  A_j^{(2f)}(\omega_{1,2}) \ 
  {\rm Tr} \Big[ \frac{P_v}{2}\, \overline \Upsilon^{(2f)}_{j\mu} \,
  \frac{\nslash}{2}\,  \Gamma^{(0)}_{j'\nu} \pm \mbox{h.c.} \Big] P_i^{\mu\nu} 
  \,,\nn
\end{eqnarray}
where $\omega_1=z_1 \bn\mcdot p$ and $\omega_2=(1\!-\!z_1)\bn\mcdot p$.

Putting all the pieces together we have
\begin{eqnarray} \label{W2ci}
W_i^{(2c)} 
 &=& \sum_{m=3,4} 
  \frac{h_i^{[2c]m+8}(\bn\mcdot p)}{m_b} \:
  \int_{0}^{p_X^+} \! \!\! dk^+\: 
   {\cal J}^{(2)}_m(p_X^-\, k^+ )\: 
   f^{(0)}(k^+  \!+\!\overline\Lambda\!-\! p_X^+) 
 \\  
 &+& \sum_{m=5}^{10} \int\!\! dz_1  \:
  \frac{h_i^{[2c]m+8}(z_1,\bn\mcdot p)}{m_b} \:
  \int_{0}^{p_X^+} \! \!\! dk^+\: 
   {\cal J}^{(2)}_m(z_{1},p_X^-\, k^+ )\: 
   f^{(0)}(k^+  \!+\!\overline\Lambda\!-\! p_X^+) \,.\ \ \phantom{x}
  \nn
\end{eqnarray}

\subsubsection{Calculation of the remaining $\hat T^{(2)}$'s}

The remaining $T^{(2)}$'s for which we have not yet defined hard, jet and soft
functions include $T^{(2La)}$, $T^{(2Lb)}$, $T^{(2LL)}$ and $T^{(2Ga)}$.
These time-ordered products all have jet functions that start at one-loop order
and, unlike $T^{(2b,2c)}$, they also induce new shape functions, $g_j$. 
Factorization formulae can be derived for these contributions by following 
similar steps to the previous cases. Rather than going through this exercise, 
we instead list some of the soft operators and shape functions that would be 
required. 

For $T^{(2La)}$  we have 
\begin{eqnarray}
 P^\beta_{8\lambda}(\ell_{1,2}^+)
 &=& \int\!\! \frac{dx^-dy^-}{\!32\pi^2}\, e^{-\frac{i}{2}\ell_2^+ x^-} 
   e^{-\frac{i}{2}(\ell^+_1 - \ell^+_2) y^-}
  \Big[ 
  \overline {\cal H}_v(\tilde x)   
     i\! {\overrightarrow {\cal D}_{\!us}^{\beta}}(\tilde y)
   \gamma^T_{\lambda}\gamma_5 {\cal H}_v(0) 
    \Big] \,, \nn\\
%
 P^\beta_{9\lambda}(\ell_{1,2}^+)
 &=& \int\!\! \frac{dx^-dy^-}{\!32\pi^2}\, e^{-\frac{i}{2}\ell_2^+ x^-} 
   e^{-\frac{i}{2}(\ell^+_1 - \ell^+_2) y^-}
  \Big[ 
  \overline {\cal H}_v(\tilde x)   T^A
     i\! {\overrightarrow {\cal D}_{\!us}^{\beta}}(\tilde y) T^A
   \gamma^T_{\lambda}\gamma_5 {\cal H}_v(0) 
    \Big]\nn\,,\\[5pt]
&& \hspace{-2.2cm}
\langle \bar B_v | P_{8,9}^{\beta\lambda}(\ell^+_{1,2}) | \bar B_v\rangle 
    = \: \epsilon_\perp^{\beta\lambda} \:
     g_{11,12}^{(2)}(\ell_1^+,\ell_2^+) \,,
\end{eqnarray}
while for $T^{(2Lb)}$ we find
\begin{eqnarray}
    O_{8}^{\beta}(\ell_{1,2}^+)\: 
 &=& \int\!\! \frac{dx^-dy^-}{\!32\pi^2}\,  e^{-\frac{i}{2}\ell_2^+ x^-}
   e^{-\frac{i}{2}(\ell^+_1 - \ell^+_2) y^-}
  \Big[ 
   \overline {\cal H}_v(\tilde x)   
   i\! {\overrightarrow {\cal D}_{\!us}^{\beta}}(\tilde y) {\cal H}_v(0) 
    \Big] 
   \,, \nn \\ 
%
  O_{9}^{\beta}(\ell_{1,2}^+)\: 
 &=& \int\!\! \frac{dx^-dy^-}{\!32\pi^2}\,  e^{-\frac{i}{2}\ell_2^+ x^-}
   e^{-\frac{i}{2}(\ell^+_1 - \ell^+_2) y^-}
  \Big[ 
   \overline {\cal H}_v(\tilde x)  T^A 
   i\! {\overrightarrow {\cal D}_{\!us}^{\beta}}(\tilde y) T^A {\cal H}_v(0) 
    \Big] 
   \nn\,, \\[5pt]
&& \hspace{-2.2cm}
 \langle \bar B_v | \bn_\beta\, O_{8,9}^{\beta}(\ell_{1,2}^+) | \bar B_v\rangle 
    =    g_{13,14}^{(2)}(\ell^+_1,\ell_2^+)  \,.
\end{eqnarray}
For both $T^{(2La)}$ and $T^{(2Lb)}$, the induced jet functions are of order
$\lambda^0$.

For $T^{(2LL)}$ the necessary soft operators are
\begin{eqnarray}
O_{10}^{\alpha\beta}(\ell_{1,2,3}^+) &=&
  i\!
  \int\!\! \frac{dx^-  dy^- dz^-}{128\pi^3} 
    e^{-\frac{i}{2} \{\ell_3 x^- + (\ell^+_2 - \ell^+_3) y^- +
    (\ell^+_1 - \ell^+_2) z^- \} }
   \big[\overline {\cal H}_v(\tilde x) i \cD_{us}^{\perp\alpha}(\tilde y) i
   \cD_{us}^{\perp\beta}(\tilde z) 
   {\cal H}_v(0) \big] \nn\,,
\\
P_{10\lambda}^{\alpha\beta}(\ell_{1,2,3}^+) &=&
  i\!
  \int\!\! \frac{dx^-  dy^- dz^-}{128\pi^3} 
    e^{-\frac{i}{2} \{\:\, \cdots \: \} }
   \big[\overline {\cal H}_v(\tilde x) i \cD_{us}^{\perp\alpha}(\tilde y) i
   \cD_{us}^{\perp\beta}(\tilde z) \gamma^T_\lambda\gamma_5\,
   {\cal H}_v(0) \big] \,,
\end{eqnarray}
together with operators $O_{11-15}^{\alpha\beta}(\ell_{1,2,3}^+)$ and
$P_{11-15\lambda}^{\alpha\beta}(\ell_{1,2,3}^+)$ that have different color
contractions between the gauge-invariant objects $\bar {\cal H}_v$,
$i\cD^{\perp\alpha}$, $i\cD^{\perp\beta}$, and ${\cal H}_v$. The shape functions
are
\begin{eqnarray}
  \langle \bar B_v | O_{10-15}^{\alpha\beta}(\ell^+_{1,2,3}) | \bar B_v\rangle 
    &=& \: g_\perp^{\alpha\beta} \:
     g_{15-20}^{(4)}(\ell_1^+,\ell_2^+,\ell_3^+) \,,\nn\\[3pt] 
  \langle \bar B_v | P_{10-15\lambda}^{\alpha\beta}(\ell^+_{1,2,3}) | \bar B_v\rangle 
    &=& \:-\epsilon_\perp^{\alpha\beta} 
     \Big(v_\lambda \!-\! \frac{n_\lambda}{n\mcdot v}\Big) \:
     g_{21-26}^{(4)}(\ell_1^+,\ell_2^+,\ell_3^+) \,.
\end{eqnarray}
For $T^{(2LL)}$ the jet functions are ${\cal O}(\lambda^{-2})$.

Finally, for $T^{(2Ga)}$ the time-ordered product has a structure that will
produce the same soft operators, $O_3^{\alpha\beta}$ and
$P_{4\lambda}^{\alpha\beta}$, as $T^{(2L)}$ and thus has shape functions
$f_3^{(4)}(\ell_1^+,\ell_2^+)$ and $f_4^{(4)}(\ell_1^+,\ell_2^+)$.  The jet
functions for $T^{(2Ga)}$ are ${\cal O}(\lambda^{-2})$.

\subsection{Summary of hard coefficients}
\label{sectHard}

To determine $h_i^{1f-6f}$ we need to compute the traces in
Eqs.~(\ref{h2ab}), (\ref{h2L}) and (\ref{h2q}). Since there is no possibility
of confusion, we shall now drop the bracketed factors, $[2-]$, in the 
superscripts. 
For $B\to X_u e\bar\nu$ ($f=u$), expanding to leading order in 
$\bar y_H \gg u_H$ gives
\begin{eqnarray}
  \label{eq:NLOhu}
h_1^{1u} \!\!& = &\!\! \frac{1}{8} \,, \quad\ \:
h_2^{1u}   =   -\frac{(2 \!-\! \bar y_H)}{2 \bar y_H^2} \,, \quad\ \ 
h_3^{1u}   =   \frac{1}{4 m_B \bar y_H} \,, \quad\!\!
h_4^{1u}   =  -\frac{1}{m_B^2 \bar y_H^2} \,, \quad\!\!\!
h_5^{1u}   =   \frac{4-\bar y_H}{4m_B \bar y_H^2} \,, 
  \nn \\[5pt]
h_1^{2u} \!\!& = &\!\! \frac{-1}{4} \,, \quad\!
h_2^{2u}   =   \frac{ 4 \bar y_H \!-\!\bar y_H^2 \!-\!2}{\bar y_H^3}, \quad\!\!
h_3^{2u}   =   \frac{-1}{2 m_B \bar y_H} , \quad
h_4^{2u}  =  \frac{2 y_H\!-\!2}{m_B^2 \bar y_H^3} , \quad
h_5^{2u}   =   \frac{\bar y_H^2 \!-\! 6\bar y_H \!+\! 4}{2m_B \bar y_H^3} , 
  \nn \\[5pt]
h_1^{3u} \!\!& = &\!\! \frac{-1}{4 \bar y_H} \,, \quad\!\!
h_2^{3u}   =   -\frac{1}{\bar y_H^2} \,, \quad\ \ \ \ \ \ \ \  
h_3^{3u}   =  \frac{-1}{2 m_B \bar y_H^{\,2}} , \quad 
h_4^{3u}  = 0\,,\quad \ \ \ \  \ \ \ 
h_5^{3u}   =   \frac{1}{2 m_B \bar y_H^2} \,, 
  \nn \\[5pt]
h_1^{4u} \!\!& = &\!\! \frac{1}{4 \bar y_H} \,, \quad\!\!
h_2^{4u}   =   \frac{\bar y_H\!-\!2}{\bar y_H^3} \,, \quad\ \ \ \ \ 
h_3^{4u}   =   \frac{1}{2 m_B \bar y_H^{\,2}} \,, \quad 
h_4^{4u}  =  \frac{-2}{m_B^2 \bar y_H^3} \,, \quad\ 
h_5^{4u}   =   \frac{4\!-\!\bar y_H}{2 m_B \bar y_H^{\,3} } \,, 
  \nn \\[5pt]
h_1^{5u} \!\!& = &\!\! \frac{-1}{2} \,, \quad
h_2^{5u}   = \frac{2-2\bar y_H}{\bar y_H^2}  \,, \quad\ \ \ 
h_3^{5u}   =  \frac{-1}{m_B \bar y_H} \,, \quad \: 
h_4^{5u}  =  \frac{2}{m_B^2 \bar y_H^2} \,, \quad\ \,
h_5^{5u}   =   -\frac{(2-\bar y_H)}{m_B \bar y_H^2} \,, 
  \nn \\[5pt]
h_1^{6u} \!\!& = &\!\! 0 \,, \quad\  \ \ 
h_2^{6u}   =  \frac{2}{\bar y_H^2} \,, \quad\ \ \ \ \ \ \ \ \ \:
h_3^{6u}   =  0 \,,\quad \ \ \ \ \ 
h_4^{6u}  =  \frac{2}{m_B^2 \bar y_H^2} \,, \quad\ \ \ 
h_5^{6u}   =   \frac{-2}{m_B \bar y_H^2} \,.  
\end{eqnarray}
For $B\to X_s\gamma$ $(f=s)$ we expand around $x_H=1$, which gives
\begin{eqnarray}
   \label{eq:NLOhs}
h_1^{1s} & = & -\frac{m_B^2}{8} 
   \,, \quad
h_2^{1s}   =  0\,, \quad
h_3^{1s}   =   -\frac{m_B}{4} 
   \,, \quad 
h_4^{1s}  =  \frac{5}{8}\,, \quad\quad
h_5^{1s}   =   -\frac{m_B}{4} 
   \,,  \nn \\
h_1^{2s} & = & \frac{m_B^2}{4} 
   \,, \quad\quad
h_2^{2s}   =   0\,, \quad
h_3^{2s}   =   \frac{m_B}{2} 
   \,, \quad\ \  
h_4^{2s}  =  -\frac{7}{4}\,, \quad
h_5^{2s}   =   \frac{m_B}{2} 
   \,, \nn \\
h_1^{3s} & = & -\frac{m_B^2}{4}
   \,, \quad
h_2^{3s}   =   0\,, \quad
h_3^{3s}   =   -\frac{m_B}{2} 
   \,, \quad 
h_4^{3s}  =  \frac{3}{4}\,, \quad\quad
h_5^{3s}   =   -\frac{m_B}{2} 
   \,, 
 \nn \\
h_1^{4s} & = & \frac{m_B^2}{4} 
  \,, \quad\ \ 
h_2^{4s}   =   0\,, \quad\ \ 
h_3^{4s}   =   \frac{m_B}{2} 
  \,,\quad 
h_4^{4s}  =  -\frac{5}{4}\,, \quad
h_5^{4s}   =   \frac{m_B}{2} 
  \,, \nn \\
h_1^{5s} & = & -\frac{m_B^2}{2} 
  \,, \quad
h_2^{5s}   =   0\,, \quad
h_3^{5s}   =   -m_B 
   \,,\quad 
h_4^{5s}  =  2\,, \quad\ \ 
h_5^{5s}   =   -m_B 
  \,,  \nn \\
h_1^{6s} & = & 
h_2^{6s}   =   
h_3^{6s}   =   0\,, \quad 
h_4^{6s}   =   \frac{1}{2}\,, \quad
h_5^{6s}   =   0\,. 
\end{eqnarray}



\section{Summary of Decay Rates to NLO} \label{sect_Summary}


\subsection{Discussion of NLO results} \label{sect_NLOdiscussion}

To facilitate the computation of the NLO corrections to the decay rates, they
were divided into several pieces, as discussed at the beginning of
Sec.~\ref{sect_NLO}. The full results at order $\Lambda/m_b$ and all orders in
$\alpha_s$ are quite complicated. However, now that factorization for
the NLO rates has been achieved, we are able to expand consistently in factors
of $\alpha_s$ evaluated at perturbative scales. In this section, we discuss 
which terms are kept for phenomenological purposes.

The most complicated contributions are the NLO correction to the $W_i$,
namely $W_i^{(2)f}$, where $f=s$ for $B\to X_s\gamma$ and $f=u$ for $B\to
X_u\ell\bar\nu$.  From Eq.~(\ref{Wfact00b}) in Sec.~\ref{sect_NLO}, the NLO
factorization theorem is
\begin{eqnarray} \label{Wfact00c}
W_i^{(2)f}  &=& 
  \frac{h_i^{0f}(\bn\mcdot p)}{2m_b} \: 
  \int_{0}^{p_X^+} \!\! dk^+\ 
  {\cal J}^{(0)}(\bn\mcdot p\, k^+,\mu )\:
   f^{(2)}_0\big(k^+ + r^+,\mu\big) \,
  \nn\\
 &+&
  \sum_{r=1}^2
  \frac{h_i^{rf}(\bn\mcdot p)}{m_b} \: 
  \int_{0}^{p_X^+} \!\! dk^+\ 
  {\cal J}^{(0)}(\bn\mcdot p\, k^+,\mu )\:
   f^{(2)}_r\big(k^+ + r^+,\mu\big) \,
  \nn\\
 &+& 
  \sum_{r=3}^4
  \frac{h_i^{rf}(\bn\mcdot p)}{m_b} 
  \int\!\! dk_1^+\: dk_2^+ 
   {\cal J}^{(-2)}(\bn\mcdot p\, k_j^+,\mu )\:
   f^{(4)}_r\big(k_j^+ + r^+,\mu\big) \,
 \nn\\
 &+&
  \sum_{r=5}^6
    \frac{h_i^{rf}(\bn\mcdot p)}{\bn\mcdot p} 
  \int\!\! dk_1^+ dk_2^+ dk_3^+  \ 
 {\cal J}^{(-4)}(\bn\mcdot p\, k_{j'}^+,\mu )\:
   f^{(6)}_r\big(k_{j'}^+ + r^+,\mu\big) 
 \nn\\
 &+& \ldots \,,
\end{eqnarray}
where $j = 1,2$ and $j' = 1,2,3$.  The ellipses denote terms given in
Eq.~(\ref{Wfact00b}) that have jet functions ${\cal J}$ that start at one-loop
order. These additional terms are of order 
\begin{eqnarray} \label{drop}
 \frac{\alpha_s(\mu_0)}{\pi} \ \frac{\Lambda}{m_b} \,,
\end{eqnarray}
where $\mu_0\simeq \sqrt{m_b\Lambda_{\rm QCD}}$,
so they are suppressed relative to the terms displayed in Eq.~(\ref{Wfact00c}).
They also induce dependence on new unknown shape functions, $g_{0-26}$,
which makes it prohibitive to include them phenomenologically. 
Thus, throughout this section we shall stick to the terms displayed in 
Eq.~(\ref{Wfact00c}), which enter at order $\Lambda/m_b$ ($r=0$--$4$) and 
$4\pi\alpha_s(\mu_0)\Lambda/m_b$ ($r=5,6$).  
For convenience we drop the subscript $1$ on ${\cal J}_1^{(-2,-4)}$
in this section, since there is no possibility of confusion.  At tree level, the
jet functions were computed in Sec.~\ref{sectJet2} and are
\begin{eqnarray}
 && {\cal J}^{(0)}(k^+) = \delta(k^+) \,,\qquad 
  {\cal J}^{(-2)}(k_j^+)=\frac{\delta(k_1^+)-\delta(k_2^+)}{k_2^+ - k_1^+}
   \,,\\
 && {\cal J}^{(-4)}(k_j^+) =  
  {4\pi\alpha_s(\mu_0)}
  \left[\frac{\delta(k_1^+)}{(k_2^+)(k_3^+)} 
      + \frac{\delta(k_2^+)}{(k_1^+)(k_3^+)} 
      + \frac{\delta(k_3^+)}{(k_1^+)(k_2^+)} 
      - \pi^2 \delta(k_1^+)\delta(k_2^+)\delta(k_3^+)
      \right]  .\nn
\end{eqnarray}
The result for ${\cal J}^{(0)}$ at one-loop order can be found in 
Eq.~(\ref{J01loop}).
Examples of diagrams that contribute to Eq.~(\ref{Wfact00c}) can be found in
Table~\ref{table_T}.  Results for the hard functions $h_i^{0f}$ were given in
Eqs.~(\ref{hiuLO}) and (\ref{hisLO}) and for $h_i^{1f-6f}$ in
Eqs.~(\ref{eq:NLOhu}) and (\ref{eq:NLOhs}).  The running of the $h_i^{rf}$'s in
Eq.~(\ref{Wfact00c}) is the same as the running of the $h_i^{0f}$'s in
Eq.~(\ref{hiuLO}), and the $h_i^{0f}$'s depend only on the running of the
$C_i$'s computed in Ref.~\cite{Bauer:2000yr}.  The shape functions
$f_{0-6}^{(n)}$ are defined by Eqs.~(\ref{defnfs1}) and (\ref{defnfs2}).

More ${\cal O}(\lambda^2)$ power corrections occur when one switches
to hadronic variables in the LO factorization theorem; these were given in
Eq.~(\ref{Wi0rNLO}). For phenomenological purposes terms that start suppressed
by the factor in Eq.~(\ref{drop}) are not included, so we include the
$(\lambda_1+3\lambda_2)$ term but not the $\overline \Lambda$ term.

Finally, the simplest source of power corrections is the kinematic expansion of
the prefactors in the decay-rate formulae, which depend on which rate we
consider (see Eqs.~(\ref{dGamma3t})-(\ref{dGamma6u})).\footnote{If one desires,
  he or she can straightforwardly treat $x_H$ in the doubly and triply
  differential rates in a different manner from what we have done here, by using
  Eqs.~(\ref{dGamma3u}) and (\ref{dGamma4u}) instead of Eqs.~(\ref{dGamma6u})
  and (\ref{dGamma5u}) and making the desired expansions.} In addition
there are corrections from the expansion of $n\mcdot q$ in the $h_i$'s; these
give the $h_i^{0'f}$'s in Eqs.~(\ref{hiuLOp}) and (\ref{hisLOp}).  Both of these
sources involve the same jet function and shape function as the LO contributions
but different hard coefficients, which we shall denote by $G^S$, $G^T_{1a,1b}$,
and $G^D$ in the next section.

\subsection{NLO Results for the Endpoint Decay Rates} \label{sectKin}

In this section we combine the pieces to arrive at the NLO results for
decay rates. The hadronic dimensionless variables that we use are
\begin{eqnarray}
  x_H^\gamma \!= \frac{2 E_\gamma}{m_B} , \quad
  x_H \!= \frac{2 E_{\ell}}{m_B} , \quad
  y_H \!= \frac{q^2}{m_B^2}, \quad 
  s_H \!= \frac{m_X^2}{m_B^2} ,\quad
  \overline y_H \!= \frac{\bn\mcdot p_X}{m_B} ,\quad
  u_H \!= \frac{n\mcdot p_X}{m_B}  ,
\end{eqnarray}
where only the first one is for $B\to X_s\gamma$ and the rest are for $B\to
X_u\ell\bar\nu$. Note that $s_H=u_H \overline y_H$ and $y_H =(1\!-\!
u_H)(1\!-\!\overline y_H)$, so these variables are not independent.  The decay
rates derived with SCET are valid for spectra dominated by the endpoint or SCET
region, in which $u_H/\overline y_H\lesssim \lambda_H^2$ and
$1-x_H^\gamma\lesssim \lambda_H^2$, where $\lambda_H^2$ is a small expansion
parameter which one can choose to be $\sim 0.2$ or $\sim\Lambda/m_b\simeq 0.1$.
In this region we write the decay rates as
\begin{eqnarray} \frac{1}{\Gamma} \frac{d\Gamma}{dZ} &=&
  \bigg( \frac{1}{\Gamma} \frac{d\Gamma}{dZ} \bigg)^{\! LO} + \bigg(
  \frac{1}{\Gamma} \frac{d\Gamma}{dZ} \bigg)^{\! NLO} + \ldots \,,
\end{eqnarray}
where $Z$ denotes a generic choice of phase-space variables. The doubly
differential and triply differential decay rates that we consider were discussed
in Secs.~\ref{sect_decay} and \ref{sect_kin}. We caution that if the SCET
expansion is used for decay rates integrated over a larger region of phase space
than the SCET region then it is important to check that sufficient smearing has
occured so that the contributions outside the SCET region do not cause the power
expansion to break down.  In Eq.~(\ref{combine}) we proposed a method one could
use to combine results from the local OPE with those from SCET even if a larger
region of phase space was desired (which become relevant when radiative
corrections are included).  Note that the $\lambda_H$ parameter provides a means
of testing for cases where a pure SCET expansion is valid. Here we present
results for arbitary $\lambda_H$, and leave the investigation of the expansion
in different decay rates to future work.

For convenience, we define a shorthand notation for the convolutions that appear
at LO and NLO:
\begin{eqnarray}\label{J0f0}
  \big[{\cal J}\otimes f\big](p_X^+,p_X^-) 
   &=& \int \!\! dk^+_j
   \ 
   {\cal J}(p_X^-\: k^+_j,\mu ) 
   \
   f\big(k_j^+ +\overline \Lambda \!-\! p_{X}^+,\mu\big)\,.
\end{eqnarray}
At LO we simply combine the $h_i^{0s}$'s and $h_i^{0u}$'s from
Eqs.~(\ref{hiuLO}) and (\ref{hisLO}) and find
\begin{eqnarray} \label{dGammaLO}
\bigg( \frac{1}{\Gamma_0^s}\: \frac{d\Gamma^s}{  dx_H^\gamma} \bigg)^{LO}
    \! &=&\! H^{S}\ 
   \big[{{\cal J}^{(0)}\otimes f^{(0)}}\big]\big(m_B(1\!-\! x_H^\gamma),m_B\big)
  \,,  \\[5pt]
 \bigg(\frac{1}{\Gamma_0^u}\: 
 \frac{d^3\Gamma^u}{ dx_H^{cut}\, d\overline y_H\, du_H}  \bigg)^{LO}
    \! &=& \!  H^{T}(\overline y_H)\
   \big[{{\cal J}^{(0)}\otimes f^{(0)}}\big](m_B u_H, m_B \overline y_H)
   \,,\nn\\[5pt]
 \bigg(\frac{1}{\Gamma_0^u}\: 
 \frac{d^2\Gamma^u}{  d\overline y_H\, du_H}  \bigg)^{LO}
    &=&  H^{D}(\overline y_H)\ 
   \big[{{\cal J}^{(0)}\otimes f^{(0)}}\big](m_B u_H, m_B \overline y_H)
    \,, \nn
\end{eqnarray}
where the convolution of the LO jet function and shape function is given by
Eq.~(\ref{J0f0}) (with one integration variable $k^+$) and the hard coefficients
are
\begin{eqnarray}
  H^{S}(\mu) \!\! &=&\!\! {m_B}
 \big[ C_1^{(t)}  \!-\!\frac12\, C_2^{(t)}\!-\!C_3^{(t)} \big]^2 \,,
  \\
  H^{T}(\mu,\overline y_H)
   \!\! &=&\!\! {12m_B}\: \overline y_H (1 \!-\! \overline y_H)
   \big[C_1^{(v)}\: \big]^2 \,,
  \nn\\
  H^{D}(\mu,\overline y_H)
   \!\! &=&\!\! {2m_B\:  \overline y_H^{\,2}}\, \, 
   \Big\{ (3\!-\!2 \overline y_H)(C_1^{(v)})^2 
      \!+\! 2 C_1^{(v)}\big(C_3^{(v)} \!+\! \frac{\overline y_H}{2} C_2^{(v)}\big)
     \!+\! \big(C_3^{(v)} \!+\! \frac{\overline y_H}{2} C_2^{(v)}\big)^2 \Big\}
   \nn \,.
\end{eqnarray}
Note that we have suppressed the functional dependence of the coefficients,
which once perturbative corrections are included is $C^{(t)}_i(\mu,m_B)$ and
$C^{(v)}_i(\mu,{m_B},\overline y_H)$.  Our $W_i^{(0)}$'s for both $B\to
X_u\ell\bar\nu$ and $B\to X_s\gamma$, and hence the corresponding LO triply and
singly differential rates, respectively, agree with Ref.~\cite{Bauer:2003pi}.
These results also match those of Ref.~\cite{Bosch:2004th}.

At NLO, combining all contributions, as discussed in the
Sec.~\ref{sect_NLOdiscussion}, leads to the following expressions:
\begin{eqnarray} 
\bigg( \frac{1}{\Gamma_0^s}\: \frac{d\Gamma^s}{  dx_H^\gamma} \bigg)^{\!\! NLO}
    \!\!
         &= & -\,\frac{(\lambda_1+3\lambda_2)}{2m_B} \: H^{S}(m_b)   
   \   \big[{{\cal J}^{(0)}\otimes \widetilde{f^{(0)}}}\big]
    (m_B(1\!-\!x_H^\gamma),m_B)
  \, \nn \\
   &+& (1\!-\!x_H^\gamma)\ G^{S}\   
      \big[{{\cal J}^{(0)}\otimes f^{(0)}}\big](m_B(1\!-\!x_H^\gamma),m_B)
  \, \nn \\
   &+&  \sum_{j=0}^{6} H_j^{S}(m_B)
   \   \big[{\cal J}^{(n_j)}\otimes f_j\big](m_B(1\!-\!x_H^\gamma),m_B)
   \,, \label{dGammaNLOa} \\[5pt]
 \bigg(\frac{1}{\Gamma_0^u}\: 
 \frac{d^3\Gamma^u}{ dx_H^{cut}\, d\overline y_H\, du_H}  \bigg)^{\!\! NLO}
    \!\!\!\! 
  &=& -\, \frac{(\lambda_1+3\lambda_2)}{2m_B} H^{T}(\overline y_H)\   
   \   \big[{{\cal J}^{(0)}\otimes \widetilde{f^{(0)}}}\big](m_B u_H,m_B\overline y_H)
  \, \nn \\
  &&\hspace{-1cm}+ \Big[u_H\: G_{1a}^{T}(\overline y_H)
   \!+\! (u_H\!+\!x_H\!-\!1)\: G_{1b}^{T}(\overline y_H)  \Big] 
  \big[{{\cal J}^{(0)}\otimes f^{(0)}}\big](m_B u_H,m_B\overline y_H)
   \, \nn \\
  &&\hspace{-1cm} +   \sum_{j=0}^{6} H_{j}^{T}(\bar y_H)
   \   \big[{\cal J}^{(n_j)}\otimes f_j\big](m_B u_H,m_B\overline y_H)
   \,, \label{dGammaNLOb}\\[5pt]
 \bigg(\frac{1}{\Gamma_0^u}\: 
 \frac{d^2\Gamma^u}{  d\overline y_H\, du_H}  \bigg)^{\!\! NLO}
    \!\! 
      &=& -\, \frac{(\lambda_1+3\lambda_2)}{2m_B} H^{D}(\overline y_H)\   
   \   \big[{{\cal J}^{(0)}\otimes \widetilde{f^{(0)}}}\big]
     (m_B u_H,m_B\overline y_H)
  \, \nn \\
     &+&  u_H\: G^{D}(\overline y_H)\ 
   \big[{{\cal J}^{(0)}\otimes f^{(0)}}\big](m_B u_H,m_B\overline y_H)
   \, \nn \\
         &+& \sum_{j=0}^{6} H_{j}^{D}(\bar y_H)
   \   \big[{\cal J}^{(n_j)}\otimes f_j \big](m_B u_H,m_B\overline y_H)
     \,, \label{dGammaNLOc} 
\end{eqnarray}
where $n_0=n_1=n_2=0$, $n_3=n_4=-2$, and $n_5=n_6=-4$, and the notation
$\big[{{\cal J}^{(n_j)}\otimes f_j}\big]$ denotes the convolutions displayed in
Eq.~(\ref{Wfact00c}) (see also Eq.~(\ref{J0tildef0})).  
From the NLO phase-space factors in Eqs.~(\ref{dGamma3t}), (\ref{dGamma6u}) 
and (\ref{dGamma5u}), and from the NLO terms in the expansion of the $h_i$, 
i.e. Eqs.~(\ref{hisLOp}) and (\ref{hiuLOp}), we find that
\begin{eqnarray}
  G^{S}(m_b)
  &=& -{3\, m_B} 
  \big( C_1^{(t)} \!-\!\frac12\, C_2^{(t)}\!-\!C_3^{(t)}\big)^2 \,,
  \\
  G_{1a}^{T}(\overline y_H) &=& {12m_B} (\overline y_H^{\,2}-1) (C_1^{(v)})^2
  \,, \nn\\
  G_{1b}^{T}(\overline y_H) &=& {12m_B} 
    \Big\{ 2(1\!-\!\overline y_H ) (C_1^{(v)})^2 
   - \big[ C_1^{(v)} \!+\! C_3^{(v)} \!+\! \frac{\overline y_H}{2} 
   C_2^{(v)} \big]^2 \Big\}\,,
  \nn\\
  G^{D}(\overline y_H) &=& {2m_B}\, \overline y_H  \Big[ 
   (2\overline y_H^{\,2}\!-\!6)(C_1^{(v)})^2 
   - \overline y_H (\overline y_H\!+\!3)(C_1^{(v)}\!+\! C_3^{(v)})\, C_2^{(v)}
 \nn\\
  &&\quad - \overline y_H^2 (C_2^{(v)})^2 
    -2(1\!+\!\overline y_H) (2 C_1^{(v)}\!+\!C_3^{(v)})C_3^{(v)}
  \Big] \,.
  \nn
\end{eqnarray}
By using Eqs.~(\ref{eq:NLOhu}) and (\ref{eq:NLOhs}), we find that the hard
coefficients for the terms with subleading shape functions are
\begin{eqnarray}
H_0^{S} & = & \frac{H^S}{2 m_B}
  \,, \quad
H_1^{S}   = -\frac{1}{2} \,, \quad
H_2^{S}   =  1  \,, \quad
H_3^{S}  =  -1 \,, \quad
H_4^{S}   =    1 \,, \quad
H_5^{S}   =   -2 \,, \quad
H_6^{S}   =    0\,, \nn \\[5pt]
H_{0}^{T} & = & \frac{H^T}{2 m_B}
  \,, \qquad 
H_{1}^{T}   =   6 \overline y_H (1\!-\!\overline y_H)\,, \qquad
H_{2}^{T}   =   -12\overline y_H (1\!-\!\overline y_H)\,, \qquad
H_{3}^{T}   =   -12(1\!-\!\bar y_H) \,, \nn \\
 && \hspace{-0.7cm} H_{4}^{T}  =  12(1\!-\!\bar y_H) \,, \qquad
H_{5}^{T}   =   -24(1\!-\!\overline y_H) \,, \qquad
H_{6}^{T}   =    0 \,, \nn \\[5pt]
H_{0}^{D} & = & \frac{H^D}{2 m_B}
 \,, \qquad
H_{1}^{D}  =  \overline y_H^2(1-2\overline y_H)  , \quad
H_{2}^{D}   =   2\overline y_H (2 \overline y_H^2 +\overline y_H -2) , \quad 
H_{3}^{D}   =   4\overline y_H^2-6\overline y_H\,, \nn \\
&& \hspace{-0.7cm} H_{4}^{D} = 2 \overline y_H -4\overline y_H^2\,, \qquad
H_{5}^{D}   =   8\overline y_H^2-8\overline y_H\,, \qquad
H_{6}^{D}   =    4 \overline y_H \,.
\end{eqnarray}
The $H_{1-6}$ factors displayed here will be corrected by $\alpha_s(m_b)/\pi$
terms at one-loop order.  Note that the shape functions $f_{5,6}$ depend on the
flavor of the light quark in the four-quark operator and therefore differ for
$B\to X_s$ and $B\to X_u$.

Equation~(\ref{dGammaNLOa}) describes $B\to X_s\gamma$ in the endpoint region.
In Eqs.~(\ref{dGammaNLOb}) and (\ref{dGammaNLOc}), convergence of the SCET
expansion requires $u_H \ll \bar y_H$. For Eq.~(\ref{dGammaNLOb}), we have in
addition made a cut on $x_H$ such that $1-x_H \sim \lambda^2$.  In contrast, for
Eq.~(\ref{dGammaNLOc}) the full range of $x_H$ has been integrated over.  We can
also straightforwardly obtain the rate $d^2\Gamma^u/dq^2 dm_X^2$, or 
equivalently $d^2\Gamma^u/d s_H d y_H$, by changing variables from $\{\overline
y_H,u_H\}$ to $\{y_H,s_H\}$ in the $W_i$ and re-expanding. This is done by using
$\bar y_H = \zeta - s_H/\zeta - \ldots$ and $u_H = s_H/\zeta + \ldots$, where
$\zeta=1-y_H+s_H$ and the ``$\ldots$'' terms are not needed at NLO in the power
expansion. The result may then be substituted into Eq.~(\ref{dGamma5sy}).
Integration over $y_H$ then gives the $m_X^2$ spectrum, $d\Gamma^u/ds_H$, for
which restriction to the endpoint region requires $cut(s_H,y_H)\equiv
\{s_H/\zeta^2\le \lambda_H^2/(1+\lambda_H^2)^2\}$.


\subsection{Singly Differential 
Spectra} \label{sect_singly}

In this section we give results for singly differential spectra at NLO that are
useful for measurements of $|V_{ub}|$. Since our goal is to explore the effect
of the subleading shape functions, in this section we shall work at lowest order
in $\alpha_s$ for the hard and jet functions, both at LO and NLO. For the shape
functions we shall use $f_1^{(2)}(\omega) = 2\omega
f^{(0)}(\omega)$~\cite{Bauer:2001mh} and define
\begin{eqnarray} \label{defnF0}
  F(p^+) & = & f^{(0)}(\bar\Lambda-p^+) 
   + \frac{1}{2m_B}\, f_0^{(2)}(\bar\Lambda-p^+) 
   - \frac{\lambda_1+3\lambda_2}{2m_B} f^{(0)\,\prime}(\bar\Lambda-p^+)\,,
 \\
  F_2(p^+) & = & f_2^{(2)}(\bar\Lambda-p^+)   \,, \nn
\end{eqnarray}
where a prime denotes a derivative,
as well as
\begin{eqnarray}
 F_{3,4}(p^+) \!\!& = &\!\! \int\! dk_1^+ \, dk_2^+ 
  \left[\frac{\delta(k_1^+)-\delta(k_2^+)}{k_2^+ - k_1^+} \right] 
  f_{3,4}^{(4)}(k_j^+ +\bar\Lambda-p^+ ), \\
 F_{5,6}(p^+) \!\! & = &\!\! \int\! dk_1^+ \, dk_2^+ \, dk_3^+ 
  \left[\frac{\delta(k_1^+)}{(k_2^+)(k_3^+)} 
      \!+\! \frac{\delta(k_2^+)}{(k_1^+)(k_3^+)} 
      \!+\! \frac{\delta(k_3^+)}{(k_1^+)(k_2^+)} 
      \!-\! \pi^2 \delta(k_1^+)\delta(k_2^+)\delta(k_3^+)
      \right]\nn\\
 && \quad\times
  f_{5,6}^{(6)}(k_j^+ + \bar\Lambda-p^+) \,. \nn
\end{eqnarray}
We shall put an additional superscript $s$ or $u$ on $F_{5,6}(p^+)$ to
distinguish the original four-quark operator with strange- and up-type quarks.
For $B\to X_s\gamma$ the rate $d\Gamma^s/dx_H^\gamma$ in the endpoint region is
equivalent to making a cut on $x_H^\gamma$. The necessary results already appear
in Eqs.~(\ref{dGammaLO}) and (\ref{dGammaNLOa}) and combining them gives
\begin{eqnarray} \label{eq:dGsdx}
 \frac{1}{\Gamma_0^s} \frac{d\Gamma^s}{dx_H^\gamma}\bigg|_{x_H^\gamma> x_H^c} 
 \!\!\! &=&  m_B \big[ 1 - 3 (1\!-\!x_H^\gamma)\big] 
    F\big(m_B(1\!-\! x_H^\gamma)\big) 
  + \big[m_B(1\!-\!x_H^\gamma)\!-\!\overline\Lambda\,\big]
    F\big(m_B(1\!-\! x_H^\gamma)\big) \nn \\
    & &  
         + F_2\big(m_B(1\!-\!x_H^\gamma)\big)
         - F_3\big(m_B(1\!-\!x_H^\gamma)\big) 
        +F_4\big(m_B(1\!-\!x_H^\gamma)\big) \nn\\
    & &  - 8\pi\alpha_s(\mu_0)\ F_5^s\big(m_B(1\!-\!x_H^\gamma)\big)
    , 
\end{eqnarray}
where $1-x_H^c \sim \lambda^2$.  Here the $-3(1-x_H^\gamma)$ term comes from
expanding the kinematic prefactors, while the term that depends on
$[m_B(1-x_H)-\overline\Lambda]$ comes from reducing $f_1^{(2)}$ to $f^{(0)}$
(and we write it in terms of $F$ for convenience, even though this includes
some formally higher-order terms).  The rate in Eq.~(\ref{eq:dGsdx}) for
$B\rightarrow X_s\gamma$ has been computed previously at NLO in
Ref.~\cite{Bauer:2001mh}. The relation of our shape functions to those in
Ref.~\cite{Bauer:2001mh} is shown in Table~\ref{table:notatn}. Our result in
Eq.~(\ref{eq:dGsdx}) agrees with theirs, up to the $F_5^{(6)}$ term, which was
computed here for the first time.\footnote{In making this comparison, note that
  Ref.~\cite{Bauer:2001mh} gave their result with partonic variables, whereas we
  have expressed ours in terms of hadronic variables here.}


To compute $d\Gamma^u/dx_H$ for $B \rightarrow X_u l \bar\nu$, we can use
Eq.~(\ref{dGamma3u}) and case i) of Table~\ref{table_limits2}. Treating
$r_\pi^2=0.026\sim \lambda^4$, we can set $r_\pi=0$ even at NLO.  To ensure that
the SCET expansion converges requires that one makes a cut on $u_H$. First
consider the $x_H$ spectrum. Since $u_H\le 1-x_H$, making a cut $x_H > x_H^c$
restricts $u_H$ to the desired small values. However, at NLO accuracy this is
not equivalent to our $u_H \le \lambda_H^2 \overline y_H$ definition of the SCET
region of phase space, and an additional term, $K(\lambda_H,x_H)$, is added to
correct for this. Depending on the choice of other cuts, the error in including
a larger region of phase space may be power suppressed.  The parameter
$\lambda_H$ provides us with a way of testing this by comparing $\lambda_H=0.2$
and $\lambda_H=1$.  We present our final results in a manner that makes it easy
to take the $\lambda_H\to 1$ limit for situations where a large enough region
has been smeared over that this is the case.  For the $x_H$ spectrum the result is
\begin{eqnarray} \label{eq:dGudx}
\frac{1}{\Gamma_0^u}\ 
 \frac{d\Gamma^u}{dx_H} \bigg|_{\substack{x_H > x_H^c\\ 
  \frac{u_H}{\overline y_H} < \lambda_H^2}} 
  &=& {2}\mspace{-50.0mu}\int\limits_0^{\mspace{70.0mu}\min\{1-x_H,\lambda_H^2
 \}}\!\!\!\!\! d u_H\, \Big\{
    m_B (1-4 u_H)F(m_B u_H) 
    + (\overline\Lambda\!-\!m_B u_H)\, F(m_B u_H)
  \nn \\
  &-& 
    {F_2(m_B u_H)} - {3}  F_3(m_B u_H) + {3} F_4(m_B u_H) 
  -24\pi\alpha_s(\mu_0) \: {F_5^u(m_B u_H)} \Big\} \nn\\[3pt]
 &+& K(\lambda_H, x_H) 
  \,,
\end{eqnarray}
where
\begin{eqnarray} \label{eq:K}
  K(\lambda_H, x_H) & = & 
   2\mspace{-50.0mu} \int\limits_{\qquad (1-x_H)\lambda_H^2}^{\qquad\quad
  \min\{1-x_H,\lambda_H^2\}}
  \!\!\!\!\! d u_H\, \bigg\{
  \frac{u_H^2}{\lambda_H^6}\: (2u_H\!-\! 3\lambda_H^2)\:
\Big[m_B F(m_B u_H) + (\overline\Lambda\!-\!m_B u_H)F(m_B u_H)
\nn\\ 
& &\mspace{-60.0mu}
 - F_2(m_B u_H) \Big]
  -\frac{u_H}{\lambda_H^4}\: (u_H\!-\! 2\lambda_H^2)\:  
\Big[3 F_3(m_B u_H)- 3 F_4(m_B u_H) + 24\pi\alpha_s(\mu_0) F_5^u(m_B u_H)\Big]
\nn\\
  && \mspace{-60.0mu} - \frac{2 u_H (3 u_H (1\!-\!x_H) \lambda_H^2 \!+\! u_H^3 
  \!-\! 3 (1\!-\!x_H) \lambda_H^4 \!-\! 3 u_H^2 \lambda_H^2 )}{\lambda_H^6} 
  \Big[m_B F(m_B u_H)\Big] \bigg\} \,.
\end{eqnarray}
The $\lambda_H$-dependent term, $K(\lambda_H, x_H)$, arises when one writes the
integral over $\bar y_H$ as $\int_{1-x_H}^{1}\! d \bar y_H\, +
\int_{u_H/\lambda_H^2}^{1-x_H}\! d \bar y_H\,
\theta\left(\frac{u_H}{\lambda_H^2} -1 + x_H\right) $, which is done to ensure
that $u_H \le \lambda_H^2 \overline y_H$.  After inputing models for the shape
functions one can use the $K(\lambda_H,x_H)$ term to check the consistency of
the operator product expansion with this level of smearing.

The rate $d\Gamma^u/dx_H$ has been computed at NLO in Ref.~\cite{Bauer:2002yu}
and we can compare our result with theirs. To do this we take $\lambda_H\to 1$,
which sets $K(\lambda_H,x_H)\to 0$, since the restrictions we imposed on the
phase space were not considered there. We also must convert back to partonic
variables, which means dropping the $(\lambda_1 \!+\! 3\lambda_2)$ term in
Eq.~(\ref{defnF0}).  After doing this, we find agreement with their
Eq.~(35)~\footnote{We found that they have an overall 2 typographical error in
  this equation (see also~\cite{Neubert:2002yx}).} on the coefficients of the LO
$F$ term and the $F_2$ and $F_4$ terms, but we disagree on the $u_H F$ term and
the $F_3$ term. (Again the $F_5$ term is computed for the first time here, so no
cross-check on this is possible.) The coefficients of our $F_3$ term also
disagrees with the RPI constraint derived in Ref.~\cite{Campanario:2002fy},
which predicted that they occur in the combination $m_B F - F_3$.\footnote{Note
  that, from the point of view of the factorization theorem, such a
  reparameterization constraint would be very interesting, since it would give a
  relation between the jet functions ${\cal J}^{(-2)}$ and ${\cal J}^{(0)}$ to
  all orders in $\alpha_s$, even though these operators appear to be defined by
  unrelated matrix elements, Eqs.~(\ref{Jmelt}) and (\ref{Jetm2}).} We found
that this combination occurs for $B\to X_s\gamma$, but not for $B\to
X_u\ell\bar\nu$.  In the next section, we present a non-trivial cross-check of
the coefficients of our result in Eq.~(\ref{eq:dGudx}), namely that when
re-expanded it correctly reproduces terms in the local OPE up to $1/m_b^3$.

Next we consider the $p_X^-$ spectrum, namely $d\Gamma/d\overline y_H$.
Integrating the doubly differential rate from Eqs.~(\ref{dGammaLO}) and
(\ref{dGammaNLOc}) with a cut on $u_H$, i.e.\ $u_H \leq \lambda_H^2 \bar y_H$,
restricts one to the triangular SCET region shown in Fig.~\ref{fig:phase} and
gives
\begin{eqnarray} \label{eq:dGdyb}
 \frac{1}{\Gamma_0^u}\:
  \frac{d\Gamma^u}{d\bar y_H}\bigg|_{\bar u_H \le {\lambda_H^2}\overline y_H}
   &=& {2}
   \int_{0}^{\lambda_H^2\bar y_H}\!\!\! du_H\:  \bigg\{
    m_B \big[ (3\bar y_H^2\!-\!2\bar y_H^3) 
    + 2 u_H  (\bar y_H^3\! -\! 3\bar y_H)\big]
   F(m_B u_H)  \nn\\
 &&  + (\overline\Lambda\!-\!m_B u_H)  ({\bar y_H^2}\!-\!2\bar y_H^3) 
  {F(m_B u_H)}
  + (2\bar y_H^3\! +\! \bar y_H^2 \!-\! 2{\bar y_H}) F_2(m_B u_H)\nn
    \\[5pt]
   & &  + (2\bar y_H^2\!-\!3\bar y_H) F_3(m_B u_H) 
        + (\bar y_H\!-\!2\bar y_H^2) F_4(m_B u_H)
   \nn \\[3pt]
   & &   +16\pi\alpha_s(\mu_0)(\bar y_H^2\!-\!\bar y_H) 
      F_5^u(m_B u_H)
      + {8\pi\alpha_s(\mu_0)\bar y_H} F_6^u(m_B u_H)  
   \bigg\}
   \,.
\end{eqnarray}
Another possibility is to consider the $p_X^+$ spectrum, which is $d\Gamma/du_H$,
where we now integrate with the cut on $\bar y_H$, obtaining
\begin{eqnarray} \label{eq:dGdu}
 \frac{1}{\Gamma_0^u}\:
 \frac{d\Gamma^u}{d u_H}\bigg|_{\overline y_H \ge \frac{u_H}{\lambda_H^2}}
 & = & 
   \bigg\{ m_B\big[ (1 \!-\! 2 {\tilde u}^3 \!+\! \tilde u^4) 
  - u_H (5 \!-\! 6 \tilde u^2 \!+\! \tilde u^4) \big] \, F(m_B u_H)
 \\
& & \hspace{-0.8cm} - \frac{1}{3} (\overline\Lambda \!-\!m_B u_H)
   (1\!+\!2 \tilde u^3\! -\!3\tilde u^4)
  \, F(m_B u_H)
 - \frac{1}{3} (1\!-\! 6 \tilde u^2 \!+\! 2 \tilde u^3 \!+\! 3\tilde u^4)
 \, F_2(m_B u_H) 
\nn \\
& &  \hspace{-0.8cm}- \frac{1}{3} 
 (5 \!-\! 9\tilde u^2 \!+\! 4 \tilde u^3)\, F_3(m_B u_H)
 - \frac{1}{3} (1 \!+\! 3 \tilde u^2 \!-\! 4 \tilde u^3)\, F_4(m_B u_H)
\nn \\
& &  \hspace{-0.8cm} 
   -\frac{16\pi\alpha_s(\mu_0)}{3}(1\!-\! 3 \tilde u^2\!+\! 2 \tilde u^3)\, 
      F_5^u(m_B u_H)
    +8\pi\alpha_s(\mu_0) (1\!-\!\tilde u^2)\, F_6^u(m_B u_H) 
  \bigg\}, 
  \nn
\end{eqnarray}
where $\tilde u = u_H/\lambda_H^2$ is a parameter of order 1 for $\lambda_H\sim
0.2$. Given sufficient smearing, one can take $\lambda_H\to 1$ and the $\tilde
u$ terms become subleading.  The subleading terms in $d\Gamma^u/d\overline u_H$
provide power corrections to the phenomenological analysis using this spectrum
in Ref.~\cite{Bosch:2004bt}. We leave the derivation of $d\Gamma^u/ds_H$ with
phase-space cuts to future work, and so have not compared this rate with
Ref.~\cite{Burrell:2003cf}.



\subsection{Comparison with the local OPE} \label{sect_localOPE}

A non-trivial check on our power-correction results can be obtained by  
expanding the subleading shape functions in a manner appropriate to the case  
where the local OPE is valid.  Essentially this means expanding
\begin{eqnarray}
   \delta(\ell^+-in\mcdot D) = \delta(\ell^+) - \delta'(\ell^+) in\mcdot D + 
   \ldots \,.
\end{eqnarray}
The comparison will be made at the level of the partonic $d\Gamma^u/dx$  
decay rate, and we define $w=m_B u_H-\overline\Lambda$. From  
Eq.~(\ref{eq:dGudx}) we have
\begin{eqnarray} \label{eq:dGudxp}
   \frac{1}{\hat\Gamma_0^u} \frac{d\Gamma^u}{dx}
   &=& {2}\Big(\frac{m_B}{m_b}\Big)^4
   \int_0^{1-x_H}\!\!\! d u_H\, \Big\{
     m_B (1-4 u_H) f^{(0)}(\overline\Lambda\!-\!m_B u_H)
     + \frac12 f_0^{(2)}(\overline\Lambda\!-\!m_B u_H)
   \\[3pt]
   &-&\frac{(\lambda_1\!+\!3\lambda_2)}{2}  
f^{(0)'}(\overline\Lambda\!-\!m_B u_H)
   + (\overline\Lambda\!-\!m_B u_H)\, f^{(0)}(\overline\Lambda\!-\!m_B  u_H)
   -{f_2^{(2)}(\overline\Lambda\!-\!m_B u_H)}  \nn\\[3pt]
   &-& {3}  F_3(m_B u_H)   + {3} F_4(m_B u_H)
     -24\pi\alpha_s(\mu_0) \: {F_5^u(m_B u_H)} \Big\}
    + K(\lambda_H, x_H)   \nn\\
   &&\hspace{-1cm}= {2}\Big(\frac{m_B}{m_b}\Big)^4
   \int_{-\overline\Lambda}^{m_b (1-x)}\!\!\! dw\, \Big\{
     \Big[1- \frac{4w}{m_b}-\frac{4\overline\Lambda}{m_b} \Big]  
f^{(0)}(-w)
     + \frac{1}{2m_b} f_0^{(2)}(-w) - \frac{w}{m_b} f^{(0)}(-w)
   \nn \\[3pt]
   && \hspace{-1cm} -
   \frac{1}{m_b} {f_2^{(2)}(-w)}  - \frac{3}{m_b}  F_3(w\!+\!\overline\Lambda)
   + \frac{3}{m_b} F_4(w\!+\!\overline\Lambda)
    - \frac{24\pi\alpha_s(\mu_0)}{m_b} \: {F_5^u(w\!+\!\overline\Lambda)} \Big\}
    + K(\lambda_H, x_H) ,  \nn
\end{eqnarray}
where $x=2E_\ell/m_b$ and $\hat\Gamma_0^u= m_b^5/m_B^5\: \Gamma_0^u$. In  
writing the second equality, the $(\lambda_1+3\lambda_2)$ term was cancelled 
by the change in the upper limit of integration.  The $-4\overline\Lambda/m_b$ 
term cancels the leading term in the expansion of $(m_B/m_b)^4$, while the 
terms that cancel higher terms are beyond order $\lambda^2$ in the SCET result. 
The $F_5$ term gives an $\alpha_s/m_b^3$ term, which we shall drop below.  
In expanding the shape functions, we obtain singular functions peaking at 
$x=1$, so it is safe to take $\lambda_H=1$ and drop $K(\lambda_H,x_H)$. For 
the remaining shape functions, the local expansion gives~\cite{Bauer:2002yu}
\begin{eqnarray}
   f^{(0)}(-w) &=& \delta(w) - \frac{\lambda_1}{6}\, \delta''(w)
     - \frac{\rho_1}{18}\, \delta'''(w) + \ldots \,,\nn\\
   f_0^{(2)}(-w) &=& -(\lambda_1+3\lambda_2)\, \delta'(w) +  
\frac{\tau}{2}\,
   \delta''(w) + \ldots \,,\nn\\
   f_2^{(2)}(-w) &=& \lambda_2 \,\delta'(w) + \frac{\rho_2}{2}\,  
\delta''(w)
   + \ldots\,,\nn\\
   F_3(\overline \Lambda\!+\!w) &=& -\frac{2}{3} \lambda_1 \,  
\delta'(w)
   + \ldots\,,\nn\\
   F_4(\overline \Lambda\!+\!w) &=& - \lambda_2 \, \delta'(w)
   + \ldots\,,
\end{eqnarray}
where $\lambda_1$, $\lambda_2$, $\rho_1$, $\rho_2$ are the standard HQET
parameters, which are matrix elements of local operators, $\tau$ is a
combination of time-ordered products of HQET operators, and terms beyond
$1/m_b^3$ have been dropped. Substituting this result into our  
Eq.~(\ref{eq:dGudxp}) and integrating by parts using the 
relations~\cite{Bauer:2002yu} $\int dw [w \delta''(w)] = \int 
dw [-2 \delta'(w)]$ and $\int dw [w \delta'''(w)] =  \int dw
[-3 \delta''(w)]$, which are valid with smooth test functions, we find
\begin{eqnarray} \label{localterms}
   \frac{1}{\hat \Gamma_0^u} \frac{d\Gamma^u}{dx} &=&
      2 \theta(1-x) - \frac{\lambda_1}{3 m_b^2}\,\delta'(1-x)
      - \frac{\lambda_1}{3m_b^2}\,\delta(1-x) -
      \frac{11\lambda_2}{m_b^2}\,\delta(1-x) \\
  &&    + \frac{\tau}{2m_b^3} \, \delta'(1-x) -\frac{\rho_1}{9m_b^3}\,
  \delta''(1-x) -\frac{5\rho_1}{3 m_b^3}\, \delta'(1-x)  
-\frac{\rho_2}{m_b^3}
  \, \delta'(1-x) \,.\nn
\end{eqnarray}
This agrees exactly with the result obtained from the local OPE in
Refs.~\cite{Manohar:1994qn,Gremm:1997df}. (Note that we do not compare  
the $1/m_b^3$ annihilation term, which does not arise from one of the shape
functions appearing at order $\lambda^2$ in SCET.)

Of the terms in Eq.~(\ref{localterms}), it is those proportional to $\lambda_1$
and $\rho_1$ that test the difference between our results and those in
Ref.~\cite{Bauer:2002yu}. Ref.~\cite{Bauer:2002yu} also obtained the $\lambda_1$
result in Eq.~(\ref{localterms}), and even though we disagree on the $u_H
f^{(0)}$ and $ F_3$ terms, the combination of the two gives the same $\lambda_1$
result.  For the $\rho_1$ terms, their $\rho_1\delta''$ term agrees with
Eq.~(\ref{localterms}), but the $\rho_1\delta'$ term does not.  In the very
recent paper~\cite{Mannel:2004as} it was pointed out that from the local OPE the
coefficient of the $\rho_1\delta'$ term should be $-5/3$, as in
Eq.~(\ref{localterms}), rather than the $-1/3$ quoted in~\cite{Bauer:2002yu}.

Ref.~\cite{Mannel:2004as} went further to advocate a different approach  
to the shape-function region that involves using an unexpanded $b$-quark  
field, and doing this obtained a result with subleading shape functions whose  
expansion is consistent with the local OPE in Eq.~(\ref{localterms}). However,  
their result for the power-suppressed $d\Gamma^u/dx$ decay rate does not agree 
with  the rate obtained here. In particular, our $-4 u_H F$ term is not present
there, and instead of our $-3 F_3$ term they have the result 
``$4 F_1 +2 G_2 -3 G_3$''.  Their ``$G_2$'' term is defined by operators that, 
in our analysis, can only show up suppressed by at least one factor of 
$\alpha_s$ through jet  functions. In fact, the operator structure of our 
result actually agrees with the  original one in Ref.~\cite{Bauer:2002yu}, 
rather than the one in  Ref.~\cite{Mannel:2004as}.
No proof of factorization has yet been achieved for this approach with  
the unexpanded $b$-quark field, and it is conceivable that this may help to  
explain why a different structure of operators was obtained.




\section{Conclusions and Discussion}\label{sect_concl}

In this paper, we have computed a factorization theorem for the leading-order
power corrections to inclusive $B\to X_s\gamma$ and $B\to X_u\ell\bar\nu$ 
decays in the endpoint region, where the $X$ is jet-like. In particular, we 
have shown that these power corrections can be fully categorized and thus 
treated in a systematic fashion using the Soft-Collinear Effective Theory. 
A main result of our analysis is that perturbative power corrections to the 
decay rates can be systematically computed, and our result explicitly 
disentangles hard factors of $\alpha_s(m_b^2)$, collinear jet-induced factors of
$\alpha_s(m_X^2)$, and soft (``$\alpha_s(\Lambda^2)$'') non-perturbative QCD
effects.

In addition, our results can be used as a starting point for the systematic
resummation of Sudakov double logarithms in the power corrections. To achieve
this, one needs to compute the anomalous dimensions of all the operators we have
defined that appear in the subleading factorization theorem. Some of the terms
here are already known. In the body of the paper, we have shown that if we 
consider only subleading terms with non-vanishing jet functions at lowest order in
$\alpha_s$, then the logarithms that can be resummed into the hard function in
these NLO contributions are {\em identical} to the analogous logarithms in the
LO result. (These logs can be thought of as occurring between the scales $m_b^2$
and $m_b\Lambda$.) There are additional logarithms that are sensitive to the
split between the jet and soft functions (logs between $m_b\Lambda$ and
$\Lambda^2$), which require knowledge of the anomalous dimensions of 
subleading soft operators. The latter are very unlikely to be universal, and 
have not been computed here.

Our main final decay-rate formulae have been collected in
Sec.~\ref{sect_Summary}. At lowest order in $\alpha_s$, they include a derivation
of the power corrections for the triply differential $B\to X_u\ell\bar\nu$ rate.
Results have been derived in the literature for $d\Gamma/dE_\gamma$ in 
$B\to X_s\gamma$~\cite{Bauer:2001mh} and the singly differential 
$B\to X_u\ell\bar\nu$ rate 
$d\Gamma/dE_\ell$~\cite{Leibovich:2002ys,Bauer:2002yu,Mannel:2004as} (and
$d\Gamma/dm_X^2$~\cite{Burrell:2003cf}), and a comparison was given in
Sec.~\ref{sect_singly}. Agreement was found for $B\to X_s\gamma$, but for
$d\Gamma/dE_\ell$ we found disagreement on two terms at subleading order. (A
check on our $d\Gamma/dE_\ell$ result was obtained by expanding it to compare
with the local OPE, and full agreement was found up to $1/m_b^3$, as discussed 
in Sec.~\ref{sect_localOPE}.)  Using our results, we found it straightforward to
present the power corrections to doubly differential rates, as well as other
singly differential rates such as $d\Gamma/dp_X^-$ and $d\Gamma/dp_X^+$.

On the phenomenological side, our most important result is the identification of
two new shape functions, which involve four-quark operators and have not been
previously considered in the literature. They are denoted by $f_{5,6}$
($F_{5,6}$), and definitions can be found in Eq.~(\ref{defnfs2}).  In the
endpoint decay rates they induce power-suppressed terms, which are quite large,
of order
\begin{eqnarray} \label{estimate}
  4\pi \alpha_s \frac{\Lambda}{m_b} \,.
\end{eqnarray}
Since $4\pi \alpha_s\sim 4$, these power corrections might numerically dominate
over those that are simply of order $\alpha_s^0\, \Lambda/m_b$. We have given
results for the effect of these shape functions in all the considered decay
rates.  In our results for the decay rates, the numerical prefactors for
$f_{5,6}$ turned out to be sizeable (e.g.\ $24\pi\alpha_s$ for $f_5$ in
$d\Gamma/dE_\ell$), which justifies including the factor of $4$ in
Eq.~(\ref{estimate}).  For the extraction of $|V_{ub}|$ from $d\Gamma/dE_\ell$,
the important thing to consider is the difference between how these new shape
functions affect $B\to X_u\ell\bar\nu$ and $B\to X_s\gamma$. In this case,
comparing the combinations of $F$ and $F_5$ in Eqs.~(\ref{eq:dGsdx}) and
(\ref{eq:dGudx}), we see that the mismatch is
\begin{eqnarray}
  \simeq -8\pi\alpha_s(\mu_0) \big[ 3 F_5^{u} - F_5^{s} \big]\,,
\end{eqnarray}
where the index $u$ or $s$ denotes the fact that these shape functions involve
different flavors of light quark.  To obtain a
numerical estimate we approximate $F_5^{u}/(m_B F) \simeq F_5^{s}/(m_B F)\sim
\Lambda/m_b\simeq 0.1 \epsilon'$ and find that they can cause a deviation of
\begin{eqnarray}
  -16\pi\alpha_s(\mu_0)\: 0.1 \epsilon'\simeq  (180\%) \epsilon' \,,
\end{eqnarray} 
where $\epsilon'$ denotes any additional dynamical suppression from the
non-perturbative functions.  Even for $\epsilon' \sim 0.1-0.3$ these terms
provide a sizeable new uncertainty for the cut spectrum
$d\Gamma/dE_\ell$ approach to measuring $|V_{ub}|$.
From Eq.~(\ref{eq:dGdu}) we see that the situation is only slightly better for a
cut $d\Gamma/dp_X^+$ spectrum. Our conclusion is that, until theoretical methods
are developed to control these new subleading shape-function effects or better
bounds on these contributions are obtained, it will be challenging to argue that
methods sensitive to the shape-function region are trustworthy for inclusive
determinations of $|V_{ub}|$ at the $\lesssim$ 10\% level.  One possible future
direction is to derive experimental bounds on the new subleading shape-function
effects by comparing endpoint-dependent methods with different spectra and
different cuts on the phase space. It would also be useful to find
model-independent ways of determining the size of the subleading shape functions
that go beyond the simple dimensional analysis used here.

Theoretically, there are several avenues for future work on $B\to X_s\gamma$ and
$B\to X_u\ell\bar\nu$. These include the calculation of perturbative corrections
in the factorization theorems at subleading order, as well as a complete
resummation of Sudakov logarithms. It would also be interesting to consider the
structure of the subleading factorization theorems in moment space, as opposed
to the momentum-space version considered here.  Starting with our triply
differential $B\to X_u\ell\bar\nu$ result, one could derive other doubly and
singly differential decay spectra and consider their phenomenological
implications. More formally, it remains to be checked that the convolutions that
appear in our subleading factorization theorems actually converge when the
functional forms of the jet functions are considered at higher orders in
$\alpha_s$. From a formal standpoint this is necessary for a complete ``proof''
of these results as factorization theorems. However, from a pragmatic standpoint
this can be checked as each new phenomenologically relevant term is computed. We
are not aware of any factorization formulae where convergence problems occur at
higher orders in the perturbative expansion of the kernels when they are not
present in the leading non-vanishing kernel results (the convergence of which we
have checked).  Finally, it should also be possible to extend the
techniques used here to closely related physical cases such as deep inelastic
scattering for $x\to 1$ (i.e. Bjorken $x\sim 1-\Lambda/Q$).

Note added: Recently the paper~\cite{Neubert:2004dd} appeared, in which a
different type of power-suppressed correction was studied near the endpoint
region, namely those corrections suppressed by $\Lambda/\Delta$, where $\Delta$
depends on the cut on the photon energy spectrum.  These corrections involve
using the leading-order shape function and give a measure of the effect of
extrapolating away from the SCET expansion region we discussed. The results
derived here should enable this analysis to be further extended to include the
power corrections suppressed by a factor of $\Lambda/m_b$.

\acknowledgments The authors thank C.~Bauer, B.~Lange, and D.~Pirjol for their
comments.  This work was supported in part by the U.S.  Department of Energy
(DOE) under the cooperative research agreement DF-FC02-94ER40818. I.S.  was also
supported in part by a DOE Outstanding Junior Investigator award.

\appendix

\section{Expansion of the heavy quark field and derivation of the currents}
\label{app_J}

To derive the power expansion of the heavy-to-light current $J =\bar
\psi^{(q)}\Gamma \psi^{(b)}$ at tree level we just need the expansion of the
light- and heavy-quark full-theory fields, $\psi^{(q)}$ and $\psi^{(b)}$, in
terms of SCET collinear fields ($\xi_n$) and soft heavy fields ($h_v$) 
respectively.
$\psi^{(q)}$ also contains terms with usoft light quark fields, but they do
not contribute for inclusive processes until higher order, and so we neglect
these terms here.  For the light quark field we have~\cite{Manohar:2002fd}
\begin{eqnarray}
  \psi^{(q)} &=& \bigg[ 1 + \frac{1}{i\bn\mcdot D_c+ Wi\bn\mcdot D_{us} W^\dagger}
   (i \DSppP + W i \DSppPus W^\dagger) \frac{\bnslash}{2} \bigg] \xi_n \\
   &=& \xi_n + \frac{1}{i\bn\mcdot D_c} i \DSppP \frac{\bnslash}{2} \xi_n 
    + W \frac{1}{\bnP} i \DSppPus \frac{\bnslash}{2} W^\dagger \xi_n 
    - W \frac{i\bn\mcdot D_{us}}{\bnP^2} W^\dagger i\DSppP \frac{\bnslash}{2}
    \xi_n  + {\cal O}(\lambda^4) \,.\nn
\end{eqnarray}
Here we have used the collinear fields~\cite{Bauer:2003mg}
\begin{eqnarray}
 && in\mcdot D=in\mcdot \partial \!+\! gn\mcdot A_c \!+\! gn\mcdot A_{us}
  \,, \qquad  
  i D^\perp = i D_c^\perp \!+\! W i D_{us}^\perp W^\dagger \,, \nn\\
 && i \bn\mcdot D = i\bn\mcdot D_c \!+\! W i \bn\mcdot D_{us} W^\dagger \,,
   \label{eq:fullD} 
\end{eqnarray}
which obviously will give an expansion with terms that are individually usoft
and collinear gauge covariant at each order in $\lambda$.

For the heavy quark fields we start with the QCD Lagrangian ${\cal L}=
\bar\psi^{(b)} (i\Dslash-m_b) \psi^{(b)}$. We then divide the quark fields into
on-shell and off-shell terms, $\psi^{(b)} = \exp(-imv\mcdot x) (h_v + \psi_B)$,
where $\vslash h_v = h_v$.  Unlike in HQET, the off-shell field $\psi_B$ does not
satisfy $\vslash \psi_B = -\psi_B$, since collinear gluons give off-shell quark
fields that have quark components, $\vslash \psi_B = \psi_B$.  This gives the
Lagrangian 
\begin{eqnarray}
  {\cal L} = \bar h_v i v\mcdot D_{us} h_v 
   + \bar\psi_B [ i\Dslash + m_b (\vslash-1) ]\psi_B 
   + \bar h_v ( i\Dslash - \cPslash) \psi_B 
   + \bar \psi_B ( i\Dslash - \cPslash) h_v \,,
\end{eqnarray}
where the subtraction in $(i\Dslash-\cPslash)$ removes the terms that vanish by
momentum conservation. Varying with respect to $\bar\psi_B$ gives
\begin{eqnarray} \label{eom}
  [ i\Dslash + m_b(\vslash-1)] \psi_B = - (i\Dslash-\cPslash) h_v \,,
\end{eqnarray}
which is a higher-order version of the LO equation for $\psi_B$ derived in
Ref.~\cite{Bauer:2001yt}.  To solve for $\psi_B$ in terms of $h_v$ at higher
orders we use a strategy proposed in Ref.~\cite{Beneke:2002ph}, namely expanding
$\psi_B=\psi_B^{(3)}+\psi_B^{(4)}+\ldots$, and considering the solutions order
by order in $\lambda$ (noting that $h_v\sim \lambda^3$).  To facilitate this we
rewrite Eq.~(\ref{eom}) as
\begin{eqnarray}
  && \Big[ \Delta^{(0)} + i\Dslash_c^\perp + \Delta^{(2)} \Big] (\psi_B^{(3)} +
  \psi_B^{(4)}+\ldots) = 
 - \Big[ \frac{\nslash}{2} g\bn\mcdot A_c + g\Aslash_c^\perp + \Delta^{(2)}
 \Big] h_v\,,\\
 && \Delta^{(0)} = \frac{\nslash}{2} i\bn\mcdot D_c \!+\! m_b (\vslash\!-\!1) 
   \,,\qquad
  \Delta^{(2)} =  W \big( \frac{\nslash}{2} i\bn\mcdot D_{us}
    \!+\! i\Dslash_{us}^\perp \big) W^\dagger 
    \!+\! \frac{\bnslash}{2} in\mcdot D\,.\nn
\end{eqnarray}
The complete set of equations to solve is then 
\begin{eqnarray}
  \Delta^{(0)} \psi_B^{(3)} &=& -\frac{\nslash}{2} g \bn\mcdot A_c h_v \,,\\
  \Delta^{(0)} \psi_B^{(4)} &=&  - i\Dslash_\perp^c \psi_B^{(3)}
       - g\Aslash^\perp_c h_v \,, \nn\\
  \Delta^{(0)} \psi_B^{(5)} &=& - i\Dslash_\perp^c  \psi_B^{(4)} 
       - \Delta^{(2)} (\psi_B^{(3)}  + h_v) \,, \nn\\
  \Delta^{(0)} \psi_B^{(n)} &=&  - i\Dslash_\perp^c  \psi_B^{(n-1)} 
       - \Delta^{(2)} \psi_B^{(n-2)} \,, \nn
\end{eqnarray}
where $n\ge 6$. Obviously, the crucial point is to invert the operator
$\Delta^{(0)}$. When acting on a collinear field, the inverse
$[\Delta^{(0)}]_c^{-1}$ is simply given by
\begin{eqnarray} \label{Dinvc}
  [\Delta^{(0)}]_c^{-1}
  &=&  \frac{\nslash}{2m_b\,n\mcdot v } 
  + \frac{(1+\vslash)}{n\mcdot v\,  i\bn\mcdot D_c} \,.
\end{eqnarray}
Owing to the $1/(i\bn\mcdot D_c)$, this solution is ill-defined when acting on an
ultrasoft field and will not suffice to solve for $\psi_B^{(n\ge 5)}$. For the
solution to the first equation we find
\begin{eqnarray}
  \psi_B^{(3)} = - \frac{1}{i\bn\mcdot D_c} g\bn\mcdot A_c h_v = (W-1) h_v \,,
\end{eqnarray}
which, as expected, is in agreement with the LO solution in the appendix of
Ref.~\cite{Bauer:2001yt}.  Thus, at lowest order the solution for the
full-theory field is $\psi=h_v +\psi_B^{(3)} = W h_v$. At the next order, we
find that
\begin{eqnarray}
  \psi_B^{(4)} &=& -[\Delta^{(0)}]_c^{-1} (i\Dslash^\perp_c W-\cPslash_\perp)  h_v
     = -[\Delta^{(0)}]_c^{-1} \frac{1}{i\bn\mcdot D_c} ig \Bslash_\perp^c W h_v
  \\
  &=& -\frac{1}{m_b\,n\mcdot v} \frac{\nslash}{2}  W 
    \Big[ \frac{1}{\bnP} W^\dagger ig\Bslash_\perp^c W \Big] h_v
   - \frac{2}{n\mcdot v} W\: \Big[ \frac{1}{\bnP^2} W^\dagger ig v\mcdot
   B_\perp^c W \Big] h_v \,, \nn
\end{eqnarray} 
where
\begin{equation}
   ig B_\perp^{c \mu} = [i\bn\mcdot D^c,i D_c^{\perp \mu}].
      \label{eq:BM1}
\end{equation}
The result agrees with Refs.~\cite{Beneke:2002ph,Pirjol:2002km}, except that we
have written the last term as $v\mcdot B_\perp^c$ to emphasize that it must
start with at least one collinear gluon.

To proceed we let 
\begin{eqnarray}
 \psi_B^{(n)} = W \tilde\psi_B^{(n)} \,,\qquad
 \Delta^{(0)} W  =W \tilde \Delta^{(0)} \,,\quad
\end{eqnarray}
and then write the remaining equations for $\tilde\psi_B^{(n)}$.  Using the
equation of motion for $W$ gives
\begin{eqnarray} \label{tDelta}
    \tilde \Delta^{(0)} = \frac{\nslash}{2} \bnP \!+\! m_b (\vslash\!-\!1)
    \,,\qquad 
 [\tilde\Delta^{(0)}]_c^{-1} =
    \frac{\nslash}{2m_b\,n\mcdot v } 
    + \frac{(1+\vslash)}{n\mcdot v\, \bnP } \,,
\end{eqnarray}
and so, after we multiply on the left by $W^\dagger$, the remaining equations 
become
\begin{eqnarray} \label{neweqtns}
  \tilde\Delta^{(0)} \tilde \psi_B^{(5)} &=&
       -\tilde\Delta^{(1)} \tilde \psi_B^{(4)} 
       - \tilde \Delta^{(2)} h_v  - i\Dslash_{us} h_v \,, \\
  \tilde \Delta^{(0)} \tilde \psi_B^{(n)} &=&  
       - \tilde\Delta^{(1)} \tilde \psi_B^{(n-1)} 
       - \tilde \Delta^{(2)} \tilde \psi_B^{(n-2)}
       - i\Dslash_{us} \tilde \psi_B^{(n-2)} \,, \nn\\
  \tilde \Delta^{(1)} &=& (W^\dagger i\Dslash_\perp^c W)
       \,, \qquad 
   \tilde \Delta^{(2)} = \frac{\bnslash}{2} (W^\dagger in\mcdot D W -in\mcdot D_{us})
      \,, \nn
\end{eqnarray}
where $n\ge 5$. 

To solve Eqs.~(\ref{neweqtns}) we invert $\tilde \Delta^{(0)}$. For terms with 
at least one collinear gluon field, such as $\tilde \Delta^{(1,2)}$, we can use
the inverse $[\tilde \Delta^{(0)}]_c^{-1}$ from Eq.~(\ref{tDelta}).  
Only terms that are purely usoft, including $i\Dslash_{us} h_v$ and possibly 
$i\Dslash_{us} \tilde \psi_B^{(n-2)}$, need to be handled with care. 
For these terms a purely usoft $\psi_B^{(n)}$ suffices, in which case 
$\tilde \Delta^{(0)}= m(\vslash-1)$. The
subset of purely usoft terms in Eqs.~(\ref{neweqtns}) is given uniquely by the
terms $m(\vslash-1) \psi_B^{(5)}= - i\Dslash_{us} h_v$ and $m(\vslash-1)
\psi_B^{(n)}= - i\protect{\Dslash_{us}} \psi_B^{(n-2)}$, which are the same as
we would find in pure HQET. Thus, the solution for the pure usoft terms is
\begin{eqnarray}
  \psi_B^{us} = \frac{1}{2m_b+iv\mcdot D_{us}} i \Dslash_{us}^T h_v \,,\qquad
   [\tilde \Delta^{(0)}]^{-1}_{us} = -\frac{1}{2m_b} \frac{(1-\vslash)}{2} \,,
\end{eqnarray}
where it is sufficient to use this $[\tilde \Delta^{(0)}]^{-1}_{us}$ since
$(1-\vslash)/2\, \psi_B^{us} = \psi_B^{us}$. From Eq.~(\ref{neweqtns}), the 
solution for $\psi_B^{(5)}$ is then
\begin{eqnarray}
  \psi_B^{(5)} 
 &=&  W \frac{i\Dslash^T_{us}}{2m_b} h_v 
     -W [\tilde \Delta^{(0)}]^{-1}_{c} \tilde\Delta^{(1)}\tilde\psi_B^{(4)} 
     - W[\tilde \Delta^{(0)}]^{-1}_{c} \tilde \Delta^{(2)} h_v  
       \\
 &=& W \frac{i\Dslash^T_{us}}{2m_b} h_v  
   - W  \frac{\nslash\bnslash}{4m_b\,n\mcdot v }
     \Big[ \frac{1}{\bnP} W^\dagger i g n\mcdot B_c W \Big] h_v 
   - W  \frac{\bn\mcdot v}{n\mcdot
     v\,\bnP}   \Big[ \frac{1}{\bnP} W^\dagger i g n\mcdot B_c W \Big] h_v 
  \nn\\
 && - W   \frac{1}{m_b\,n\mcdot v}
   \Big[ \frac{1}{\bnP} W^\dagger i\Dslash_c^\perp W \Big]
    \Big[ \frac{1}{\bnP} W^\dagger ig\Bslash_\perp^c W \Big] h_v \nn\\
 && + W   \frac{\nslash}{m_b\,(n\mcdot v)^2}
   \Big[ \frac{1}{\bnP} W^\dagger iv\mcdot D_c^\perp W \Big]
    \Big[ \frac{1}{\bnP} W^\dagger ig\Bslash_\perp^c W \Big] h_v \nn\\
&& + W \frac{\nslash}{m_b\,(n\mcdot v)^2 }
   \Big[ \frac{1}{\bnP} W^\dagger ig \Bslash^c_\perp W \Big]
    \: \Big[ \frac{1}{\bnP^2} W^\dagger ig v\mcdot
   B_\perp^c W \Big] h_v \nn\\
&& + W \frac{4}{(n\mcdot v)^2}
   \Big[\frac{1}{\bnP} W^\dagger iv\mcdot D_c^\perp W \Big]
    \: \Big[ \frac{1}{\bnP^2} W^\dagger ig v\mcdot
   B_\perp^c W \Big] h_v 
   \nn  \,.
\end{eqnarray}

From the expansion of the fields we obtain the expansion of the currents at tree
level,
\begin{eqnarray}
  J &=& \big[ \bar\xi_n +  \bar\psi_L^{(2)}+ \bar\psi_L^{(3)}+ \ldots ] 
   \:\Gamma\: [ W h_v + \psi_B^{(4)} + \psi_B^{(5)} \ldots ]  \\
  &=& [ \bar\xi_n \Gamma W h_v ] 
  + [ \bar\xi_n \Gamma \psi_B^{(4)} + \bar\psi_L^{(2)}\Gamma W h_v ] 
  +[ \bar\xi_n \Gamma \psi_B^{(5)} + \bar\psi_L^{(3)}\Gamma W h_v 
    + \bar\psi_L^{(2)}\Gamma \psi_B^{(4)} ]  + \ldots \,. \nn
\end{eqnarray}
The leading-order SCET heavy-to-light current is
\begin{eqnarray}
  J^{(0)}(\omega) &=&   (\bar\xi_n W)_\omega \Gamma h_v \,.
\end{eqnarray}
For the currents suppressed by $\lambda$ we obtain
\begin{eqnarray}
  J^{(1)} &=&  -\, \bar\xi_n \DSppPl \frac{\bnslash}{2}
   \frac{1}{i \bn\mcdot\lD} \Gamma W h_v
    -\, \frac{1}{n\mcdot v m}\bar\xi_n \Gamma 
    \frac{\nslash}{2} \frac{1}{i \bn\mcdot D_c} ig \Bslash_\perp^c W h_v  \nn\\
 && -\, \frac{2}{n\mcdot v}\: \bar\xi_n \Gamma\:
\frac{1}{(i\bn\mcdot D_c)^2} ig v \mcdot B_\perp^c W h_v \,. \nn
\end{eqnarray}
Making the field redefinition and putting in the most general $\omega_i$
dependence consistent with RPI gives~\cite{Pirjol:2002km}\footnote{Note that for
  $J^{(1c)}$ we find a field strength $B_\perp^c$ rather than a $v\mcdot
  D_\perp^c$ acting on $W$.} 
\begin{eqnarray}
  J^{(1a)}(\omega) &=&  \frac{1}{\omega} \big( \bar\xi_n \DSppPl W \big)_{\omega} 
    \frac{\bnslash}{2} \Gamma (Y^\dagger h_v) \,, \\
  J^{(1b)}(\omega_1,\omega_2) 
    &=& -\, \frac{1}{m_b\: n\mcdot v} (\bar\xi_n W)_{\omega_1} 
    \Gamma \frac{\nslash}{2} 
    \: (ig {\Bcslash}_c^\perp)_{\omega_2}\: (Y^\dagger h_v) \,, \nn\\
  J^{(1c)}(\omega) &=& -\, \frac{2}{n\mcdot v}\: \Big(\bar\xi_n 
\frac{1}{(i\bn\mcdot D_c)^2} ig v \mcdot B_\perp^c W\Big)_{\omega} \Gamma\:
 (Y^\dagger h_v) \,, \nn
\end{eqnarray}
where ${\cal B}_c^\perp$ is defined in Eq.~(\ref{calB}). For the currents
suppressed by $\lambda^2$ we find
\begin{eqnarray}
  J^{(2)} &=& \bar\xi_n \Gamma W \frac{i\Dslash_{us}}{2m} h_v
     - \frac{1}{n\mcdot v} \bar\xi_n \Gamma 
     \left(\frac{\nslash\bnslash}{4m} + \frac{\bn\mcdot v}{i\bn\mcdot D_c} \right) 
     \frac{1}{i\bn\mcdot D_c} ig n\mcdot M W h_v  \\
  &+& \frac{1}{n\mcdot v} \bar\xi_n \DSppPl 
     \frac{\bnslash}{2} \frac{1}{i \bn\mcdot\lD} \Gamma
     \left( \frac{\nslash}{2m} \frac{1}{i\bn\mcdot D_c} ig \Bslash_\perp^c 
     + \frac{2}{(i\bn\mcdot D_c)^2} ig v \mcdot B_\perp^c \right) W h_v \nn \\ 
  &-& \bar\xi_n W i{\overleftarrow \Dslash}_{us}^\perp W^\dagger \frac{\bnslash}{2} 
     \frac{1}{i \bn\mcdot\lD} \Gamma W h_v 
   -  \frac{1}{n\mcdot v m}\bar\xi_n \Gamma
     \frac{1}{i\bn\mcdot D_c} i\Dslash_\perp^{c} \frac{1}{i\bn\mcdot D_c}
     ig \Bslash_\perp^c W h_v \nn\\ 
  &+&  \frac{1}{(n\mcdot v)^2 m} \bar\xi_n \Gamma \frac{1}{i\bn\mcdot D_c}
     \DvppP \nslash \frac{1}{i\bn\mcdot D_c} ig \Bslash_\perp^c W h_v \nn\\
  &+&  \frac{4}{(n\mcdot v)^2} \bar\xi_n \Gamma \frac{1}{i\bn\mcdot D_c} \DvppP 
     \frac{1}{(i\bn\mcdot D_c)^2} i g v \mcdot B_\perp^c W h_v \nn\\
  &+&  \frac{1}{(n\mcdot v)^2 m} \bar\xi_n \Gamma \nslash
      \left( i\Dslash_\perp^{c} \frac{1}{i\bn\mcdot D_c} 
      - \frac{1}{i\bn\mcdot D_c}  i\Dslash_\perp^{c} \right) 
      \frac{1}{i\bn\mcdot D_c} ig v \mcdot B_\perp^c W h_v \,, \nn
\end{eqnarray}
where 
\begin{equation} 
   ig n\mcdot M = [i\bn\mcdot D^c,in\mcdot D].
      \label{eq:BM2}
\end{equation}

It will be convenient for us to label the terms as follows:
\begin{eqnarray}
 J^{(2a)} &=& \bar\xi_n \Gamma W \frac{i\Dslash_{us}}{2m} h_v
   \,,\qquad
 J^{(2b)} = -\bar\xi_n W i{\overleftarrow \Dslash}_{us}^\perp W^\dagger 
    \frac{\bnslash}{2} \frac{1}{i \bn\mcdot\lD} \Gamma W h_v
   \,,  \nn\\
J^{(2c)} &=& - \frac{\bn\mcdot v}{n\mcdot v} \bar\xi_n \Gamma 
    \frac{1}{(i\bn\mcdot D_c)^2} ig n\mcdot M W h_v
    \,, \nn\\
 J^{(2d)} &=& - \frac{1}{n\mcdot v \,m} \bar\xi_n \Gamma \frac{\nslash\bnslash}{4}
     \frac{1}{i\bn\mcdot D_c} ig n\mcdot M W h_v 
    \,, \nn\\
 J^{(2e)} &=&  \frac{1}{n\mcdot v \,m} \bar\xi_n \DSppPl
     \frac{\bnslash}{2} \frac{1}{i \bn\mcdot\lD} \Gamma
     \frac{\nslash}{2} \frac{1}{i\bn\mcdot D_c} ig \Bslash_\perp^c W h_v
    \,, \nn \\  
 J^{(2f)} &=& -  \frac{1}{n\mcdot v \,m}\bar\xi_n \Gamma
     \frac{1}{i\bn\mcdot D_c} i\Dslash_\perp^{c} \frac{1}{i\bn\mcdot D_c}
     ig \Bslash_\perp^c W h_v  
    \,, \nn\\ 
 J^{(2g)} &=& \frac{1}{(n\mcdot v)^2 \,m} \bar\xi_n \Gamma \frac{1}{i\bn\mcdot D_c}
     \DvppP \nslash \frac{1}{i\bn\mcdot D_c} ig \Bslash_\perp^c W h_v  
    \,, \nn\\
 J^{(2h)} &=&  \, \frac{1}{(n\mcdot v)^2 \,m} 
   \bar\xi_n \Gamma \nslash
   \frac{1}{i\bn\mcdot D_c} ig\Bslash_\perp^{c} 
   \frac{1}{(i\bn\mcdot D_c)^2} ig v \mcdot B_\perp^c W h_v \,, \nn\\
 J^{(2i)} &=&  \frac{2}{n\mcdot v} \bar\xi_n \DSppPl
     \frac{\bnslash}{2} \frac{1}{i \bn\mcdot\lD} \Gamma
     \frac{1}{(i\bn\mcdot D_c)^2} ig v \mcdot B_\perp^c W h_v 
    \nn \,, \\  
 J^{(2j)} &=& \frac{4}{(n\mcdot v)^2} \bar\xi_n \Gamma \frac{1}{i\bn\mcdot D_c} 
   \DvppP \frac{1}{(i\bn\mcdot D_c)^2} i g v \mcdot B_\perp^c W h_v
      \, .
\end{eqnarray}

The terms $J^{(2a)}$ to $J^{(2f)}$ agree with those found in
Refs.~\cite{Beneke:2002ph,Beneke:2002ni}, which use the position-space
formulation of SCET.  $J^{(2a)}$, $J^{(2b)}$ and $J^{(2c)}$ correspond to the
third, last and fourth terms, respectively, of Eq.~(41) in
Ref.~\cite{Beneke:2002ni}, while $J^{(2d)}$, $J^{(2e)}$ and $J^{(2f)}$
correspond to the second, last and third terms, respectively, of their Eq.~(43).
The rest of their second-order terms correspond to Lagrangian insertions in our
SCET calculation.  The terms $J^{(2g)}$ to $J^{(2j)}$ do not appear in
Refs.~\cite{Beneke:2002ph,Beneke:2002ni}, because they set $v_\perp^\mu \sim
\lambda$ (more generally, the assignment $v_\perp^\mu \sim 1$ used here is
allowed). We follow the common practice of dropping the $v_\perp$ terms in our
analysis of the factorization theorems by picking a frame where $v_\perp=0$.
\OMIT{(Besides this, in Eqs.~(41) and (43) they have specialized to the frame
  where $v^\mu = (n^\mu + \bn^\mu)/2$.)}


\section{Reparameterization invariance for the currents}
\label{app_RPI}

The choice of the light-cone vectors $n$ and $\bn$ must satisfy
$n^2 = \bn^2 = 0, \;\; n\mcdot\bn = 2$, but otherwise is arbitrary.
Therefore one can make the following reparameterization (RPI) transformations
\cite{Manohar:2002fd,Pirjol:2002km}:
\begin{eqnarray}\label{repinv}
\mbox{(I)}
\left\{
\begin{tabular}{l}
$n_\mu \to n_\mu + \Delta_\mu^\perp$ \\
$\bn_\mu \to \bn_\mu$
\end{tabular}
\right.\qquad
\mbox{(II)}
\left\{
\begin{tabular}{l}
$n_\mu \to n_\mu$ \\
$\bn_\mu \to \bn_\mu + \varepsilon_\mu^\perp$
\end{tabular}
\right.\qquad
\mbox{(III)}
\left\{
\begin{tabular}{l}
$n_\mu \to (1+\alpha)\, n_\mu$ \\
$\bn_\mu \to (1-\alpha)\, \bn_\mu$
\end{tabular}
\right. \,,
\end{eqnarray}
where $\Delta^\perp\sim\lambda$, while $\varepsilon^\perp\sim \alpha \sim
\lambda^0$. 

It is useful to separate the RPI transformations according to the
power suppression in $\lambda$ that they cause. We shall denote this
by a bracketed superscript.
For type-I transformations we have, at the same order in $\lambda$,
\begin{eqnarray}
&& \delta_{\rm I}(\bn\mcdot D) = \delta_I(W) = 0 \,, 
\nonumber \\
&& \delta_{\rm I}^{(\lambda^0)}(n \mcdot D) = \Delta^\perp \mcdot D_c^\perp \,, 
\qquad
\delta_{\rm I}^{(\lambda^0)}(D_{c \mu}^\perp) = 
    - \frac{\Delta_\mu^\perp}{2} \bn\mcdot D_c \,, 
\nonumber \\
&& \delta_{\rm I}^{(\lambda^0)}(\xi_n) = 0 \,. 
\end{eqnarray}
Type-I transformations one order higher in $\lambda$ are
\begin{eqnarray}
&& \delta_{\rm I}^{(\lambda^1)} (D_{c \mu}^\perp) + 
   \delta_{\rm I}^{(\lambda^0)}(W D_{us, \mu}^\perp W^\dagger) = 
   -\frac{\bn_\mu}{2} \Delta^\perp \mcdot D_c^\perp \,,
\nonumber \\
&& \delta_{\rm I}^{(\lambda^1)} (\bar\xi_n) 
    = \bar\xi_n \frac{\bnslash\Delslash^\perp}{4}\,. 
\end{eqnarray}
For type-II transformations we have, at the same order in $\lambda$,
\begin{eqnarray}
 \delta_{\rm II} (n\mcdot D) & = & 0 \,,
\nonumber \\
 \delta_{\rm II}^{(\lambda^0)} (D_{c \mu}^\perp) & = &
   -\frac{n_\mu}{2} \varepsilon^\perp \mcdot D_c^\perp \,,\nn\\
 \delta_{\rm II}^{(\lambda^0)} (\xi_n) &=&
 \delta_{\rm II}^{(\lambda^0)} (W) =0 \,.
\end{eqnarray}
Type-II transformations one order higher in $\lambda$ are
\begin{eqnarray}
&& \delta_{\rm II}^{(\lambda^1)} (D_{c \mu}^\perp) +
   \delta_{\rm II}^{(\lambda^0)}(W D_{us, \mu}^\perp W^\dagger) =
   -\frac{\varepsilon_\mu^\perp}{2} n \mcdot D
   -\frac{n_\mu}{2} W \varepsilon^\perp \mcdot D_{us}^\perp W^\dagger \,,
\nonumber \\
&& \delta_{\rm II}^{(\lambda^1)}(\bar\xi_n) =
   \bar\xi_n \overleftarrow \Dslash_c^\perp 
   \frac{1}{\bn\mcdot \overleftarrow {D}_c}\frac{\vepslash_\perp}{2}\,,
\qquad \delta_{\rm II}^{(\lambda^1)}(W) = 
 \left[ - \frac{1}{\bn\mcdot D_c}\: \varepsilon^\perp\mcdot D_c^\perp W
 \right] \,,
\end{eqnarray}
where the differential operators do not act outside the square brackets.
Furthermore,
\begin{eqnarray}
&& \delta_{\rm II}^{(\lambda^2)} (\bar\xi_n) = 
      \bar\xi_n W \overleftarrow \Dslash_{us}^\perp W^\dagger
   \frac{1}{\bn\mcdot \overleftarrow {D}_c}\frac{\vepslash_\perp}{2} \,.
\end{eqnarray}
Note that Table~I of Ref.~\cite{Manohar:2002fd} seems to suggest that
the transformation of $W$ involves the full covariant derivative with
both collinear and usoft parts. However, the Wilson line referred to in 
that table is constructed from $\bn \mcdot (A_c + A_{us})$;
for the Wilson line built out of $\bn \mcdot A_c$ alone, the equation
of motion is $i \bn \mcdot D_c W = 0$, and hence
$\delta_{\rm II}(W) = \delta_{\rm II}^{(\lambda)}(W)$.
Nevertheless, there is a potential complication related to the choice of 
the full covariant derivative. One possible reparameterization-invariant
choice is 
$i\bn\mcdot D = i\bn\mcdot D_c + i\bn\mcdot D_{us}$,
$i D^\perp = i D_c^\perp + i D_{us}^\perp$.
With this choice, one must redefine $\bn\mcdot A_c$ and 
$A_c^\perp$ if one wants the Lagrangian to be invariant order by
order in $\lambda$ under collinear gauge transformations 
\cite{Bauer:2003mg}. This turns out to be equivalent to the choice
given by Eq.~(\ref{eq:fullD}), with the covariant derivatives 
defined in terms of the new gauge field. Furthermore, the Wilson
line built out of the new gauge field has the same equation of
motion as previously, and hence the same transformation law.

Using the above transformations and the definitions (\ref{eq:BM1})
and (\ref{eq:BM2}), we obtain
\begin{eqnarray}
\delta_{\rm I}^{(\lambda^1)} (ig B_\perp^{c \mu})
& = & -\frac{1}{2} \bn^\mu \Delta_\alpha^\perp 
 [i\bn\mcdot D_c, iD_c^{\perp\alpha}] 
+ [ \delta_{\rm I}^{(\lambda^0)}(W iD_{us}^{\perp\mu} W^\dagger),
   i\bn\mcdot D_c]
\nonumber \\
& = & -\frac{1}{2} \bn^\mu ig \Delta^\perp \mcdot B_\perp^c \,, 
\end{eqnarray}
since the second commutator in the first line above equals
$[W \delta_{\rm I}^{(\lambda^0)}(iD_{us}^{\perp\mu}) W^\dagger, 
W \bnP W^\dagger] 
= W [\delta_{\rm I}^{(\lambda^0)}(iD_{us}^{\perp\mu}), \bnP] W^\dagger = 0$.
Similarly, we find that
\begin{equation}
\delta_{\rm I}^{(\lambda^0)} (ig B_\perp^{c \mu}) = 0 \,, \qquad
\delta_{\rm I}^{(\lambda^0)} (ig n \mcdot M) 
   = ig \Delta^\perp \mcdot B_\perp^c \,.
\end{equation}
Then, it is straightforward to show that
\begin{eqnarray} \label{typeI}
  \delta_{\rm I}^{(\lambda^1)} J^{(1a)}  
  + \delta_{\rm I}^{(\lambda^0)} J^{(2b)} & = & 0 \,, 
\nonumber \\
  \delta_{\rm I}^{(\lambda^1)} J^{(1b)} 
  + \delta_{\rm I}^{(\lambda^0)} J^{(2d)}
  + \delta_{\rm I}^{(\lambda^0)} J^{(2e)}
  + \delta_{\rm I}^{(\lambda^0)} J^{(2f)}
  + \delta_{\rm I}^{(\lambda^0)} J^{(2g)} & = & 0 \,,
\nonumber \\
  \delta_{\rm I}^{(\lambda^1)} J^{(1c)}  
  + \delta_{\rm I}^{(\lambda^0)} J^{(2c)} 
  + \delta_{\rm I}^{(\lambda^0)} J^{(2i)} 
  + \delta_{\rm I}^{(\lambda^0)} J^{(2j)} & = & 0 \,. 
\end{eqnarray}
It follows that 
$\delta_{\rm I}^{(\lambda^1)} J^{(1)} 
+ \delta_{\rm I}^{(\lambda^0)} J^{(2)} = 0$.

Repeating the process for type-II transformations, we obtain
\begin{eqnarray}
\delta_{\rm II}^{(\lambda^1)} (ig B_\perp^{c \mu})
& = & \varepsilon_\nu^\perp [iD_c^{\perp \nu}, i D_c^{\perp \mu}]
     - \frac{\varepsilon_\perp^\mu}{2} [i\bn\mcdot D_c, in\mcdot D]
\nonumber \\
&& \qquad
     - \frac{n^\mu}{2}[i\bn\mcdot D_c, 
           W i \varepsilon^\perp \mcdot D_{us}^\perp W^\dagger]
     - [i\bn\mcdot D_c, \delta_{\rm II}^{(\lambda^0)}
           (W iD_{us}^{\perp \mu} W^\dagger)] 
\nonumber \\
& = & ig \varepsilon_\nu^\perp \mcdot G_{c\perp}^{\mu\nu}
      - \frac{\varepsilon_\perp^\mu}{2} ig n \mcdot M \,,
\end{eqnarray}
where 
$g G_{c\perp}^{\mu\nu} = i[iD_c^{\perp\mu}, iD_c^{\perp\nu}]$.
Similarly, we find that
\begin{equation}
\delta_{\rm II}^{(\lambda^0)} (ig B_\perp^{c \mu})
   = \frac{n^\mu}{2} ig \varepsilon^\perp \mcdot B_\perp^c \,, \qquad
\delta_{\rm II}^{(\lambda^0)} (ig n \mcdot M) = 0 \,.
\end{equation}
Then,
\begin{eqnarray}
   \lefteqn{
      \delta_{\rm II}^{(\lambda^2)} J^{(0)}
      + \delta_{\rm II}^{(\lambda^1)} J^{(1a)}  
      + \delta_{\rm II}^{(\lambda^1)} J^{(1c)}  
      + \delta_{\rm II}^{(\lambda^0)} J^{(2b)}
      + \delta_{\rm II}^{(\lambda^0)} J^{(2c)} 
      + \delta_{\rm II}^{(\lambda^0)} J^{(2i)} 
      + \delta_{\rm II}^{(\lambda^0)} J^{(2j)}} 
\nonumber \\
    & = & -\, \bar\xi_n \DSppPl \frac{1}{i \bn\mcdot\lD} 
   \frac{\vepslash_\perp}{2} \DSppPl \frac{\bnslash}{2}
   \frac{1}{i \bn\mcdot\lD} \Gamma W h_v
    + \bar\xi_n \frac{\vepslash_\perp}{2} i n\mcdot{\overleftarrow D}
     \frac{\bnslash}{2}\frac{1}{i \bn\mcdot\lD} \Gamma W h_v
\nonumber \\
   & &
    + \, \bar\xi_n \DSppPl \frac{\bnslash}{2}
      \frac{1}{i \bn\mcdot\lD} i \varepsilon^\perp \mcdot \lD
      \frac{1}{i \bn\mcdot\lD} \Gamma W h_v
    + \bar\xi_n \DSppPl \frac{\bnslash}{2} \frac{1}{i \bn\mcdot\lD}
    \Gamma \left[ \frac{1}{i\bn\mcdot D_c}\: 
     i\varepsilon^\perp\mcdot D_c^\perp W \right] h_v
\nonumber \\
   & &
   + \, \frac{2}{n\mcdot v} \bar\xi_n \Gamma \frac{1}{(i \bn\mcdot D_c)^2} 
    \left(
    i\varepsilon^\perp\mcdot D_c^\perp \frac{1}{i \bn\mcdot D_c}
    ig v\mcdot B_\perp^c W h_v
   - v_\mu \varepsilon^\perp_\nu ig G_{c\perp}^{\mu\nu} W h_v
    \right. 
\nonumber \\
   & &
   \left. + \, ig v\mcdot B_\perp^c \left[ \frac{1}{i\bn\mcdot D_c}\:
     i\varepsilon^\perp\mcdot D_c^\perp W \right] h_v
   \right)
   - \frac{2}{n\mcdot v}\bar\xi_n \Gamma \frac{1}{i \bn\mcdot D_c}
     i v\mcdot D_c^\perp \frac{1}{(i \bn\mcdot D_c)^2}
    ig \varepsilon^\perp\mcdot B_\perp^c W h_v
\nonumber \\
   & &
   - \, \bar\xi_n \DSppPl \frac{\bnslash}{2} \frac{1}{i \bn\mcdot\lD}
    \Gamma \frac{1}{(i \bn\mcdot D_c)^2} 
    ig \varepsilon^\perp\mcdot B_\perp^c W h_v \,.
   \label{eq:rpiII}
\end{eqnarray}
Now, since 
$(i\bn\mcdot D_c)^{-1} ig \varepsilon^\perp\mcdot B_\perp^c W h_v
= \varepsilon^\perp\mcdot (iD_c^\perp W - \ppP) h_v $, 
we can write
\begin{equation}
\left[ \frac{1}{i\bn\mcdot D_c}\:
     i\varepsilon^\perp\mcdot D_c^\perp W \right] h_v
   = \frac{1}{i\bn\mcdot D_c} \varepsilon^\perp\mcdot 
         (iD_c^\perp W - \ppP) h_v
   = \frac{1}{(i\bn\mcdot D_c)^2} 
     ig \varepsilon^\perp\mcdot B_\perp^c W h_v \,.
   \label{eq:epsB}
\end{equation}
Substituting this into Eq.~(\ref{eq:rpiII}) and combining the first
and third terms, we obtain
\begin{eqnarray}
   \lefteqn{
      \delta_{\rm II}^{(\lambda^2)} J^{(0)}
      + \delta_{\rm II}^{(\lambda^1)} J^{(1a)}  
      + \delta_{\rm II}^{(\lambda^1)} J^{(1c)}  
      + \delta_{\rm II}^{(\lambda^0)} J^{(2b)}
      + \delta_{\rm II}^{(\lambda^0)} J^{(2c)} 
      + \delta_{\rm II}^{(\lambda^0)} J^{(2i)} 
      + \delta_{\rm II}^{(\lambda^0)} J^{(2j)}} 
\nonumber \\
   & = &
  \bar\xi_n \DSppPl \frac{1}{i \bn\mcdot\lD} \DSppPl
  \frac{\vepslash_\perp}{2} \frac{\bnslash}{2} 
  \frac{1}{i \bn\mcdot\lD} \Gamma W h_v
    + \bar\xi_n \frac{\vepslash_\perp}{2} i n\mcdot{\overleftarrow D}
     \frac{\bnslash}{2}\frac{1}{i \bn\mcdot\lD} \Gamma W h_v
\nonumber \\
   & & 
  + \frac{2}{n\mcdot v} \bar\xi_n \Gamma \frac{1}{(i \bn\mcdot D_c)^2}
    \Big( i \varepsilon^\perp\mcdot D_c^\perp \, 
     v\mcdot (i D_c^\perp W - \ppP) h_v  
   \nonumber \\
   & & \qquad
   - \big\{ i\varepsilon^\perp\mcdot D_c^\perp \,
     v\mcdot (i D_c^\perp W - \ppP)
    -  (iv\mcdot  D_c^\perp  i \varepsilon^\perp\mcdot D_c^\perp W 
       -  i \varepsilon^\perp\mcdot D_c^\perp v\mcdot \ppP) 
     \big\} h_v
\nonumber \\
    & & \qquad
   - iv\mcdot  D_c^\perp \, \varepsilon^\perp\mcdot (iD_c^\perp W - \ppP) h_v
     \Big)
\nonumber \\
   & = &
  - \bar\xi_n \left( \DSppPl\frac{1}{i \bn\mcdot\lD} \DSppPl +
      i n\mcdot{\overleftarrow D} \right) \frac{\bnslash\vepslash_\perp}{4}
        \frac{1}{i \bn\mcdot\lD} \Gamma W h_v 
\nonumber \\
   & &
  + \frac{2}{n\mcdot v} \bar\xi_n \Gamma \frac{1}{(i \bn\mcdot D_c)^2}
    (i v\mcdot D_c^\perp \varepsilon^\perp\mcdot\ppP 
      - i \varepsilon^\perp\mcdot D_c^\perp v\mcdot\ppP) h_v
\nonumber \\
   & = & 0 \,.
\end{eqnarray}
The first term of the second-last equality above is zero, owing to the equation 
of motion obtained from Eq.~(\ref{L0}). In the second term, the expression in
brackets begins at one collinear gluon, ensuring that the $1/(i \bn\mcdot D_c)$
does not cause any singularities.  Thus the label operators make this term zero.

Using completely analogous manipulations, we also find the closure relation that
involves the $J^{(1b)}$ current. Thus the type-II constraints are
\begin{eqnarray}\label{typeII}
    &&  \delta_{\rm II}^{(\lambda^2)} J^{(0)}
      + \delta_{\rm II}^{(\lambda^1)} J^{(1a)}  
      + \delta_{\rm II}^{(\lambda^1)} J^{(1c)}  
      + \delta_{\rm II}^{(\lambda^0)} J^{(2b)}
      + \delta_{\rm II}^{(\lambda^0)} J^{(2c)} 
      + \delta_{\rm II}^{(\lambda^0)} J^{(2i)} 
      + \delta_{\rm II}^{(\lambda^0)} J^{(2j)}  =  0\,,
\nonumber \\
&&   \delta_{\rm II}^{(\lambda^1)} J^{(1b)} 
  + \delta_{\rm II}^{(\lambda^0)} J^{(2d)}
  + \delta_{\rm II}^{(\lambda^0)} J^{(2e)}
  + \delta_{\rm II}^{(\lambda^0)} J^{(2f)}
  + \delta_{\rm II}^{(\lambda^0)} J^{(2g)}
  + \delta_{\rm II}^{(\lambda^0)} J^{(2h)}  = 0\,.
\end{eqnarray}
It then follows that $\delta_{\rm II}^{(\lambda^2)} J^{(0)} +\delta_{\rm
  II}^{(\lambda^1)} J^{(1)} + \delta_{\rm II}^{(\lambda^0)} J^{(2)} =
0$. Dropping the $J^{(2g)-(2j)}$ currents, these
relations agree with Ref.~\cite{Beneke:2002ph}.

From the results in Eqs.~(\ref{typeI}) and (\ref{typeII}) we see that the
currents split into two sets:
\begin{eqnarray}
 && J^{(0)} + J^{(1a)} + J^{(1c)} + J^{(2b)} + J^{(2c)} + J^{(2i)} + J^{(2j)} 
   \,,\nn\\
 && J^{(1b)} + J^{(2d)} + J^{(2e)} + J^{(2f)} + J^{(2g)} + J^{(2h)} \,.
\end{eqnarray}
When we add non-trivial Wilson coefficients to all the currents, the number of
$\omega_i$ parameters they depend on is restricted by RPI, i.e.\ in the set
with $J^{(0)}$ they have one parameter $\omega$ and in the second set with
$J^{(1b)}$ they have two parameters, $\omega_{1,2}$. In the second set the
combination of fields restricted to have momentum $\omega_1$ and $\omega_2$ is
determined by the manner in which the terms in the RPI-transformed currents
cancel.  Making the $Y$ field redefinition in Eq.~(\ref{fr}) and putting in the
most general $\omega_i$ dependence consistent with RPI, we find
\begin{eqnarray}
 J^{(2a)}(\omega) &=&  \frac{1}{2m_b}\: (\bar\xi_n W)_{\omega} \Gamma  
   \Big( Y^\dagger  {i\Dslash_{us}^{\,T} } h_v\Big)  
   \,,\nn\\
 J^{(2b)}(\omega) &=& \frac{1}{\omega}\: (\bar\xi_n W)_{\omega}
   \Big(Y^\dagger \DSppPlus     \frac{\bnslash}{2} 
   \Gamma  h_v\Big) \,,  \nn\\
J^{(2c)}(\omega) &=& - \frac{\bn\mcdot v}{n\mcdot v} \Big(\bar\xi_n \Gamma 
   \frac{1}{(i\bn\mcdot D_c)^2}  ig n\mcdot B_c W\Big)_{\omega} (Y^\dagger h_v) 
    \,, \nn\\
 J^{(2d)}(\omega_1,\omega_2) &=& - \frac{1}{m_b\,n\mcdot v} 
   \big( \bar\xi_n W\big)_{\omega_1} \Gamma \frac{\nslash\bnslash}{4} 
   \Big( \frac{1}{\bnP} W ig n\mcdot B_c W\Big)_{\omega_2} (Y^\dagger h_v) \,, 
   \nn\\
 J^{(2e)}(\omega_1,\omega_2) &=& - \frac{1}{m_b\, n\mcdot v\,\omega_1} 
   \big( \bar\xi_n \DSppPl W\big)_{\omega_1}
   \frac{\bnslash}{2} \Gamma \frac{\nslash}{2}\:
   ( ig {\Bcslash}_c^\perp)_{\omega_2} (Y^\dagger h_v) \,, \nn \\  
 J^{(2f)}(\omega_1,\omega_2) &=& -\, \frac{1}{m_b\, n\mcdot v} 
   \big(\bar\xi_n W)_{\omega_1}\Gamma 
   \Big(\frac{1}{\bnP} W^\dagger i\Dslash_\perp^{c} \frac{1}{i\bn\mcdot D_c}
   ig \Bslash_\perp^c W \Big)_{\omega_2} (Y^\dagger h_v) \,, \nn\\ 
 J^{(2g)}(\omega_1,\omega_2) &=&
   \, \frac{1}{m_b (n\mcdot v)^2} \big( \bar\xi_n W\big)_{\omega_1} \Gamma\nslash
    \Big( \frac{1}{\bnP} W^\dagger
   \DvppP  \frac{1}{i\bn\mcdot D_c} ig \Bslash_\perp^c W  \Big)_{\omega_2}
   (Y^\dagger h_v)   \,, \nn\\
 J^{(2h)}(\omega_1,\omega_2) &=&  \, \frac{1}{m_b\,(n\mcdot v)^2} 
   \big(\bar\xi_n W)_{\omega_1}\Gamma \nslash
   \Big(\frac{1}{\bnP} W^\dagger ig\Bslash_\perp^{c} 
   \frac{1}{(i\bn\mcdot D_c)^2} ig v \mcdot B_\perp^c W\Big)_{\omega_2} 
   (Y^\dagger h_v) \,, \nn\\
 J^{(2i)}(\omega) &=& \frac{2}{n\mcdot v} \Big( \bar\xi_n \DSppPl 
   \frac{1}{i\bn\mcdot\lD} 
    \frac{1}{(i\bn\mcdot D_c)^2} ig v \mcdot B_\perp^c W \Big)_{\omega}
   \frac{\bnslash}{2} \Gamma (Y^\dagger h_v) \nn \,, \\  
 J^{(2j)}(\omega) &=&  \frac{4}{(n\mcdot v)^2} 
   \Big( \bar\xi_n \Gamma \frac{1}{i\bn\mcdot D_c} \DvppP 
   \frac{1}{(i\bn\mcdot D_c)^2} i g v \mcdot B_\perp^c W\Big)_{\omega} 
    (Y^\dagger h_v)  \, ,
\end{eqnarray}
where 
\begin{eqnarray}
  ig B_{c\perp}^\mu &=& [i\bn\mcdot D_c , iD_{c\perp}^\mu] \,,
\qquad
  ig n\mcdot B_c = [ i\bn\mcdot D_c , in\mcdot \partial+gn\mcdot A_n ]\,.
\end{eqnarray}


\bibliography{klis}

\end{document}